\documentclass[letterpaper,12pt,titlepage,oneside,printwatermark]{book}
\usepackage{lipsum}
\usepackage{scrhack} 
\usepackage{epigraph}
\usepackage[
  inner	=	4.0cm, 
  outer	=	3.0cm, 
  top	=	3.0cm, 
  bottom=	3.0cm, 
  includeheadfoot, 
]{geometry}
\usepackage{lscape}
\usepackage{pdfpages}
\usepackage{url}
\usepackage{graphicx, graphics}
\usepackage{bm}
\usepackage{amssymb}
\usepackage{yfonts,color}
\usepackage{amsmath}
\usepackage{amstext}
\usepackage{amsfonts}
\usepackage{latexsym}
\usepackage[pdftex]{hyperref}
\usepackage{cite}
\usepackage{enumerate}
\usepackage{quotchap}
\usepackage{lettrine}
\usepackage[numbers]{natbib}
\usepackage{erewhon}
\usepackage{GoudyIn}
\usepackage{textcomp}

\usepackage[nopostdot,nogroupskip,style=super,nonumberlist,toc,automake,order=letter]{glossaries}
 \usepackage{glossaries-extra}
\usepackage{appendix}

\hypersetup{
    colorlinks=true,       
    linkcolor=[rgb]{0.0,0.2,0.8},          
    citecolor=[rgb]{1.0,0.0,0.0},       
    filecolor=magenta,      
    urlcolor=[rgb]{0.0,0.0,1.0}          
}

\graphicspath{{Include/Figures/}}

\definecolor{red1}{rgb}{0.7, 0.11, 0.11}

\newcommand{\pac}[1]{ \left\{ #1 \right\} }
\newcommand{\pap}[1]{\left( #1 \right)}
\newcommand{\pas}[1]{\left[#1 \right]}

\newcommand{\tr}[1]{\mathrm{tr}\left\{ #1 \right\}}

\def\ii{{\rm i}}
\def\ee{{\rm e}}
\def\l{{\it  l}}

\def\la{\langle}
\def\ra{\rangle}

\def\tr{\rm{Tr}}

\newcommand{\beq}{\begin{equation}}
\newcommand{\eeq}{\end{equation}}
\newcommand{\beqa}{\begin{eqnarray}}
\newcommand{\eeqa}{\end{eqnarray}}

\newcommand{\Ho}{\hat{H}}

\newcommand{\Jo}{\hat{J}}

\newcommand{\rhoo}{\hat{\rho}}
\newcommand{\lam}{\lambda}
\newcommand{\ome}{\omega}
\newcommand{\eps}{\epsilon}

\newcommand{\ups}{\upsilon}

\DeclareMathOperator*{\ketGS}{\left|\psi_{0}\right\rangle}
\DeclareMathOperator*{\braGS}{\left\langle \psi_{0}\right|}

\DeclareMathOperator*{\ketC}{\left|0\right\rangle}
\DeclareMathOperator*{\ket1}{\left|1\right\rangle}

\DeclareMathOperator*{\sx}{\hat{\sigma}^{x}}
\DeclareMathOperator*{\sy}{\hat{\sigma}^{y}}
\DeclareMathOperator*{\sz}{\hat{\sigma}^{z}}

\makeglossaries

\newglossaryentry{AGW}{name={AGW},description={\textbf{A}garwal-\textbf{W}igner \textbf{F}unction}}
\newglossaryentry{AXY-TFIM}{name={AXY-TFIM},description={\textbf{A}nisotropic \textbf{XY} \textbf{I}sing model in a \textbf{T}ransversal magnetic \textbf{F}ield}}
\newglossaryentry{LGI}{name={LGI},description={\textbf{L}eggett-\textbf{G}arg \textbf{I}nequalities}}
\newglossaryentry{QPT}{name={QPT},description={\textbf{Q}uantum \textbf{P}hase \textbf{T}ransition}}
\newglossaryentry{STC}{name={STC},description={\textbf{S}ingle-site \textbf{T}wo-time \textbf{C}orrelations}}
\newglossaryentry{MFC}{name={MFC},description={\textbf{M}ajorana \textbf{F}ermion \textbf{C}hains}}
\newglossaryentry{ZEM}{name={ZEM},description={\textbf{Z}ero \textbf{E}nergy \textbf{M}ajorana \textbf{M}odes}}
\newglossaryentry{KC}{name={KC},description={\textbf{K}itaev \textbf{C}hain}}
\newglossaryentry{TFI}{name={TFI},description={\textbf{T}ransverse \textbf{F}ield \textbf{I}sing}}
\newglossaryentry{OTOC}{name={OTOC},description={\textbf{O}ut-of-\textbf{T}ime-\textbf{O}rdered \textbf{C}orrelations}}
\newglossaryentry{BCH}{name={BCH},description={\textbf{B}aker-\textbf{C}ampell-\textbf{H}ausdorff formula}}
\newglossaryentry{NP}{name={NP},description={\textbf{N}arayana \textbf{P}olynomials}}
\newglossaryentry{DM}{name={DM},description={\textbf{D}icke \textbf{M}odel}}
\newglossaryentry{LZM}{name={LZM},description={\textbf{L}andau-\textbf{Z}ener \textbf{M}odel}}
\newglossaryentry{QCP}{name={QCP},description={\textbf{Q}uantum \textbf{C}ritical \textbf{P}oint}}
\newglossaryentry{LZS}{name={LZS},description={\textbf{L}andau-\textbf{Z}ener \textbf{S}t\"uckelberg}}
\newglossaryentry{TL}{name={TL},description={\textbf{T}hermodynamic \textbf{L}imit}}
\newglossaryentry{QED}{name={QED},description={\textbf{Q}uantum \textbf{E}lectro\textbf{D}ynamics}}
\newglossaryentry{BEC}{name={BEC},description={\textbf{B}ose-\textbf{E}instein \textbf{C}ondensate}}
\newglossaryentry{KZM}{name={KZM},description={\textbf{K}ibble-\textbf{Z}ureck \textbf{M}echanims}}
\newglossaryentry{IKZM}{name={IKZM},description={\textbf{I}nhomogeneous  \textbf{K}ibble-\textbf{Z}ureck \textbf{M}echanims}}
\newglossaryentry{TFQIM}{name={TFQIM},description={ \textbf{T}ransversal \textbf{F}ield  \textbf{Q}uantum \textbf{I}sing \textbf{M}odel}}
\newglossaryentry{STD}{name={STD},description={\textbf{S}tatistics of \textbf{T}opological \textbf{D}efects}}
\newglossaryentry{DMRG}{name={DMRG},description={\textbf{D}ensity \textbf{M}atrix \textbf{R}enormalization \textbf{G}roup}}
\newglossaryentry{TEBD}{name={TEBD},description={\textbf{D}ime-\textbf{E}volving \textbf{B}lock \textbf{D}ecimation}}

\definecolor{Uniandes}{rgb}{1.0,1.0,0}	
\newgeometry{ignoreall,top=2cm,outer=2cm,inner=2cm}
\begin{document}
\begin{titlepage}
\newlength{\centeroffset}
\setlength{\centeroffset}{-0.5\oddsidemargin}
\addtolength{\centeroffset}{0.5\evensidemargin}
\thispagestyle{empty}

\pagecolor{Uniandes}

\begin{center}
\begin{figure}
\begin{center}
\includegraphics[scale=0.7]{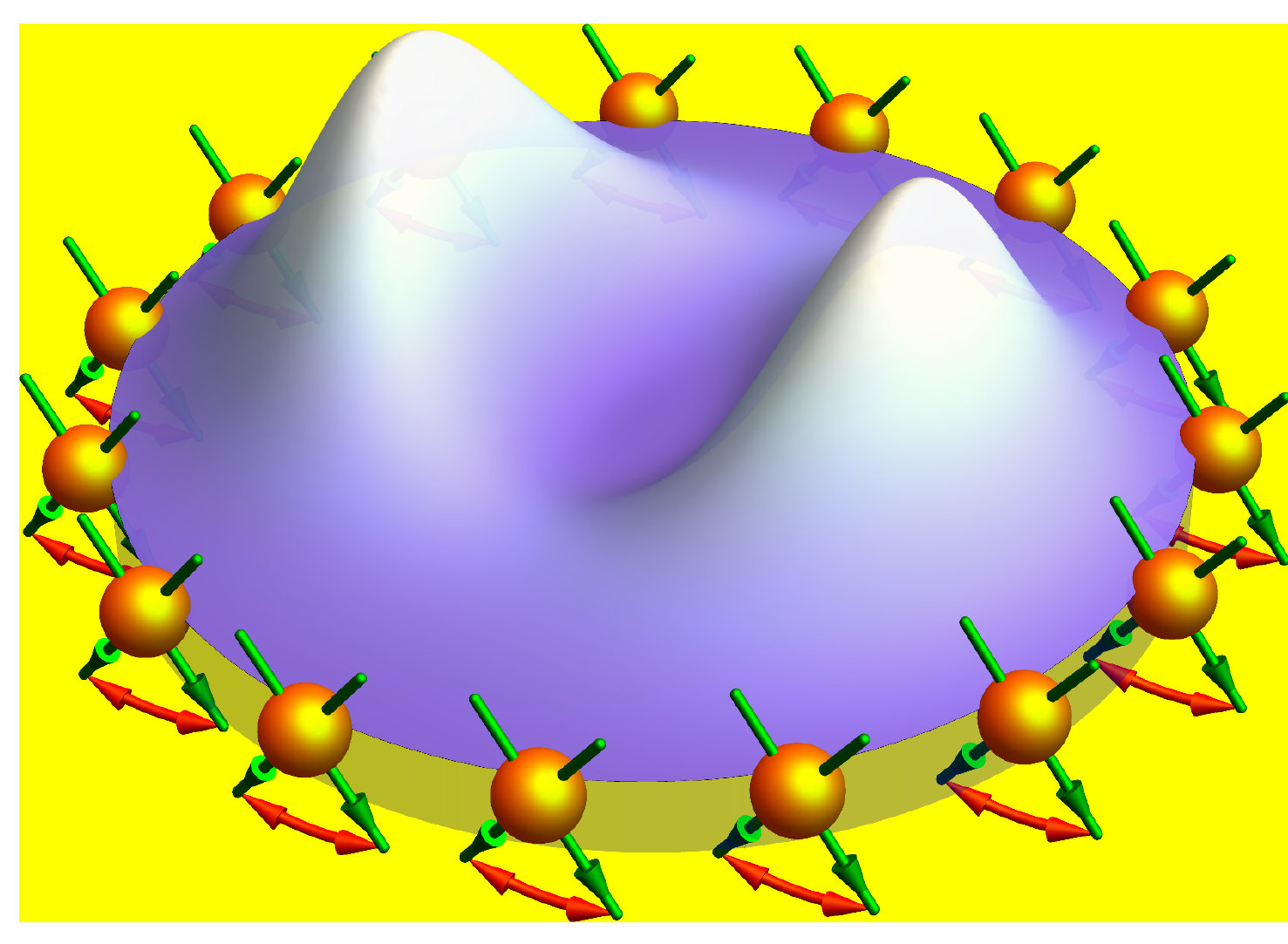}
\end{center}
\end{figure}
\end{center}
\hfill\linebreak\\
\vspace{0.5cm}
\begin{center}
\textbf{\Huge{Quantum Non-equilibrium Many-Body Spin-Photon Systems}}
\end{center}
\hfill\linebreak\\
\begin{center}
\textit{\Large{by:}}\\
\Large{Fernando Javier G\'omez-Ruiz}
\end{center}
\vfill
\begin{center}
\begin{figure}[h!]
\begin{center}
\includegraphics[scale=1.0]{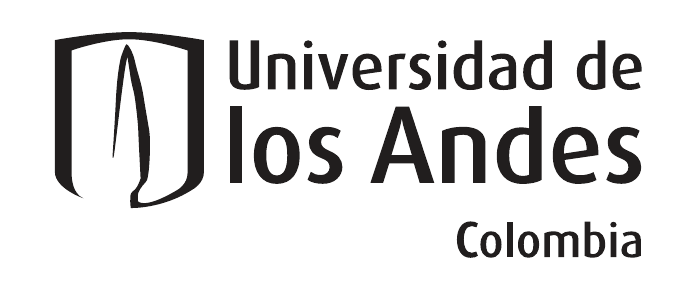}
\end{center}
\end{figure}
\end{center}
\end{titlepage}
\pagecolor{white}
\thispagestyle{empty}

\frontmatter
\thispagestyle{empty}
\hfill\linebreak\\
\begin{center}
\href{https://uniandes.edu.co}{\Huge{Universidad de los Andes}}
\end{center}
\hfill\linebreak\\
\vspace{-0.1cm}
\begin{center}
{\huge Doctoral Thesis}
\end{center}
\hfill\linebreak\\
\vspace{-1.0cm}
\begin{center}
 \hrule width \hsize \kern 1mm \hrule width \hsize height 2pt 
 \hfill\linebreak\\
 \textbf{\Huge{Quantum Non-equilibrium Many-Body Spin-Photon Systems}}
 \hfill\linebreak\\
 \hrule width \hsize \kern 1mm \hrule width \hsize height 2pt 
\end{center}
 \hfill\linebreak\\
\begin{center}
\begin{tabular}{p{6cm} p{2cm} p{6cm}}
\emph{Author:}&&\emph{Advisor:}\\
    \href{}{Fernando Javier G\'omez-Ruiz}&&\href{http://fisica.uniandes.edu.co/index.php/personal/profesores-de-planta/ferney-javier-rodriguez-duenas}{Ferney Rodr\'iguez Due\~nas}
   \end{tabular}
\end{center}
 \hfill\linebreak\\
\begin{center}
\emph{A thesis submitted in fulfillment of the requirements}\\
\emph{for the degree of Doctor in Science-Physics}\\
\emph{in the research group:}\\
 \hfill\linebreak\\
\href{https://fimaco.uniandes.edu.co/index.php/en/}{\large{Grupo de F\'isica de la Materia Condensada, FIMACO}}
 \hfill\linebreak\\
\href{https://fisica.uniandes.edu.co/en/}{\large{Departamento de F\'isica}}
\end{center}
\begin{center}
\begin{figure}[h!]
\begin{center}
\includegraphics[scale=0.7]{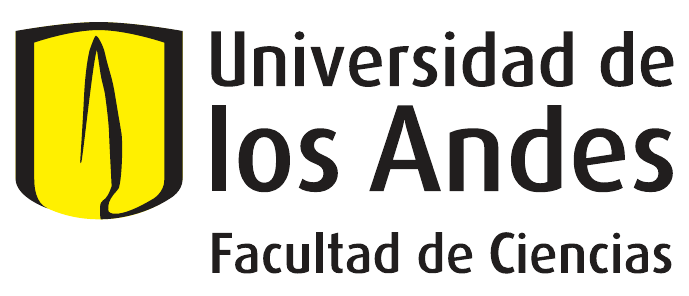}
\end{center}
\end{figure}
\end{center}
\hfill
\begin{center}
Bogot\'a D.C., Colombia\\
May 2019. 
\end{center}

\newpage
\thispagestyle{empty}
 \hfill\linebreak\\
\begin{center}
\textcopyright\,Copyright by Fernando Javier G\'omez-Ruiz, 2019.\\
All rights reserved.
\end{center}
\newpage
\setcounter{page}{1}
\addcontentsline{toc}{chapter}{Declaration of Authorship}
 \hfill\linebreak\\
 \vspace{3cm}
 \begin{center}
 \Huge{\textbf{Declaration of Authorship}}
 \end{center}
 \hfill\linebreak\\
 \begin{flushleft}
 I, Fernando Javier G\'omez-Ruiz, declare that this thesis titled, ``Quantum Non-equilibrium Many-Body Spin-Photon Systems" and the work presented in it are my own. I confirm that:
 
\begin{itemize}
\item This work was done wholly or mainly while in candidature for a research degree at this
University.
\item Where any part of this thesis has previously been submitted for a degree or any other
qualification at this University or any other institution, this has been clearly stated.
\item Where I have consulted the published work of others, this is always clearly attributed.
\item Where I have quoted from the work of others, the source is always given. With the
exception of such quotations, this thesis is entirely my own work.
\item I have acknowledged all main sources of help.
\item Where the thesis is based on work done by myself jointly with others, I have made clear
exactly what was done by others and what I have contributed myself.
 \end{itemize}
 Signed:\\
 \vspace{0.2cm}
\hspace{2cm} \rule{9cm}{1pt}\\
 Date:\\
 \vspace{0.2cm}
\hspace{2cm} \rule{9cm}{1pt}
\end{flushleft}
\newpage
\thispagestyle{empty}
 \hfill\linebreak\\
 \begin{center}
\begin{figure}[h!]
\begin{center}
\includegraphics[scale=0.7]{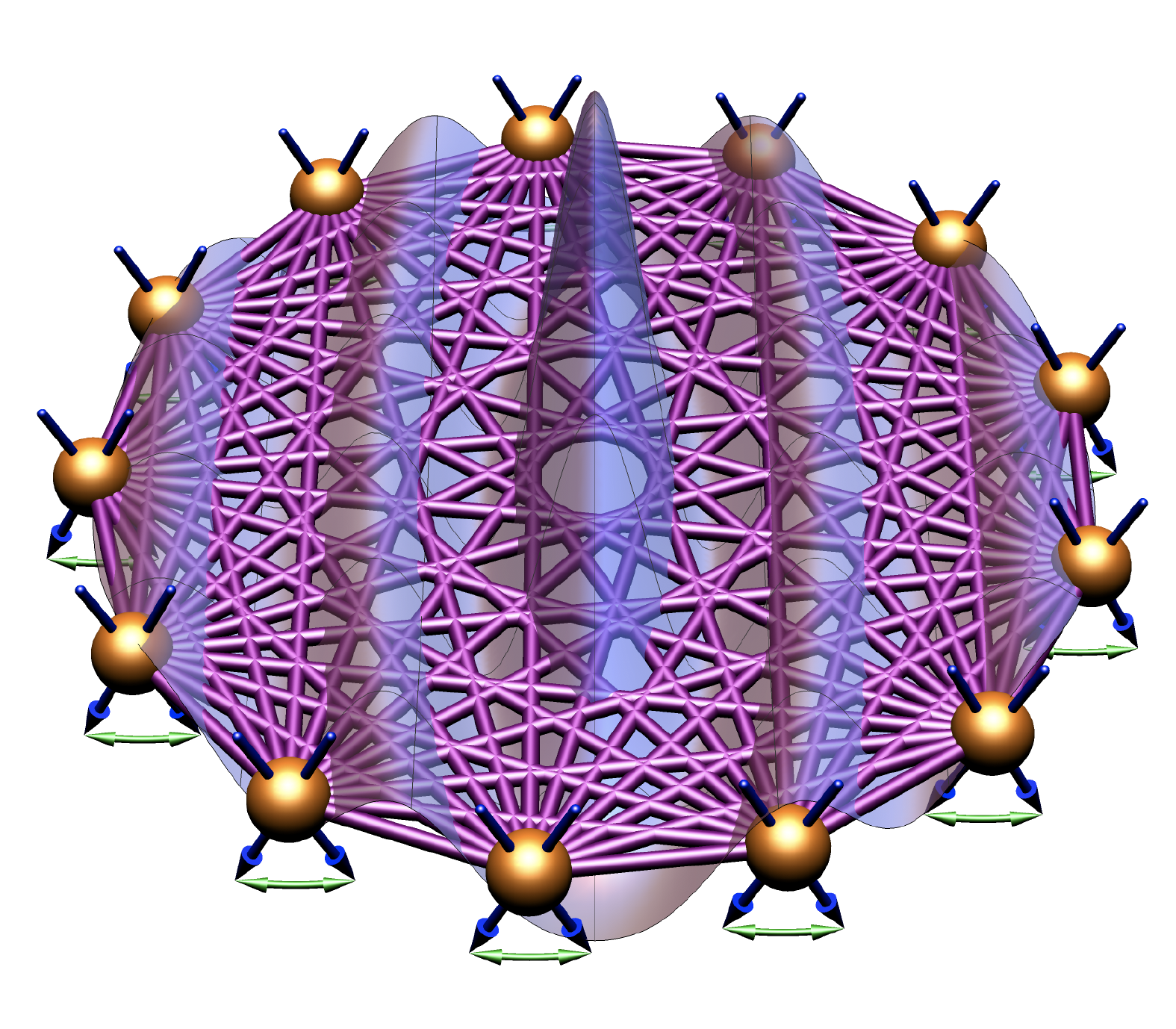}
\end{center}
\end{figure}
\end{center}
 \vspace{1.5cm}
 \begin{flushleft}
\emph{``Trying to find a computer simulation of physics seems to me to be an excellent program to follow out...  the real use of it would be with quantum mechanics... Nature isn\textquotesingle t classical... and if you want to make a simulation of Nature, you\textquotesingle d better make it quantum mechanical, and by golly it is a wonderful problem, because it doesn\textquotesingle t look so easy."}\footnote{Feynman, R. Simulating Physics with Computers. Keynote address delivered at the MIT Physics of Computation Conference. Published in \emph{Int. J. Theor. Phys. 21 (6/7), 1982. (Excerpts reprinted with permission from the International Journal of Theoretical Physics.)}}
\end{flushleft} 
 \hfill\linebreak\\
 \begin{flushright}
 \textit{--- Richard P. Feynman}
\end{flushright}
\newpage
\addcontentsline{toc}{chapter}{Abstract}
\hfill\linebreak\\
\begin{center}
\href{https://uniandes.edu.co}{\Huge{Universidad de los Andes}}
\end{center}
\hfill\linebreak\\
\vspace{-1cm}
\begin{center}
\emph{\Huge{Abstract}}
\end{center}
\begin{center}
\href{https://ciencias.uniandes.edu.co}{\large{Facultad de Ciencias}}\\
\vspace{0.3cm}
\href{https://fisica.uniandes.edu.co/en/}{\large{Departamento de F\'isica}}\\
\vspace{0.5cm}
\emph{Doctor in Science-Physics}\\
\vspace{0.5cm}
\textbf{Quantum Non-equilibrium Many-Body Spin-Photon Systems}\\
\vspace{0.3cm}
by Fernando Javier G\'omez-Ruiz
\end{center}

\lettrine{\color{red1}{\GoudyInfamily{T}}}{his} thesis dissertation concerns the quantum dynamics of strongly-correlated quantum systems in out-of-equilibrium states. The research is neither restricted to static properties or long-term relaxation evolutions, nor does it neglect effects on any relevant subsystem as is frequently done with the environment in master equations approaches.The focus of this work is to explore different quantum systems during severals regimes of operations, then discover results that might be of interest to quantum control, and hence to quantum computation and quantum information processing. Our main results can be summarized as follows in three parts.\\
\\
{\em Signature of Critical Dynamics.---} We thoroughly investigate the fingerprint of equilibrium quantum phase transitions through the single-site two-time correlations and violation of Leggett-Garg inequalities in spins-$1/2$ and Majorana Fermion systems. By means of simple analytical arguments for a general spin-$1/2$ Hamiltonian, and matrix product simulations of one-dimensional XXZ and anisotropic XY models, we argue that finite-order quantum phase transitions can be determined by singularities of the time correlations or their derivatives at criticality. The same features are exhibited by corresponding Leggett-Garg functions, which noticeably indicate violation of the Leggett-Garg inequalities for early times and all the Hamiltonian parameters considered~\cite{Gomez_PRB2016}. Moreover, we propose that two-time correlations of Majorana edge localized fermions constitute a novel and versatile toolbox for assessing the topological phases of 1D open lattices. Using analytical and numerical calculations on the Kitaev model, we uncover universal relationships between the decay of the short-time correlations and a particular family of out-of-time-ordered correlators, which provide direct experimental alternatives to the quantitative analysis of the system regime, either normal or topological~\cite{Gomez_PRB2018}.\\
\\
{\em Driven Dicke Model as a Test-bed of Ultra-Strong Coupling.---} In this part, we investigate the non-equilibrium quantum dynamics of a canonical light-matter system, namely the Dicke model, when the light-matter interaction is ramped up and down through a cycle across the quantum phase transition. Our calculations reveal a rich set of dynamical behaviors determined by the cycle times, ranging from the slow, near adiabatic regime  through to the fast, sudden quench regime. As the cycle time decreases, we uncover a crossover from an oscillatory exchange of quantum information between light and matter that approaches a reversible adiabatic process, to a dispersive regime that generates large values of light-matter entanglement~\cite{Gomez_Entropy2016} and we show that a pulsed stimulus can be used to generate many-body quantum coherences in light-matter systems of general size~\cite{Gomez_Frontiers2018}. Additionally, our results reveal that both types of quantities indicate the emergence of the superradiant phase when crossing the quantum critical point. In addition, at the end of the pulse light and matter remain entangled even though they become uncoupled, which could be exploited to generate entangled states in non-interacting systems~\cite{Gomez_JPB2017}. Our findings show robustness to losses and noise, and have potential functional implications at the systems level for a variety of nanosystems, including collections of $N$ atoms, molecules, spins, or superconducting qubits in cavities -- and possibly even vibration-enhanced light-harvesting processes in macromolecules.\\
\\
\noindent {\em Beyond the Kibble-Zurek Mechanism.---}  The Kibble-Zurek mechanism is a highly successful paradigm to describe the dynamics of both thermal and quantum phase transitions. It is one of the few theoretical tools that provide an account of nonequilibrium behavior in terms of equilibrium properties. It predicts that in the course of a phase transition topological defects are formed. In this work, we elucidate the emergence of adiabatic dynamics in an inhomogeneous quantum phase transition. We show that the dependence of the density of excitations with the quench rate is universal and exhibits a crossover between the standard KZM behavior at fast quench rates, and a steeper power-law dependence for slower ramps.  Local driving of quantum critical systems thus leads to a much more pronounced suppression of the density of defects, that constitute a testable prediction amenable to a variety of platforms for quantum simulation including cold atoms in optical gases, trapped ions and superconducting qubits. Our results establish the universal character of the critical dynamics across an inhomogeneous quantum phase transition, that we proposed for favoring adiabatic dynamics. Our results should prove useful in a variety of contexts including the preparation of phases of matter in quantum simulators and condensed matter systems, as well as the engineering of inhomogeneous schedules in quantum annealing~\cite{Gomez_PRL2019}. Moreover, using a trapped-ion quantum simulator, we experimentally probe the the kink distribution resulting from driving  a one-dimensional quantum Ising chain through the paramagnet-ferromagnet quantum phase transition. Our results establish that  the universal character of  the critical dynamics can be extended beyond the paradigmatic Kibble-Zurek mechanism, that accounts for  the mean kink number, to characterize the full  probability distribution of topological defects~\cite{Cui_Arxiv2019}. Finally, we argue that our results have a broad impact as our experiment has manifold applications. Knowledge of the distribution of defects proves useful in the characterization in adiabatic quantum computers.  Assessing the statistics of topological defects and its universal character is bound to motivate new experiments across the plethora of experimental platforms where topological defect formation has been explored, e.g., in the light of the Kibble-Zurek mechanism. 

\newpage
\thispagestyle{empty}
\hfill\linebreak\\
\newpage
\addcontentsline{toc}{chapter}{Acknowledgements}
\hfill\linebreak\\
\vspace{2cm}
\begin{flushright}
\emph{\Huge{Acknowledgements}}
\end{flushright}

\vspace{1.5cm}
\lettrine{\color{red1}{\GoudyInfamily{T}}}{his} work could not have been performed without the support and contributions of a number of people to whom I am particularly grateful:
\begin{itemize}
\item Prof. Ferney Rodr{\'i}guez, the director of my thesis, for his friendly and helpful guidance. His generous advice highly added value to this work and my professional life. Thank you, Ferney for continuously believing in me and understanding in me a five-year proximity effect, which continued far past my critical temperature. Additionally, I am also thankful to Ferney for his financial support with several research projects. I know that in Colombia it is difficult to find funding for Ph.D. students, but Ferney's passion, intuition, and energy helped procure funding and also taught and motivated me to find funding opportunities as well.  I really appreciate that he told me several times ``You already have "no" as an answer, why not try it?". Ferney has been my mentor in many aspects of my life.      
\item Prof. Luis Quiroga, leader of the FIMACO group, for his great supervision. Luis has provided an amazing example of how to perform the highest quality and most rigorous science. His knowledge and advice have considerably contributed to this work. I benefited from numerous productive discussions and ideas to overcome the scientific difficulties of this thesis. One of Luis’s most distinguishing qualities as a mentor is that he truly cares about the success and training of his mentees, and I have been lucky to train under him.
\item Prof. Adolfo del Campo, for his great supervision and friendship in Boston. Thank you, Adolfo, for giving me the opportunity to work in your theoretical lab, where I enjoyed every moment and learned so much about physics. I was shown much hospitality and benefited greatly from fruitful discussions with your group. Adolfo's dedication to science is also inspiring and I hope to emulate that throughout my career. When I arrived in Boston, I was emotionally broken; nonetheless, Adolfo's passion and energy were a light in my way. Boston was for me the best experience in my Ph.D.   
\item Over the course of my research I have had the opportunity to collaborate with many colleagues on a number of interesting projects. In particular I want to thank Dr. Juan Jos\'e Mendoza-Arenas (Postdoctoral Fellow Uniandes), Dr. Oscar Acevedo, Dr. Luis Pedro Garc\'ia-Pintos (Postdoctoral Fellow Umass-Boston), Dr. Jin-Ming Cui (Postdoctoral Fellow University of Science and Technology of China), Prof. Neil Johnson (George Washington University), Prof. Zhen-Yu Xu (Soochow University), and Prof. Hidetoshi Nishimori (Tokyo Institute of Technology). 
\item I must also express my sincerest gratitude to Universidad de Los Andes for not only providing a wonderfully stimulating academic and social environment over the last 4 years but also for allowing me to grow as a person and physicist.
\item Finally, and most importantly, I would like to thank my family - my mom Luz Linda Ruiz, my dad Teofilo G\'omez Chaves, my sisters Lorena and Maribel who throughout all my life have been a light in my way. Additionally, I would like to thank my nephew and nieces, they give me support and an infinite number of beautiful moments. Day to day my family provides me with the energy and power necessary to continue this beautiful adventure in the knowledge. Along my path, I had the fortune to know my second family, Cardona-Rodr\'iguez Family -Escilda, Marcos, Alex, and Giovanny- thank you for your great support. Finally, I want to say thank you to my girlfriend Ershela Durresi who in the last months of this journey has been a very important person in my life. Shela every day has add value to my life and I grow in several aspects of my life. Boston not only provided me very nice academic moments, but also gave me the opportunity to get to know this beautiful woman. I want to continue walking together side by side in our next adventures.                   
\end{itemize}
\hfill\linebreak\\
\newpage
\thispagestyle{empty}
\hfill\linebreak\\
\vspace{7cm}
\begin{center}
\emph{I would like to dedicate this thesis to my loving family.}\\
\emph{Their support has always been priceless.}
\end{center}
\vspace{3cm}
\begin{center}
\emph{In Memory to Marcos Cardona}\\
\emph{who passed away shortly before I began this work.}
\end{center}
\newpage
\thispagestyle{empty}
\tableofcontents
\addcontentsline{toc}{chapter}{Contents}
\listoffigures	
\addcontentsline{toc}{chapter}{List of Figures}
\listoftables
\addcontentsline{toc}{chapter}{List of Tables}
\printglossary[title={List of Abbreviations}]

\begin{savequote}[45mm]
``Study hard what interests you the most in the most undisciplined, irreverent and original manner possible."
\qauthor{Richard P. Feynman}
\end{savequote}
\chapter[Preface]{Preface}
\begin{center}
\begin{tabular}{p{15cm}}
\vspace{0.1cm}
\lettrine{\color{red1}{\GoudyInfamily{T}}}{his} thesis is based upon the work I performed during my graduate studies and is made of three parts, corresponding to the three different projects I have conducted as a PhD student. The first one is titled Signature of equilibrium properties in spin$-1/2$ systems. This project has been carried out mainly at the Universidad de los Andes - Bogot\'a, under the guidance of Prof. Luis Quiroga and Ferney J. Rod\'iguez and with the collaboration of Dr. Juan J. Mendoza-Arenas (post-doctoral fellow). The second one is titled Driven Dicke Model as a Test-bed of Ultra-Strong Coupling. This fruitful project has been developed in collaboration with Prof. Neil Johnson from George Washington University, USA. The third and final part of this thesis is based on the project I have conducted at the University of Massachusetts Boston on beyond the Kibble-Zurek Mechanisms with Prof. Adolfo del Campo. I spent one year of my doctorate as a graduate visiting student at the University of Massachusetts Boston and am very grateful for the generosity and support of Dr. del Campo, which made this year incredibly productive and allowed for an amazing collaboration.      
\end{tabular}
\end{center}
\newpage
\thispagestyle{empty}
\hfill\linebreak
\mainmatter
\begin{savequote}[45mm]
``Physics is like sex: sure, it may give some practical results, but that's not why we do it."
\qauthor{Richard P. Feynman}
\end{savequote}
\chapter{Introduction}
\lettrine{\color{red1}{\GoudyInfamily{I}}}{n} the last two decades, there have been several breakthroughs in the experimental realization of systems that mimic specific many-body quantum models~\cite{Han2013}. This is especially true in systems involving aggregates of real or artificial atoms in cavities and superconducting qubits~\cite{ExpDicke1,ExpDicke2}, as well as trapped ultra-cold atomic systems~\cite{bloch2012nat,schneider2012rpp,georgescu2014rmp}.  These advances have stimulated a vast among of theoretical research on a wide variety of phenomena exhibited by these systems, such as quantum phase transitions (\gls{QPT}s)~\cite{sachdev, aeppli}, the collective generation and propagation of entanglement~\cite{amico2002nature,Acevedo2015NJP,AcevedoPRA2015}, and critical universality~\cite{Acevedo2014PRL}. \\
\\
Future applications in the area of quantum technology will involve exploiting -- and hence fully understanding -- the {\em non-equilibrium} quantum properties of such many-body systems. Radiation-matter systems are of particular importance: not only because optoelectronics has always been the main platform for technological innovations, but also in terms of basic science because the interaction between light and matter is a fundamental phenomenon in nature. On a concrete level, light-matter interactions are especially important for most quantum control processes, with the simplest manifestation being the non-trivial interaction between a single atom and a single photon~\cite{AgarwalPRL1984}. One of the key goals of experimental research is to improve both the intensity and tunability of the atom-light interaction~\cite{WillPRL2016, BegleyPRL2016}.\\
\\
In particular, we are interested on two models with a  wide range of applications on Condensed Matter Physics as well as in Atomic, Molecular and Optical Physics.These are  the driven Dicke Model (\gls{DM}) and one-dimensional spin$-1/2$ XYZ chain. Both models have been extensively studied  in many circumstances before, for static and dynamical properties. \\
\\ 
In the case of the \gls{DM} despite significant accumulated knowledge regarding its static properties across its parameter space, with the emergence of a quantum phase transition, little is known about the dynamical crossing of this transition. In previous work, we advanced on the understanding of the \gls{DM}'s dynamics by exploring the effects of crossing the \gls{QPT} using a tuned interaction, hence taking the system in a single sweep from a non-interacting regime into one where strong correlations within and between the matter and light subsystems play an essential role~\cite{AcevedoPRA2015,Acevedo2015NJP,Acevedo2014PRL}. Our previous analyses also revealed universal dynamical scaling behavior for a class of models concerning their near-adiabatic behavior in the region of a \gls{QPT}, in particular the Transverse-Field Ising model, the \gls{DM} and the Lipkin-Meshkov-Glick model. These findings, which lie beyond traditional critical exponent analysis like the Kibble-Zurek mechanism~\cite{Zurek96a, Kibble76b} and adiabatic perturbation approximations, are valid even in situations where the excitations have not yet stabilized -- hence they provide a time-resolved understanding of \gls{QPT}s encompassing a wide range of near adiabatic regimes. Additionally, we analyze the effects of driving the system through a round trip across the \gls{QPT}, by successively ramping up and down the light-matter interaction so that the system passes from the non-interacting regime into the strongly interacting region and back again. We restrict ourselves to the case of a closed \gls{DM} such that a description of the temporal evolution using unitary dynamics is sufficient. Depending on the time interval within which the cycle is realized, we find that the system can show surprisingly strong signatures of \emph{quantum hysteresis}, i.e. different paths in the system's quantum state evolution during the forwards and backwards process, and that these memory effects vary in a highly non-monotonic way as the round-trip time changes~\cite{Gomez_Entropy2016}. The adiabatic theorem ensures that if the cycle is sufficiently slow, the process will be entirely reversible. In the other extreme, where the round-trip ramping is performed within a very short time, the total change undergone by the system is negligible. However in between these two regimes, we find a remarkably rich set of behaviors.\\  
\\
On the other hand, one the most important is the one-dimensional spin$-1/2$ XXZ model, this spin model describe the competition between hopping and interactions of excitations. It has become one of the most studied models in Condensed Matter Physics, and is usually considered as a testbed to identify general many-body phenomena applicable beyond magnetic systems. The phase diagram for the XXZ model has been extensively studied and characterized under equilibrium conditions. Actually, one of the most active research fields in condensed matter physics corresponds to the properties of this system in ou­t-of-equilibrium states. On the one hand, state-of­-the-art experimental techniques have allowed researchers to manipulate severals degrees of freedom of complex systems up to unprecedented levels, in both condensed matter\cite{Cavalleri_Nat2016} and ultracold atomic systems\cite{Bloch_PRM}.   
In addition, several theoretical studies in non-­equilibrium systems have shown intriguing physics such as spontaneous emergence of long-range order in boundary-driven spin chains\cite{prosen2010prl}, existence of phase transitions absent in equilibrium\cite{ajisaka2014njp}, etc. We introduce a new way of \gls{QPT}s of equilibrium many-body systems, by mean of local time correlations and Leggett-Garg inequalities. \\
\\
Additionally, The development of new methods to induce or mimic adiabatic dynamics is essential to the progress of quantum technologies. In many-body systems, the need to develop new methods to approach adiabatic dynamics is underlined for their potential application to quantum simulation and adiabatic quantum computation. The Kibble-Zurek mechanism (\gls{KZM}) is a highly successful paradigm to describe the dynamics of both thermal and quantum phase transitions~\cite{Zurek96a, Kibble76b}. It is one of the few theoretical tools that provide an account of nonequilibrium behavior in terms of equilibrium properties. It predicts that in the course of a phase transition topological defects are formed. We elucidate the emergence of adiabatic dynamics in an inhomogeneous quantum phase transition. We show that the dependence of the density of excitations with the quench rate is universal and exhibits a crossover between the standard \gls{KZM} behavior at fast quench rates, and a steeper power-law dependence for slower ramps.  Local driving of quantum critical systems thus leads to a much more pronounced suppression of the density of defects, that constitute a testable prediction amenable to a variety of platforms for quantum simulation including cold atoms in optical gases, trapped ions and superconducting qubits.\\
\\
Our results establish the universal character of the critical dynamics across an inhomogeneous quantum phase transition, that we proposed for favoring adiabatic dynamics. Our results should prove useful in a variety of contexts including the preparation of phases of matter in quantum simulators (e.g., with trapped ions, superconducting qubits) and condensed matter systems, as well as the engineering of inhomogeneous schedules in quantum annealing.\\
\\ 
\begin{savequote}[45mm]
``Trying to understand the way nature works involves a most terrible test of human reasoning ability. It involves subtle trickery, beautiful tightropes of logic on which one has to walk in order not to make a mistake in predicting what will happen. The quantum mechanical and the relativity ideas are examples of this."
\qauthor{Richard P. Feynman}
\end{savequote}

\chapter{Theoretical Background}\label{Cap2}
\begin{center}
\begin{tabular}{p{15cm}}
\vspace{0.1cm}
\quad \lettrine{\color{red1}{\GoudyInfamily{T}}}{his} chapter is a preamble containing essential ingredients that will play a major role in the next chapters, and gives the first impression of the kind of many-body equilibrium and non-equilibrium  situation that has been at the center of this doctoral thesis.
\end{tabular}
\end{center}

\section{Quantum Phase Transitions in 1-Dimensional Spin-$\frac{1}{2}$ models}
For several decades, models of interacting spins have attracted a lot of attention due
to their fascinating physical properties. The most prominent example is undoubtedly the spin$-1/2$ Heisenberg model, which was recognized  from the early days of quantum mechanics as a key element to understand the origin of the ferromagnetic behavior of several systems. This model corresponds to an effective Hamiltonian for a lattice of interacting electrons with overlapping wave functions, and arises from the Pauli exclusion principle and the Coulomb repulsion. The effective spin coupling resulting from the combination of these two effects, known as the exchange interaction, features a very simple form. Nevertheless, extracting information from it has proven to be a quite challenging task, and several analytical and numerical methods have been developed over the years for that purpose. Even now, open questions exist regarding its physical properties, and a lot of effort is still performed to  find an answer to them. \\
\\
We principal focus on a one dimensional spin$-­1/2$ chain described by an general Hamiltonian of the form
\begin{equation} \label{Hami_general}
\hat{H}=\sum_{\alpha}\sum_{i,j} J_{\alpha}^{i,j}\hat{\sigma}_i^{\alpha}\hat{\sigma}_j^{\alpha}+\sum_{\alpha}\sum_{i}B_{\alpha}^{i}\hat{\sigma}_i^{\alpha}.
\end{equation}
Here $\hat{\sigma}_i^{\alpha}$ denotes the Pauli operators at site $i$ ($\alpha=x,y,z$), $J_{\alpha}^{i,j}$ is the coupling between spins at sites $i$ and $j$ along direction $\alpha$, $B_{\alpha}^{i}$ is the magnetic field at site $i$ along direction $\alpha$, and $\hbar=1$. No restrictions on the dimensionality of the system or the range of the interactions are in principle required. we restrict our numerical studies to two particular testbed Hamiltonians of condensed matter physics, namely the 1D XXZ and anisotropic XY models with nearest-neighbour interactions. These systems have been extensively studied in the literature, and their ground-state phase diagrams are very well known~\cite{takahashi,giamarchi,Sutherland}. In this section we briefly describe the \gls{QPT}s featured by these models.

\subsection{Spin-$\frac{1}{2}$ XXZ Model}
\noindent We first consider a 1D system in which $1/2$  spins are coupled through an anisotropic Heisenberg interaction. This case, known as the XXZ model, corresponds to $J_{x}^{i,j}=J_{y}^{i,j}=J$, $J_{z}^{i,j}=J\Delta$, and $B_{\alpha}^i=0$. Thus it is described by the Hamiltonian
\begin{equation} \label{hami_xxz}
\hat{H}=J\sum_{i}\left[\hat{\sigma}_{i}^{x}\hat{\sigma}_{i+1}^{x} + \hat{\sigma}_{i}^{y}\hat{\sigma}_{i+1}^{y} +\Delta \hat{\sigma}_{i}^{z}\hat{\sigma}_{i+1}^{z}\right].
\end{equation}		
\noindent Here the coupling $J > 0$ represents the exchange interaction between nearest neighbors, and $\Delta$ is the dimensionless anisotropy along the $z$ direction~\footnote{Alternatively, the 1D XXZ model can be mapped to a chain of spinless fermions by means of a Jordan-Wigner transformation~\cite{takahashi,giamarchi}, where $J$ corresponds to the hopping to nearest neighbors and $J \Delta$ to a density-density interaction.}. This model can be exactly solved by means of the Bethe ansatz~\cite{takahashi,giamarchi,Sutherland}, and possesses several symmetries. Namely, it features a continuous $U(1)$ symmetry due to the conservation of the total magnetization in the $z$ direction for any $\Delta$ and an additional $SU(2)$ symmetry at $\Delta = \pm 1$ due to the conservation of the total magnetization along the $x$ and $y$ directions. Furthermore, the Hamiltonian is invariant under transformations $\hat{\sigma}_{i}^{z}\to -\hat{\sigma}_{i}^{z}$, thus having $\mathbb{Z}_{2}$ symmetry.\\

\begin{figure}[h!]
\begin{center}
\includegraphics[scale=1.2]{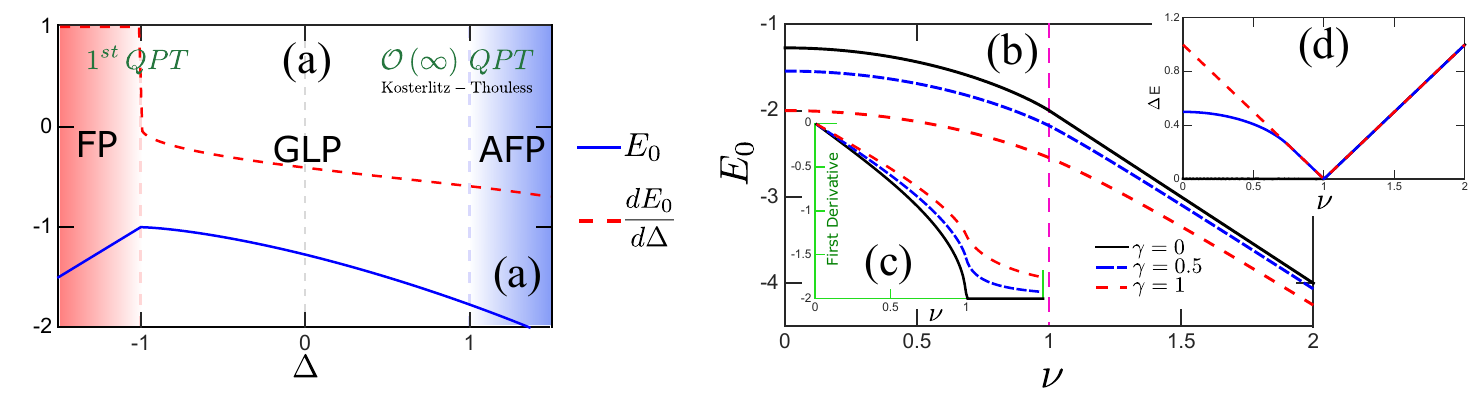}
\caption[Exact results for ground state energy XXZ and anisotropic XY Models.]{\label{ground} {\bf (a)} {\it Left panel:} Exact result for ground state XXZ model~\cite{takahashi} and first derivative ground state as a function $\Delta$, additionally we showed three different phases: ferromagnetic phase (FP), gapless phase (GLP) and antiferromagnetic phase (AFP) with the order of transition.  {\it Right  panel:} {\bf (b)} Exact ground state for XY model as a function of $\nu=2B_{z}/J$. Insets: {\bf (c)} First derivative ground state energy as a function $\nu$ and {\bf (d)} Gap energy for XY model. }
\end{center}
\end{figure}

\noindent The model presents three different phases. First, for $\Delta<-1$ the ground state consists of a fully polarized configuration along the $z$ direction, i.e. it corresponds to a ferromagnetic state (FP). In the intermediate regime $-1 < \Delta < 1$ the system is in a gapless phase (GLP), which can be shown to correspond to a Luttinger liquid in the continuum limit~\cite{giamarchi}. Finally, for $\Delta > 1$, the ground state corresponds to an antiferromagnetic configuration (AFP). The ferromagnetic and gapless states are separated by a first-order \gls{QPT} at $\Delta = -1$, while the gapless and antiferromagnetic states are separated by a (infinite-order) Kosterlitz-Thouless \gls{QPT} at $\Delta=1$ (see panel left Fig.~\ref{ground}).

\subsection{Spin-$\frac{1}{2}$  XY Model}	
\noindent We now describe the anisotropic 1D XY Hamiltonian for spins $1/2$. It corresponds to $J_{x}^{i,j}=\frac{1}{2}J(1+\gamma)$, $J_{y}^{i,j}=\frac{1}{2}J(1-\gamma)$, $B_z^i=B_z$, $J_{z}^{i,j}=B_x^i=B_y^i=0$ and is given by
\begin{equation} \label{hami_xy}
\hat{H}=J\sum_{i}\left[\frac{1+\gamma}{2} \hat{\sigma}_{i+1}^{x}\hat{\sigma}_{i}^{x} + \frac{1-\gamma}{2}\hat{\sigma}_{i+1}^{y}\hat{\sigma}_{i}^{y}\right]+B_{z}\sum_{i}\hat{\sigma}_{i}^{z}.
\end{equation}
Here $J > 0$ represents the exchange interaction between nearest neighbors, $\gamma > 0$ is the anisotropy parameter in the XY plane and $B_{z} > 0$ is the magnetic field along the $z$ direction. The limiting value  $\gamma = 1$ corresponds to the Ising model in a transverse magnetic field, which possesses a $\mathbb{Z}_2$ symmetry, and the limit $\gamma = 0$ is the isotropic XY model. In the thermodynamic limit $N \to \infty$, the anisotropic XY model can be exactly diagonalized by means of Jordan-Wigner and  Bogolyubov transformations~\cite{barouch1970pra,barouch1971pra}.\\

\noindent For the anisotropic case $0 < \gamma \leq 1$ the model belongs to the Ising universality class, and its phase diagram is determined by the ratio $\nu = 2B_z/J$.  When $\nu > 1$ the magnetic field dominates over the nearest-neighbor coupling, polarizing the spins along the $z$ direction. This corresponds to a paramagnetic state, with zero magnetization in the $xy$ plane. On the other hand, when $0 \leq \nu < 1$ the ground state of the system corresponds to a ferromagnetic configuration with polarization along the $xy$ plane. These phases are separated by a second-order \gls{QPT} at the critical point $\nu = 1$. Finally, for the isotropic case $\gamma = 0$, a \gls{QPT} is observed between gapless ($\nu < 2$) and ferromagnetic ($\nu > 2$) phases.
\section{Majorana Fermion Chain}
We focus on a concrete realization of a Majorana fermion chain in terms of the Kitaev model~\cite{kitaev}. It is described by the Hamiltonian:
\begin{equation}\label{Hkitaev}
\hat{H}=-\frac{\mu}{2}\sum_{j=1}^{N} \left(2\hat{n}_{j}-1\right)-\omega\sum_{j=1}^{N-1}\left(\hat{c}_{j}^{\dagger}\hat{c}_{j+1}+\hat{c}_{j+1}^{\dagger}\hat{c}_{j}\right)+\Delta \sum_{j=1}^{N-1}\left(\hat{c}_{j}\hat{c}_{j+1}+\hat{c}_{j+1}^{\dagger}\hat{c}_{j}^{\dagger}\right),
\end{equation}
representing a system of non-interacting  spinless fermions on an open end chain of $N$ sites labeled by $j=1,\ldots, N$. The single site fermion occupation operator is denoted by $\hat{n}_{j}=\hat{c}_{j}^{\dagger}\hat{c}_{j}$, the chemical potential is $\mu$, taken as uniform along the chain, $\omega$ is the hopping amplitude between nearest-neighbor sites (we assume $\omega\geq0$ without loss of generality because the case with $\omega\leq0$ can be obtained by a unitary transformation: $\hat{c}_{j}\to-\ii\left(-1\right)^{j}\hat{c}_{j}$) and $\Delta$ is the $p-$wave paring gap, which is assumed to be real and $\Delta\geq 0$ (the case $\Delta\leq0$ can be obtained by transformation $\hat{c}_{j}\to \ii\,\hat{c}_{j}$ for all $j$). This model captures the physics of a 1-D topological superconductor with a phase transition between topological and nontopological (trivial) phases at $\mu=2\Delta$, for $\Delta=\omega$. Notice that for this symmetric hopping-pairing Kitaev Hamiltonian, i.e. $\omega=\Delta$, a Jordan-Wigner transformation leads directly into the transverse field Ising model~\cite{leenjp16}. Thus, from now on we will refer as Majorana fermion chain either the Kitaev chain or the transverse field Ising model.\\

Let us introduce Majorana operators $\hat{\gamma}_{j}$ to express the real space spinless fermion annihilation and creation operators, as:
\begin{equation}\label{Majo_op}
\hat{c}_j=\frac{1}{2}\left ( \hat{\gamma}_{2j-1}+\ii\hat{\gamma}_{2j} \right ), \qquad \hat{c}_j^{\dagger}=\frac{1}{2}\left ( \hat{\gamma}_{2j-1}-\ii\hat{\gamma}_{2j} \right ).
\end{equation}
These are Hermitian operators $(\hat{\gamma}_j=\hat{\gamma}_j^{\dagger})$, satisfy the property $\left(\hat{\gamma}_{j}\right)^{2}=\left(\hat{\gamma}_{i}^{\dagger}\right)^{2}=1$, and obey the modified anticommutation relations $\lbrace\hat{\gamma}_{i},\hat{\gamma}_{j}\rbrace=2\delta_{i,j}$, with $i,j=1,\dots,2N$. From the definition of Majorana operators~\eqref{Majo_op} it is evident that for each spinless fermion on site $j$, two Majorana fermions are assigned to that site, which are denoted by $\hat{\gamma}_{2j-1}$ and $\hat{\gamma}_{2j}$.  They allow the Kitaev Hamiltonian in Eq.~\eqref{Hkitaev} to be written in the equivalent form:
\begin{equation}\label{Hmajo1}
\hat{H}=-\ii \frac{\mu}{2}\sum_{j=1}^{N}\hat{\gamma}_{2j-1}\hat{\gamma}_{2j}
+\frac{\ii}{2}\sum_{j=1}^{N-1}\left[(\omega+\Delta)\;\hat{\gamma}_{2j}\hat{\gamma}_{2j+1}-(\omega-\Delta)\;\hat{\gamma}_{2j-1}\hat{\gamma}_{2j+2}\right].
\end{equation}
The parameters $\mu$, $\Delta$ and $\omega$ induce relative complex interactions between the Majorana modes. Now we briefly explain two limit cases of the KM Hamiltonian.
\begin{figure}[h!]
\begin{center}
\includegraphics[scale=0.7]{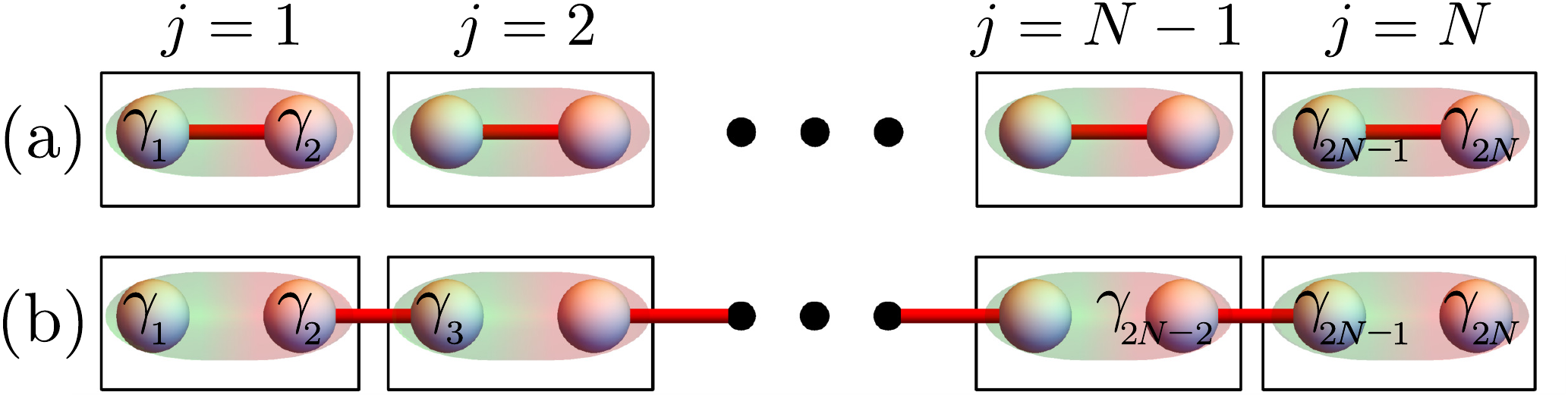}
\caption[Schematic illustration of the Kitaev-Majorana Hamiltonian.]{\label{Apar_1} Schematic illustration of the Kitaev-Majorana Hamiltonian, {\bf (a)} in the trivial limit, and {\bf (b)} in the nontrivial limit with coupling between Majorana fermions $\hat{\gamma}_{2j}$ and $\hat{\gamma}_{2j+1}$ only. The solid spheres represent the Majorana fermions $\gamma_{2j-1}$ and $\gamma_{2j}$ making up each physical $j$ site in the Kitaev chain. In the nontrivial phase, the zero energy Majorana Modes (ZEM) are present at the left and right boundaries of the lattice, which are illustrated by the two unpaired spheres.}
\end{center}
\end{figure}

{\em First limit case:} We start with the simplest case $\Delta=\omega=0$ with $\mu<0$, yielding to a Kitaev state in the so-called topologically trivial phase. Therefore, the KM Hamiltonian of Eq.~\eqref{Hmajo1} takes the trivial form
\begin{align}\label{Hmajo}
\hat{H}_{\text{Trivial}}=-\ii\frac{\mu}{2}\sum_{j=1}^{N}\hat{\gamma}_{2j-1}\hat{\gamma}_{2j}.
\end{align}
Here only the first term of equation~\eqref{Hmajo1} is different from zero, leaving a coupling only between Majorana modes $\hat{\gamma}_{2j-1}$ and $\hat{\gamma}_{2j}$ at the same lattice site $j$, as Fig.~\ref{Apar_1}(a) schematically illustrates. This leads to a ground state with all occupation numbers equal to $0$.\\

{\em Second limit case:} We now consider the Hamiltonian of Eq.~\eqref{Hmajo1} with $\mu=0$, namely
\begin{equation*}
\hat{H}=\frac{\ii}{2}\sum_{j=1}^{N-1}\left[\left(\Delta+\omega\right)\hat{\gamma}_{2j}\gamma_{2j+1}-\left(\omega-\Delta\right)\hat{\gamma}_{2j-1}\hat{\gamma}_{2j+2}\right].
\end{equation*}
This last form simplifies even more when $\Delta=\omega>0$ to the compact expression $\hat{H}=\Delta\ii\sum_{j=1}^{N-1}\hat{\gamma}_{2j}\hat{\gamma}_{2j+1}$, indicating that Majorana operators from neighboring sites are paired together, so that the even numbered $\hat{\gamma}$ at site $j$ is coupled to the odd numbered $\hat{\gamma}$ at site $j+1$, as depicted in Fig.~\ref{Apar_1}(b). The first and last Majorana fermions are thus left unpaired, corresponding to the zero energy Majorana Modes (ZEM)~\cite{Leijnse,AguadoR}.

\subsection{Exact diagonalization of the Kitaev Hamiltonian: Bogoliubov-de Gennes approach}
Since the Kitaev Hamiltonian is quadratic in fermionic operators $\hat{c}_{j}$ and $\hat{c}_{j}^{\dagger}$, its exact diagonalization via a Bogoliubov-de Gennes transformation is always feasible~\cite{Olesia}. The matrix representation of the Kitaev Hamiltonian given by Eq.~\ref{H_num}  

\begin{equation*} \label{Matrix_Kitaev}
\hat{H}=\frac{1}{2}\begin{pmatrix}
\hat{c}_1^{\dagger} & \hat{c}_{1} & \cdots & \hat{c}_N^{\dagger} & \hat{c}_{N}
\end{pmatrix}\left(\begin{array}{cccccccccc}
-\mu & 0 & -\Delta & -w & 0 & 0 & 0 & \cdots  &0& 0\\
0 & \mu & w & \Delta & 0 & 0 & 0 &\cdots  & 0 & 0\\
-\Delta & w & -\mu & 0 & -\Delta & -w & 0 &  \cdots & 0 & 0\\
-w &\Delta & 0 &\mu &w & \Delta &0 &\cdots & 0 & 0\\
0 & 0 &-\Delta & w & \ddots & \ddots & \ddots&\ddots  &\vdots &\vdots \\
0 & 0 & -w & \Delta & \ddots & \ddots & \ddots & \ddots  & \ddots&\vdots \\
0 & 0 & 0 & 0 & \ddots  &\ddots & \ddots & \ddots &   \ddots & \vdots\\
\vdots&\vdots&\vdots&\vdots&\vdots&\vdots&\vdots&\vdots&\vdots&\vdots\\
0&0&0&0&\cdots&0&-\Delta&w&-\mu&0\\
0&0&0&0&\cdots&0&-w&\Delta&0&\mu
\end{array}
\right)\begin{pmatrix}
\hat{c}_1\\
\hat{c}_{1}^{\dagger}\\
\vdots\\
\vdots\\
\vdots\\
\vdots\\
\vdots\\
\hat{c}_N\\
\hat{c}_{N}^{\dagger}
\end{pmatrix}.
\end{equation*}
In order to put the Majorana fermion Hamiltonian in Eq.~\eqref{Hkitaev} (or equivalently in Eq.~\eqref{Hmajo1}) in diagonal form, a standard Bogoliubov transformation is performed:
\begin{align}\label{BGT}
\hat{c}_{j}^{\dagger} &= \sum_{k=1}^{N} \left ( u_{2k,j} \hat{d}_{k} + v_{2k,j}\hat{d}^{\dagger}_{k}\right ), 
&\hat{c}_{j} &= \sum_{k=1}^{N} \left ( u_{2k,j} \hat{d}^{\dagger}_{k} + v_{2k,j}\hat{d}_{k}\right ),
\end{align}
where $k$ denotes a single fermion mode, $u_{2k,j}$ and $v_{2k,j}$ are real numbers, and the canonical fermion anticomutation relations for the new operators $\hat{d}_{k}$, $\hat{d}^{\dagger}_{k}$ remain true, that is $\left\{\hat{d}_{k},\hat{d}^{\dagger}_{k'}\right\}=\delta_{k,k'}$, $\left\{\hat{d}^{\dagger}_{k},\hat{d}^{\dagger}_{k'}\right\} =\left\{\hat{d}_{k},\hat{d}_{k'}\right\}=0$.
Thus the exact diagonalization of the Kitaev Hamiltonian in Eq.~\eqref{Hkitaev}, in terms of the new independent fermion mode operators $\hat{d} \pap{\hat{d}^{\dagger}}$, leads to:
\begin{equation}\label{H_num}
\hat{H}=\sum_{k=1}^{N}\eps_{k}\pas{\hat{d}^{\dagger}_{k}\hat{d}_{k}-\frac{1}{2}},
\end{equation}
where the new fermion mode energies $\eps_{k}\geq 0$ are to be numerically calculated for a Kitaev chain with open ends (although analytical exact results may be found in some cases, see~\cite{narozhny}).
The central $2N\times 2N$ matrix in Eq.~\eqref{Matrix_Kitaev}, which we denote by $\hat{\mathcal{H}}$, can be rendered to a diagonal form $\hat{\mathcal{H}}_D=\hat{D}^{-1}\hat{\mathcal{H}}\hat{D}$ by a unitary matrix such as:
\begin{equation}\label{duv1}
\hat{D}=\left(\begin{array}{cccccccccc}
u_{1,1} & u_{2,1} & u_{3,1} & u_{4,1} & \cdots & \cdots & \cdots & \cdots  &u_{2N-1,1}& u_{2N,1}\\
v_{1,1} & v_{2,1} & v_{3,1} & v_{4,1} & \cdots & \cdots & \cdots & \cdots  &v_{2N-1,1}& v_{2N,1}\\
u_{1,2} & u_{2,2} & u_{3,2} & u_{4,2} & \cdots & \cdots & \cdots & \cdots  &u_{2N-1,2}& u_{2N,2}\\
v_{1,2} & v_{2,2} & v_{3,2} & v_{4,2} & \cdots & \cdots & \cdots& \cdots  &v_{2N-1,2}& v_{2N,2}\\
\vdots & \vdots &\vdots & \vdots & \ddots & \ddots & \ddots&\ddots  &\vdots &\vdots \\
\vdots & \vdots &\vdots & \vdots & \ddots & \ddots & \ddots&\ddots  &\vdots &\vdots \\
u_{1,N} & u_{2,N} & u_{3,N} & u_{4,N} & \cdots & \cdots & \cdots & \cdots  &u_{2N-1,N}& u_{2N,N}\\
v_{1,N} & v_{2,N} & v_{3,N} & v_{4,N} & \cdots & \cdots & \cdots & \cdots  &v_{2N-1,N}& v_{2N,N}
\end{array}
\right).
\end{equation}
Since $u_{2q-1,j}=v_{2q,j}$ and $v_{2q-1,j}=u_{2q,j}$, the unitary property of matrix $\hat{D}$ implies that
\begin{align}\label{diag1}
\sum_{q=1}^{N}\pas{u_{2q,i}u_{2q,j}+v_{2q,i}v_{2q,j}}=\delta_{i,j},\qquad\qquad\sum_{q=1}^{N}u_{2q,i}v_{2q,j}=0
\end{align}
for every site $j=1,...,N$. The diagonal matrix $\hat{\mathcal{H}}_D$ is ordered as
\begin{equation}
\hat{\mathcal{H}}_D=\left(\begin{array}{cccccccccc}
-\epsilon_1 & 0 & \cdots & 0 & 0\\
0 & \epsilon_1 & \cdots & 0 & 0\\
\vdots & \vdots & \ddots  &\vdots & \vdots\\
0 & 0 & \cdots & -\epsilon_N & 0\\
0 & 0 & \cdots & 0 & \epsilon_N
\end{array}
\right)
\end{equation}
where, for $\omega=\Delta$, the positive energies are given by:
\begin{equation}\label{diag2}
\epsilon_k=-\mu\sum_{j=1}^N\left [ u_{2k,j}^2-v_{2k,j}^2 \right ]-2\Delta \sum_{j=1}^{N-1}\left [ u_{2k,j}-v_{2k,j} \right ]\left [ u_{2k,j+1}+v_{2k,j+1} \right ].
\end{equation}
From the entries of matrix $\hat{D}$ in Eq.~\eqref{duv1} the standard  Bogoliubov-de Gennes transformation given by Eq.~\eqref{BGT}.

\section{Dynamics of the Dicke Model across its quantum phase transition}\markright{Dynamics of the Dicke Model across its \gls{QPT}}{}
\noindent The \gls{DM} of a many-body light-matter interacting quantum system, features a set of $N$ identical two-level systems (commonly referred to as qubits) each of which is coupled to a single radiation mode. It can be described by the following microscopic Hamiltonian:
\begin{equation}\label{hdic}
\hat{H}(t)=\frac{\epsilon}{2}\sum_{i=1}^{N}\hat{\sigma}_{z}^{i} + \omega \hat{a}^{\dagger}\hat{a} +\frac{\lambda(t)}{\sqrt{N}}\left(\hat{a}^{\dagger}+\hat{a}\right)\sum_{i=1}^{N}\hat{\sigma}_{x}^{i}\:.
\end{equation}
We purposely avoid common approximations such as the rotating-wave approximation. Here $\sigma_{\alpha}^{i}$ denotes the Pauli operators for qubit $i$ $\left(\alpha=x,z\right)$;  $\hat{a}^{\dag} \pap{\hat{a}}$ is the creation (annihilation) operator of the radiation field. $\epsilon$ and $\omega$ represent the qubit and field transition frequencies respectively; and $\lambda(t)$ represents the strength of the radiation-matter interaction at time $t$ which can be varied over time. In many situations such as those we consider here in our work, the dynamics of the \gls{DM} do not require the consideration of the entire $2^{\otimes N}\otimes \mathbb{N}$ dimensional Hamiltonian. Instead, $SU(2)$ collective operators $\hat{J}_{\alpha}=\frac{1}{2}\sum_{i=1}^{N}\hat{\sigma}^{i}_{\alpha}$ can be used. The Hamiltonian~\eqref{hdic} can then be written in the following form:
\begin{equation}\label{Hdicke}
\hat{H}(t)=\epsilon \hat{J}_{z} + \omega \hat{a}^{\dagger}\hat{a} +\frac{2\lambda(t)}{\sqrt{N}}\hat{J}_{x}\left(\hat{a}^{\dagger}+\hat{a}\right)\ \ .
\end{equation}

\begin{figure}[h!]
\begin{center}
\includegraphics[scale=0.93]{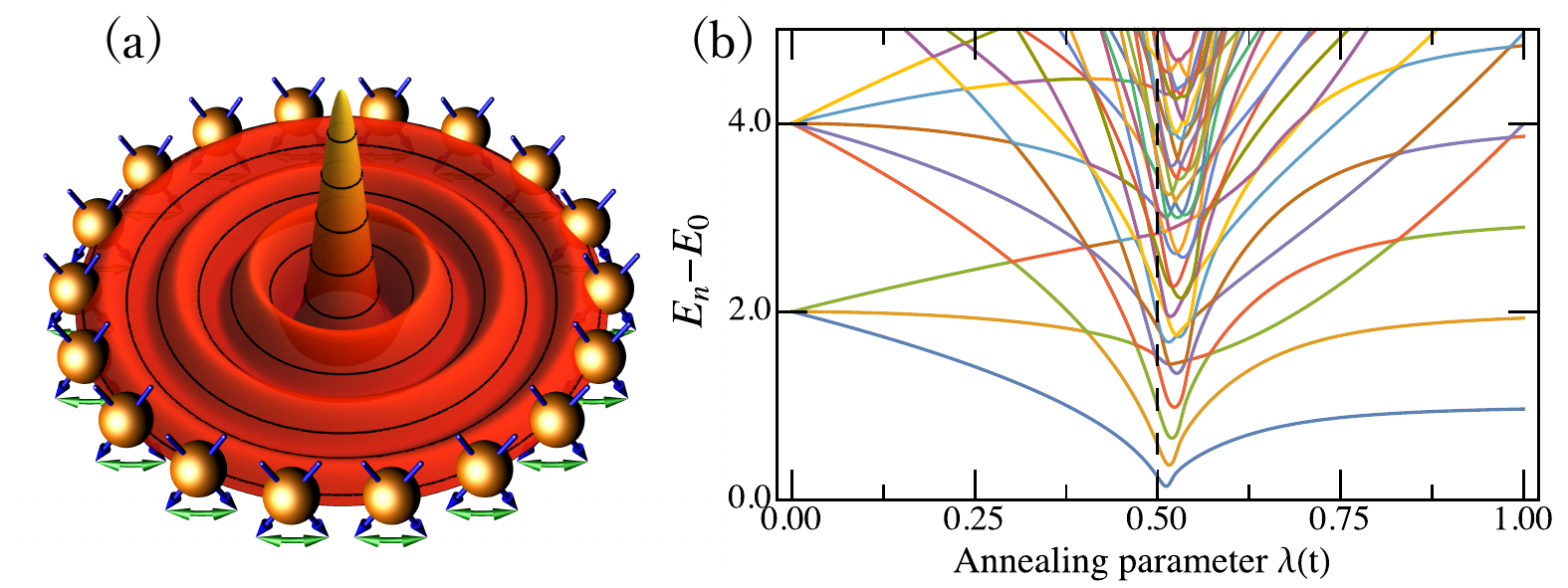}
\caption[Dicke Model energy spectrum.]{\label{Espectro} {\bf (a)} Illustration of the Dicke model (\gls{DM}) network of spins, or equivalently qubits. The electromagnetic field mode mediates the all-to-all interaction among these qubits. {\bf (b)} Energy spectrum of the \gls{DM} as a function of the interaction parameter $\lam(t)$ for a finite number of qubits $N=257$. The numerical results in this chapter are generated under the condition of resonance between the qubits and the radiation frequency: $\epsilon=\omega=1$ in Eq.~\eqref{hdic}. All energies are measured with respect to the ground state energy $E_{0}$. In the thermodynamic limit (\gls{TL}), a second order \gls{QPT} occurs at value $\lam_c= 0.5$ at which point  the energy gap vanishes. As can be seen in (b) for finite $N$, a finite-size version of the \gls{QCP} with minimum gap arises near to $\lam_c$. Only the even parity sector of the model is depicted here.}
\end{center}
\end{figure}

The static properties of Dicke Model have been widely studied and characterized in the last two decades~\cite{Nagy2010prl,Das2016njp}. It is well known that, in the thermodynamic limit $N\to\infty$, it exhibits a second-order \gls{QPT}~\cite{HioePRA} at $\lam_{c}=\sqrt{\eps\ome}/2$ with order parameter $\hat{a}^{\dag}\hat{a}/J$, separating the normal phase at $\lam_{c}<\sqrt{\eps\ome}/2$ from the superradiant phase in which there is a finite value of the macroscopic order parameter, e.g. finite boson expectation number \cite{Acevedo2014PRL}. When the static coupling parameter is above this critical value $\lam_c$, the ground state of the system is characterized by a non-zero expectation value of the excitation operators,   
\begin{equation}
\bigl\langle \hat{N}_{b} \bigr\rangle\equiv \bigl\langle \hat{a}^{\dagger}\hat{a}\bigr\rangle \qquad \text{and} \qquad \bigl\langle \hat{N}_{q} \bigr\rangle\equiv \biggl \langle \hat{J}_{z} -\frac{N}{2}\biggr\rangle \:.
\end{equation}
When $\lambda < \lambda_{c}$, the order parameters are zero. Because of this, the region when $\lambda > \lambda_{c}$ is called the ordered or superradiant phase, while the region when $\lambda < \lambda_{c}$ is called the normal phase. Near the phase-boundary in the vicinity of this superradiant phase, there is a dependence of the order parameter as follows: $\bigl\langle \hat{N}_{b} \bigr\rangle\propto\left(\lambda -\lambda_{c}\right)$ and  $\bigl\langle \hat{N}_{q} \bigr\rangle\propto\left(\lambda -\lambda_{c}\right)^{1/2}$~\cite{Emary_PRE}. This power-law behavior is typical of second-order phase transitions where the critical exponents are characteristic of the universality class to which the model belongs. In the ordered quantum phase $\left(\lambda >\lambda_c\right)$, the $\mathbb{Z}_{2}$ symmetry related to parity is spontaneously broken, which originates from the fact that the TL ground state is two-fold degenerate corresponding to the two different eigenvalues of $\hat{P}$. Also at the \gls{QPT}, the \gls{DM} presents an infinitely-degenerate vanishing energy gap~\cite{Emary_PRE}, as shown in Fig.~\ref{Espectro}. 

In this thesis, now we study its dynamical properties under the time-dependent model in Eq.~\ref{Hdicke}. We obtain the full driven \gls{DM} instantaneous state $\left|\psi\pap{t}\right.\rangle$ by numerically solving the time-dependent Schr\"odinger equation. Our numerical solution of the driven \gls{DM} profits from the fact that the operator $\hat{{\bf J}}^{2}=\sum_{\alpha}\hat{J}_{\alpha}^{2}$ is a constant of motion with eigenvalue $J\left(J+1\right)$, and that the parity operator $\hat{\mathcal{P}}=\exp\left(\imath \pi\left[\hat{a}^{\dag}\hat{a} +\Jo_{z}+J\right]\right)$ is also conserved and commutes with $\hat{{\bf J}}^{2}$. Since we are seeking results that have general validity, we avoid making the rotating-wave approximation that is commonly used to solve the static version of the driven \gls{DM} and which makes it Bethe ansatz integrable~\cite{Gaudin_JPF1976}. The general structure of the state $\left|\psi\pap{t}\right.\rangle$ at any time $t$ is given by
\begin{equation}\label{state}
\left|\psi\pap{t}\right.\rangle=\sum_{m_{z}=-N/2}^{N/2}\sum_{n=0}^{\chi}C_{n,m_{z}}\pap{t}\left| m_{z},n\right.\rangle.
\end{equation}
Here $\chi$ is the truncation parameter of the size of the bosonic Fock space, whose value we choose to be large enough to ensure that the numerical results converge  \cite{Acevedo2014PRL}. The basis states $\left|m_{z},n\right.\rangle=\left|m_{z}\right.\rangle\otimes\left|n\right.\rangle$ are defined such that $\left|m_{z}\right.\rangle$ is an eigenvector of $J_{z}$ in the subspace of even parity with eigenvalue $m_z$, and $\left|n\right.\rangle$ is a bosonic Fock state with occupation $n$. The initial state of the dynamics at $t=0$, with negligible light-matter coupling $\lambda(t)=0$, is the non-interacting ground state $\left|\psi(0)\right.\rangle=\bigotimes_{i=1}^{N}\left|\downarrow\right.\rangle\otimes\left|n=0\right.\rangle=\left|-\frac{N}{2},0\right.\rangle$, where both the matter and light subsystems have zero excitations. All qubits are polarized in the state with $\langle\sigma_z\rangle=-1$, and the field is in the Fock state of zero photons. Since the total angular momentum and parity are conserved quantities, we can without loss of generality restrict our study to the maximum angular momentum sector $J=N/2$ and $\mathcal{P}=1$.\\
\\
\noindent The manifestations of the \gls{QPT} in static properties of the \gls{DM} at finite values of $N$ makes us to also expect effects of it in the dynamical aspects when the Hamiltonian varies on time through variations of its parameters. This question about the dynamical aspects of the \gls{DM} has been quite scarcely addressed, and this contrasts with the big amount of literature that has been written about the static properties. In fact, easier ways to explore the static aspects are still being researched. On the other hand, among the few works we have found to explore the issue of the dynamical evolution of the parameters, all exploit the thermodynamic limit at some level, and either resort to a semiclassical approximation or other perturbative approaches, or open system's Langevin equations that fail to cross the phase-boundary. In our approach we expect to tackle the problem at the very quantum level. However, we will also limit the issue in some sense, since we will restrict the attention on manageable values of $N$ and neglect the open system situation that it is unavoidably present in all current experimental realizations of the model. However, we expect to effectively explore large enough sizes of the system to be able to make plausible extrapolations to the thermodynamic limit; and results from unitary evolutions of the total system are by no means out of interest, neither theoretically, nor experimentally. From the theoretical point of view, the closed \gls{DM} is already a non-trivial many-body system that can produce interesting insights about the collective behavior of quantum systems. Also, the results would not be out of empirical verification, since ultrafast spectroscopical techniques keep being improved, then leaving the possibility of operation times being short enough to rule out any significant  decoherence effect from external in influences. 

\subsection{Experimental Platform to Driving Dicke Model}
To date, several experimental scenarios have shown efficient and effective ways to simulate radiation-matter interaction systems with time-varying couplings. Among the experimental possibilities that have attracted most attention, with many being built around implementations in circuit \gls{QED} where superconducting qubits play the role of the matter subsystem~\cite{ExpDicke1, ExpDicke2, ExpDicke3}. Additionally to date, important experimental scenarios have shown efficient and effective ways to simulate radiation-matter interaction systems with time-varying couplings. Some of the most promising experimental possibilities are thermal gases of atoms~\cite{BlackPLR}, and \gls{BEC}s using momentum~\cite{ExpDicke4} and hyperfine states~\cite{BaumannPRL}. However, we believe that the branch of experiments deserving most attention is that demonstrating \gls{DM} superradiance in ultra-cold atom optical traps, especially $^{87}$Rb Bose-Einstein condensates~\cite{KlinderPNC2015, ExpDicke4, ExpDicke5, ExpDicke6}. 
Indeed, these atom-trap experiments are so promising that a brief description of them is pertinent here, following Ref.~\cite{KlinderPNC2015}. \\
\\
 We now proceed to briefly review how the essential physics from those experiments get captured in our \gls{DM} Hamiltonian. The involved degrees of freedom can be described by the creation (annihilation) operators $\hat{a}\,\left(a^{\dagger}\right)$, the corresponding matter wave field operator $\hat{\psi}\left(r\right)$, and the pumping laser amplitude $\alpha_{p}\left(t\right)$ simulated by a c-number. The Hamiltonian then reads\footnote{C. J. Pethick and H. Smith, Bose-Einstein Condensation in Dilute Gases (Cambridge University Press, 2008).},
\begin{equation}
\begin{split}
\hat{H}_{\text{BEC}}=\omega_{0}\hat{a}^{\dagger}\hat{a} +\int {\rm d}^{2}r \,&\hat{\psi}^{\dagger}\left(r\right)\left[-\frac{\nabla^{2}_r}{2m_a}+\right.\\
&\Delta_{0}\left(\hat{a}^{\dagger}\cos kz +\alpha^{*}_{p}(t)\cos ky\right)\left(\hat{a}\cos kz +\alpha^{}_{p}(t)\cos ky\right)\left.\right]\hat{\psi}\left(r\right)
\end{split}
\end{equation}
\begin{figure}[t]
\begin{center}
\includegraphics[scale=1.0]{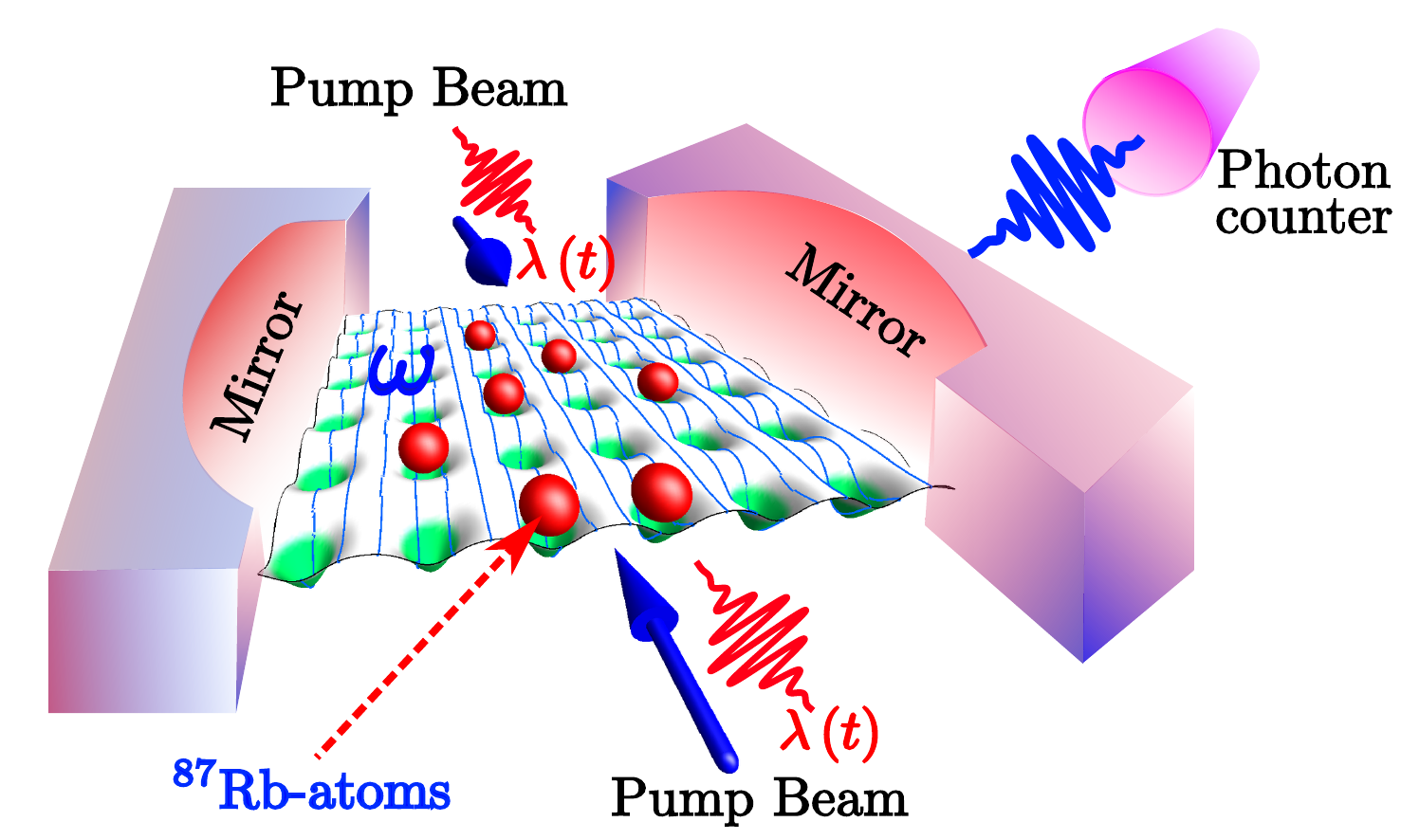}
\caption[Schematic representation of the Driven Dicke Model]{Schematic of recent successful realizations of the Dicke Model in a \gls{BEC} of $^{87}$Rb atoms confined by a magneto-optical trap. The atoms (qubits) are made to interact with the  field mode of a high  finesse cavity by means of a pumped transverse field.~\cite{KlinderPNC2015} }\label{fig_1a}
\end{center}
\end{figure}
where $m_a$ is the mass of the atoms, $\Delta_0$ is the light shift per intra-cavity photon, and $\omega_0$ is the radiation mode frequency. Now, we approximate the field operator by a Fourier series in the grid plane (which we suppose is the $zy$ plane) discarding global phases; we go to the rotating frame; and we use Dicke states such as
\begin{equation}
\left|\right.m\rangle=\frac{1}{\sqrt{m!\left(N-m\right)!}}\left(\hat{c}^{\dagger}_{e}\right)^{m}\left(\hat{c}^{\dagger}_{g}\right)^{m}\left|\right.0\rangle\ .
\end{equation}
As a result, we obtain the Hamiltonian in Eq.~\eqref{Hdicke} with an explicit time-dependent light-matter coupling. Then, there exists the following equivalences among parameters:
\begin{align}
\epsilon = \frac{k^{2}}{2m_a}, && \omega = \delta_{c} -\frac{1}{2}\Delta_0 N, && \lambda(t)=\frac{\sqrt{N}}{4}\Delta_0 \left|\alpha_{p}(t)\right|
\end{align}
where $\delta_c$ is the the atom-laser frequency detuning. We note how the laser becomes the actual
control parameter, the knob for driving the \gls{DM} across parameter space. Hence every time we mention in the chapter a time-modulation in the coupling $\lambda(t)$, we are in fact implementing a real physical process such as changing the pumping laser power. Figure~\ref{fig_1a} depicts a schematic of the main components of the atom-trap \gls{DM} realization. It corresponds to an ultracold cloud of $N\sim 10^{5}$ $^{87}$Rb atoms confined by a magneto-optical trap inside a high  finesse Fabri-Perot cavity. The cloud is driven by a transverse pump laser whose wavelength is the same as that of the fundamental mode of the cavity. The combined cavity and pump laser setting produces an optical lattice potential that affects the motion of the atoms in the cloud through coupling with far-detuned atomic resonances. This coupling causes the atoms to interact with each other through the mediating presence of the radiation mode. At low intensity pump power $\epsilon_{p}$, the \gls{BEC} remains in its (almost spatially uniform) translational ground state. However, when a critical value of $\epsilon_p$ is reached, the ground state becomes a grid-like matter wave as the one shown in Fig~\ref{fig_1}. This change of configuration constitutes the \gls{QPT} whose spontaneous symmetry breaking is caused by the fact that two matter wave configurations, which are distinguishable only by a phase difference of $\pi$, have the same lowest possible energy. The effective two-level (qubit) system is composed of the ground \gls{BEC} translational state and the fundamental grid-like matter wave state for each atom. There are several ways to monitor the system. The two most fundamental are (1) addressing the radiation  field by coupling one of the (unavoidably) leaky walls of the cavity to a detector, and (2) using time-of-flight methods to measure the matter wave modes. Note that each technique measures the state of one of the two main components of the \gls{DM}, i.e. light and matter respectively. 
\begin{savequote}[45mm]
``There is no authority who decides what is a good idea."
\qauthor{Richard P. Feynman}
\end{savequote}
\chapter[\gls{QPT}s Detected by \gls{STC} and Violations of \gls{LGI}s]{Time correlations and Leggett-Garg Inequalities as local probes of quantum phase transitions in interacting spin systems}\label{Cap3}
\begin{center}
\begin{tabular}{p{15cm}}
\vspace{0.1cm}
\quad \lettrine{\color{red1}{\GoudyInfamily{T}}}{his} chapter focus the study of fingerprint of critical dynamic  properties is spins latices systems. We discuss an alternative form to identify quantum phase transitions of many-body systems. Namely we show that local time correlations and Leggett-Garg inequalities help locate the quantum critical points of finite- and infinite-order transitions. By means of an analytical calculation on a general spin-$1/2$ Hamiltonian, and matrix product simulations of one-dimensional XXZ and anisotropic XY models, we argue that finite-order \gls{QPT}s can be determined by singularities of the time correlations or their derivatives at criticality. The same features are exhibited by corresponding Leggett-Garg functions, which remarkably indicate violation of the Leggett-Garg inequalities for early times and all the Hamiltonian parameters considered. In addition, we find that the infinite-order transition of the XXZ model at the isotropic point can be revealed by the maximal violation of the Leggett-Garg inequalities. We thus show that quantum phase transitions can be identified by purely local measurements, and that many-body systems constitute important candidates to observe the violation of Leggett-Garg inequalities.\\
\\
This chapter is published in reference~\cite{Gomez_PRB2016}: {\bf F. J. G\'omez-Ruiz}, J. J. Mendoza-Arenas, F. J. Rodr\'iguez, C. Tejedor, and L. Quiroga. {\it Quantum phase transitions detected by a local probe using time correlations and violations of Leggett-Garg inequalities}. Phys. Rev. B, {\bf 93}, 035441 (2016).
\end{tabular}
\end{center}

\section{Introduction}
 In recent years, quantum phase transitions (\gls{QPT}s) of many-body systems have been the object of intense research~\cite{sachdev,aeppli}. This is the case not only due to the intrinsic interest that critical phenomena exhibit, but also because the understanding and development of new states in condensed matter or atomic systems may have prominent applications in areas such as high-temperature superconductivity~\cite{lee2006rmp} and quantum computation~\cite{nielsen}. The seminal recent advances on quantum simulation schemes~\cite{georgescu2014rmp} in systems such as cold atoms in optical lattices~\cite{Bloch_PRM} and trapped ions~\cite{Monroe11,schneider2012rpp} constitute fundamental steps in this direction.

Usually finite-order \gls{QPT}s of a particular system are characterized by discontinuities of its ground state energy, or singularities of its derivatives with respect to the parameter that drives the transitions. Besides the determination of order parameters, quantities such as gaps, spatial correlation functions and structure factors are commonly used to determine the quantum critical points of several models. Remarkably, a few years ago it was realized that entanglement plays a fundamental role in critical phenomena, and that different measures of entanglement can be used to determine the location of several types of \gls{QPT}s~\cite{amico2002nature,gu2003pra,latorre2003prl,wu2004prl,mosseri2004pra,Reslen2005epl,laflorencie2006prl,zanardi2006njp,amico2008rmp,buonsante2007prl,canovi2014prb,hofmann2014prb}. Furthermore, the relation of Bell inequalities and criticality has been recently explored~\cite{justino2012pra,sun2014pra,sun2014_2pra}.

Since non-local measurements are not always accessible, in this work we propose an alternative form to characterize \gls{QPT}s by exploiting single-site protocols to obtain bulk properties of many-body systems~\cite{Gessner2011PRL,Gessner2013PRA,Gessner2014epl,Gessner2014Nat}. We argue that local time correlations can indicate the location of critical points for finite-order \gls{QPT}s, in a similar way to measures of bipartite entanglement such as concurrence and negativity~\cite{wu2004prl}. This is exemplified by numerical simulations, based on tensor network algorithms, of time correlations of one-dimensional (1D) spin-$1/2$ lattices described by XXZ and anisotropic XY Hamiltonians, which correspond to exhaustively-studied models of condensed matter physics. The first- and second-order transitions of these models are determined by nonanalyticities of the time correlations and their first derivative, respectively. We also relate \gls{QPT}s to a different characterization of quantumness of a system, namely the violation of Leggett-Garg inequalities (\gls{LGI})~\cite{lgi_original,emary,huelga1996pra,castillo2013pra,bell_leggett_garg_2015,kofler2007prl}, which indicates the absence of macroscopic realism and non-invasive measurability. We show that by maximizing the violation of these inequalities along all possible directions, the infinite-order \gls{QPT} of the XXZ model can be identified. Given that the models considered in our work describe several condensed-matter systems~\cite{aeppli} and can be implemented in a variety of quantum simulators~\cite{georgescu2014rmp}, our analysis places them as interesting many-body scenarios for the experimental observation of the violation of Leggett-Garg inequalities.
\section{Single-site Two-time correlations} \label{section_time_correl}
We discuss how single-site two-time correlations (\gls{STC}) can indicate different types of quantum phase transitions. We consider the symmetrized temporal correlation $C(t)$ for a single-site operator $A$, given by

\begin{equation}
C(t)=\frac{1}{2}\braGS \lbrace A(t),A(0)\rbrace \ketGS,
\end{equation}	
with $\ketGS$ the ground state of the time-independent Hamiltonian of interest $H$, $\lbrace .,.\rbrace$ the anticommutator between two operators, and $A (t)$ the operator at time $t$,
\begin{equation}\label{eq:Correl}
A(t)=e^{iHt}A(0)e^{-iHt}.
\end{equation}
For simplicity, we consider that $A(0)$ corresponds to one of the Pauli operators of a particular site $k$ ($A(0)=\sigma_k^{\mu}$, $\mu=x,y,z$). First note that the time correlations can be rewritten as
\begin{align} \label{time_correl}
\begin{split}
C(t)=Re\left[e^{iE_{0}t}\braGS A(0)e^{-iHt}A(0)\ketGS\right],
\end{split}
\end{align}
with $E_0$ the ground-state energy. To proceed, we expand the time evolution operator, as
\begin{equation}
e^{-iHt}=\sum_{l=0}^{\infty} \frac{\left(-it\right)^{l}}{l!}H^{l}
\end{equation} 
So,  the operators product $A(0)e^{-iHt}A(0)$ can be written as
\begin{align}
A(0)e^{-iHt}A(0)&=A(0)\sum_{l=0}^{\infty} \frac{\left(-it\right)^{l}}{l!}H^{l}A(0)\notag\\
&= \sum_{l=0}^{\infty} \frac{\left(-it\right)^{l}}{l!}A(0)H^{l}A(0)\notag\\
A(0)e^{-iHt}A(0)&=\sum_{l=0}^{\infty} \frac{\left(-it\right)^{l}}{l!}A(0)\underbrace{H\cdot H\cdot\dotsm\cdot H\cdot H}_{l-\text{factors}}A(0)\notag
\end{align}
Considering that $A(0)A(0)=\left(\sigma_k^{\gamma}\right)^2=\mathcal{I}$, then
\begin{align}
A(0)e^{-iHt}A(0)&=\sum_{l=0}^{\infty} \frac{\left(-it\right)^{l}}{l!}\underbrace{A(0)H\mathcal{I} H\mathcal{I}\dotsm\mathcal{I}H\mathcal{I}HA(0)}_{l-\text{factors}}\notag\\
&=\sum_{l=0}^{\infty} \frac{\left(-it\right)^{l}}{l!}\left(A(0)HA(0)\right)^{l}\notag\\
A(0)e^{-iHt}A(0)&=e^{-it A(0)HA(0)}.
\end{align}  
 Here we restrict our calculations to spin-$1/2$ Hamiltonians written in the form
\begin{equation}
H=\sum_{\alpha}\sum_{i,j} J_{\alpha,\alpha}^{i,j}\sigma_i^{\alpha}\sigma_j^{\alpha}+\sum_{\alpha}\sum_{i}B_{\alpha}^{i}\sigma_i^{\alpha},
\end{equation}
where $\sigma_i^{\alpha}$ denotes the Pauli operators at site $i$ ($\alpha=x,y,z$), $J_{\alpha,\alpha}^{i,j}$ is the two-site coupling between sites $i$ and $j$ along direction $\alpha$, and $B_{\alpha}^{i}$ is the magnetic field at site $i$ along direction $\alpha$. Restricting to nearest-neighbor homogeneous couplings, this Hamiltonian includes the $XX$ model in a transversal magnetic field ($J_{x,x}^{i,j}=J_{y,y}^{i,j}=J$, $J_{zz}^{i,j}=0$), the anisotropic XY model ($J_{x,x}^{i,j}=1+\gamma$ and $J_{y,y}^{i,j}=1-\gamma$, $J_{zz}^{i,j}=0$, with $\gamma$ the anisotropy parameter), and the XXZ model ($J_{x,x}^{i,j}=J_{y,y}^{i,j}=J$, $J_{zz}^{ij}=J\Delta\neq0$, $B_{\alpha}^i=0$). No restriction to the dimensionality of the system is made here. Also, the couplings are considered to be time-independent. In addition, observe that the Hamiltonian can be written in the following form
\begin{equation}
H=\sum_{\alpha}\sum_{i,j}J_{\alpha,\alpha}^{i,j}\frac{\partial H}{\partial J_{\alpha,\alpha}^{i,j}}+\sum_{\alpha}\sum_{i}B_{\alpha}^i\frac{\partial H}{\partial B_{\alpha}^i},
\end{equation}
given that
\begin{equation} \label{derivatives_hamiltonian}
\frac{\partial H}{\partial J_{\alpha,\alpha}^{i,j}}=\sigma_i^{\alpha}\sigma_j^{\alpha}\quad\text{and}\quad\frac{\partial H}{\partial B_{\alpha}^i}=\sigma_i^{\alpha}.
\end{equation}
Therefore, the product of operators in the exponent can be written as		
\begin{align*}
A(0)HA(0)&=\sigma_k^{\gamma}\left(\sum_{\alpha}\sum_{i,j} J_{\alpha,\alpha}^{i,j}\sigma_i^{\alpha}\sigma_j^{\alpha}+\sum_{\alpha}\sum_{i}B_{\alpha}^{i}\sigma_i^{\alpha}\right)\sigma_k^{\gamma}\\
&=\sum_{\alpha}\sum_{i,j} J_{\alpha,\alpha}^{i,j}\sigma_k^{\gamma}\sigma_i^{\alpha}\sigma_j^{\alpha}\sigma_k^{\gamma}+\sum_{\alpha}\sum_{i}B_{\alpha}^{i}\sigma_k^{\gamma}\sigma_i^{\alpha}\sigma_k^{\gamma}.
\end{align*}		
Different cases emerge within this operation. First note that when $k\neq i,j$, the operators at site $k$ commute with those of sites $i$ and $j$, leading to an unchanged term (because $(\sigma_k^{\gamma})^2=\mathcal{I}$). Second, if $k=i$ or $k=j$ and $\alpha=\gamma$, we have products of three operators of the same kind on the same site, which result in a single operator of the same kind (i.e. $(\sigma_k^{\gamma})^3=\sigma_k^{\gamma}$)). Finally, if $k=i$ or $k=j$ but $\alpha\neq\gamma$, we have
\begin{equation}
\sigma_k^{\gamma}\sigma_k^{\alpha}\sigma_k^{\gamma}=-\sigma_k^{\alpha}.
\end{equation}	
So we have that
\begin{align*}
A(0)HA(0)=\sum_{\alpha}\sum_{i,j}\mathcal{S}^{\alpha\gamma}_{ijk}J_{\alpha,\alpha}^{i,j}\sigma_i^{\alpha}\sigma_j^{\alpha}+\sum_{\alpha}\sum_{i}\mathcal{T}^{\alpha\gamma}_{ik}B_{\alpha}^i\sigma_i^{\alpha},
\end{align*}		
where we have defined the tensors $\mathcal{S}$ and $\mathcal{T}$ by the following rules:
\begin{equation}
\mathcal{S}^{\alpha\gamma}_{ijk}=
\begin{cases}
1 & \text{if}\quad \begin{cases} 
	k=i\,\text{or}\,k=j,\,\text{and}\,\alpha=\gamma\\
	k\neq i,j
      \end{cases}\\
-1 & \text{if} \qquad k=i\,\text{or}\,k=j,\,\text{and}\,\alpha\neq\gamma\\    
\end{cases}
\end{equation}
and		
\begin{equation}
\mathcal{T}^{\alpha\gamma}_{ik}=
\begin{cases}
1 & \text{if}\quad \begin{cases} 
	k=i\,\text{and}\,\alpha=\gamma\\
	k\neq i
      \end{cases}\\
-1 & \text{if} \qquad k=i,\,\text{and}\,\alpha\neq\gamma\\    
\end{cases}
\end{equation}
Separating the cases in which $\alpha=\gamma$ and $\alpha\neq\gamma$, we obtain
\begin{align*}
A(0)HA(0)=&\sum_{i,j}J_{\gamma,\gamma}^{i,j}\sigma_i^{\gamma}\sigma_j^{\gamma} +\sum_{i}B_{\gamma}^i\sigma_i^{\gamma}+\sum_{\alpha\neq\gamma}\sum_{i,j}\mathcal{S}^{\alpha\gamma}_{ijk}J_{\alpha,\alpha}^{i,j}\sigma_i^{\alpha}\sigma_j^{\alpha}+\sum_{\alpha\neq\gamma}\sum_{i}\mathcal{T}^{\alpha\gamma}_{ik}B_{\alpha}^i\sigma_i^{\alpha}
\end{align*}
Now, developing further the second line of the previous equation, we get
\begin{align*}
A(0)HA(0)=&\sum_{i,j}J_{\gamma,\gamma}^{i,j}\sigma_i^{\gamma}\sigma_j^{\gamma} +\sum_{i}B_{\gamma}^i\sigma_i^{\gamma}+ \sum_{\alpha\neq\gamma}\sum_{i,j\neq k}J_{\alpha,\alpha}^{i,j}\sigma_i^{\alpha}\sigma_j^{\alpha}
+\sum_{\alpha\neq\gamma}\sum_{i\neq k}B_{\alpha}^i\sigma_i^{\alpha}\\
&- \sum_{\alpha\neq\gamma}\sum_{j}J_{\alpha,\alpha}^{k,j}\sigma_k^{\alpha}\sigma_j^{\alpha}-\sum_{\alpha\neq\gamma}\sum_{i}J_{\alpha,\alpha}^{i,k}\sigma_i^{\alpha}\sigma_k^{\alpha} -\sum_{\alpha\neq\gamma}B_{\alpha}^k\sigma_k^{\alpha}.
\end{align*}
Finally, adding and subtracting the negative terms of the previous equation, we obtain
\begin{align} \label{AHA}
A(0)&HA(0)=H-2\sum_{\alpha\neq\gamma}\sum_{j>k} J_{\alpha,\alpha}^{j,k}\sigma_k^{\alpha}\sigma_j^{\alpha}-2\sum_{\alpha\neq\gamma}\sum_{j<k} J_{\alpha,\alpha}^{k,j}\sigma_j^{\alpha}\sigma_k^{\alpha}-2\sum_{\alpha\neq\gamma}B_{\alpha}^k\sigma_k^{\alpha}.
\end{align}
For simplicity in further calculations, we rewrite Eq.~\eqref{AHA} in the form
\begin{align}
A(0)HA(0)&=H - f_{k}^{\gamma}.
\end{align}
where the second term explicitly depends on site $k$ where the correlations are calculates, namely 
\begin{align}
f_{k}^{\gamma}&=2\sum_{\alpha\neq\gamma}\left(\sum_{j>k}J_{\alpha,\alpha}^{k,j}\sigma_{k}^{\alpha}\sigma_{j}^{\alpha} + \sum_{j<k}J_{\alpha,\alpha}^{j,k}\sigma_{j}^{\alpha}\sigma_{k}^{\alpha}+B_{\alpha}^{k}\sigma_{k}^{\alpha}\right)\notag\\
&=2\sum_{\alpha\neq\gamma}\left(\sum_{j}J_{\alpha,\alpha}^{j,k}\sigma_{k}^{\alpha}\sigma_{j}^{\alpha} +B_{\alpha}^{k}\sigma_{k}^{\alpha}\right)
\end{align}
where in the second line of the previous equation we have used that $J_{\alpha,\alpha}^{j,k}=J_{\alpha,\alpha}^{k,j}$, and thus the sum over $j$ gives contributions from all sites $j$ coupled to site $k$. Therefore, the unequal time correlation is given by
\begin{equation} \label{correlation}
C(t)=e^{i E_{0}t}\braGS e^{-it\left(H-f_{k}^{\gamma}\right)}\ketGS.
\end{equation}
In this way, using the Hellmann-Feynman relations
\begin{equation}
\left\langle\psi_{0}\left|\frac{\partial H}{\partial\lambda}\right|\psi_{0}\right\rangle=\frac{\partial E_0}{\partial\lambda},
\end{equation}
with $\lambda$ any parameter of the Hamiltonian~\cite{griffiths}, we obtain that up to first order in the expansion of the time evolution operator, the time correlations are given by
\begin{align} \label{time_correl_final}
C(t)\approx & \cos\left(E_{0}t\right)+\sin\left(E_{0}t\right)\bigg[E_{0}t-2t\sum_{\alpha\neq\mu}\left(\sum_{j}J_{\alpha}^{j,k}\frac{\partial E_{0}}{\partial J_{\alpha}^{j,k}} +B_{\alpha}^{k}\frac{\partial E_{0}}{\partial B_{\alpha}^{k}}\right)\biggr]
\end{align}
Thus we have obtained that the \gls{STC} are proportional (apart from a structureless term $E_{0}t$)  to the first derivatives of the ground-state energy of the system, which show a discontinuity at the critical point of a first-order \gls{QPT}~\cite{wu2004prl}. This means that first-order \gls{QPT}s are directly identified by discontinuities of the \gls{STC} as a function of Hamiltonian parameters.\\
\\
Now we consider the first derivative of Eq.~\eqref{time_correl_final} with respect to some Hamiltonian parameter, e.g. $J_{\beta}^{m,n}$. We obtain
\begin{align} \label{1der_time_correl_final}
\begin{split}
\frac{\partial C(t)}{\partial J_{\beta}^{m,n}}&\approx t\cos\left(E_{0}t\right)\frac{\partial E_{0}}{\partial J_{\beta}^{m,n}}\bigg[E_{0}t-2t\sum_{\alpha\neq\mu}\left(\sum_{j}J_{\alpha}^{j,k}\frac{\partial E_{0}}{\partial J_{\alpha}^{j,k}} +B_{\alpha}^{k}\frac{\partial E_{0}}{\partial B_{\alpha}^{k}}\right)\biggr]\notag\\
&-2t\sin\left(E_{0}t\right)\sum_{\alpha\neq\mu}\Biggr[\sum_{j}\Biggr(\delta_{\alpha,\beta}\delta_{m,j}\delta_{n,k}\frac{\partial E_{0}}{\partial J_{\alpha}^{j,k}}+J_{\alpha}^{j,k}\frac{\partial^{2}E_{0}}{\partial J_{\beta}^{m,n}\partial J_{\alpha}^{j,k}}\Biggr)+B_{\alpha}^{k}\frac{\partial^{2}E_{0}}{\partial J_{\beta}^{m,n}\partial B_{\alpha}^{k}}\Biggr]
\end{split}
\end{align}
This means that the first derivative of the \gls{STC} with respect to a Hamiltonian parameter is proportional to the second derivative of the ground-state energy with respect to the same parameter,
\begin{equation} \label{1der_time_correl_final_prop}
\frac{\partial C(t)}{\partial J_{\beta}^{m,n}}\propto t\frac{\partial^2 E_0}{\partial (J^{m,n}_{\beta})^2}\quad\text{and}\quad\frac{\partial C(t)}{\partial J_{\beta}^{m,n}}\propto E_0t.
\end{equation}
As a result of the first proportionality relation of Eq.~\eqref{1der_time_correl_final_prop}, the derivatives of unequal-time correlations indicate second-order \gls{QPT}s by means of a discontinuity or divergence at the corresponding quantum critical points.

The previous results indicate that, in general, a finite-order \gls{QPT} can be identified by the properties of the \gls{STC} of the system. Namely, given that
\begin{align} \label{p_der_time_correl_final_prop}
\begin{split}
&\frac{\partial^{p-1} C(t)}{\partial (J_{\beta}^{m,n})^{p-1}}\propto t\frac{\partial^p E_0}{\partial (J^{m,n}_{\beta})^p},\\
\text{and}\quad&\frac{\partial^{p-1} C(t)}{\partial (J_{\beta}^{m,n})^{p-1}}\propto t\frac{\partial^{p-1} E_0}{\partial (J^{m,n}_{\beta})^{p-1}},\\
&\quad\quad\quad\quad\quad\quad\vdots\\
\text{and}\quad&\frac{\partial^{p-1} C(t)}{\partial (J_{\beta}^{m,n})^{p-1}}\propto E_0t,
\end{split}
\end{align}
the $(p-1)$th derivative of the \gls{STC} with respect to some Hamiltonian parameter is a function of all $q$th derivatives of the ground state energy with respect to the same parameter, for $0\leq q\leq p$. Thus a $p$th order \gls{QPT}, which corresponds to a discontinuity or divergence of the $p$th derivative of the ground-state energy, can be identified by the $(p-1)$th derivative of the \gls{STC}.\\

The validity of this result is not affected by taking the expansion of the exponential of Eq.~\eqref{correlation} to higher orders. Consider, for instance, the second-order correction $C^{(2)}(t)$, which adds the terms
\begin{align} \label{second_order}
\begin{split}
C^{(2)}(t)=&-\frac{t^2}{2}\cos(E_0t)\left(E_0^2-2E_0\braGS f_{k}^{\lambda}\ketGS + \braGS (f_{k}^{\lambda})^2\ketGS\right)
\end{split}
\end{align}
to the time correlations in Eq.~\eqref{time_correl_final}. The term $\braGS f_{k}^{\lambda}\ketGS$ has the form already displayed in Eq.~\eqref{time_correl_final}. The third component $\braGS (f_{k}^{\lambda})^2\ketGS$ results in more complicated (up to three-site) expectation values in addition to more terms $\partial E_0/\partial J_{\alpha}^{j,k}$ and $\partial E_0/\partial B_{\alpha}^k$. These elements will continue appearing in higher-order expansions, either separately or in expectation values $\braGS f_{k}^{\lambda}\ketGS$. So these expansions would lead to the observation of finite-order \gls{QPT}s as previously discussed for the first-order case. The last results shown that for Hamiltonian~\eqref{Hami_general}, the \gls{STC} of Eq.~\eqref{time_correl} and their derivatives constitute appropriate quantities to determine the location of finite-order \gls{QPT}s.  In particular we obtain that the $(p-1)$-th derivative of the \gls{STC} is a function of $E_0$ and its first $p$ derivatives, so it can be written in the form
\begin{equation} \label{p_der_time_correl_final_prop_main}
\frac{\partial^{p-1} C(t)}{\partial \mu^{p-1}} =\mathcal{F}\left(E_{0},\frac{\partial E_0}{\partial \mu},\ldots,\frac{\partial^p E_0}{\partial \mu^p},t\right),
\end{equation}
where $\mu$ can be any Hamiltonian parameter, such as $\Delta$ for the XXZ model and $\nu$ for the XY model. This means that, in general, a $p$th order \gls{QPT}, which corresponds to a discontinuity or divergence of the $p$th derivative of the ground-state energy with respect to some Hamiltonian parameter, can be identified by the $(p-1)$th derivative of the \gls{STC} with respect to the same parameter. Thus, a first-order \gls{QPT} should in principle be identified by a discontinuity of the time correlations $C(t)$, at any time $Jt>0$, as a function of the parameter driving the transition. Similarly, a second-order \gls{QPT} should be recognized by a discontinuity or divergence of the first derivative of the time correlations with respect to the driving parameter. Note that this result is similar to the observation of finite-order \gls{QPT}s by measures of bipartite entanglement~\cite{wu2004prl}. Here, however, we are able to determine transitions by looking at a purely local (single-site) quantity.\\

To provide stronger evidence that this is in fact the case, we calculate the time correlations for the XXZ and anisotropic XY models, and examine their behavior at the corresponding quantum critical points. Even though both models are exactly solvable, obtaining their physical properties is a very challenging task. For example, for zero temperature exact time correlations are only known for $\Delta=0$ in the XXZ model (equivalent to the limit $\gamma=0$ and $B_z=0$ in the XY model) and for $A(0)=\sigma^z$~\cite{katsura1970physica}. Calculations based on a mean-field approach fail to reproduce the time correlations correctly. Furthermore, exact diagonalization methods are restricted to small lattices. Thus to obtain quantitatively-correct results for much longer systems we perform numerical simulations based on tensor-network algorithms. Namely, we first obtain the ground state of both models for several parameters by means of the density matrix renormalization group algorithm~\cite{white1992prl}, using a matrix product state description~\cite{schollwock2011ann}. Subsequently we simulate the time evolution described in Eq.~\eqref{time_correl} by means of the time evolving block decimation~\cite{vidal2004prl}. These methods allow us to carry out our simulations efficiently, for lattices of several sites. In particular, we consider systems of $N=100$ spins (unless stated otherwise) with open boundary conditions, described by matrix product states with bond dimensions of up to $\chi=400$. Our implementation of the algorithms is based on the open-source Tensor Network Theory (TNT) library~\cite{tnt}.

\subsection{Fingerprint of First-order \gls{QPT}}

\begin{figure}[h!]
\begin{center}
\includegraphics[scale=0.43]{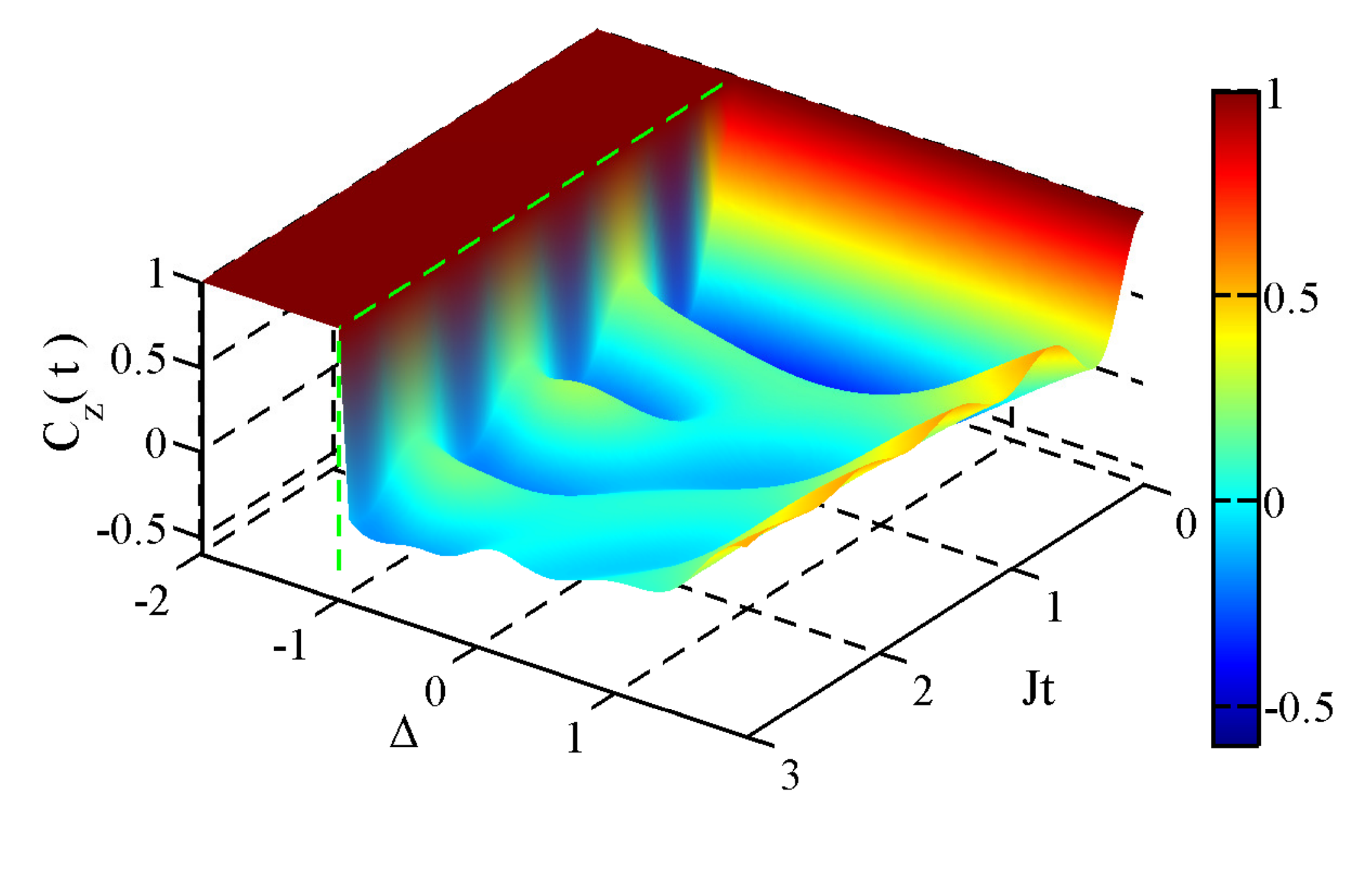}
\includegraphics[scale=1]{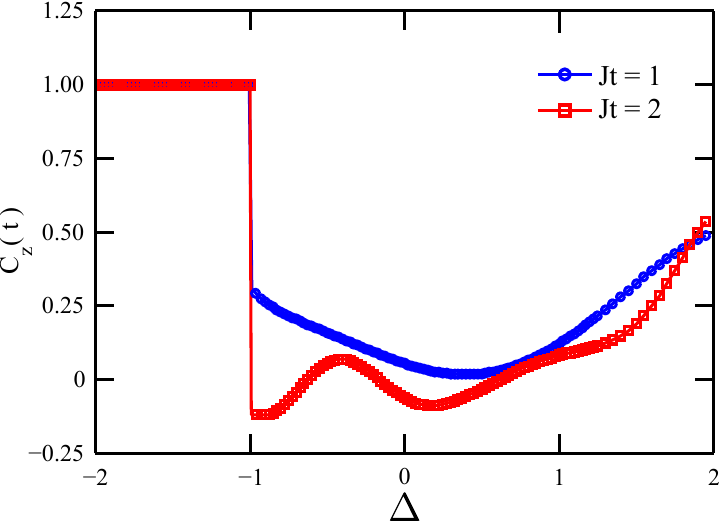}
\caption[Single two-time correlation along $z$ direction of the XXZ model.]{\label{correl_z_xxz_2D} {\it Left panel:} \gls{STC} along $z$ direction of the XXZ model, as a function $Jt$  and anisotropy parameter $\Delta$. The dashed green lines indicate the critical point $\Delta=-1$. {\it Right panel:} \gls{STC} along $z$ direction of the XXZ model, as a function of $\Delta$, for  $Jt=1$ and $Jt=2$.}
\end{center}
\end{figure}

We start by observing the \gls{STC} for the XXZ model, and focus on the transition between ferromagnetic and gapless states at $\Delta=-1$. All the results to be presented are in a time scale from $0$ to $3$ in units of $1/J$. Since recent experiments on non-equilibrium spin models implemented in ultracold-atom quantum simulators have been performed for similar ($J/\hbar=2\pi\times 8.6Hz$)~\cite{Fukuhara}  and even longer time scales ($J/\hbar=14.1Hz$)~\cite{Hildprl}, the effects we will present are in a time scale perfectly observable with current technology. In the left panel of Fig.~\ref{correl_z_xxz_2D} we show the correlations $C_z(t)$ (i.e. for $A(0)=\sigma_k^z$), evaluated at site $k=50$, as a function of $\Delta$ and $t$. Additionally, in the right panel of Fig.~\ref{correl_z_xxz_2D} we plot the corresponding correlations at times $Jt=1$ and $Jt=2$ as a function of $\Delta$.\\

First, note that since the ground state of the system is ferromagnetic for $\Delta<-1$, so $\sigma_k^z\ketGS=\pm\ketGS$ in this regime, with the sign depending on the direction of polarization along the $z$ axis. Furthermore, this state remains unchanged under magnetization-conserving time evolution, such as that of the XXZ model. Thus the time correlations remain constant, with value $C_z(t)=1$. For $\Delta>-1$ this is no longer the case. Since in this regime the states $\sigma_k^z\ketGS$ are not fully polarized, they are strongly affected by time evolution. More importantly, when the system crosses the quantum critical point $\Delta=-1$ and enters the gapless state, the correlations exhibit a discontinuous jump to values $C_z(t)<1$ at any finite time $Jt>0$, as depicted in Fig.~\ref{correl_z_xxz_2D}.\\

\begin{figure}[h!]
\begin{center}
\includegraphics[scale=0.5]{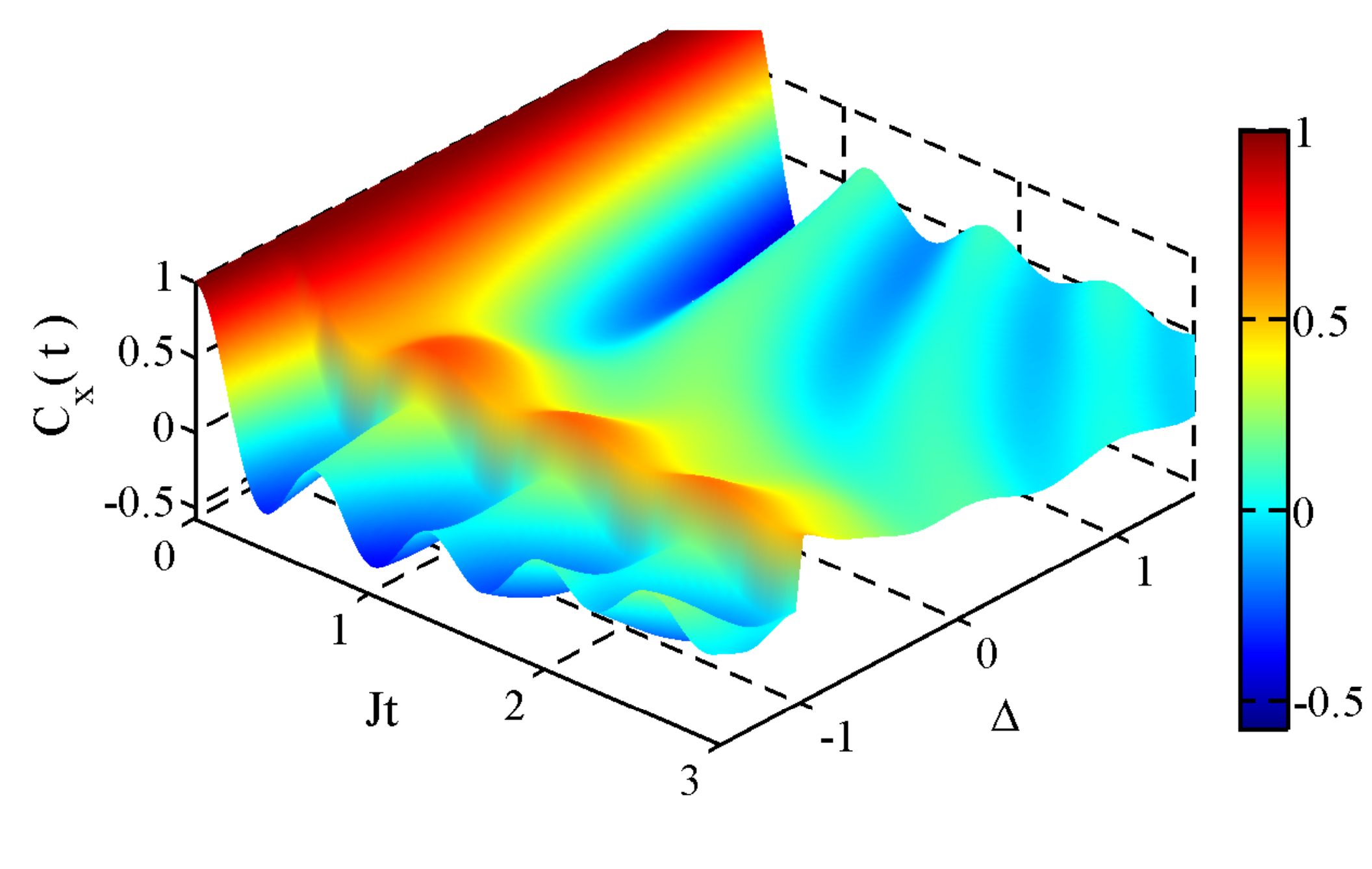}
\includegraphics[scale=1]{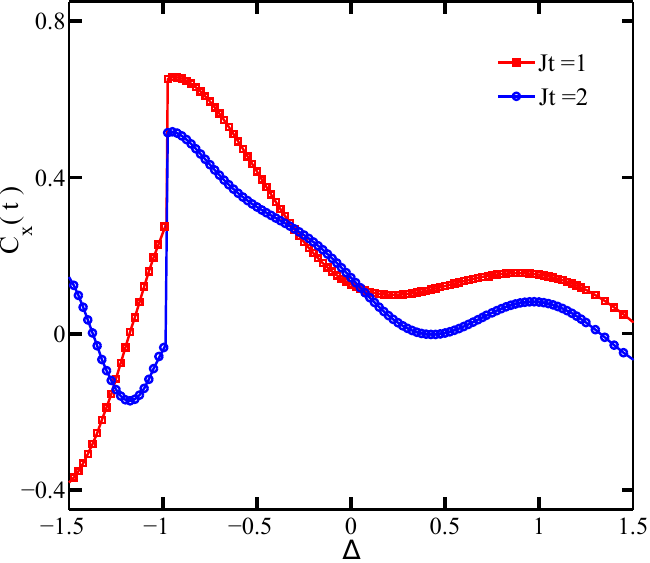}
\caption[Single two-time correlation along $x$ direction of the XXZ model.]{\label{correl_x_xxz_2D} {\it Left panel:} \gls{STC} along $x$ direction of the XXZ model, as a function of  $Jt$ and anisotropy parameter $\Delta$. {\it Right panel:} \gls{STC} along $x$ direction of the XXZ model, as a function of $\Delta$, for $Jt=1$ and $Jt=2$.}
\end{center}
\end{figure}

A similar result is obtained when calculating the \gls{STC} $C_x(t)$, i.e. with $A(0)=\sigma_k^x$, which give identical results to the correlations along $y$ direction due to the symmetry of the Hamiltonian~\eqref{hami_xxz}. These are shown in the left panel of Fig.~\ref{correl_x_xxz_2D} as a function of $\Delta$ and $Jt$, and in the right panel of Fig.~\ref{correl_x_xxz_2D} for two specific times, namely $Jt=1$ and $Jt=2$. In contrast to $C_z(t)$, the correlations along $x$ direction do not remain constant in the ferromagnetic regime $\Delta<-1$, since $\sigma_k^x$ flips the spin at site $k$ and thus induces dynamics on the system. However, the correlations $C_x(t)$ also show a discontinuity at $\Delta=-1$. Thus as expected from Eq.~\eqref{p_der_time_correl_final_prop_main}, the different \gls{STC} indicate the first-order \gls{QPT} of the XXZ model by means of a discontinuity as a function of $\Delta$ at the quantum critical point.\\

Note that neither $C_z(t)$ nor $C_x(t)$, or any of their derivatives, indicate the existence of the Kosterlitz-Thouless QPT at $\Delta=1$,  given that it is of infinite order. In this way, we conclude that the observation that a first-order quantum phase transition can be directly identified by means of a discontinuity of the single-site two-time correlations $C(t)$ at any time $t$. Moreover, our density matrix renormalization group (\gls{DMRG}) + time-evolving block decimation (\gls{TEBD}) simulations correspond to the first calculation of this kind, given the difficulty to correctly obtain these results from analytical calculations.\\

In the next subsection we discusses how finite-order \gls{QPT}s can be identified from local time correlations, providing examples for the spin models previously described.

\newpage
\subsection{Fingerprint of Second-order \gls{QPT}}
Now we consider the transition between ferromagnetic and paramagnetic phases of the anisotropic XY model. In particular, we illustrate the transition for two cases, namely the limit $\gamma=1$, which corresponds to the Ising model with a transverse magnetic field, and the intermediate case $\gamma=0.5$.  We verified that the \gls{STC} along any direction $ \alpha = x, y, z $ give the same qualitative information regarding the \gls{QPT}, which is in agreement with Eq.~\eqref{p_der_time_correl_final_prop_main}, so we focus on $ C_z (t) $ and do not show the results of the other directions.
\begin{figure}[h!]
\begin{minipage}{0.5\textwidth} 
\centering \includegraphics[scale=0.39]{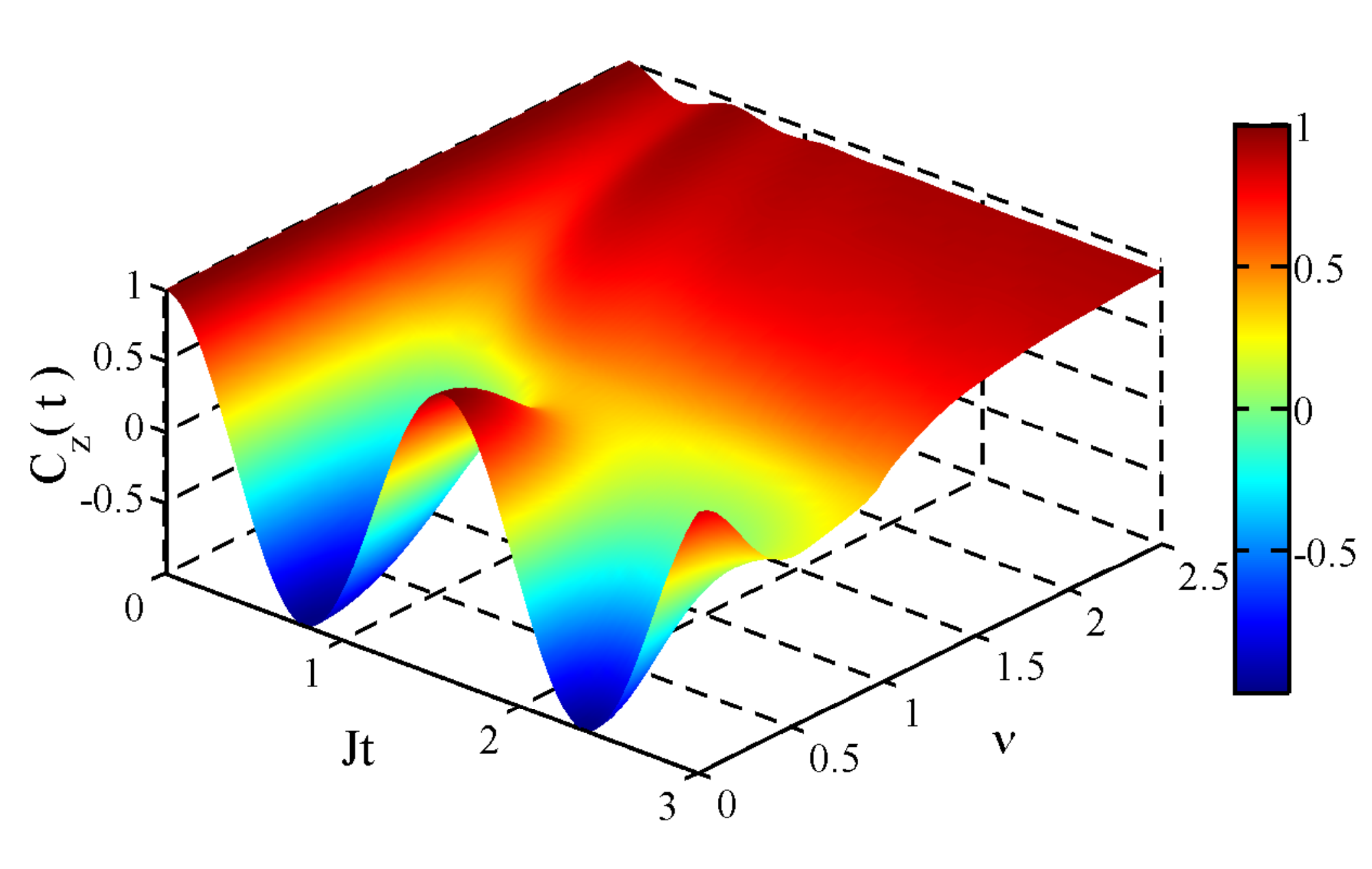}
		\caption[Single two-time correlation along $z$ direction of the Ising model.]{\label{correl_z_xy_3D} \gls{STC} along $z$ direction of the anisotropic XY model for $\gamma=1$ (Ising model), as a function of  $Jt$ and $\nu$.}
\end{minipage}
\hfill
\begin{minipage}{0.43\textwidth}
In Fig.~\ref{correl_z_xy_3D} we show the $z$ time correlations as a function of $Jt$ and $\nu$, for $\gamma=1$; the results for $\gamma=0.5$ are qualitatively similar. In addition, we depict in the left panel of Fig.~\ref{correl_z_xy_2D} the correlations for times $Jt=1$ and $Jt=2$ as a function of the magnetic field, for both $\gamma=1$ and $\gamma=0.5$. In contrast to the XXZ case, here the correlations are continuous for the whole range of values of $\nu$ considered. However, the first derivative with respect to $\nu$ is not a well-behaved function. As exemplified in the lower panel of Fig.~\ref{correl_z_xy_2D} for two particular times, $\frac{dC_z(t)}{d\nu}$ shows a sharp maximum at the quantum critical point $\nu=1$.					
\end{minipage}				
\end{figure}

\begin{figure}[h!]
\begin{center}
\includegraphics[scale=0.4]{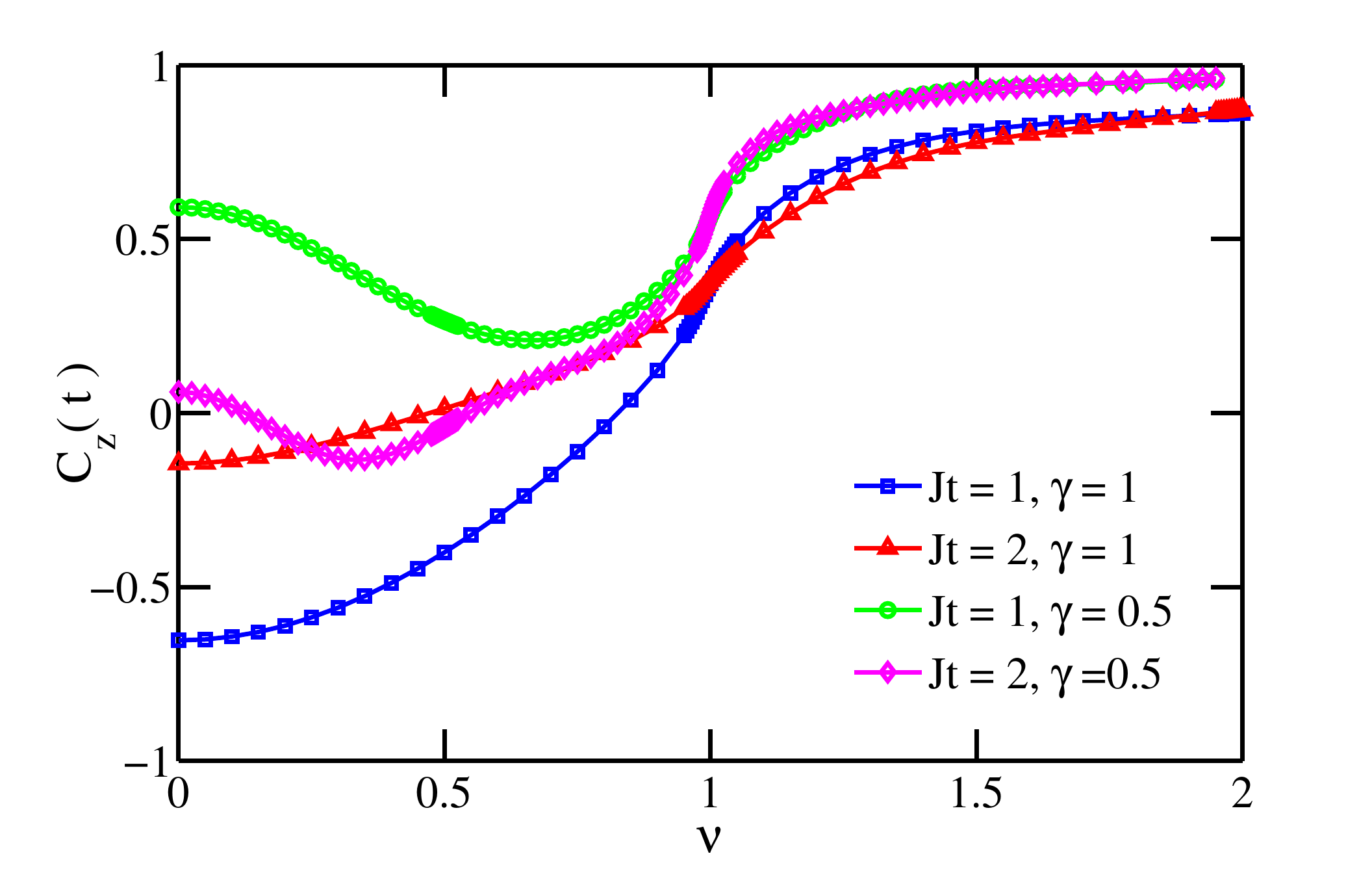}
\includegraphics[scale=0.4]{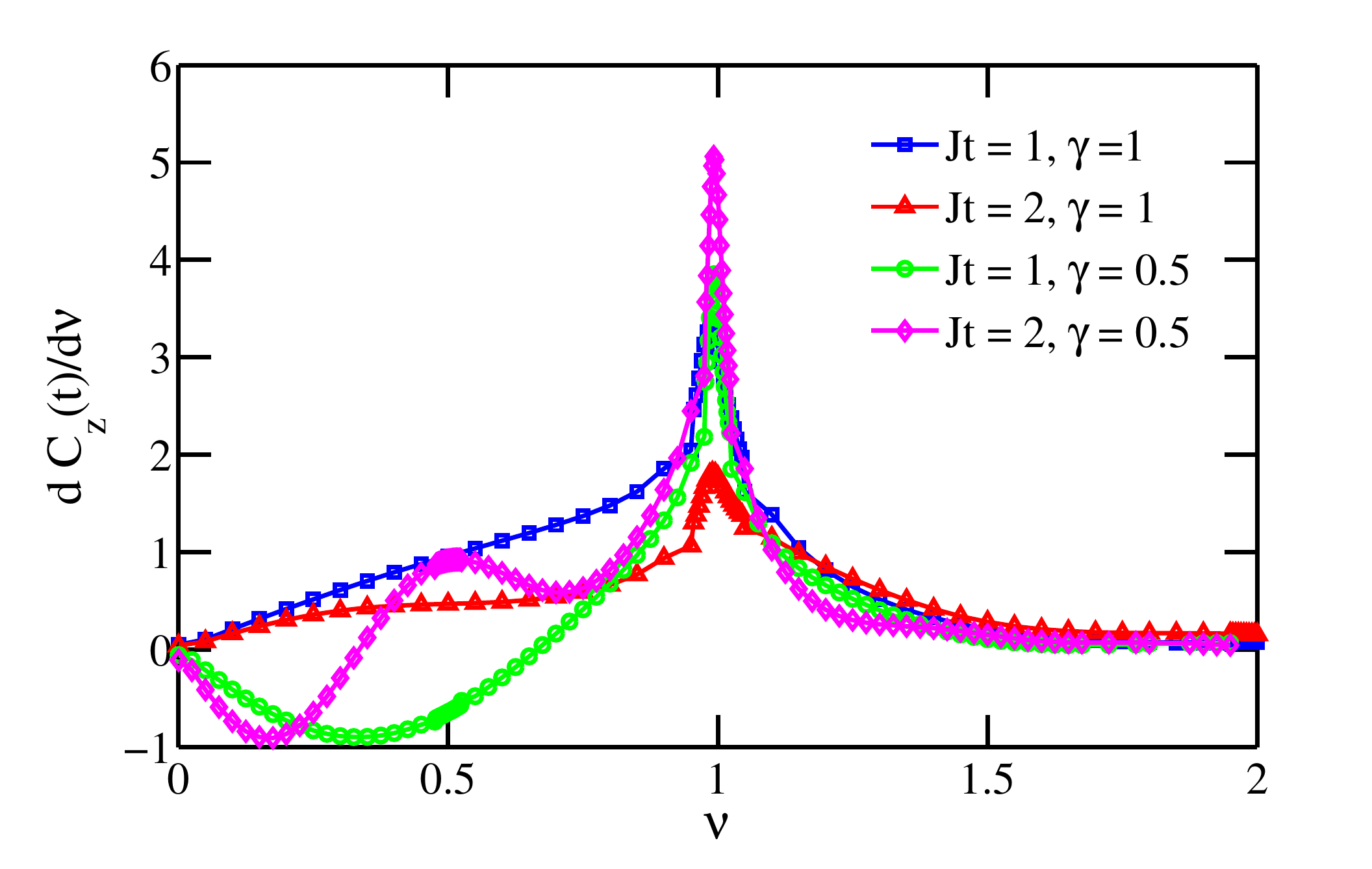}
\caption[Single two-time correlation along $z$ direction of the Ising model, for  $Jt=1$ and $Jt=2$.]{\label{correl_z_xy_2D}  \gls{STC} ({\it left panel}) and first derivative ({\it right panel}) along $z$ direction of the anisotropic XY model ($\gamma=0.5,1$), as a function of $\nu$, for  $Jt=1$ and $Jt=2$.}
\end{center}
\end{figure}
Thus, in accordance with Eq.~\eqref{p_der_time_correl_final_prop_main}, the second-order QPT of the model can be identified by means of a singularity in the first derivative of the local time correlations with respect to the Hamiltonian parameter which drives the transition, which in this case is $\nu$.\\

\section{Leggett-Garg inequalities} \label{sect_lgi}

Since the birth of quantum mechanics, its non-deterministic nature and nonlocal structure have motivated many theoretical debates that have recently moved to the experimental field. In particular, Bell inequalities establish a natural border to the spatial quantum correlations in separate systems. Leggett and Garg~\cite{lgi_original} in 1985 showed that the temporal correlations obey similar inequalities.\\

In our intuitive view of the world, probabilities are due to our uncertainty about the state of a system, but they are not a fundamental description of it. For example, when we toss a coin to the air, it has probability one half of landing tails or heads. We also assume that if we had the precise knowledge of its position and momentum, and enough computational power, we would be able to determine on which side the coin will land. We do not think that the coin is in a superposition of states, such a Schr\"odinger's cat. This is known as macroscopic realism. In addition, we assume that making measurements on a system does not modify its present state, in the way projective quantum measurements do. This is referred to as non-invasive measurability. Based on these two principles Leggett and Garg obtained a set of inequalities, which are consistent with the macroscopic intuition. One form these \gls{LGI} can take is

\begin{equation}\label{LGI}
C\left(t_{1},t_{3}\right)-C\left(t_{1},t_{2}\right)-C\left(t_{2},t_{3}\right)\geq -1
\end{equation}
where $C\left(t_{i},t_{j}\right)=\frac{1}{2}\langle \lbrace Q(t_{i}),Q(t_{j})\rbrace\rangle$ is the two-time correlation of a dichotomic observable	$Q$ (with eigenvalues $q=\pm1$) between times $t_i$ and  $t_j$, and $t_1 < t_2 < t_3$. On the other hand, if the correlation functions $C\left(t_{i},t_{j}\right)$ are stationary, i.e. they only depend on the time difference $\tau=t_i -t_j$, then the Leggett-Garg inequality~\eqref{LGI} can be written as~\cite{huelga1995pra}
\begin{equation}\label{eq9}
K_{-}\left(\tau\right) \equiv C\left(2\tau \right) - 2C\left(\tau\right)\geq-1,
\end{equation}
which defines the Leggett-Garg functions $K_-(\tau)$ for time $\tau$. Just as with Bell inequalities, any system that violates inequality~\eqref{eq9} shows some behavior that is essentially nonclassical. This is why violations of \gls{LGI} are used as a measure of quantumness~\cite{HuelgaPRL2015}. In the following we discuss different Leggett-Garg functions $K_{-}^{\alpha}(t)$, corresponding to measurements of spin components along $\alpha$ direction, and see whether they can give information about the \gls{QPT}s previously discussed.

\subsection{Leggett-Garg Inequalities and finite-order \gls{QPT}s}
\begin{figure}[h!]
\begin{center}
\includegraphics[scale=0.8]{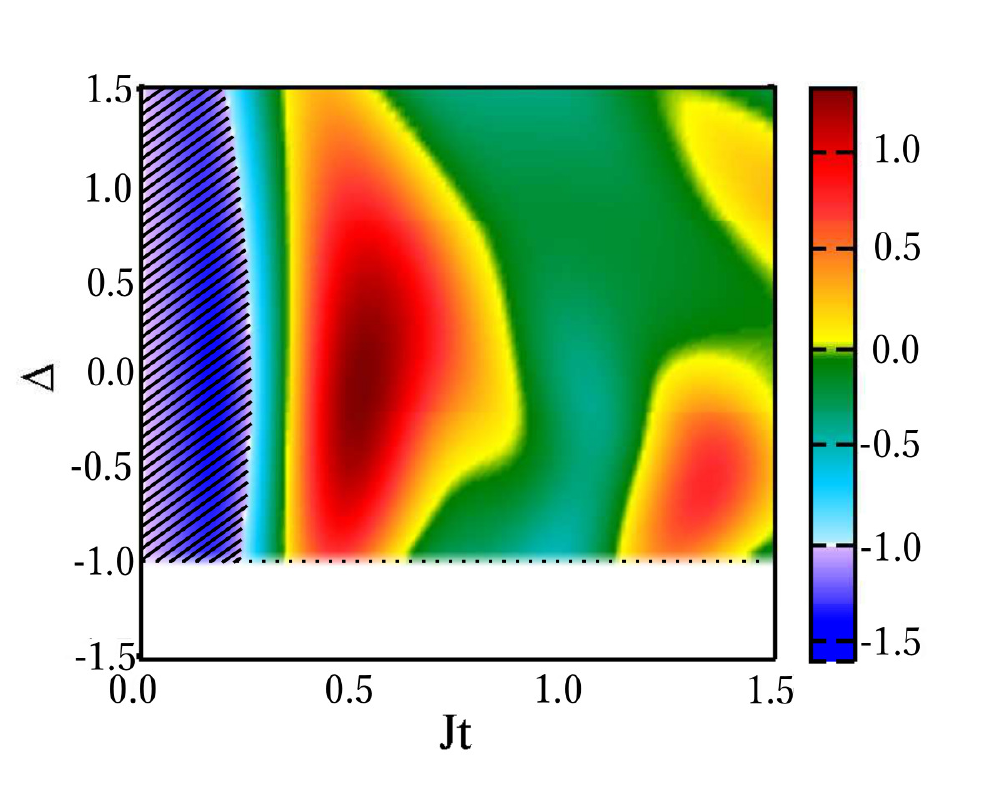}
\includegraphics[scale=0.77]{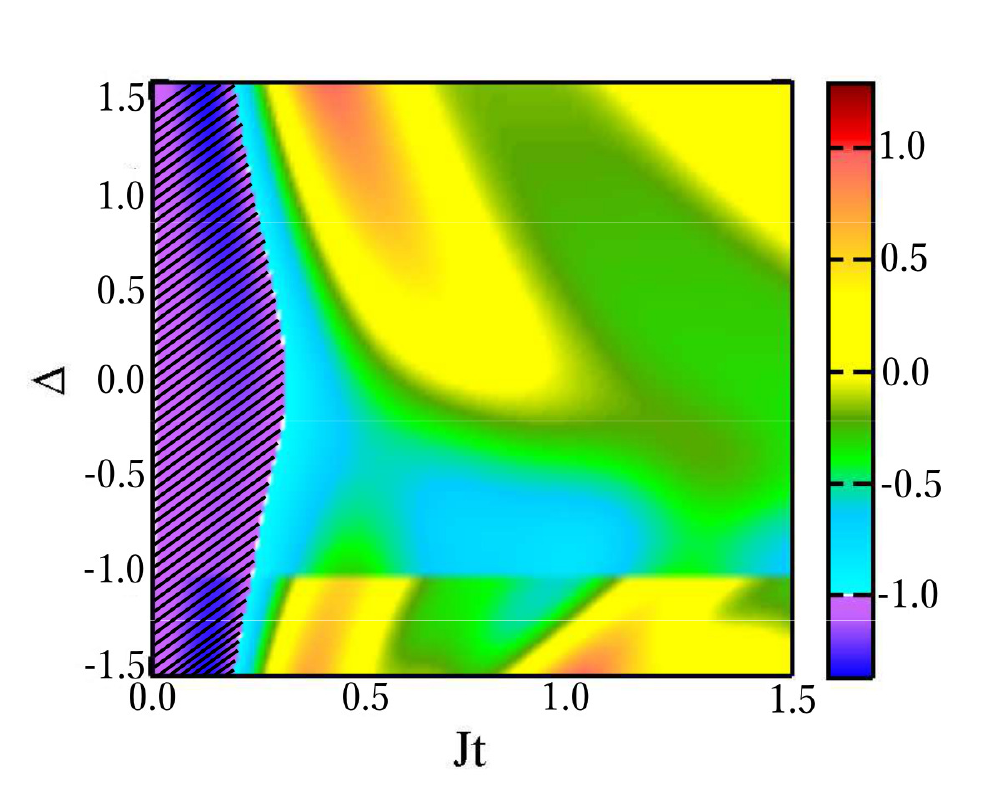}
\caption[Leggett-Garg functions for the XXZ model.] {\label{lgi_z_and_x_xxz_3D} Leggett-Garg functions for the XXZ model. {\it Left panel:} $K_{-}^{z}(t)$. The regions with diagonal lines corresponds to the regime of anisotropies $\Delta$ and  $Jt$ in which the $x$ Leggett-Garg inequalities are violated. The white region below $\Delta=-1$ indicates that there the time correlations remain constant. {\it Right panel}: $K_{-}^{x}(t)$. The  regions with diagonal lines region corresponds to the regime of anisotropies $\Delta$ and  $Jt$ in which the $x$ Leggett-Garg inequalities are violated.}
\end{center}
\end{figure}

We start by showing how the Leggett-Garg functions $K_{-}^{\alpha}(t)$ signal the finite-order QPTs discussed in Section~\ref{section_time_correl}. In Fig.~\ref{lgi_z_and_x_xxz_3D} we depict both $K_{-}^{z}(t)$ (left panel) and $K_{-}^{x}(t)$ (right panel) for the $XXZ$ model as a function of $\Delta$  and time. Regarding the results along $\alpha=z$ direction, we first note that for $\Delta<-1$ the value of the Leggett-Garg function remains equal to $K_{-}^{z}(t)=-1$ for any time. Thus in the ferromagnetic phase the corresponding LGIs are never violated. This is clearly a direct consequence of the constant value of the time correlations previously discussed (see Fig.~\ref{correl_z_xxz_2D}), and manifests the classical nature of the ferromagnetic state when undisturbed. The situation is entirely different for $\Delta>-1$. Not only $K_{-}^{z}(t)$ varies on time, but indicates a violation of the Leggett-Garg inequalities for early times. As to the results along $\alpha=x$ direction, all the values of $\Delta$ considered show a violation of the inequalities for early times. In addition, the violation lasts longer as $|\Delta|$ decreases.
\begin{figure}[h!]
\begin{minipage}{0.5\textwidth} 
\centering \includegraphics[scale=1.1]{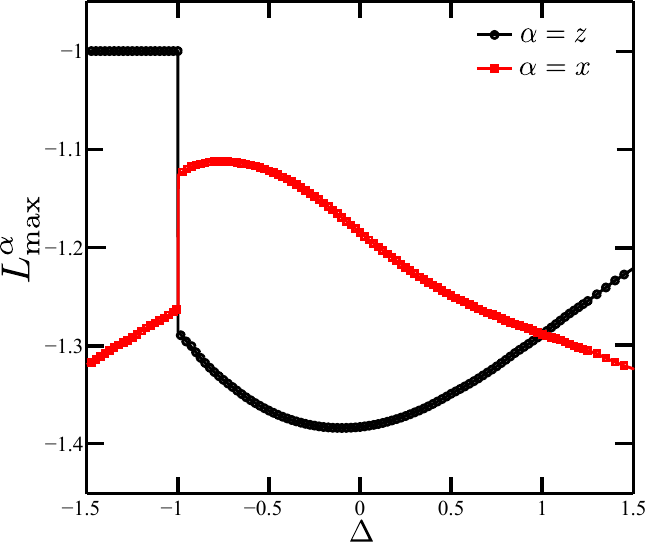}
		\caption[Maximum violation of Leggett-Garg inequalities for the XXZ model.]{\label{lgi_max_xxz_x_z} Maximum violation of Leggett-Garg inequalities for the XXZ model, along both $z$ and $x$ directions.}
\end{minipage}
\hfill
\begin{minipage}{0.43\textwidth}
In Fig.~\ref{lgi_max_xxz_x_z} we plot the maximum value $L_{\text{max}}^{\alpha}$ we obtain for the violation of the Leggett-Garg inequalities as defined by
\begin{equation}
L_{\text{max}}^{\alpha}=\min_{t}K_{-}^{\alpha}(t),
\end{equation}
for both directions $\alpha=z,x$ as a function of $\Delta$. This clearly shows that similarly to time correlations, the first-order \gls{QPT} of the XXZ model can be identified by a discontinuity of the $L_{\text{max}}^{\alpha}$ function at the critical point. Also, the maximal violation occurs along $z$ direction, close to the noninteracting limit $\Delta=0$. In contrast, for the magnetically-ordered phases, the maximal violation occurs along $x$ direction.
\end{minipage}				
\end{figure}

\begin{figure}[h!]
\begin{center}
\includegraphics[scale=0.79]{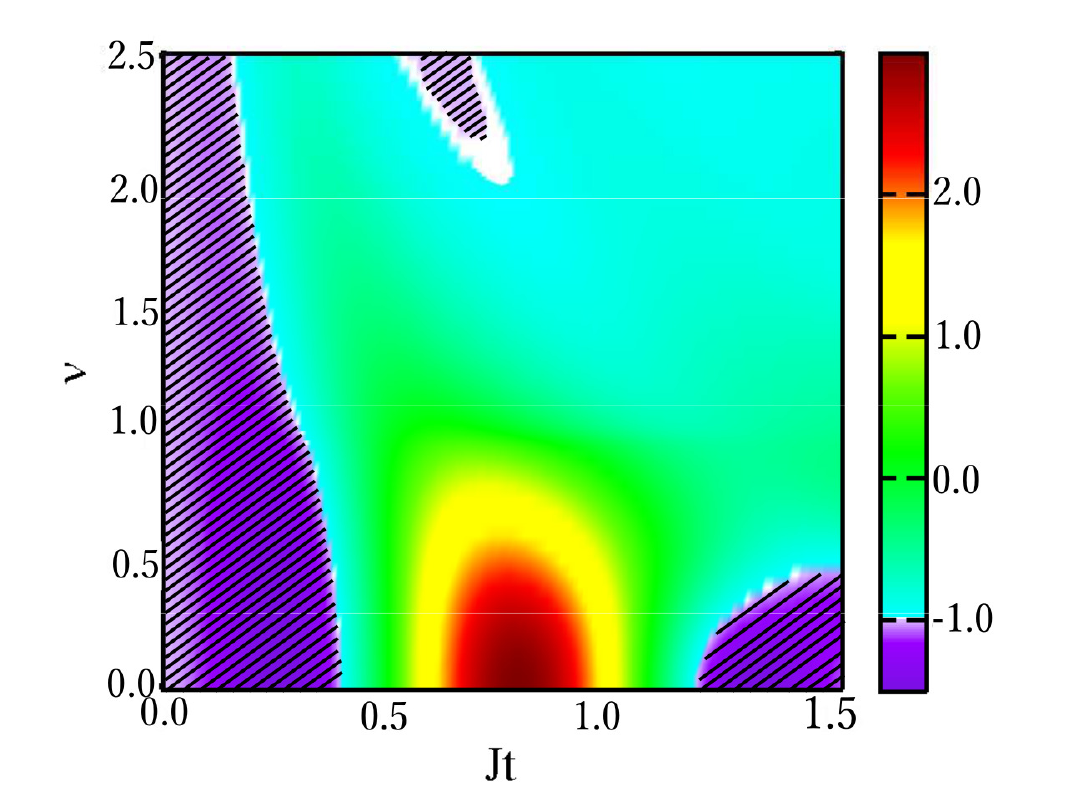}
\includegraphics[scale=0.79]{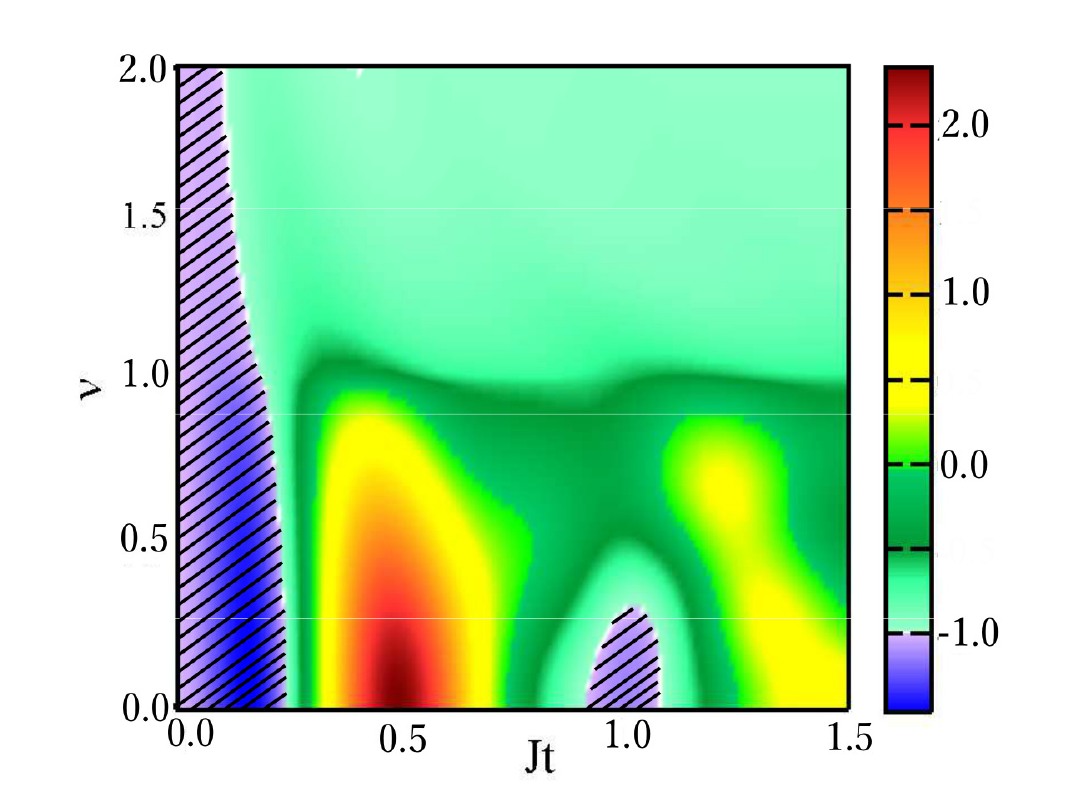}
\caption[Leggett-Garg functions $K_{-}^{z}(t)$ for the anisotropic XY model.]{\label{lgi_z_gamma_1_and_05_xy_3D} Leggett-Garg functions $K_{-}^{z}(t)$ for the anisotropic XY model. {\it Left panel}: $\gamma=1$. {\it Right panel}: $\gamma=0.5$. The regions with diagonal lines corresponds to the regime of parameter  $\nu$ and  $Jt$ in which the $z$ Leggett-Garg inequalities are violated.}
\end{center}
\end{figure}

The Leggett-Garg functions also help determine the second-order QPT of the anisotropic XY model. In Fig.~\ref{lgi_z_gamma_1_and_05_xy_3D} we show $K_{-}^{z}(t)$ for several values of $\nu$ as a function of time, for $\gamma=1$ (left panel) and $\gamma=0.5$ (right panel). Notably, for all the values of $\nu$ considered, the system features violation of the inequalities. Initially, for $Jt<0.5$, the violation of the inequalities lasts longer as $\nu$ decreases. And interestingly, for longer times, revivals of the violations are seen for low values of $\nu$. Thus weak magnetic fields favor the observation of the violation of the Leggett-Garg inequalities along $z$ direction. \\

\begin{figure}[h!]
\begin{center}
\includegraphics[scale=1.4]{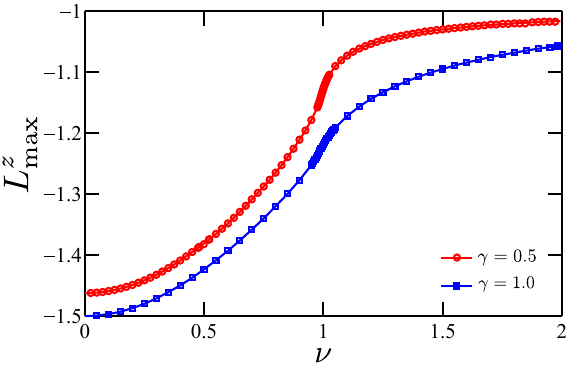}
\includegraphics[scale=1.4]{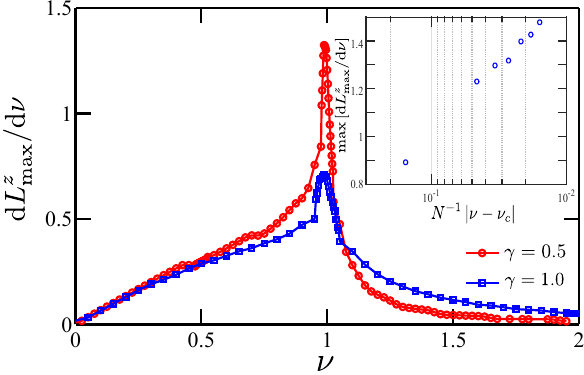}
\caption[Maximum violation of Leggett-Garg inequalities for the anisotropic XY model.]{\label{lgi_z_max_xy} {\it Left panel}: Maximum violation of Leggett-Garg inequalities for the anisotropic XY model ($\gamma=1,0.5$) along $z$ direction, as a function of $\nu$. {\it Right panel}: First derivative with respect to $\nu$.  Inset:  scaling of  maximum of the derivative with respect to $\nu$} 
\end{center}
\end{figure}

Just as the time correlations, the Leggett-Garg functions $K_{-}^{z}(t)$ and the maximal violation functions $L_{\text{max}}^{z}$ (see upper panel of Fig.~\ref{lgi_z_max_xy}) are continuous in the whole parameter regime. However, their first derivative tends to diverge at the quantum critical point $\nu=1$ as the size of the system increases (see inset lower panel of Fig.~\ref{lgi_z_max_xy}). This is shown in the lower panel of Fig.~\ref{lgi_z_max_xy} for $L_{\text{max}}^{z}$, and both $\gamma=1$ and $\gamma=0.5$. As expected, the behaviour of the time correlations is translated to the Leggett-Garg functions, and they are able to signal the second-order QPT of the anisotropic XY model by means of a singularity in their first derivative.\\

\begin{figure}[h!]
\begin{center}
\includegraphics[scale=1.1]{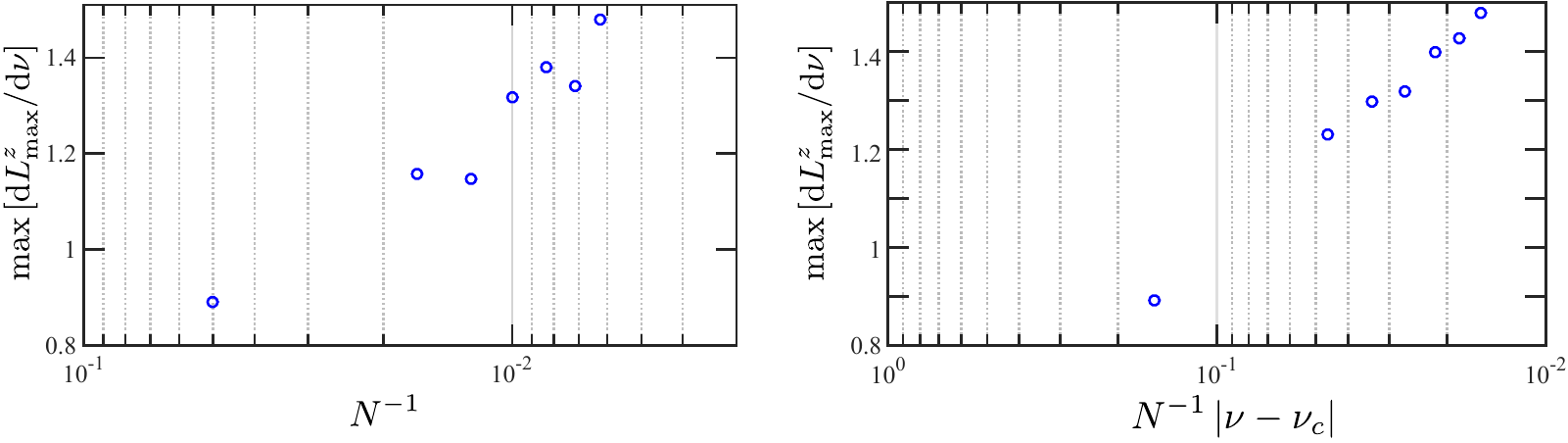}
\caption[Scaling maximum violation of Leggett-Garg inequalities for the anisotropic XY model.]{\label{Scaling} Left panel. Maximum value of the derivative function  $L_{\max}^{z}$ versus the inverse of the size of the system $N$. Right panel. Maximum value of the derivative of the function $L_{\max}^{z}$ versus the difference of theoretical value $\nu_{c}$ for the CP and numerical value $\nu$, for the anisotropic XY model with $\gamma = 0.5.$}
\end{center}
\end{figure}

For each of the models considered in the present work, we perform an analysis of convergence for the results, both of the bond link used in the MPS processes $\chi$ and system size $N$. For the Ising and anisotropic XY model as the critical point (CP) is tracked with  discontinuity of the first derivative of the function violation maximum $L_{\max}^{\alpha}$ respect to the parameter  control $\nu$, we performed a fine sweep near the CP for different sizes showing that the point of discontinuity of the derivative diverges as the size increases, see Fig.~\ref{Scaling} of this resubmission letter. Finally we chose a size of $N = 100$ for the results presented, since the difference of the known theoretical value $\nu_{c}$ for the CP with the value found numerically disagree on the order of $10^{-2}$.

\subsection{Infinite-order \gls{QPT} of the XXZ model} \label{lgi_kt}
We have observed that finite-order QPTs can in principle be determined by means of a singular behaviour of local unequal-time correlations and Leggett-Garg functions, or of their derivatives. However, this form is not suitable to identify infinite-order transitions. In fact, the results shown so far do not feature any singular property at the quantum critical point $\Delta=1$ of the Kosterlitz-Thouless transition of the one-dimensional XXZ model. However, it is possible to locate this transition from Leggett-Garg functions, as we discuss in the present Section. This is very similar to the observation of the transition from Bell inequalities~\cite{justino2012pra}, with the notable difference that here we actually have violation of the respective inequalities, and thus we can perceive the quantumness of the system.\\

\begin{figure}[h!]
\begin{minipage}{0.5\textwidth} 
\centering \includegraphics[scale=1]{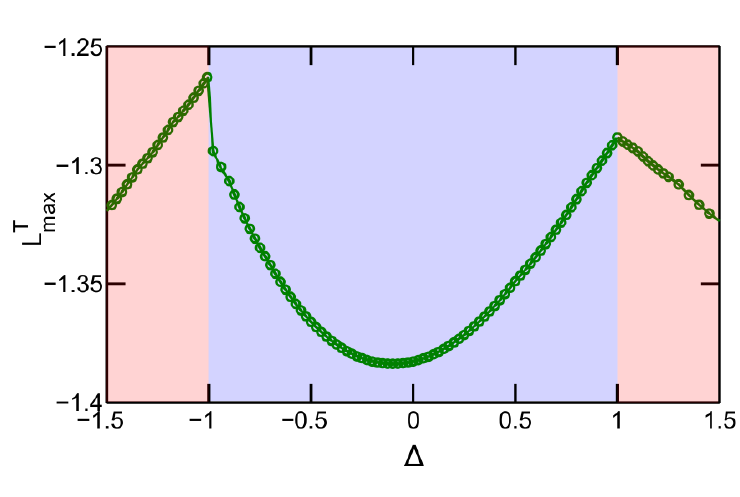}
		\caption[Maximizes over all times and directions the violation of the Leggett-Garg inequalities for the XXZ model.]{\label{lgi_max_xxz_tot} Total maximum violation of Leggett-Garg inequalities for the XXZ model as a function of $\Delta$.}
\end{minipage}
\hfill
\begin{minipage}{0.43\textwidth}
In the Fig~\ref{lgi_max_xxz_tot}, the light-red zones indicate the regimes in which the maximal violations comes from inequalities along $x$ direction, while the light-blue zone in between shows the regime in which the maximal violations occur along $z$ direction; see Fig.~\ref{lgi_max_xxz_x_z}.
\end{minipage}				
\end{figure}

The first point to note is that to actually establish that a violation of the inequalities exists, and also when the maximal violation occurs, we must consider all the possible directions $\alpha$ of evaluation of time correlations. For the XXZ model, this corresponds to $\alpha=z$ and $\alpha=x$, the latter giving the same results than for $\alpha=y$ due to the symmetry of the Hamiltonian. For this we define the function
\begin{equation}
L_{\text{max}}^{\text{T}}=\max_{\alpha=z,x}L_{\text{max}}^{\alpha},
\end{equation}
which maximizes over all times and directions the violation of the Leggett-Garg inequalities. We show $L_{\text{max}}^{\text{T}}$ as a function of $\Delta$ in Fig.~\ref{lgi_max_xxz_tot}; note that it indicates the first-order QPT at $\Delta=-1$ by means of a discontinuity.

The second point to note, responsible for the observation of the Kosterlitz-Thouless transition by means of Bell inequalities, is that at the isotropic point of the Hamiltonian there is a change in the largest type of spatial correlation~\cite{justino2012pra}. Namely, for $|\Delta|>1$ and spins separated by $r$ lattice sites, $|\langle\sigma_i^x\sigma_{i+r}^x\rangle|\leq|\langle\sigma_i^z\sigma_{i+r}^z\rangle|$, while for $-1<\Delta<1$ we have that $|\langle\sigma_i^x\sigma_{i+r}^x\rangle|>|\langle\sigma_i^z\sigma_{i+r}^z\rangle|$. This behaviour is translated to the local time correlations and related functions. In fact, as seen in Fig.~\ref{lgi_max_xxz_x_z}, $L_{\text{max}}^{x}<L_{\text{max}}^{z}$ for the ordered phases, while $L_{\text{max}}^{x}>L_{\text{max}}^{z}$ for the gapless regime. As shown in Fig.~\ref{lgi_max_xxz_tot}, a sharp local maximum appears at $\Delta=1$ in the $L_{\text{max}}^{\text{T}}$ function, and a singularity of its first derivative results. Thus, by means of a function characterizing the total maximal violation of Leggett-Garg inequalities, we are able to locate the infinite-order Kosterlitz-Thouless transition of the XXZ model. It would be interesting to observe whether Kosterlitz-Thouless transitions for other quantum systems can be identified in this form.
\section{Conclusions}
We have discussed whether single-site time correlations and Leggett-Garg inequalities allow the identification of \gls{QPT}s in many-body quantum systems. By means of efficient matrix product simulations and analytical arguments, we have answered this question in the affirmative for different spin-$1/2$ models, for both finite- and infinite-order \gls{QPT}s. Thus we have shown that \gls{QPT}s can be detected by purely-local measurements.\\

Initially, by means of a first-order approximation for a general spin-$1/2$ Hamiltonian, we argued that a $p$th order QPT can be located by a singular behaviour of the $(p-1)$th derivative of the local time correlations at the quantum critical point. Thus, these correlations indicate quantum criticality in a form similar to different measures of bipartite entanglement~\cite{wu2004prl}. Furthermore, this behaviour is directly transferred to the corresponding Leggett-Garg functions.\\

To support this general result, we calculated several time correlations for large one-dimensional XXZ and anisotropic XY spin systems, using the density matrix renormalization group and time evolving block decimation methods. In particular, we showed that the first-order ferromagnetic-gapless \gls{QPT} of the XXZ model is manifested as a discontinuity of the correlations at $\Delta=-1$, along any possible direction and for any finite time. Subsequently we showed that the second-order paramagnetic-ferromagnetic \gls{QPT} of the anisotropic XY model is observed by means of a divergence of the first derivative of the correlations with respect to the magnetic field at $\nu=1$.\\

We also showed that the Leggett-Garg functions can help identify finite-order \gls{QPT}s in a similar fashion. More importantly, we found that at least for one direction, the Leggett-Garg inequalities are violated for early times and the whole regime of parameters considered, in contrast to Bell inequalities~\cite{justino2012pra}. Furthermore, the maximization of this violation allowed us to identify the infinite-order Kosterlitz-Thouless \gls{QPT} of the XXZ model at $\Delta=1$, which was not possible from the separate observation of time correlations along each direction. Given the large amount of materials described by the testbed models discussed in our work~\cite{aeppli,blundell_magnetism}, and the seminal advances on their implementation in quantum simulators~\cite{georgescu2014rmp}, we expect that our results extend the range of systems in which the violation of Leggett-Garg inequalities can be observed experimentally~\cite{emary}.
\begin{savequote}[45mm]
``I learned very early the difference between knowing the name of something and knowing something"
\qauthor{Richard P. Feynman}
\end{savequote}
\chapter[\gls{STC} and \gls{LGI} by Majorana Fermion qubits]{Time correlations and Leggett-Garg inequalities for probing the topological phase transition in the Kitaev chain}
\begin{center}
\begin{tabular}{p{15cm}}
\vspace{0.1cm}
\quad \lettrine{\color{red1}{\GoudyInfamily{I}}}{n}  this chapter, we continue exploring the signatures of critical dynamic properties in strongly correlated many-body quantum systems, thought, two-time correlations. Now, we focus in topological states in the Kitaev chain. Topological states have shown as robust quantum information entities with potential applications in topological quantum computation protocols. A major challenge in these new proposals is the control of both the autonomous as well as directed time evolution of total system, an issue rather unexplored up to now. We evaluate the interplay between time dependent quantum correlations and nonlocal quantum objects such as Majorana based qubits. We use \gls{STC} and \gls{LGI} for identifying the transition between normal and the topological phase in a Kitaev chain. \gls{STC} and \gls{LGI} of dichotomic quantum observables associated with fermion occupation number of both local as well as nonlocal qubit operators (formed by pairing local and non-local Majorana fermions) are analyzed for different chain lengths and chemical potentials. In order to gain further insight on the physical properties of the system's dynamics, violations of \gls{LGI} are also evaluated for different string order parameter qubits. We obtain analytical results which allow us to understand the fundamental aspects of \gls{STC} in topological Kitaev chains.
\end{tabular}
\end{center}
\newpage
\begin{center}
\begin{tabular}{p{15cm}}
\vspace{0.1cm}
This chapter is published in reference~\cite{Gomez_PRB2018}: {\bf F. J. G\'omez-Ruiz}, J. J. Mendoza-Arenas, F. J. Rodr\'iguez, C. Tejedor, and L. Quiroga. {\it Universal two-time correlations, out-of-time-ordered correlators, and Leggett-Garg inequality violation by edge Majorana fermion qubits}. Phys. Rev. B, {\bf 97}, 235134 (2018).
\end{tabular}
\end{center}

\section{Introduction}
 In the last few years, the development of new quantum devices has fueled the search for novel materials and control mechanisms to engineer unprecedented technologies. Along this path, topological systems have been identified as robust entities with potential applications in quantum computation and information processing due to their unusual braiding properties~\cite{alisea1,dassarma1,Sau}. Candidates for topological qubits include chains of magnetic atoms on top of a superconducting surface~\cite{Nadj}, hybrid systems between $s-$wave superconductors and topological insulators~\cite{HaltermanPRB}, $p-$wave superconductors~\cite{TanakaPRB}, fractional quantum Hall systems~\cite{MooreNP} and 1D semiconductor-superconductor heterostructure based quantum wires~\cite{RomanPRL,Mourik1003,Das}. Notably, the latter have aroused great interest given their high experimental accessibility and controllability~\cite{DumitrescuPRB}. In addition, edge-localized Majorana zero modes, expected to be robust against dephasing and dissipation~\cite{elliott,dassarma2,AguadoR,Albrecht}, have been predicted to exist in these systems. The search of new topological configurations allowing for Majorana zero modes has also been extended to Josephson junction based nanostructures~\cite{KuertenPRB,KalcheimPRB,AliPRB,MohaPRB,Alidoust}. Moreover, Zero energy Majorana modes (\gls{ZEM}) are expected to be robust against dephasing and dissipation~\cite{elliott,dassarma2}, but topological information protocols require an ingredient which remains to be fully explored: the control of the dynamics of each component of the physical system. Deeply related to this, as well as a fundamental problem in quantum physics, is the detection of correlations beyond the scope of classical physics. A large number of protocols have been proposed to this end, and a particularly important subset are those based on spatial non-local correlations as embodied in Bell inequalities, which have been studied vigorously in the quantum information community over the last two decades~\cite{Brunner_RMP2014}.\\
\\
Concurrently with the chase of novel materials is the search for experimentally-accessible properties to identify their truly nonclassical features, such as topological quantum phases. A large number of protocols have been proposed to this end, and a particularly important subset are those based on spatial non-local correlations as embodied in Bell inequalities~\cite{Brunner_RMP2014,drummond2014prb,dassarma2}.
More recently there has been a surge of theoretical and experimental interest in using temporal correlations instead for similar purposes, since in some scenarios nonlocal measurements are quite challenging.  Thus local measurements such as \gls{STC} can be used to gain further access to the underlying physics~\cite{Gessner2014epl,Gessner2014Nat}.\\
\\
More recently, there has been a surge of theoretical and experimental interest in temporal, in addition to spatial, quantum correlations. First, in some scenarios nonlocal measurements are quite challenging, and thus local measurements such as \gls{STC} can be used instead to access the underlying physics~\cite{Gessner2011PRL,Gessner2013PRA,Gessner2014epl,Gessner2014Nat,abdelrahman2017Nat,Cosco2017arxiv}. Second, \gls{STC} are key quantities to understand the phenomenology and control mechanisms of strongly correlated systems in and out of equilibrium~\cite{georges1996rmp,tsuji2011prl,werner2013prl,eckstein2014rmp,gramsch2014prb,JJ2017ann}. And importantly for the purposes of the present work, \gls{STC} can be used to assess the quantumness of a system, in a form similar to spatial correlations through Bell inequalities. Namely, combinations of \gls{STC} allow for \gls{LGI}~\cite{Leggett2,emary} to be tested. These inequalities are satisfied in macroscopic classical systems, characterized by macrorealism (a system is on one particular state at a time only, not in a superposition) and noninvasive measurability (a system is unaffected by a measurement). Their violation thus indicates the existence of macroscopic quantum coherence. Not only there has been an intense search for experimental schemes in which these violations can be observed~\cite{palacios,goggin,knee,athalye,waldherr,dressel,bell_leggett_garg_2016,huffman2017pra}, but also several applications for them have been proposed, including identification of quantum phase transitions in many-body systems~\cite{Gomez_PRB2016} and characterization of quantum transport~\cite{lambert2010prl}.\\
\\
One of the most important challenges to detect superposition of quantum macroscopic states is the robustness of these states against decoherence. Recent experiments~\cite{palacios,goggin,knee,jordan,athalye,xu,dressel,waldherr}  have focused on the detection including interactions with realistic reservoirs. One of the novel signatures is the emergence of non-trivial time dependent non-classical effects. In particular, in nanoresonators~\cite{odonell,teufel} or gate-spin manipulations~\cite{baudin}, the read-out scheme of qubit states, defined by the  measurement process, put new typical time scales to do it. Our main motivation here is not to propose another test of local reality by closing some loopholes. Instead, the \gls{LGI} test is used here to unambiguously establish the existence of an extremely non-classical sensitivity effect of quantum temporal correlations to topological features in a simple scenario. Our findings suggest that strong quantum spatially non-local coherences that could have been generated in \gls{MFC} experiments could have accessible signatures via edge temporal correlation measurements. Recently hybrid Bell-\gls{LGI} weak measurements have been performed for probing remote entanglement in a linear chain of qubits which could also be adapted for non-local topological set ups as the one we address in the present work~\cite{bell_leggett_garg_2016}.\\
\\
The recent interest in time correlations has not been restricted to \gls{LGI} but has been largely directed towards their second moment, the so called out-of-time-ordered correlations (\gls{OTOC}). Initially considered for analyzing superconductivity in the presence of impurities~\cite{larkin1969jetp}, and rediscovered much later in the context of chaos and quantum gravity~\cite{maldacena2016jhep,kitaev2017,hashimoto2017jhep}, \gls{OTOC} have rapidly become a valuable quantity for the analysis of many-body quantum systems~\cite{tsuji2017pra,lin2018,dora2017prb,torres2017} for several reasons. They characterize information scrambling, which refers to the spreading of quantum information over the different degrees of freedom of a system~\cite{delCampo}. They also help diagnosing the existence of quantum chaos by providing a test for the butterfly effect~\cite{aleiner2016ann,bohrdt2017njp,balazs2017prl,khemani2017}, namely that close initial conditions result in exponentially-separated dynamics. In addition, several connections to different measures of quantum correlations have been found~\cite{hosur2016jhep,hauke2017,keyserlingk2017,mezei2017jhep,fan2017scibull}. Given the well-known fundamental role of entanglement in quantum criticality, the natural possibility of observing (equilibrium and dynamical) quantum phase transitions through \gls{OTOC} has been explored with positive results, including transitions in bosonic~\cite{shen2017prb} and spin lattices~\cite{heyl2018}, in impurity systems~\cite{dora2017prb}, and many-body localization~\cite{fan2017scibull,huang2017ann,fradkin2017ann}. Furthermore, after different proposals of measurement of \gls{OTOC}~\cite{hafezi2016pra,swingle2016pra}, their experimental realization has been finally achieved in quantum simulators~\cite{garttner2017nat,li2017prx}.\\
\\
In this chapter, we consider an extension of that interest to assess the interplay between time correlations and nonlocal quantum objects in Majorana fermion chains, a situation different from any other previously considered, by focusing mainly on the Kitaev chain~\cite{kitaev}. In particular we address the open question of detecting true quantum temporal correlations in a topological quantum phase. Correlations for two types of objects are to be explored: {\bf (i)} Local Dirac fermions formed by pairing two Majorana fermions on the same edge site, and {\bf (ii)} Non-local Dirac fermions coming from the pairing of Majorana fermions located at the two opposed edge sites of the chain.  In this way we will address the pivotal role that \gls{STC}s play for detecting large memory effects of local and non-local Majorana edge qubits. In this way we will address the pivotal role that \gls{STC}s play for detecting large memory effects of spatial local and non-local edge qubits. In particular, we will show how the longtime limit of several edge \gls{STC}s provide valuable information on the specific quantum phase of the MFC. On the other hand it is also interesting to appraise the adequacy of \gls{LGI} violations to detect topological phase transitions extending that feasibility beyond the detection of the usual order-disorder quantum phase transitions~\cite{Gomez_PRB2016}. The connection between correlations in space and time domains has not been fully addressed before. Our results provide a first step for looking at a such link in a concrete topological condensed-matter set up. Moreover, we stress that all of our results remain still valid for an edge spin in the transverse field Ising (\gls{TFI}) open chain by applying a Jordan-Wigner transformation to the open \gls{KC} model.\\
\\
 Specifically we will show how the longtime limit of several boundary \gls{STC}s possesses features of an order parameter, providing information on the quantum phase transitions of the Majorana fermion system. Moreover, for the purposes of the present work, \gls{STC} can be used to assess the quantumness of a system, in a form similar to spatial correlations do through Bell inequalities. Namely, combinations of \gls{STC} allow for testing \gls{LGI}~\cite{Leggett2,emary}. These inequalities are satisfied in macroscopic classical systems, characterized by macrorealism (a system's property is well defined at every time regardless of being observed or not) and noninvasive measurability (a system is unaffected by measurements). Their violation indicates the existence of macroscopic quantum coherence.\\
\\
Not only there has been an intense search for experimental schemes in which \gls{LGI} violations can be observed~\cite{palacios,goggin,knee,athalye,waldherr,dressel,bell_leggett_garg_2016,huffman2017pra}, but also several applications for them have been proposed, including identification of order-disorder quantum phase transitions in many-body systems~\cite{Gomez_PRB2016} and characterization of quantum transport~\cite{lambert2010prl}. Indeed it is also interesting to extend the range of \gls{LGI} violation features as a detection tool of topological phase transitions. Along this line, our results provide a first step for understanding the link between correlations in space and time domains in a concrete topological condensed-matter set up. Moreover, we stress that all of our results remain still valid for an edge spin in the transverse field Ising open chain, by applying a Jordan-Wigner transformation to the open Kitaev chain model.\\
\\
\section{Short-time behavior and out-of-time-ordered correlation function}
The \gls{OTOC} are defined as~\cite{hosur2016jhep,swingle2016pra,shen2017prb,fan2017scibull,garttner2017nat,li2017prx}
\begin{equation}\label{eq:otoc}
\mathcal{T}\pap{t}=\langle \hat{O}^{\dagger}_{1}\pap{t}\hat{O}^{\dagger}_{2}\pap{0}\hat{O}_{1}\pap{t}\hat{O}_{2}\pap{0} \rangle
\end{equation}
where $\hat{O}_{1}$ and $\hat{O}_{2}$ are usually (but not necessarily) taken to be local operators and $\hat{O}_{1}\pap{t}=\ee^{\ii \hat{H}t}\hat{O}_{1}\ee^{-\ii \hat{H}t}$ is the time evolution of $\hat{O}_{1}$. For our subsequent analysis we focus on a particular family of these \gls{OTOC} functions. This family of correlations, known as multiple quantum coherences, was initially developed in the context of nuclear magnetic resonance~\cite{baum1985jcp,sanchez2009pra,alvarez2013ann,alvarez2015sci} for the characterization of many-particle coherences, and has recently received plenty of attention as it has been the center of important advances in the comprehension of information scrambling. Namely, it has been measured in a trapped-ion quantum simulator~\cite{garttner2017nat} and put forward as a probe of the build-up of multiparticle entanglement~\cite{hauke2017}. These correlations are defined through Eq.~\eqref{eq:otoc} with $\hat{O}_{1}$ and $\hat{O}_{2}$ unitary operators which commute among themselves at time zero.\\

In order to assess the sensitivity of \gls{STC} and \gls{OTOC} for detecting quantum phase transitions by looking at a single local site, we will connect the short time \gls{STC} behavior to a second-order expansion of the particular multiple quantum coherences. Let us expand up to second order in time the \gls{STC} as given by Eq.~\eqref{eq:Correl}, yielding to
\begin{equation}\label{otoc1}
\begin{split}
\mathcal{C}\pap{t}&=\frac{1}{2}\langle \lbrace \ee^{\ii \hat{H}t}\hat{Q}\ee^{-\ii \hat{H}t},\hat{Q}\rbrace \rangle\\
&\simeq 1-\frac{t^2}{2}\langle -[ \hat{H},\hat{Q} ]^2\rangle+{\cal O}(t^4)
\end{split}
\end{equation}
where $\hat{Q}$ denotes a single qubit operator (a dichotomic observable, i.e. with eigenvalues $q=\pm 1$) to be specified later, with $\left|\psi_{K}\right.\rangle$ the quantum state of the \gls{KC}, $\lbrace \bullet,\bullet\rbrace$ denotes an anticommutator. Note that the second line in the last equation holds for any single-site qubit observable $\hat{Q}$ such that $\hat{Q}^2=\hat{1}$, evolving under the action of an arbitrary (local or global) Hamiltonian $\hat{H}$. Moreover, and most interestingly, the first line in Eq.~\eqref{otoc1} is nothing but the real part of the multiple quantum coherences corresponding to a hermitian single qubit operator $\hat{O}_2=\hat{Q}_2$. Indeed, the latter single site qubit operator $\hat{O}_2$ does commute with the time-evolving operator $\hat{O}_{1}(t)=\hat{O}_1=e^{-iHt}$ at $t=0$ (i.e. with the identity).\\

First, let us consider the \gls{STC} for a single edge Majorana fermion $j=1$, i.e. $\hat{Q}=\hat{\gamma}_{1}$. By resorting to Eq.~\eqref{Hmajo1} it is easy to check that $[ \hat{H},\hat{\gamma}_{1} ]^2=-\mu^2$, a scalar quantity, thus producing for the real part of the corresponding \gls{OTOC} the simple and universal result $\langle\left.\psi_{K}\right| -[ \hat{H},\hat{\gamma}_{1} ]^2\left|\psi_{K}\right.\rangle=\mu^2$, independent of the chain size and valid for any \gls{KC} eigenstate $\left|\psi_{K}\right.\rangle$.  Note that via the Jordan-Wigner transformation, this qubit operator corresponds to $\hat{\sigma}_{1,x}=\hat{\gamma}_{1}$ for the \gls{TFI}, i.e. the $x$-spin operator of an edge site. Consequently, we rewrite the \gls{STC} in Eq.~\eqref{otoc1} as $\mathcal{C}_{1}^{(x)}\pap{t}$,
\begin{equation}\label{otoc2}
\begin{split}
\mathcal{C}_{1}^{(x)}\pap{t}&=\frac{1}{2}    \langle\left.\psi_{K}\right|  \lbrace \hat{\gamma}_1\pap{t},\hat{\gamma}_1\rbrace \left|\psi_{K}\right.\rangle\\
&\simeq 1-\frac{\mu^2}{2}t^2+{\cal O}(t^4).
\end{split}
\end{equation}

As a second case, we consider a two-Majorana edge qubit such as $\hat{Q}=2\hat{n}_1-\hat{1}=-\ii\,\hat{\gamma}_{1}\hat{\gamma}_{2}$. This qubit corresponds, via the Jordan-Wigner transformation, to the $\hat{\sigma}_{1}^{z}$ edge spin operator for the transverse field Ising model, i.e. $-\ii\,\hat{\gamma}_{1}\hat{\gamma}_{2}=\hat{\sigma}_{1}^{z}$. Now it is straightforward to show that $[ \hat{H},-\ii\,\hat{\gamma}_{1}\hat{\gamma}_{2} ]^2=-4\Delta^2$, again a scalar quantity and hence producing a result valid for any Majorana fermion chain pure state $\left|\psi_{K}\right.\rangle$ or mixed state $\hat{\rho}_K$. Thus the second derivative of the real part of the $\mathcal{T}(t)$ reduces to the universal value $\langle -[ \hat{H},-\ii\,\hat{\gamma}_{1}\hat{\gamma}_{2} ]^2\rangle=4\Delta^2$ and consequently the short time expression for $\mathcal{C}_{1}^{(z)}\pap{t}$ becomes
\begin{equation}\label{otoc3} 
\begin{split}
\mathcal{C}_{1}^{(z)}\pap{t}&=-\frac{1}{2}\big\langle \lbrace \hat{\gamma}_1\pap{t}\hat{\gamma}_2\pap{t},\hat{\gamma}_1\hat{\gamma}_2\rbrace \big\rangle \\
&\simeq 1-2\Delta^2t^2+{\cal O}(t^4).
\end{split}
\end{equation}
 As a third case, we analyze the short-time behavior of the non-local Dirac fermion formed by coupling two Majorana operators located at the two edges of the chain, $\hat{Q}_{1,N}=\ii\,\hat{\gamma}_{1}\hat{\gamma}_{2N}$. The expansion of the corresponding \gls{STC} leads to

\begin{equation}\label{otoc20}
\begin{split}
\mathcal{C}_{1,N}\pap{t}&= -\frac{1}{2}\big\langle \lbrace \hat{\gamma}_1\pap{t}\hat{\gamma}_{2N}\pap{t},\hat{\gamma}_1\hat{\gamma}_{2N}\rbrace \big\rangle\\
&\simeq 1-\mu^2\left ( 1- \langle \hat{\gamma}_1\hat{\gamma}_2\hat{\gamma}_{2N-1}\hat{\gamma}_{2N}\rangle \right )\,t^2+{\cal O}(t^4).
\end{split}
\end{equation}

It is evident that this \gls{STC} features a non-universal short-time evolution, given that it depends on the specific quantum state of the Majorana fermion chain. This is indicated by the expected value of the four Majorana operator term $\langle \hat{\gamma}_1\hat{\gamma}_2\hat{\gamma}_{2N-1}\hat{\gamma}_{2N}\rangle=\langle \hat{\gamma}_1\hat{\gamma}_2\rangle\langle \hat{\gamma}_{2N-1}\hat{\gamma}_{2N}\rangle+\langle \hat{\gamma}_1\hat{\gamma}_{2N}\rangle\langle \hat{\gamma}_2\hat{\gamma}_{2N-1}\rangle$, which for a sufficiently long it can be approximated to $\langle \hat{\gamma}_1\hat{\gamma}_2\hat{\gamma}_{2N-1}\hat{\gamma}_{2N}\rangle\simeq \langle \hat{\gamma}_1\hat{\gamma}_2\rangle^2$ in the ground state.\\

\begin{figure}[h!]
\begin{minipage}{0.5\textwidth} 
\centering \includegraphics[scale=1.1]{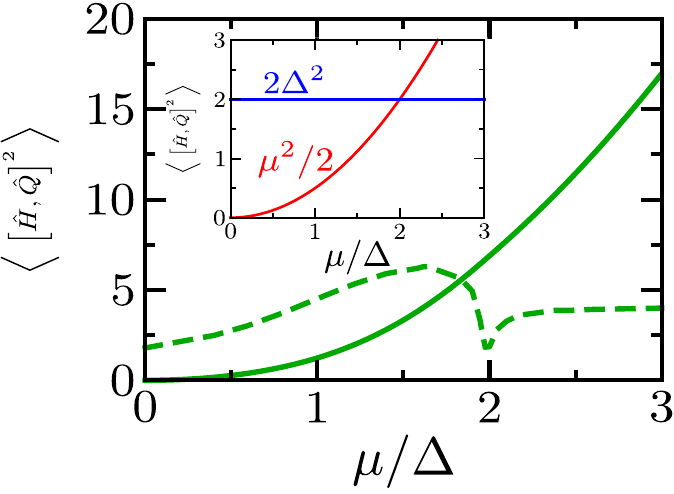}
		\caption[Short-time curvatures of the different edge \gls{STC}]{Short-time curvatures of the different edge \gls{STC}s as a function of $\mu/\Delta$. Main panel: Curvature for the non-local two-Majorana qubit $\mathcal{C}_{1,N}\pap{t}$ (green, solid line),  and its second derivative with respect to $\mu$ (green, dashed line). The latter presents a clear dip at the topological quantum critical point. Inset: \gls{STC} initial curvature for local Majorana qubits, $\mathcal{C}_{1}^{x}\pap{t}$ (red line) and $\mathcal{C}_{1}^{z}\pap{t}$ (blue line), showing a crossing just at the critical point.}\label{fig_1}
\end{minipage}
\hfill
\begin{minipage}{0.43\textwidth}
In Fig.~\ref{fig_1} the short-time curvature (second time derivative) of the edge \gls{STC}s corresponding to single- and double-Majorana fermions is depicted as a function of $\mu/\Delta$. In the main panel the non-local case of $\mathcal{C}_{1,N}\pap{t}$ is plotted (solid line), while in the inset those of $\mathcal{C}_{1}^{(x)}\pap{t}$ and $\mathcal{C}_{1}^{(z)}\pap{t}$ are depicted. Clearly, by comparing Eqs.~\eqref{otoc2}-\eqref{otoc3}, an universal crossing of initial \gls{STC} curvatures occurs for $\mu=2\Delta$, which signals the critical point for the topological-trivial phase transition in the Kitaev model, or equivalently for the ferromagnetic-paramagnetic transition in the transverse field Ising model. This remarkable universal behavior, i.e. the independence from the Majorana fermion chain quantum state, holds true only for the edge sites of both these models as realized by the Kitaev chain and transverse field Ising systems.
\end{minipage}				
\end{figure}
By contrast, the non-local $\mathcal{C}_{1,N}\pap{t}$ shows a non-universal behavior depending on the specific quantum state of the Majorana fermion chain. The results plotted in the main panel of Fig.~\ref{fig_1} have been obtained numerically, as explained below, for a Kitaev chain in the ground state. In the same panel the second derivative of the curvature with respect to $\mu$ is also plotted (dashed line), which clearly presents a dip at the critical point $\mu/\Delta=2$. Thus, we observe that the early-time correlators, both with universal and non-universal behavior, are sensitive to the topological phase transition.

\section{General two-time correlation behavior of Majorana qubits} \label{sec_4}
Having established the relevance of \gls{STC}s and a related family of out-of-time-ordered correlations for edge sites in the Majorana fermion chain, we proceed to explore the TTC behavior for qubits formed by any combination between edge and/or bulk sites, for arbitrary times. By developing the Majorana qubits in terms of Bogoliubov operators (see chapter~\ref{Cap2}) we proceed to express both the single- and double-Majorana \gls{STC} in convenient forms for numerical analysis. We follow the same notation for Bogoliubov coefficients used in Ref.~\cite{Olesia}. As we discuss below this numerical procedure is essential to further progress, except in special cases  for $\mathcal{C}_{1}^{(x)}\pap{t}$ where an exact closed form has been obtained. We first evaluate the two-time correlation as
\begin{equation}
\mathcal{C}_{1}^{(x)}\pap{t}= \frac{1}{2}\big\langle \lbrace \hat{\gamma}_1\pap{t},\hat{\gamma}_1 \rbrace\big\rangle
\end{equation}
The temporal evolution for the Majorana operator is given by $\hat{\gamma}\pap{t}=\ee^{\ii \hat{H} t} \hat{\gamma}_{1} \ee^{-\ii \hat{H} t}$, where $\hat{H}$ is the Kitaev-Majorana Hamiltonian Eq.~\eqref{Hmajo1}. We calculate this evolution using the traditional Baker-Campell-Hausdorff formula (\gls{BCH}) $\ee^{s\hat{A}}\hat{B}\ee^{-s\hat{A}}=\hat{B}+s[\hat{A},\hat{B}]+\frac{s^2}{2!}[\hat{A},[\hat{A},\hat{B}]]+...$, and by a direct substitution we evaluate each commutator as follows:
\begin{align*}
[\hat{H},\hat{\gamma}_1]&=\ii\mu\,\hat{\gamma}_2\\
[\hat{H},[\hat{H},\hat{\gamma}_1]]&=\mu^2\,\hat{\gamma}_1+2\mu \Delta\, \hat{\gamma}_3\\
[\hat{H},[\hat{H},[\hat{H},\hat{\gamma}_1]]]&=\ii\mu\left ( \mu^2+4\Delta^2 \right )\,\hat{\gamma}_2+2\ii\mu^2 \Delta\, \hat{\gamma}_4\\
[\hat{H},[\hat{H},[\hat{H},[\hat{H},\hat{\gamma}_1]]]]&=\mu^2 \left ( \mu^2+4\Delta^2 \right )\,\hat{\gamma}_1+4\mu\Delta\left (\mu^2+2\Delta^2 \right )\,\hat{\gamma}_3+4\mu^2 \Delta^2\,\hat{\gamma}_5\\
[\hat{H},[\hat{H},[\hat{H},[\hat{H},[\hat{H},\hat{\gamma}_1]]]]]&=\ii\mu \left ( \mu^4+3\mu^2 4\Delta^2+16\Delta^4 \right )\,\hat{\gamma}_2+4\ii\mu^2 \Delta\left ( \mu^2+4\Delta^2 \right )\,\hat{\gamma}_4+4\ii\mu^3 \Delta^2\,\hat{\gamma}_6\\
[\hat{H},[\hat{H},[\hat{H},[\hat{H},[\hat{H},[\hat{H},\hat{\gamma}_1]]]]]]&=\mu^2 \left ( \mu^4+12\mu^2 \Delta^2+16\Delta^4 \right )\,\hat{\gamma}_1+2\mu\Delta\left ( 3\mu^4+5\mu^2 4\Delta^2+16\Delta^4 \right )\,\hat{\gamma}_3\\
&\quad+4\mu^2 \Delta^2\left ( 3\mu^2+8\Delta^2 \right )\,\hat{\gamma}_5+8\mu^3 \Delta^3\,\hat{\gamma}_7.
\end{align*}
After evaluating these first terms of the \gls{BCH} formula, we can figure out the sequence of the emerging terms. In addition we use the fact that the Majorana operators satisfy the property $\pap{\gamma_j}^{2}=\pap{\gamma_j^{\dagger}}^{2}=\hat{1}$, and that they obey the modified anticommutation relations $\lbrace \gamma_i,\gamma_j\rbrace=2\delta_{i,j}$ with $i,j=1,...,2N$. Therefore, after a careful algebraic process, we find that
\begin{equation}\label{naraya1}
\mathcal{C}_{1}^{(x)}\pap{t}=\sum_{m=0}^{\infty}\frac{(-1)^m}{(2m)!}\left ( 2\Delta t \right )^{2m}\mathcal{N}_m(u^2),
\end{equation}
where $u=\mu/2\Delta$ and $\mathcal{N}_{m}(x)$ are the well-known Narayana polynomials (\gls{NP}). The \gls{NP} have the form
\begin{equation}
\mathcal{N}_{m}\pap{x} =\sum_{n=1}^{m}N_{m,n}x^{n}\; \qquad \text{with}\,\qquad N_{m,n}=\frac{1}{m}\binom {m}{n-1}\binom {m}{n}
\end{equation}
where $N_{m,n}$ are denoted as the Narayana numbers~\cite{Kostov,Sun}. Note that the critical point corresponds to $u=1$, for which $\mathcal{N}_m(1)=C_m=\frac{1}{m+1}\binom {2m}{m}$, the most famous Catalan numbers. Importantly Eq.~\eqref{naraya1} can be calculated in a closed form at the critical point $u=1$, yielding to the simple expression
\begin{equation}\label{naraya2}
\mathcal{C}_{1}^{(x)}\pap{t}=\frac{\mathcal{J}_1\left ( 4\Delta t \right )}{2\Delta t}
\end{equation}
in terms of the Bessel function of the first kind $\mathcal{J}_1(z)$. To the best of our knowledge this compact result has passed unnoted in the literature on both Ising and Kitaev models. We emphasize that the expressions given by Eqs.~\eqref{naraya1}-\eqref{naraya2} are always valid and thus they are of universal reach, independently of the pure or mixed state of the Majorana fermion chain. Consequently, they hold true even at infinite temperature.\\
\\
For other values of $u$ such a simple form has yet to be found. However the analytics can be developed further, leading to deeper insights on the general behavior of the \gls{STC}. First, Eq.~\eqref{naraya1} allows for establishing a link of $\mathcal{C}_{1}^{(x)}(t)$-\gls{STC} on both phases around the critical point $u=1$, which will come in handy afterwards. Since the Narayana polynomials are symmetric, the property $\mathcal{N}_m(\frac{1}{x})=\frac{1}{x^{m+1}}\mathcal{N}_m(x)$ holds. Consequently,
\begin{equation}\label{naraya3}
\mathcal{C}_{1}^{(x)}\pap{t,\frac{1}{u}}=1-\frac{1}{u^2}+\frac{1}{u^2}\,\,\mathcal{C}_{1}^{(x)}\pap{\frac{t}{u},u},
\end{equation}
indicating that the $x$-\gls{STC} behaves in one phase (reduced chemical potential $\frac{1}{u}$) as it would do in the complementary phase (reduced chemical potential $u$) but with a scaled time $\frac{t}{u}$.\\
\\
Furthermore, for numerical calculations the time evolution of a single Majorana edge fermion operator, $\hat{\gamma}_{i}(t)=\ee^{\ii \hat{H}t}\hat{\gamma}_{i}(0)\ee^{-\ii \hat{H}t}$, is found to be
\begin{equation}\label{otoc4}
\begin{split}
\hat{\gamma}_{2j-1}\pap{t}&=\sum_{m=1}^N\lbrace \hat{\gamma}_{2m-1}\,g_{m,j}^{(+,+)}(t)+\hat{\gamma}_{2m}\,h_{m,j}^{(-,+)}(t) \rbrace\\
\hat{\gamma}_{2j}\pap{t}&=\sum_{m=1}^N\lbrace \hat{\gamma}_{2m}\,g_{m,j}^{(-,-)}(t)-\hat{\gamma}_{2m-1}\,h_{m,j}^{(+,-)}(t) \rbrace,
\end{split}
\end{equation}
where
\begin{equation}\label{otoc5}
\begin{split}
g_{m,j}^{(\nu,\nu)}(t)&=\sum_{k}{\rm cos}(\epsilon_k\,t)\,(u_{2k,m}+\nu v_{2k,m})\,(u_{2k,j}+\nu v_{2k,j})\\
h_{m,j}^{(\nu,-\nu)}(t)&=\sum_{k}{\rm sin}(\epsilon_k\,t)\,(u_{2k,m}-\nu v_{2k,m})\,(u_{2k,j}+\nu v_{2k,j}),
\end{split}
\end{equation}
with $\nu=+,-$. A direct application of these relations allows us to obtain an analytical expression for the full time evolution of $\mathcal{C}_{1}^{(x)}\pap{t}$, as
\begin{equation}\label{otoc6}
\mathcal{C}_{1}^{(x)}\pap{t}=\sum_{k}{\rm cos}(\epsilon_k\,t)(u_{2k,1}+v_{2k,1})^2,
\end{equation}
where $\langle \gamma_{2i}\gamma_{2j-1}\rangle=-i\sum_{k}\left (u_{2k,i}-v_{2k,i}\right )\left (u_{2k,j}+v_{2k,j}\right )$ and $\langle \gamma_{2i}\gamma_{2j}\rangle=\langle \gamma_{2i-1}\gamma_{2j-1}\rangle=\delta_{i,j}$ have been used. By expanding Eq.~\eqref{otoc6} up to second order in time and comparing it with the universal result quoted in Eq.~\eqref{otoc2} the following identity holds true,
\begin{equation}\label{otoc7}
\sum_{k}\epsilon_k^2\,\left (u_{2k,1}+v_{2k,1}\right )^2=\mu^2,
\end{equation}
which is valid for open Kitaev and transverse field Ising models (with $\mu$ replaced by the transverse magnetic field) of arbitrary chain length. The identity given by Eq.~\eqref{otoc7} provides by itself a consistency check of numerical calculations.\\

Now let us look at the long-time limit of $\mathcal{C}_{1}^{(x)}\pap{t}$ by averaging Eq.~\eqref{otoc6} over a long time period. As the time average of $\cos(\epsilon_k\,t)$ vanishes unless some fermion mode has energy $\epsilon_M=0$, i.e. a zero energy Majorana mode exists  (for which the average is $1$), we can readily assure that for the topological regime
\begin{equation}\label{otoc11}
\lim_{t\to\infty} \mathcal{C}_{1}^{(x)}\pap{t} \simeq (u_{M,1}+v_{M,1})^2=4u_{M,1}^2,
\end{equation}
since $u_{M,1}=v_{M,1}$, i.e. the electron and hole contributions for the zero energy Majorana mode $k=M$ at site $j=1$ are the same. Consequently, we propose that a measurement of the long time saturation value of the edge $\mathcal{C}_{1}^{(x)}$-\gls{STC} provides a witness of the topological ($\neq 0$) and non-topological ($=0$) phase transition of the Majorana fermion chain systems, as it probes directly the existence of zero energy modes. Additionally, it gives direct access to the electron-hole weight of such modes.\\
\subsection{Two-time correlations of double-Majorana qubits}
We focus now on qubits formed by any pair of Majorana fermions such as $\hat{\gamma}_{2i-1}$ and $\hat{\gamma}_{2j}$. We define
\begin{equation}\label{maj1}
\hat{\theta}_{i,j}=\frac{1}{2}\left ( \hat{\gamma}_{2i-1}+\ii\,\hat{\gamma}_{2j} \right ), \,
\hat{\theta}_{i,j}^{\dagger}=\frac{1}{2}\left ( \hat{\gamma}_{2i-1}-\ii\,\hat{\gamma}_{2j} \right).
\end{equation}
Notice that $i=j$ implies that the forming Majorana modes are located on the same physical site, and the Kitaev operators in Eq.~\eqref{Hkitaev} are recovered, i.e $\hat{\theta}_{j,j}=\hat{c}_j$. On the other hand, for $i\neq j$ the Majorana fermions are located on different physical sites. It is easy to check that usual Dirac fermion relations hold true for operators $\hat{\theta}_{i,j}$ and $\hat{\theta}_{i,j}^{\dagger}$ as $\left \{\hat{\theta}_{i,j},\hat{\theta}_{i,j}^{\dagger}\right \}=1$, $\left\{\hat{\theta}_{i,j}^{\dagger},\hat{\theta}_{i,j}^{\dagger}\right\} =\left\{\hat{\theta}_{i,j},\hat{\theta}_{i,j}\right\}=0$.
Thus, we can define non-local Majorana qubits as $\hat{Q}_{i,j}=2\hat{\theta}_{i,j}^{\dagger}\hat{\theta}_{i,j}-1$, which have eigenvalues $\pm 1$.  By direct substitution of the standard Bogoliubov-de Gennes transformation (Eq.~\eqref{BGT}), we can rewrite the operator $\hat{Q}_{i,j}$ as
\begin{equation}\label{majsup2}
\begin{split}
\hat{Q}_{i,j}&=\frac{1}{2}\left ( \hat{\gamma}_{2i-1}-i\,\hat{\gamma}_{2j} \right )\left ( \hat{\gamma}_{2i-1}+i\,\hat{\gamma}_{2j} \right )-1=\frac{1}{2}\left [ \hat{c}_{i}+\hat{c}_{i}^{\dagger}, \hat{c}_{j}-\hat{c}_{j}^{\dagger}\right ]\\
\hat{Q}_{i,j}&=\frac{1}{2}\sum_{k=1}^{N}\sum_{q=1}^{N}\left ( u_{2k,i}+v_{2k,i} \right )\left ( v_{2q,j}-u_{2q,j} \right )\left [ \hat{d}_{k}+\hat{d}_{k}^{\dagger}, \hat{d}_{q}-\hat{d}_{q}^{\dagger}\right ].
\end{split}
\end{equation}

Similarly we evaluate its time-evolution $\hat{Q}_{i,j}(t)=\ee^{\ii \hat{H}t}\hat{Q}_{i,j}(0)\ee^{-\ii \hat{H}t}$, obtaining that
\begin{equation}\label{majsup3}
\hat{Q}_{i,j}(t)=\frac{1}{2}\sum_{k=1}^{N}\sum_{q=1}^{N}\left ( u_{2k,i}+v_{2k,i} \right )\left ( u_{2q,j}-v_{2q,j} \right )\left [ \ee^{-\ii \epsilon_kt}\hat{d}_{k}+\ee^{\ii \epsilon_kt}\hat{d}_{k}^{\dagger}, \ee^{-\ii \epsilon_qt}\hat{d}_{q}-\ee^{\ii \epsilon_qt}\hat{d}_{q}^{\dagger}\right ].
\end{equation}

Therefore, we explicitly calculate the symmetric \gls{STC} (see Eq.~\eqref{otoc1}). The \gls{STC} for the general non-local Majorana qubit operator $\hat{Q}_{i,j}$ is found to be
\begin{align*}
\mathcal{C}_{i,j}\pap{t}=&\frac{1}{8}\sum_{k=1}^{N}\sum_{q=1}^{N}\sum_{k'=1}^{N}\sum_{q'=1}^{N}\left ( u_{2k,i}+v_{2k,i} \right )\left ( v_{2q,j}-u_{2q,j} \right )
\left ( u_{2k',i}+v_{2k',i} \right )\left ( v_{2q',j}-u_{2q',j} \right )\\
&\bigg\langle  \biggl( \hat{d}_{k}\hat{d}_{q}\ee^{-\ii \pap{\eps_k +\eps_q}t}-\hat{d}_{k}\hat{d}_{q}^{\dagger}\ee^{-\ii \pap{\eps_k -\eps_q}t}+
\hat{d}_{k}^{\dagger}\hat{d}_{q}\ee^{\ii \pap{\eps_k -\eps_q}t}-\hat{d}_{k}^{\dagger}\hat{d}_{q}^{\dagger}\ee^{\ii \pap{\eps_k +\eps_q}t}\\
&\quad\qquad-
\hat{d}_{q}\hat{d}_{k}\ee^{-\ii \pap{\eps_k +\eps_q}t}-\hat{d}_{q}\hat{d}_{k}^{\dagger}\ee^{\ii \pap{\eps_k -\eps_q}t}+
\hat{d}_{q}^{\dagger}\hat{d}_{k}\ee^{-\ii \pap{\eps_k -\eps_q}t}+\hat{d}_{q}^{\dagger}\hat{d}_{k}^{\dagger}\ee^{\ii \pap{\eps_k +\eps_q}t}\biggr )\\
&\quad\qquad\biggl( \hat{d}_{k'}\hat{d}_{q'}-\hat{d}_{k'}\hat{d}_{q'}^{\dagger}+\hat{d}_{k'}^{\dagger}\hat{d}_{q'}-\hat{d}_{k'}^{\dagger}\hat{d}_{q'}^{\dagger}
-\hat{d}_{q'}\hat{d}_{k'}-\hat{d}_{q'}\hat{d}_{k'}^{\dagger}+\hat{d}_{q'}^{\dagger}\hat{d}_{k'}+\hat{d}_{q'}^{\dagger}\hat{d}_{k'}^{\dagger}
\biggr )\\
&\quad\qquad+\biggl( \hat{d}_{k}\hat{d}_{q}-\hat{d}_{k}\hat{d}_{q}^{\dagger}+\hat{d}_{k}^{\dagger}\hat{d}_{q}-\hat{d}_{k}^{\dagger}\hat{d}_{q}^{\dagger}
-\hat{d}_{q}\hat{d}_{k}-\hat{d}_{q}\hat{d}_{k}^{\dagger}+\hat{d}_{q}^{\dagger}\hat{d}_{k}+\hat{d}_{q}^{\dagger}\hat{d}_{k}^{\dagger}
\biggr )\\
&\quad\qquad\biggl( \hat{d}_{k'}\hat{d}_{q'}\ee^{-\ii \pap{\eps_{k'} +\eps_{q'}}t}-\hat{d}_{k'}\hat{d}_{q'}^{\dagger}\ee^{-\ii \pap{\eps_{k'} -\eps_{q'}}t}+
\hat{d}_{k'}^{\dagger}\hat{d}_{q'}\ee^{\ii \pap{\eps_{k'} -\eps_{q'}}t}-\hat{d}_{k'}^{\dagger}\hat{d}_{q'}^{\dagger}\ee^{\ii \pap{\eps_{k'} +\eps_{q'}}t}\\
&\quad\qquad-
\hat{d}_{q'}\hat{d}_{k'}\ee^{-\ii \pap{\eps_{k'} +\eps_{q'}}t}-\hat{d}_{q'}\hat{d}_{k'}^{\dagger}\ee^{\ii \pap{\eps_{k'} -\eps_{q'}}t}+
\hat{d}_{q'}^{\dagger}\hat{d}_{k'}\ee^{-\ii \pap{\eps_{k'} -\eps_{q'}}t}+\hat{d}_{q'}^{\dagger}\hat{d}_{k'}^{\dagger}\ee^{\ii \pap{\eps_{k'} +\eps_{q'}}t}\biggr )
\bigg\rangle,
\end{align*}

 In order to proceed further from last equation, we note that only elements with equal number of creation and annihilation operators are relevant since only they can produce nonvanishing expectation values for eigenstates of the system with a well defined number of elementary fermionic excitations. These nonvanishing terms turn out to be

\begin{equation}\label{nnt2}
\begin{split}
\big\langle\hat{d}_{k}\hat{d}_{q}^{\dagger}\hat{d}_{k'}\hat{d}_{q'}^{\dagger}\big\rangle&=\delta_{k,q}\delta_{k',q'}(1-n_{k})(1-n_{k'})+\delta_{k,q'}\delta_{k',q}(1-n_{k})n_{k'}\\
\big\langle\hat{d}_{k}\hat{d}_{q}^{\dagger}\hat{d}_{k'}^{\dagger}\hat{d}_{q'}\big\rangle&=\delta_{k,q}\delta_{k',q'}(1-n_{k})n_{k'}-\delta_{k,k'}\delta_{q,q'}(1-n_{k})n_{q}\\
\big\langle\hat{d}_{k}^{\dagger}\hat{d}_{q}\hat{d}_{k'}\hat{d}_{q'}^{\dagger}\big\rangle&=\delta_{k,q}\delta_{k',q'}n_{k}(1-n_{k'})-\delta_{k,k'}\delta_{q,q'}n_{k}(1-n_{q})\\
\big\langle\hat{d}_{k}^{\dagger}\hat{d}_{q}\hat{d}_{k'}^{\dagger}\hat{d}_{q'}\big\rangle&=\delta_{k,q}\delta_{k',q'}n_{k}n_{k'}+\delta_{k,q'}\delta_{q,k'}n_{k}(1-n_{q})\\
\big\langle\hat{d}_{k}\hat{d}_{q}\hat{d}_{k'}^{\dagger}\hat{d}_{q'}^{\dagger}\big\rangle&=\left (-\delta_{k,k'}\delta_{q,q'}+\delta_{k,q'}\delta_{q,k'}\right )(1-n_{k})(1-n_{q})\\
\big\langle\hat{d}_{k}^{\dagger}\hat{d}_{q}^{\dagger}\hat{d}_{k'}\hat{d}_{q'}\big\rangle&=\left (-\delta_{k,k'}\delta_{q,q'}+\delta_{k,q'}\delta_{q,k'}\right )n_{k}n_{q}.\\
\end{split}
\end{equation}
From  the expression that we obtain for $\mathcal{C}_{i,j}\pap{t}$ we develop term by term and insert therein the matrix elements as expressed by Eqs.~\eqref{nnt2}. Exchanging the final active labels $k$ and $q$ in the resulting equation, and noting that $n_k^2=n_k$, the compact expression for the \gls{STC} $\mathcal{C}_{i,j}\pap{t}$ written as
\begingroup\makeatletter\def\f@size{8}\check@mathfonts
\def\maketag@@@#1{\hbox{\m@th\large\normalfont#1}}
\begin{equation}\label{cj3}
\begin{split}
\mathcal{C}_{i,j}\pap{t}=1-\sum_{k=1}^{N}\sum_{q=1}^{N}\Biggl[&\sin^2\pap{\frac{\eps_{k}+\eps_{q}}{2}t}\left [ \left ( u_{2k,i}+v_{2k,i}\right )\left ( u_{2q,j}-v_{2q,j}\right )-\left ( u_{2q,i}+v_{2q,i}\right )\left ( u_{2k,j}-v_{2k,j}\right )\right ]^2\left [ 1-\pap{n_{q}-n_{k}}^2\right ]\\
&+\sin^2\pap{\frac{\eps_{k}-\eps_{q}}{2}t}\left [ \left ( u_{2k,i}+v_{2k,i}\right )\left ( u_{2q,j}-v_{2q,j}\right )+\left ( u_{2q,i}+v_{2q,i}\right )\left ( u_{2k,j}-v_{2k,j}\right )\right ]^2\pap{n_{q}-n_{k}}^2\Biggr].
\end{split}
\end{equation}
\endgroup
where $n_k=0$ denotes the k-th fermion mode is empty while $n_k=1$ means it is occupied. By focusing on the edge \gls{STC}, i.e. $i=j=1$ and $\mathcal{C}_{1,1}\pap{t}=\mathcal{C}_{1}^{(z)}\pap{t}$, expanding the right hand side of Eq.~\eqref{cj3} up to second order in time and comparing it to the universal result quoted in Eq.~\eqref{otoc3}, a new identity results as
\begin{equation}\label{otoc10}
\sum_{k=1}^{N}\sum_{q=1}^{N}\left ( \eps_{k}+\eps_{q}\right )^2(u_{2 k,1}v_{2 q,1}-u_{2 q,1}v_{2 k,1})^2=2\Delta^2
\end{equation}
 which is valid for both open boundary Kitaev and transverse field Ising models (with $\Delta$ replaced by the spin exchange interaction) for arbitrary chain lengths. As before the identity given by Eq.~\eqref{otoc10} turns out to be another important consistency check for numerical calculations.
 \begin{figure}[t!]
  \includegraphics[width=1 \textwidth]{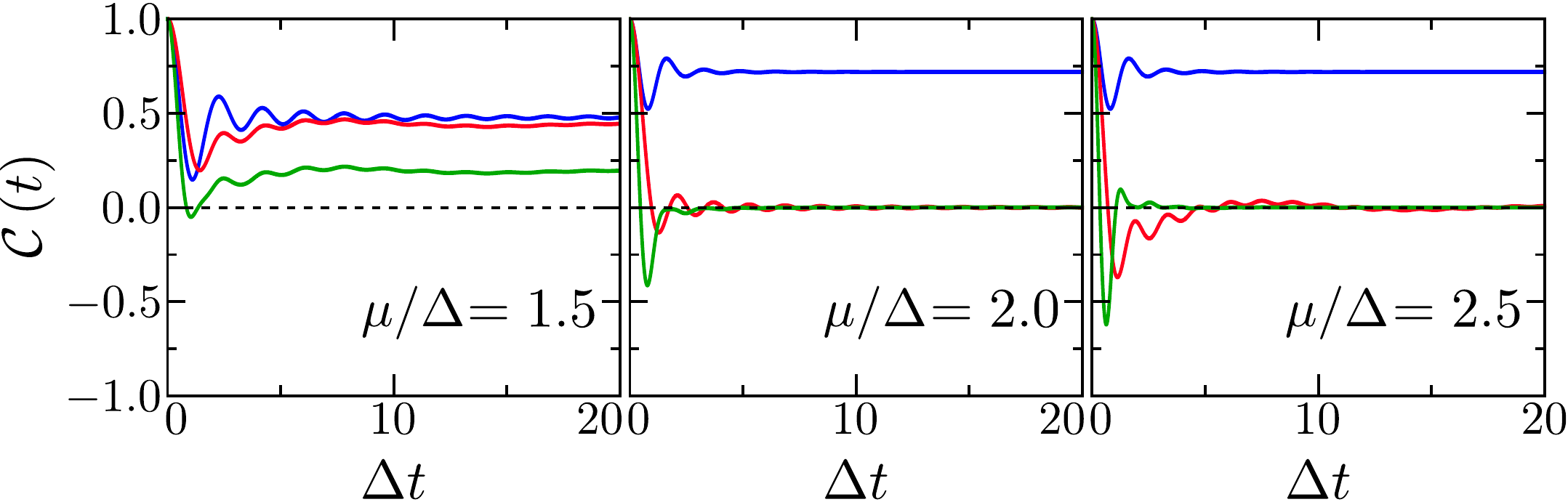}
  \caption[Edge single- and double-Majorana qubit \gls{STC}.]{Edge single- and double-Majorana qubit \gls{STC} in the topological phase ($\mu/\Delta=1.5$, left panel), at the transition point ($\mu/\Delta=2$, central panel) and in the non-topological phase ($\mu/\Delta=2.5$, right panel). In all panels the red line depicts the $\mathcal{C}_{1}^{(x)}\left(t\right)$-\gls{STC}, the blue line represents the $\mathcal{C}_{1}^{(z)}\left(t\right)$-\gls{STC}, while the green line corresponds to $\mathcal{C}_{1,N}\left(t\right)$.}
  \label{fig_2}
\end{figure}

\subsection{Numerical results for \gls{STC} in Majorana Fermion Systems}
 Now we evaluate numerically the different time correlations discussed in Sec.~\ref{sec_4}. All the results we describe below correspond to an open-ended Majorana fermion chain with $N=101$ sites in the many-body ground-state, $ \left|\psi_{K}\right.\rangle=\bigotimes_{k=1}^{N} \left|0\right.\rangle$, with symmetric hopping-pairing energies, i.e. $\omega=\Delta=1$, which also fixes the energy scale. Their inverse fixes the time scale through the dimensionless variable $\Delta t$. In Figure~\ref{fig_2} the time evolution of both single edge Majorana qubits $\mathcal{C}_{1}^{(x)}\pap{t}$ and two-Majorana edge qubits $\mathcal{C}_{1}^{(z)}\pap{t}$ and $\mathcal{C}_{1,N}\left(t\right)$  is displayed for three specific values of the chemical potential, namely $\mu/\Delta=1.5$ (left panel), $\mu/\Delta=2.0$ (central panel) and $\mu/\Delta=2.5$ (right panel). Oscillatory features are dominant for both short- and intermediate-time regimes $\Delta t < 10$, which subsequently are attenuated until the \gls{STC}s reach stationary or asymptotical values for $\Delta t > 10$. We discuss first this long-time regime.\\

It can be seen that the asymptotic behaviors of $\mathcal{C}_{1}^{(x)}\pap{t}$ and $\mathcal{C}_{1,N}\left(t\right)$-\gls{STC}s are very different from that of $\mathcal{C}_{1}^{(z)}\pap{t}$-\gls{STC} crossing the critical point to the trivial phase. We found that these three \gls{STC}s remain finite in the topological phase even at infinity time, which agrees with the numerically-based observation in Ref.~\cite{Jack} that long coherence times for edge sites in open boundary Majorana fermion chains are possible. However, the long-time limits of $\mathcal{C}_{1}^{(x)}\pap{t}$ and $\mathcal{C}_{1,N}\left(t\right)$ vanish when the system enters the non-topological or trivial phase ($\mathcal{C}_{1}^{(z)}\pap{t}$ saturates to finite values at both phases). 

\begin{figure}[h!]
\begin{minipage}{0.5\textwidth} 
\centering \includegraphics[scale=0.8]{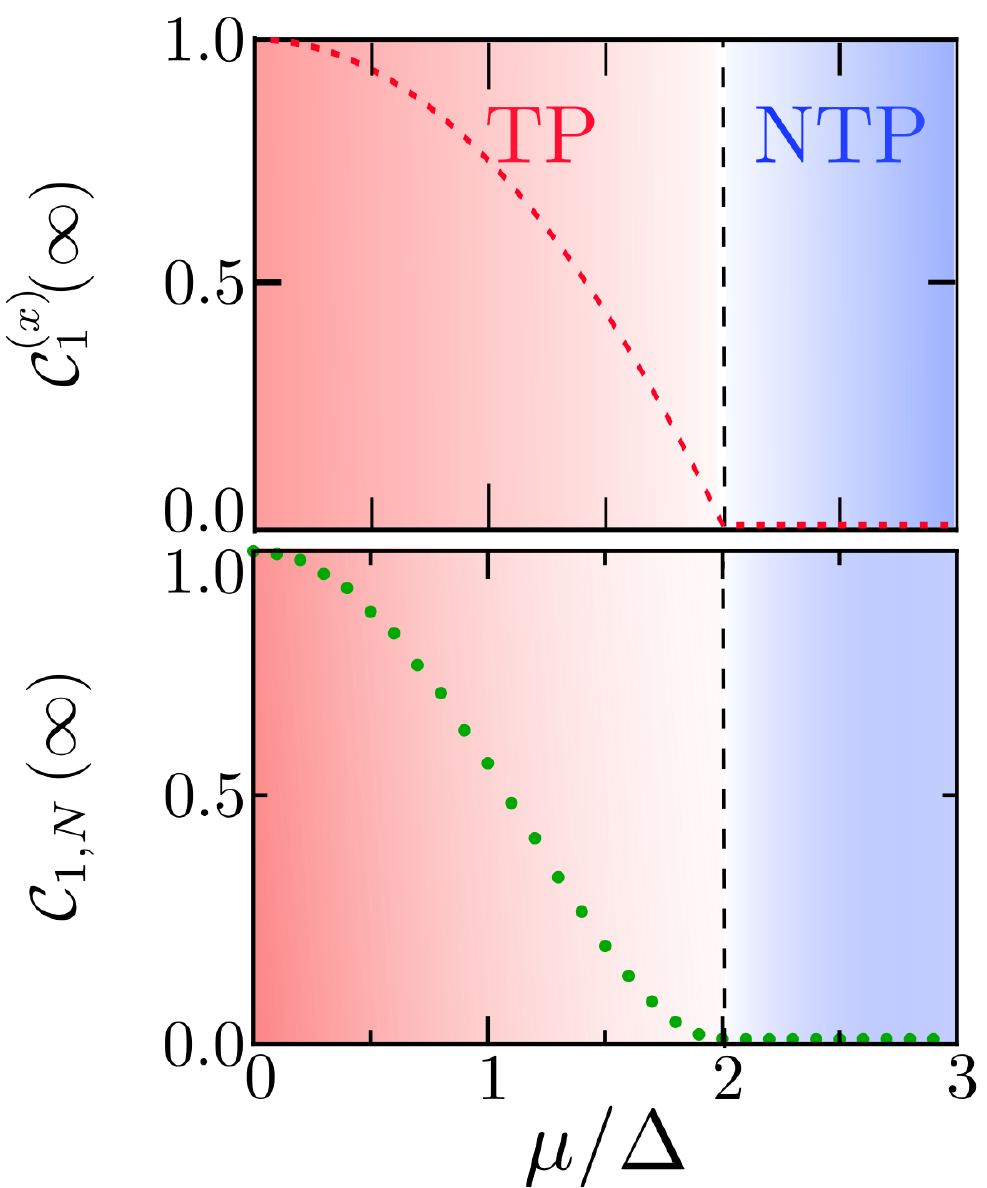}
		\caption[Long-time limits of edge \gls{STC}s as a function of $\mu/\Delta$]{Long-time limits of edge \gls{STC}s as a function of $\mu/\Delta$: top panel, single-Majorana edge $\mathcal{C}_{1}^{(x)}\left(t\right)$-\gls{STC} and bottom panel, non-local double-Majorana edge $\mathcal{C}_{1,N}\left(t\right)$-\gls{STC}. The order-parameter-like behavior exhibited by the long-time limits is evident. TP: topological phase, NTP: non-topological phase.}\label{C4_fig_3}
\end{minipage}
\hfill
\begin{minipage}{0.43\textwidth}
This order-parameter-like behavior of the \gls{STC} long-time limit is displayed in Figure~\ref{C4_fig_3}. Furthermore, by both numerical fitting as well as the exact general duality property expressed in Eq.~\eqref{naraya3}, we establish that the long-time limit of the single-Majorana edge $\mathcal{C}_{1}^{(x)}(t)$-\gls{STC} has a simple specific functional behavior given by:
\begingroup\makeatletter\def\f@size{9}\check@mathfonts
\def\maketag@@@#1{\hbox{\m@th\large\normalfont#1}}
\begin{equation}\label{orpar}
\lim_{t\to\infty}\mathcal{C}_{1}^{(x)}(t)=\begin{cases}
1-\left (\frac{\mu}{2\Delta}\right )^{2} & \text{for}\; \mu<2\Delta\\
0  &\text{for}\; \mu>2\Delta.
\end{cases}
\end{equation}
\endgroup
On the other hand, the decay of the limit value of the non-local $\mathcal{C}_{1,N}\left(t\right)$-\gls{STC} as a function of $\mu/\Delta$ has been evaluated numerically, showing a gradual transition, instead of an abrupt one, from one phase to the other. Note that these results are strictly valid for an infinitely long chain or for times below a certain limit where finite size effects could emerge, such as possible interference or revivals coming from the reflected influence of the other edge (not shown here). 
\end{minipage}				
\end{figure}
In addition, the quantum behavior of single-site \gls{STC} for the edge single- and double- Majorana qubits is similar to the $x$ and $z$ spin correlations of the transverse Ising model, and consequently its quantum critical point could also be detected by \gls{STC} measurements~\cite{Gomez_PRB2016}.
\begin{figure}[h!]
\begin{center}
  \includegraphics[width=0.9 \textwidth]{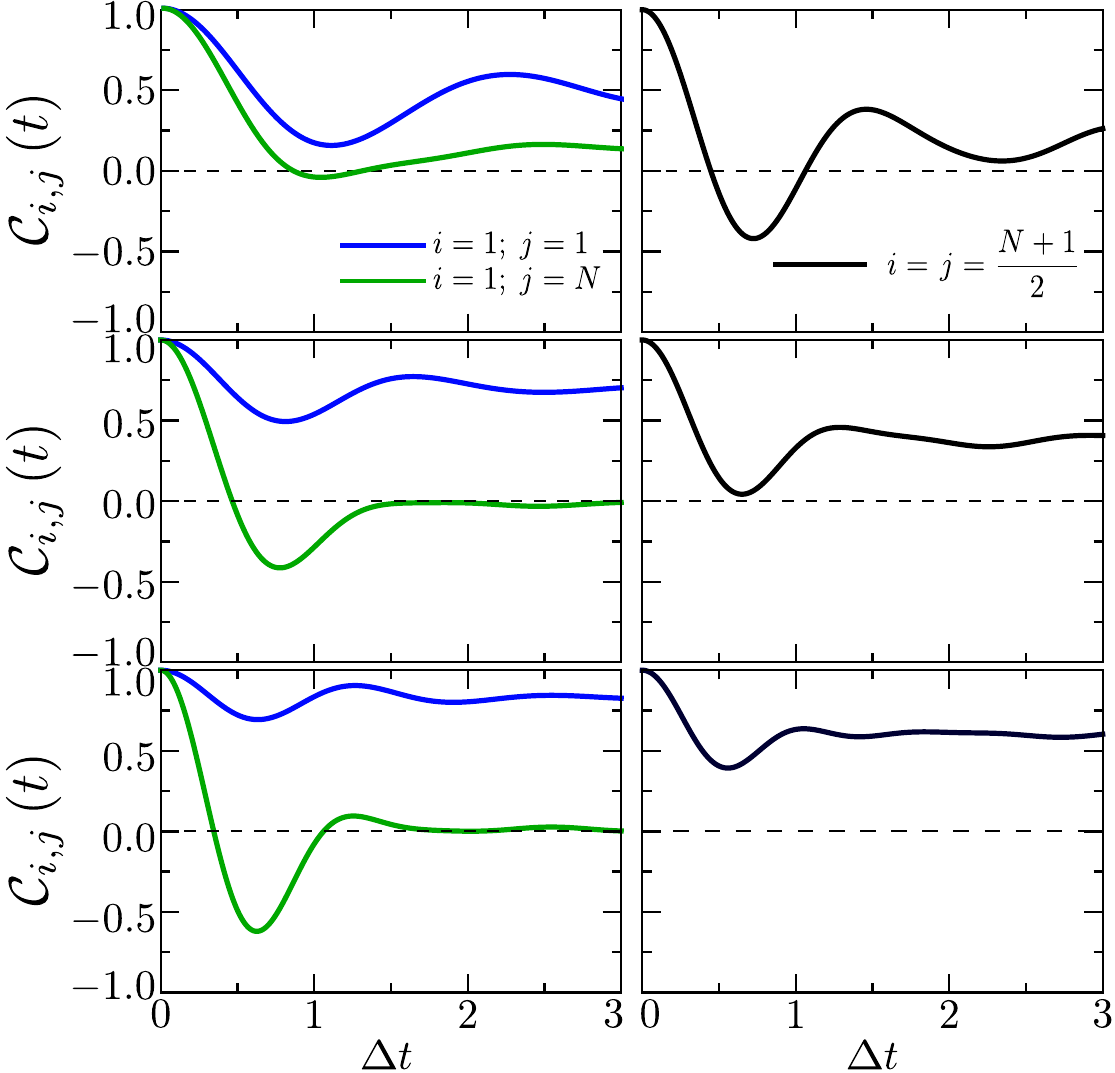}
  \caption[$\mathcal{C}_{i,j}\left(t\right)$-\gls{STC} as a function of the dimensionless time $\Delta t$ in the short- and intermediate-time regimes.]{$\mathcal{C}_{i,j}\left(t\right)$-\gls{STC} as a function of the dimensionless time $\Delta t$ in the short- and intermediate-time regimes. Left panels display the \gls{STC} for {\em two-Majorana qubits}: blue (green) line $\mathcal{C}_{1,1}\pap{t}=\mathcal{C}_{1}^{(z)}\pap{t}$ local \gls{STC} ($\mathcal{C}_{1,N}\pap{t}$ non-local \gls{STC}), respectively. Right panels illustrate the time evolution behavior of \gls{STC} for a local {\em bulk} two-Majorana qubit (middle site of the Majorana fermion chain $\mathcal{C}_{\frac{N+1}{2},\frac{N+1}{2}}\pap{t}$). The chemical potentials are $\mu/\Delta=1.5$ (upper panels), $\mu/\Delta=2.0$ (middle panels), and $\mu/\Delta=2.5$ (bottom panels). }
  \label{fig_4}
  \end{center}
\end{figure}
Finally, we end this sub-section with a comparison between edge vs. bulk \gls{STC}s. In Figure~\ref{fig_4} the short- and intermediate-time behaviors of $\mathcal{C}_{i,j}\left(t\right)$-\gls{STC} are illustrated for edge-Majorana qubits, namely the local case ${i,j}={1}$ and the non-local case ${i,j}={1,N}$, and a bulk-two-Majorana qubit ${i,j}={\frac{N+1}{2},\frac{N+1}{2}}$. We conclude that apart from a different oscillation amplitude, the local two-Majorana \gls{STC}s, either located at the edge or at a bulk site, are very similar in going to a finite long-time limit in any phase, thus not being able to detect such phase transition by looking at that specific feature. This behavior contrasts with the one offered by the two-Majorana non-local edge \gls{STC} or even, as discussed above, with that shown by the single-Majorana edge \gls{STC}. Next, we focus on the consequences of these \gls{STC}s features when assessing macroscopic quantum coherence through the Leggett-Garg inequality violations, by both local- and non-local-\gls{STC}s.
\section{Violation of \gls{LGI} in Majorana Fermion systems}
Similarly to expose in the Chapter~\ref{Cap3}, we are interesting in characterize the violation of \gls{LGI} in the Majorana Fermion context. We will focus on the following form of a \gls{LGI},
\begin{equation}
\mathcal{C}_{i,j}\left(t_{2}-t_{1}\right)+\mathcal{C}_{i,j}\left(t_{3}-t_{2}\right)-\mathcal{C}_{i,j}\left(t_{3}-t_{1}\right)\leq 1, \label{Eq:LGI1}
\end{equation}
where $\mathcal{C}_{i,j}\left(t_{\alpha},t_{\beta}\right )$ is a two-time correlation (see Eq.~\eqref{eq:Correl}) of the qubit nonlocal Majorana operator $\hat{Q}_{i,j}$ (with eigenvalues $\pm1$) between times $t_{\alpha}$ and  $t_{\beta}$, and $t_1 < t_2 < t_3$. We concentrate in the case of identical time intervals, i.e. $t_2-t_1=t_3-t_2=t$, defining a \gls{LGI} function $\mathcal{K}_{i,j}(t)$ such as \cite{Gomez_PRB2016}:
\begin{equation}\label{lgi}
\mathcal{K}_{i,j}\pap{t}=2\mathcal{C}_{i,j}\left(t\right)-\mathcal{C}_{i,j}\left(2t\right) \leq 1.
\end{equation}

\begin{figure}[h!]
\begin{minipage}{0.5\textwidth} 
\centering \includegraphics[scale=0.65]{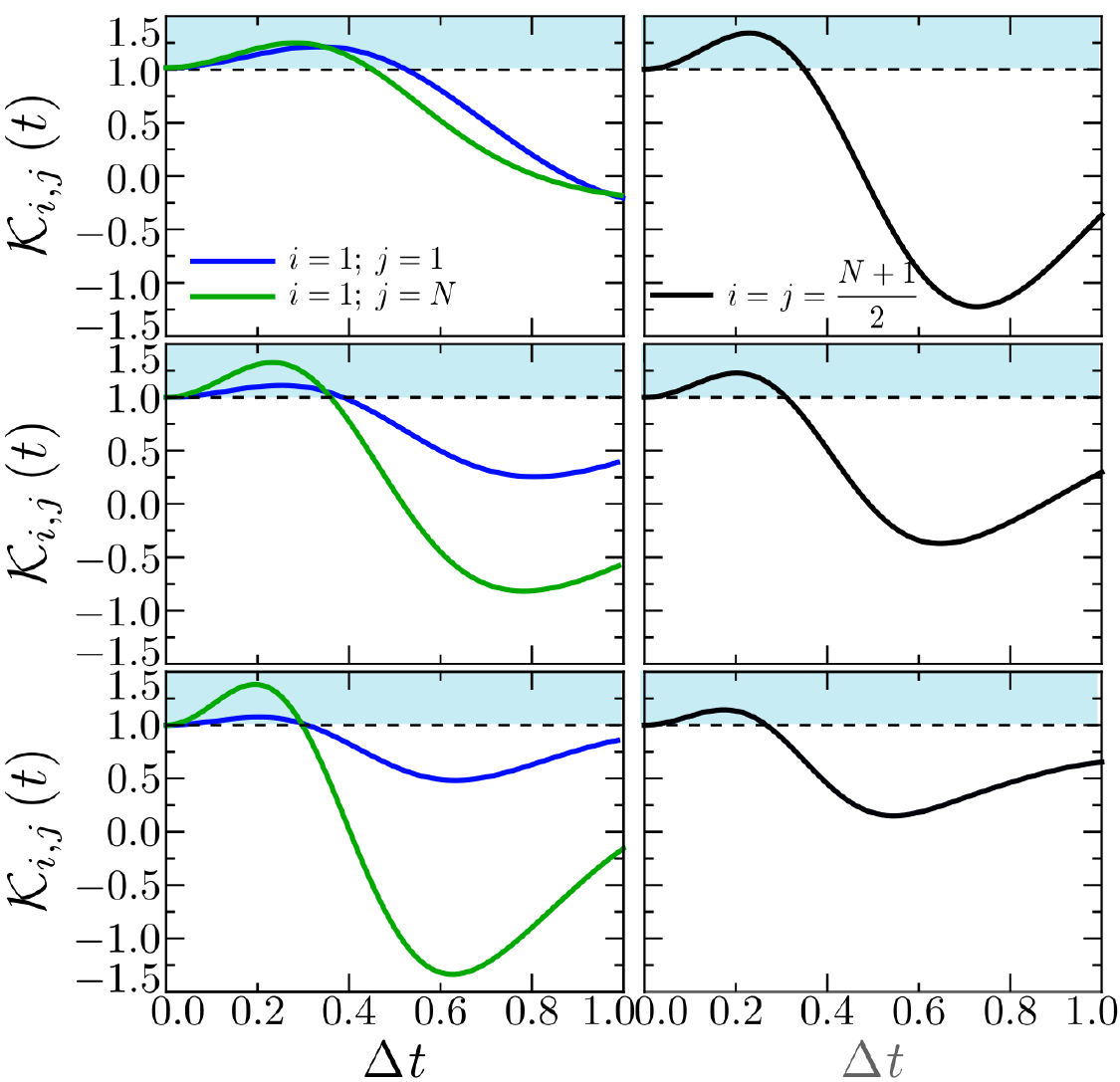}
		\caption[Two-Majorana $\mathcal{K}_{i,j}\pap{t}$ \gls{LGI} function as a function of $\Delta t$.]{Two-Majorana $\mathcal{K}_{i,j}\pap{t}$ \gls{LGI} function as a function of $\Delta t$. Panels and color lines have the same meaning as in Figure~\ref{fig_4}. The upper blue zones represent violations of the \gls{LGI} given by Eq.~\eqref{lgi}.}\label{fig_5}
\end{minipage}
\hfill
\begin{minipage}{0.43\textwidth}
Figure~\ref{fig_5} displays the evolution, as a function of $\Delta t$, of the \gls{LGI} function $\mathcal{K}_{i,j}\pap{t}$ given by Eq.~\eqref{lgi} for the same parameters as used in Figure~\ref{fig_4}. We first note that the inequality is always violated at very early times, a result that can be already understood from the $\mathcal{O}(t^2)$ expansions given in Eqs.~\eqref{otoc2} and~\eqref{otoc3}. Specifically, the $\mathcal{C}_{1}^{(x)}\pap{t}$-\gls{STC} based \gls{LGI}, denoted by $\mathcal{K}_1^{(x)}\pap{t}$, is given by
\begin{equation}
\mathcal{K}_1^{(x)}\pap{t}\simeq 1+\mu^2t^2+{\cal O}(t^4),
\end{equation}
while that based on $\mathcal{C}_{1}^{(z)}\pap{t}$, denoted by $\mathcal{K}_1^{(z)}\pap{t}$, is
\begin{equation}
\mathcal{K}_1^{(z)}\pap{t}\simeq 1+4\Delta^2t^2+{\cal O}(t^4).
\end{equation}
\end{minipage}				
\end{figure}

Thus, the initial growth of both inequality violations is captured again by the universal initial curvatures of the corresponding \gls{STC}s. Furthermore, the early-time violations for $\mathcal{K}_1^{(x)}\pap{t}$ and $\mathcal{K}_1^{(z)}\pap{t}$ become identical at $\mu=2\Delta$, i.e. the critical point. This conclusion provides an alternative route to identifying the topological phase transition.\\

\begin{figure}[h!]
\begin{center}
  \includegraphics[width=0.9 \textwidth]{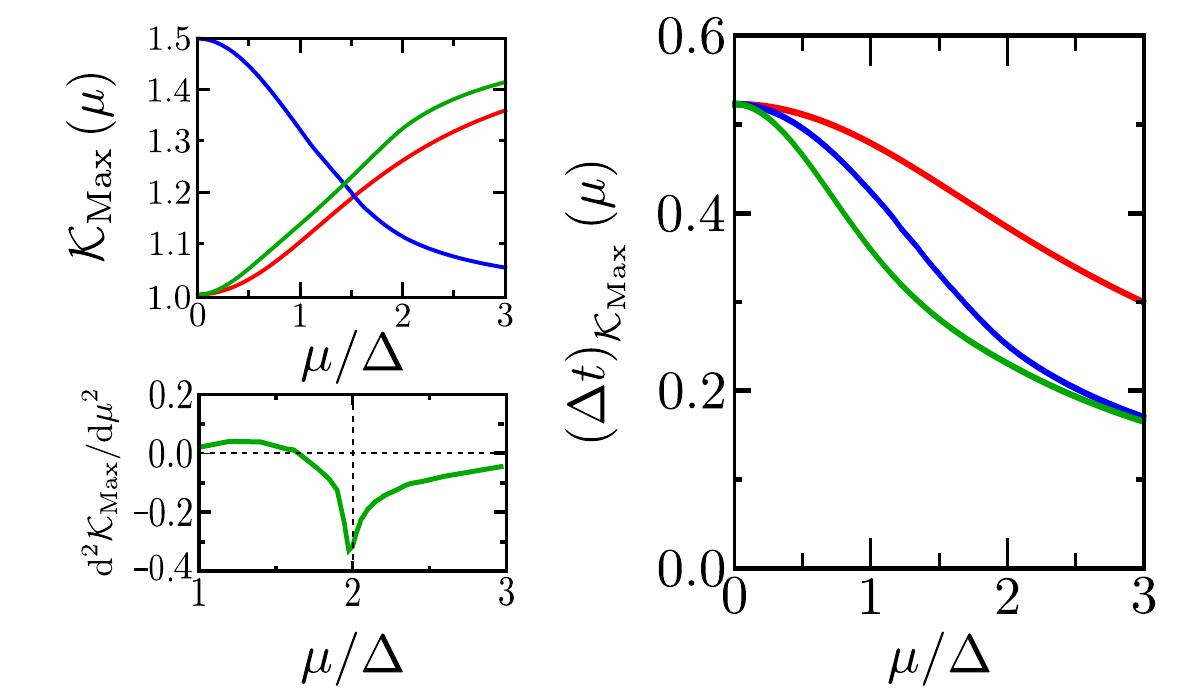}
  \caption[Critical behavior for the maximum \gls{LGI}.] {{\it Left top panel:} maximum violation of \gls{LGI} as a function of $\mu/\Delta$. {\it Right panel:} time of maximum \gls{LGI} violation as a function of $\mu/\Delta$. {\it Left bottom panel:} second derivative of the maximum \gls{LGI} violation with respect to $\mu$ showing a dip signaling the phase transition for the non-local edges two-Majorana case. Red lines depict the $\mathcal{C}_{1}^{(x)}\pap{t}$-\gls{STC} based \gls{LGI}, the blue lines represent the $\mathcal{C}_{1}^{(z)}\pap{t}$-\gls{STC} based \gls{LGI} while the green lines correspond to $\mathcal{C}_{1,N}\pap{t}$-\gls{STC} based \gls{LGI}.}
  \label{fig_6}
  \end{center}
\end{figure}

Now we consider different inequalities for longer times. It is evident that \gls{LGI} functions based on two-local-Majorana \gls{STC}s such as edge $\mathcal{C}_{1,1}\pap{t}$ and bulk $\mathcal{C}_{\frac{N+1}{2},\frac{N+1}{2}}\pap{t}$ follow a similar trend, which is very different to that of the non-local two-Majorana \gls{STC} given by $\mathcal{C}_{1,N}\pap{t}$ when crossing from one phase to the other. The local \gls{LGI} violations turn out to be stronger in the topological phase while the non-local \gls{LGI} violation increases when passing from the topological to the trivial phase. This contrasting behavior can also be seen in Figure~\ref{fig_6}, where we compare the maximum \gls{LGI} violation $\mathcal{K}_{\text{Max}}\pap{\mu}$ as a function of $\mu$ (left panel) for single- and double-Majorana qubits, as well as the times for which that maximum violation occurs $t_{\mathcal{K}_{\text{Max}}}\pap{\mu}$ for the same qubits (right panel).  Interestingly, for the non-local edges two-Majorana case, the second derivative of $\mathcal{K}_{\text{Max}}\pap{\mu}$ with respect to $\mu$ shows a dip signaling the phase transition, again an inherited feature from the corresponding time correlations $\mathcal{C}_{1,N}\pap{t}$ (see Fig.~\ref{fig_1}). Thus, we can conclude that \gls{LGI} violations by non-local Majorana qubits are sensitive to the topological features of the underlying phase, and consequently they could be explored in properly designed experimental setups.

\section{Candidates for experimental implementations} \label{sec_7}
Among the most promising candidates for experimentally detecting Majorana edge fermions in condensed matter systems are chains of magnetic atoms on superconducting surfaces~\cite{Nadj,HaltermanPRB,JianPRB} and semiconducting nanowires with large Rashba spin-orbit interaction under an applied magnetic field and induced superconductivity by proximity effects~\cite{SticletPRL,Rashba1}. Previous works have focused on local sensitive tunneling signatures of the topological phase transition in the boundary fermion occupation (Kitaev chain) or boundary spin (transverse field Ising chain).\\

In the Rashba nanowire setup Sticlet et al.~\cite{SticletPRL} define local Majorana pseudo-spins and argue that they could be measured by spin-polarized STM allowing to directly visualize the Majorana fermionic states and to test the topological character of the 1D system. On the other hand, Deng et al.~\cite{DengArxiv} reported that highly sensitive experiments have been recently conducted where the non-locality of Majorana qubits can be {\it locally} probed by a quantum dot at one end of the nanowire. These state-of-the-art experiments could evolve to develop time dependent sensitivity as required for detecting local and nonlocal \gls{STC}s. Recently, there has been great interest in contrasting distinctive signatures of spin polarization for Andreev and Majorana bound states~\cite{ZhangPRB} since when identifying topological phases effects coming from the presence of quasiparticle states inside the superconducting gap should be carefully eliminated~\cite{ChienPRB}. Thus, it is most desirable to have additional signatures available (besides tunneling conductance signatures of Majorana fermions) that would allow one to identify the topological phase transition. It has been proposed in~\cite{SticletPRL} and~\cite{DengArxiv} how to distinguish such differences between Andreev vs. Majorana signatures by accessing true nonlocal features. In this way, our results as given by the behavior of local $\mathcal{C}_{1}^{(x)}\left(t\right)$ and most importantly by nonlocal $\mathcal{C}_{1,N}\left(t\right)$, and their \gls{LGI} combinations, should be relevant for extending that kind of search of true Majorana behavior.\\

Furthermore, recently, spin noise spectroscopy has been shown as a powerful tool to experimentally accessing the autocorrelation function~\cite{NikolaiRPP,BurnellPRB2}. The universal short-time behavior described by Eqs.~\eqref{otoc1} and~\eqref{otoc2} could be exploited in spin fluctuation measurements as an alternative route to get information about the dynamics~\cite{BurnellPRB}. Such rich variety of behaviors would also permit the study of temporal effects as well as different kind of susceptibilities, through their Fourier transform equivalents, in topological quantum computing settings.\\

Therefore, in light of recent experiments, we demonstrate in the present work that \gls{STC} and \gls{LGI} behaviors exhibit a quantum-phase sensitive signature due to the appearance of zero-energy-modes in the topological phase that will manifest themselves in the long-time behavior of both local as well as nonlocal qubit \gls{STC}s. This provides an experimentally useful diagnostic tool to detect topological phase transitions.\\

\section{Conclusions} \label{sec_8}
In summary, we have provided evidence that time correlations and violations of \gls{LGI} establish new testable signatures of topological phase transitions. The behavior of that sort of inequality is a direct consequence of time correlations in local and nonlocal Majorana qubits. Specifically, we have identified signatures of the \gls{MFC} topological phase transition in any of three time domains: (i) in the short-time limit we found universal features such as the out-of-time-ordered correlation and a dip in the second $\mu$-derivative marking the phase transition; (ii) in the intermediate time region, the \gls{LGI} violations are sensitive to the quantum phase the system is; and (iii) in the long-time limit, the asymptotic values of single- and double-Majorana edge \gls{STC}s act as order-parameter-like indicators. Specifically, we propose that a measurement of the long time saturation value of the local edge $\mathcal{C}_{1}^{(x)}$-\gls{STC} as well as the non-local edge $\mathcal{C}_{1,N}$-\gls{STC} provide a witness of the topological ($\neq 0$) vs. non-topological ($=0$) phase transition of Majorana fermion chain systems, as it probes directly the existence of zero energy modes. Additionally, in the former case it gives direct access to the electron-hole weight of such modes. The results are especially relevant because the whole question of quantum coherence in complex mesoscopic systems is taking up a new impulse in the community and is of interest to researchers not only in quantum information and foundations but also in condensed matter.

\begin{savequote}[45mm]
``I... a universe of atoms, an atom in the universe."
\qauthor{Richard P. Feynman}
\end{savequote}
\chapter[Manipulating quantum coherences in driven \gls{DM}]{Manipulating quantum coherences in driven Dicke Model}
\begin{center}
\begin{tabular}{p{15cm}}
\vspace{0.1cm}
\quad \lettrine{\color{red1}{\GoudyInfamily{I}}}{n} this chapter, we begin to explore the non-equilibrium dynamics of a canonical light-matter system, namely the Dicke model, when the light-matter interaction is ramped up and down through a cycle across the quantum phase transition. Our calculations reveal a rich set of dynamical behaviors determined by the cycle times, ranging from the slow, near adiabatic regime  through to the fast, sudden quench regime. As the cycle time decreases, we uncover a crossover from an oscillatory exchange of quantum information between light and matter that approaches a reversible adiabatic process, to a dispersive regime that generates large values of light-matter entanglement. The phenomena uncovered in this work have implications in quantum control, quantum interferometry, as well as in quantum information theory.\\
\\
This chapter is published in references~\cite{Gomez_Entropy2016}:  {\bf F. J. G\'omez-Ruiz}, O. L. Acevedo, F. J. Rodr\'iguez, L. Quiroga, and N. F. Johnson. {\it Quantum hysteresis in coupled light-matter systems.} Entropy, {\bf 18(9)}, 319 (2016).
\end{tabular}
\end{center}
\section{Introduction}
In the last two decades, there have been several breakthroughs in the experimental realization of systems that mimic specific many-body quantum models~\cite{Han2013}. This is especially true in systems involving aggregates of real or artificial atoms in cavities and superconducting qubits~\cite{ExpDicke1,ExpDicke2}, as well as trapped ultra-cold atomic systems~\cite{bloch2012nat,schneider2012rpp,georgescu2014rmp}. These advances have stimulated a flurry of theoretical research on a wide variety of phenomena exhibited by these systems, such as quantum phase transitions (\gls{QPT}s)~\cite{sachdev, aeppli}, the collective generation and propagation of entanglement~\cite{amico2002nature, wu2004prl, RomeraPLA2013, Reslen2005epl, Acevedo2015NJP, AcevedoPRA2015}, the development of spatial and temporal quantum correlations~\cite{sun2014pra,FernandoPRB2016}, critical universality~\cite{Acevedo2014PRL}, and finite-size scalability~\cite{VidalEPL,Oct_PS2013,CastanoPRA2011,CastanoPRA2012}. All of these topics have implications for quantum control protocols which are in turn of interest in quantum metrology, quantum simulations, quantum computation, and quantum information processing~\cite{Gernot2006,Rey2007,Dziarmaga10,Hardal_CRP2015,NiedenzuPRE2015}. \\
\\
Future applications in the area of quantum technology will involve exploiting -- and hence fully understanding -- the {\em non-equilibrium} quantum properties of such many-body systems. Radiation-matter systems are of particular importance: not only because optoelectronics has always been the main platform for technological innovations, but also in terms of basic science because the interaction between light and matter is a fundamental phenomenon in nature. On a concrete level, light-matter interactions are especially important for most quantum control processes, with the simplest manifestation being the non-trivial interaction between a single atom and a single photon~\cite{AgarwalPRL1984}. One of the key goals of experimental research is to improve both the intensity and tunability of the atom-light interaction~\cite{WillPRL2016, BegleyPRL2016}. Unfortunately the relatively weak dipolar coupling between an atom and an electromagnetic field makes it difficult to obtain a large light-matter interaction, even when atoms are constrained to interact with a single radiation mode in a cavity. In recent years, some condensed matter systems have offered an alternative to the traditional atom-cavity implementation. Clear spectroscopic evidence has recently been presented that a charged Josephson qubit coupled to a superconductor transmission line, behaves like an atom in a cavity~\cite{XiuPRA2016} and that the dipolar coupling between these systems is $3$ to $4$ orders of magnitude greater than that in atomic systems. This type of system, known as a superconducting qubit, enables the study of effective two-level atoms interacting with a quantized single-mode electromagnetic field~\cite{StefanoPRA2016, LijunPRA2016} and allows the exploration of new regimes of strong coupling~\cite{HerreraPRL2016}. Another very successful approach to obtain strongly interacting light-matter systems has emerged in experiments involving ultra-cold trapped atoms or ions~\cite{ExpDicke1,ExpDicke2}. In this case, discrete translational degrees of freedom (vibrational modes) emerging from the optical trap are used to couple the atoms to the radiation mode. Thanks to the extremely low temperatures, the light-matter coupling effectively becomes the dominant interaction, once again allowing the exploration of a wide variety of strong-coupling phenomena.\\
\\
One important consequence of such strong coupling, is that the atom-light interaction can effectively become all-to-all, in the sense that all atoms are equally coupled to the radiation. In this case, one of the simplest and yet richest scenarios involves instances in which the Dicke model (\gls{DM}) is realized~\cite{Dicke_PR}. One of the most striking and important features of the Dicke model is the fact that it exhibits a superradiant second order \gls{QPT} in the thermodynamic limit~\cite{Lieb1973}. Despite more than sixty years of existence, this model has recently attracted renewed interest thanks to major experimental breakthroughs in terms of its realization and exploration~\cite{ExpDicke4,KlinderPNC2015,GuerinPRL2016}. This has in turn fueled a surge in theoretical investigations of the \gls{DM}, including further detailed proposals for its realization~\cite{WangPRA2016}. Regardless of this surge in theoretical interest, however, much of the focus has been on the \gls{DM}'s static properties or equilibration schemes, leaving many aspects of its non-equilibrium evolution as an open problem.\\
\\
In previous work, we had attempted to advance understanding of the \gls{DM}'s dynamics by exploring the effects of crossing the \gls{QPT} using a tuned interaction, hence taking the system in a single sweep from a non-interacting regime into one where strong correlations within and between the matter and light subsystems play an essential role~\cite{AcevedoPRA2015,Acevedo2015NJP,Acevedo2014PRL}. Our previous analyses also revealed universal dynamical scaling behavior for a class of models concerning their near-adiabatic behavior in the region of a \gls{QPT}, in particular the Transverse-Field Ising model, the \gls{DM} and the Lipkin-Meshkov-Glick model. These findings, which lie beyond traditional critical exponent analysis like the Kibble-Zurek mechanism~\cite{Zurek96a,Zurek96b,Zurek96c,Kibble76b} and adiabatic perturbation approximations, are valid even in situations where the excitations have not yet stabilized -- hence they provide a time-resolved understanding of \gls{QPT}s encompassing a wide range of near adiabatic regimes. \\
\\
In this work, instead of a single crossing of the \gls{QPT}, we analyze the effects of driving the system through a round trip across the \gls{QPT}, by successively ramping up and down the light-matter interaction so that the system passes from the non-interacting regime into the strongly interacting region and back again. We restrict ourselves to the case of a closed \gls{DM} such that a description of the temporal evolution using unitary dynamics is sufficient. Depending on the time interval within which the cycle is realized, we find that the system can show surprisingly strong signatures of \emph{quantum hysteresis}, i.e. different paths in the system's quantum state evolution during the forwards and backwards process, and that these memory effects vary in a highly non-monotonic way as the round-trip time changes. The adiabatic theorem ensures that if the cycle is sufficiently slow, the process will be entirely reversible. In the other extreme, where the round-trip ramping is performed within a very short time, the total change undergone by the system is negligible. However in between these two regimes, we find a remarkably rich set of behaviors.\\
\\

\section{Quantum Hysteresis in the Dicke Model}\label{Hyste}

Our central objective in this section is to address the effects of cyclically varying the radiation-matter interaction $\lambda(t)$ as a function of time. We will focus on a particularly simple, piecewise linear form for the time dependence:
\begin{equation}\label{ciclo}
\lambda(t)=\begin{cases}
\lambda_{1}+\frac{2\left(\lambda_2-\lambda_1\right)}{\tau}t,& t\leq \tau/2 \\
\lambda_{2}+\frac{2\left(\lambda_1-\lambda_2\right)}{\tau}t, & t>\tau/2 \:,
\end{cases}
\end{equation}
where $\lambda_1$ and $\lambda_2$ are respectively zero and one, and with $\ome=\eps=1$ in Eq.~\eqref{hdic}. Hence $\lambda(t)$ has a triangular profile. The slope of the two portions of the cycle $\pm\nu$ is characterized by a finite time $\tau$ such that $\nu= 2/ \tau$, where $\nu$ is known as the  {\it annealing velocity}. The strongly interacting regime is reached when $\lambda(t)\approx 1$. The particular choice of $\lambda(t)$ given by Eq.~\ref{ciclo}, implies that the quantum critical point is crossed twice during the cycle, first when $t\approx\tau/4$, and second when $t\approx3\tau/4$. From Eq.~\eqref{Hdicke} and Eq.~\eqref{ciclo}, we get the full \gls{DM} instantaneous state $\left|\psi(t)\right.\rangle$ by numerically solving the time-dependent Schr\"{o}dinger equation. The initial state when $t=0$ and $\lam=0$ is the non-interacting ground state, namely $\left|\psi(0)\right.\rangle=\bigotimes_{i=1}^{N}\left| \downarrow\right.\rangle\otimes\left| n=0\right.\rangle$ where both matter and light subsystems have zero excitations. All qubits are polarized in the state $\sigma_z=-1$, and the field is in the Fock state of zero photons. The \gls{DM} solution is significantly harder to obtain. Nevertheless, we have used the fact that the total angular momentum  is a conserved quantity, therefore the dynamics will lie in the $J=N/2$ subspace. The success of this solution was tested by extending the truncation limit and then checking that the results do not change (convergence test). The most naive application of this solution to the \gls{DM} would be to work with vectors of the form $\left|m\right.\rangle_{z}\otimes \left|n\right.\rangle$, where the first one is an eigenvector of $J_z$ in the subspace of even parity and the last one is a bosonic Fock state.\\
\\
To provide a first step toward understanding the complexity of the quantum hysteresis results that we generate for the \gls{DM}, there exists a model that constitutes arguably the simplest version of what happens to a quantum system when it crosses a quantum critical point driven by a time-dependent Hamiltonian. This is the Landau-Zener model (\gls{LZM}) and we will spend the rest of this section reviewing its key properties concerning the probability that the system transitions out of its ground state, together with the Landau-Zener-St\"uckelberg process which helps understand the presence of oscillatory features in our results. The Landau-Zener model is represented by a two-level system following the Hamiltonian~\cite{Landau, Zener}

\begin{equation}\label{hlandau}
\hat{H}_{{\rm LZM}}=-\frac{\Delta_{0}}{2}\hat{\sigma}_{x}+\frac{\lambda_{0}-\lambda(t)}{2}\hat{\sigma}_{z}\ \ .
\end{equation}
The energy-gap between the ground state and the excited state is $\Delta(t)=\sqrt{\Delta_{0}^{2}+\left(\lambda(t)-\lambda_{0}\right)^{2}}$. At the \gls{QCP} $\lambda=\lambda_{0}$, the system reaches its minimal energy-gap $\Delta=\Delta_{0}$. In this chapter, we are considering the light-matter interaction $\lambda(t)=\nu t$ during the ramping up, and a similar form in the ramping down, where $\nu$ is an annealing velocity and where the system starts from its ground state, i.e. $\vert0\rangle$.  It is known from previous work on the \gls{LZM} that for a two-level system starting from its ground state at $t=-\infty$, the probability of it ending in its excited state at $t=\infty$ is given by  $ P_{{\rm LZM}}=\exp\left(-\frac{\pi\Delta_{0}^{2}}{\nu}\right)$ where the ratio $\zeta=\frac{\Delta_{0}^{2}}{\nu}$ is called the adiabatic parameter. It is worth noticing that when $\nu$ is very small, we have zero probability for the system to jump to the excited state. Therefore, the limit $\nu\to 0$ corresponds to perfect adiabatic evolution. Thus the parameter $\zeta$ allows control of the probability for the system to perform either an adiabatic or diabatic transition.

\begin{figure}[t!]
\begin{center}
\includegraphics[scale=0.9]{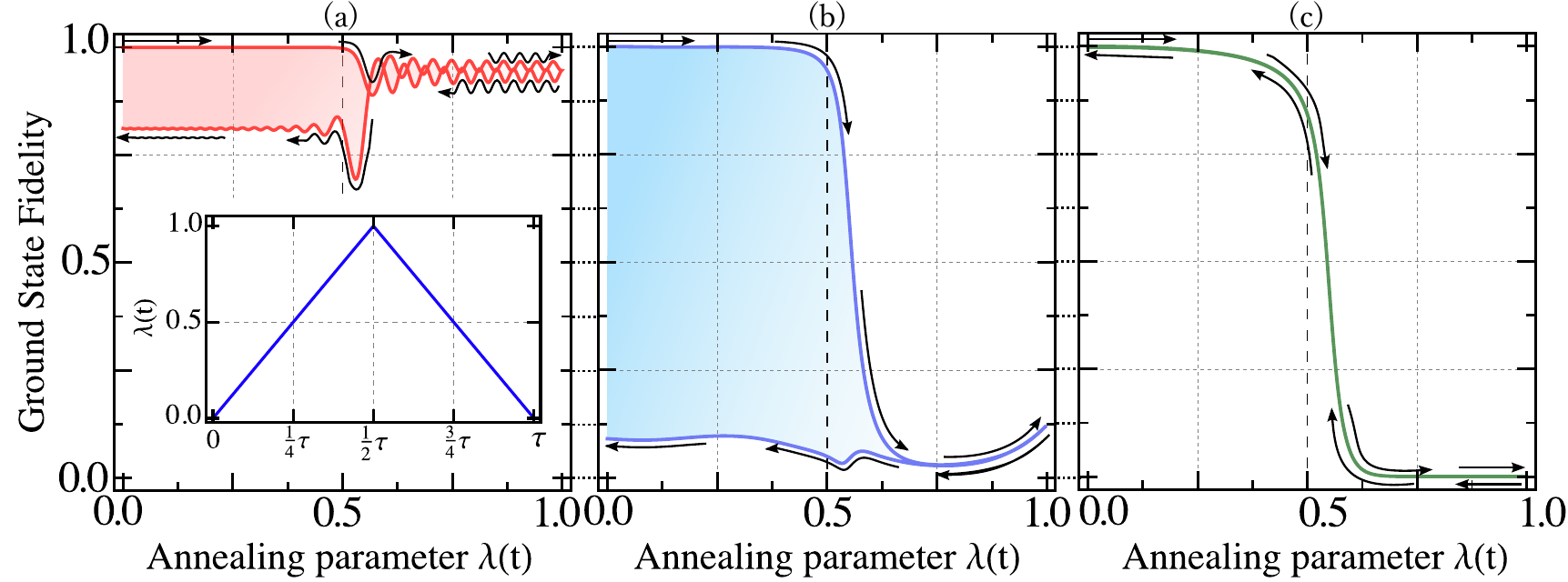}
\caption[Quantum hysteresis, in driven \gls{DM}, as measured by ground state fidelity.] {\label{Fide}  Quantum hysteresis as measured by ground state fidelity $\vert \langle \varphi^{GS}_{Ins} \left(\lambda(t)\right)\vert \psi(t)\rangle \vert$ as a function of $\lambda(t)$ when the interaction parameter performs a cycle as specified by Eq.~\eqref{ciclo}. The \gls{DM} system size is $N = 33$. The cycle time $\tau=2/\nu$ in each case is characterized by an annealing velocity $\nu$:  {\bf (a)} $Log_{2} \left(\nu\right)=-8.46$, {\bf (b)} $Log_{2} \left(\nu\right)=-4.46$, and {\bf (c)} $Log_{2}\left(\nu\right)=2.94$. Hence the annealing velocity $\nu$ increases from {\bf (a)} to {\bf (c)}. Inset in panel {\bf(a)} shows the time profile of the annealing parameter cycle $\lambda(t)$ specified by Eq.~\eqref{ciclo}.
}
\end{center}
\end{figure}

One interesting aspect of the \gls{LZM} is that much of its dynamics is determined during the short interval during which the minimum-gap is crossed. For this reason, an \gls{LZM} can be seen as analogous to a beam-splitter~\cite{HuangPRX}, since the probability $P_{{\rm LZM}}$ for the system to stay in state $\vert 0 \rangle$ can be seen as equivalent to a transmission coefficient which characterizes the probability for a beam to go through a partially reflecting mirror. The analogy with the beam-splitter can be extended to one of interest to this chapter, in which the evolution of $\lambda(t)$ is reversed and the critical gap is crossed again. This complete cycle is known as a Landau-Zener-St\"uckelberg (\gls{LZS}) process. At the end of a \gls{LZS} cycle, the probability of staying in the $\vert 0 \rangle$ state can be approximated by the following formula~\cite{HuangPRX}:
\begin{equation}\label{ProLZS}
P_{{\rm LZS}}=4P_{{\rm LZM}}\left(1-P_{{\rm LZM}}\right)\sin\left(\theta_{12}-\Phi_{S}\right),\qquad\text{with}\quad \theta_{12}=\int_{t_{1}}^{t_{2}}\Delta(t)\text{d}t.
\end{equation}
The times $t_1$ and $t_2$ define the interval between the two crossings of the gaps. The phase $\Phi_{S}$ is called the Stokes phase and it is determined entirely by the form of the minimum gap -- hence it does not depend on the annealing  velocity $\nu$. On the other hand, the dynamical phase $\theta_{12}$ is inversely proportional to $\nu$; or equivalently, it is directly proportional to the total time of the cycle $\tau \propto 2\nu^{-1}$. As a result, Eq.~\eqref{ProLZS} implies that one should expect oscillatory behavior, with respect to $\tau$, for the probability of finishing the cycle in the same state as which the cycle was started. Such oscillatory behavior is known as {\em St\"uckelberg oscillations}~\cite{Stuckel}. Despite its simplicity, the \gls{LZM} problem has found an enormous range of applications in various experimental situations. Also, some generalizations of its concepts can be performed in order  to tackle the dynamics in situations involving more than two levels~\cite{AtlandPRL, AtlandPRA, GuozhuNC}. Whether it can be extended to provide a full, formal description of the \gls{DM} hysteresis results studied here, remains an open challenge.\\
\\
Although the \gls{LZM} and \gls{LZS} can serve as a guide for understanding the full numerical results of the \gls{DM}, the complexity of the \gls{DM}'s non-equilibrium quantum dynamics involves many more than two energy levels. For any value of $\lambda(t)$, there is a set of instantaneous eigenstates $\vert\varphi_{n}\left(\lambda(t)\right)\rangle$. If $\vert\psi(t)\rangle$ represents the actual dynamical state, we can express the probability of being in the instantaneous eigenstate $n$ as follows:
\begin{equation}
P_n \left(t\right) = \left[F\left(\varphi_{n}\left(\lambda(t)\right) , \psi(t)\right)\right]^{2}
\end{equation}
where $F\left(\varphi_{n}\left(\lambda(t)\right) , \psi(t)\right)=\vert \langle \varphi_{n} \left(\lambda(t)\right)\vert \psi(t)\rangle \vert$ is the instantaneous fidelity of the $n$ state.  We consider the ground state fidelity $F(t)=\vert \langle \varphi^{GS}_{Ins} \left(\lambda(t)\right)\vert \psi(t)\rangle \vert$ as a reference quantity for characterizing the quantum hysteresis, where $\varphi^{GS}_{Ins} \left(\lambda(t)\right)$ is the ground state of the Hamiltonian that corresponds to time $t$. In Fig.~\ref{Fide} we plot the dependence of the function $F(t)$ for different values of $\nu$. Despite the fact that the finite-size \gls{DM} has no true \gls{QPT}, the system dynamics reveal significant differences between what would be the normal phase $\left(\lambda(t)<\lambda_c\right)$ and the superradiant one. This manifests itself in the curves since all functions $F(t)$ start to depart from their initial value of unity after a threshold near the \gls{QCP} is crossed. As can be deduced from Fig.~\ref{Espectro}{\bf (b)}, the critical point of the finite-size model is slightly above the TL critical value, displaced to the right in the plot. The crossing of the phase-boundary also manifests itself in the return stage of the cycle. In general, one can see that the normal phase tends to stabilize the behavior while the superradiant phase is the one in which most changes emerge.

\begin{figure}[t!]
\begin{center}
\includegraphics[scale=1.1]{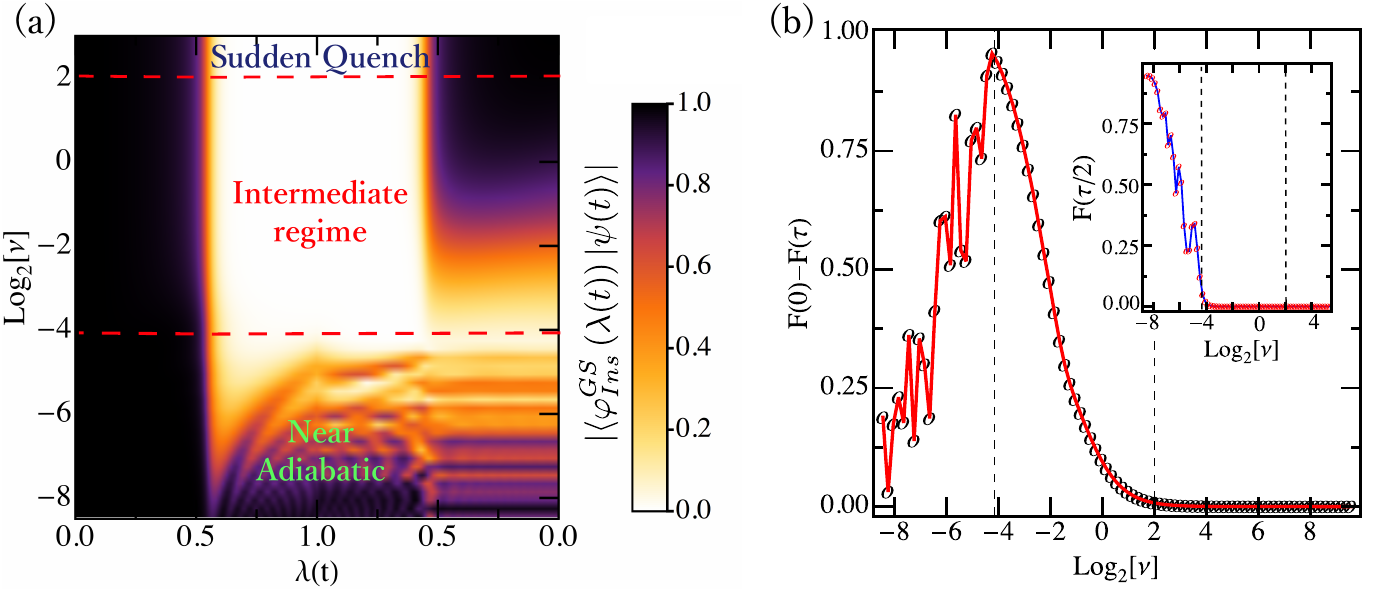}
\caption[Quantum hysteresis profiles as measured by the ground state  fidelity as a function of $\lambda(t)$.] {\label{Fide2}  {\bf (a)} Quantum hysteresis profiles as measured by the ground state  fidelity $\vert \langle \varphi^{GS}_{Ins} \left(\lambda(t)\right)\vert \psi(t)\rangle \vert$ as a function of $\lambda(t)$. The dashed lines are a guide to the eye. {\bf (b)} Difference between the values of the ground state fidelity at the start time $t=0$ and end time $t=\tau$.  Inset  in Fig.~\ref{Fide2} {\bf(b)} shows the values of the ground state fidelity at the end time $t=\tau/2$. The continuous lines are a guide to the eye. In {\bf(a)} and {\bf(b)} the interaction parameter performs the cycle specified by Eq.~\eqref{ciclo}. The profile show the existence of the three regimes of annealing velocity: {\it (i)} near adiabatic, {\it (ii)} intermediate regime, and {\it (iii)} sudden quench. The \gls{DM} system size is $N=33$.}
\end{center}
\end{figure}

Very low values of $\nu$ (or equivalently, very high values of $\tau$) correspond to almost zero probability of exciting the system. Therefore the system essentially follows the instantaneous eigenstate during its entire time evolution. This regime can hence be labelled the near adiabatic limit: see Fig.~\ref{Fide}{\bf (a)}. In the other extreme of very short cycle times, the system is simply not able to respond to the change of the Hamiltonian. Hence it remains frozen in the initial state. This is the sudden quench limit: see Fig.~\ref{Fide}{\bf (c)}. In this limit, the decrease in the fidelity is due to the differences  between the instantaneous ground state and the starting one as predicted by the \gls{QPT}. Hence this change is concentrated around the \gls{QCP}. In between these two limits, strong memory effects occur as shown in Fig.~\ref{Fide}{\bf (b)}. The trajectories enclose an area that can be referred to as a signature of quantum hysteresis behavior. Recent experimental realizations of the \gls{DM}~\cite{KlinderPNC2015} show results of such possible effects.\\
\\
In order to better understand the transition between the near adiabatic and sudden quench limits, we present in Fig.~\ref{Fide2}{\bf (a)} a dynamical profile of the time evolution of the ground state fidelity for a wide range of annealing velocities. It is clear that there is a non-monotonic path between the $\nu \to 0$ limit (i.e. logarithm of $\nu$ is negative) and $\nu \to \infty$ limit (i.e. logarithm of $\nu$ is positive) . In Fig.~\ref{Fide2}{\bf (b)} we are able to clearly distinguish three regimes based on the behavior of the final fidelity as a function of $\nu$. These are {\it (i)} near adiabatic, {\it(ii)} intermediate regime, and {\it (iii)} sudden quench. Complex oscillations arise as a function of $\nu$ in the near adiabatic regime. This is a many-body version of the St\"uckelberg oscillations as defined in Eq.~\eqref{ProLZS}. The amplitude of being either in the ground or the first excited state, accumulates a dynamical phase until the system returns to the \gls{QCP}, at which point these two channels interfere with each other and hence form the oscillatory pattern seen in the left part of Fig.~\ref{Fide2}{\bf (b)}. As is also evidenced in Fig.~\ref{Fide}{\bf (a)}, the time interval $\tau/4 <t <3\tau/4$ (i.e. when the system is in the superradiant phase) is dominated by oscillations that tend to disappear as the annealing velocity increases. These oscillations are restricted to the superradiant phase because it is only in this interval that there is a non-negligible transition amplitude between the ground and first excited state, due to a significant change of the ground state as a function of $\lam$. The near adiabatic region is the closest to a \gls{LZS} cycle in the sense that only the ground state and first excited state of the \gls{DM} are significantly excited, and hence a two-level approximation is feasible. That is why both the St\"uckelberg oscillations and the superradiant phase oscillations are only relevant for slower cycles. For faster cycles, part of the evolution information leaks to higher excited states so that the simplified \gls{LZS} scenario is no longer valid.\\
\\
Notwithstanding the oscillatory behavior, the near-adiabatic regime has a general tendency to show an increase in memory effects as the cycles get faster, which is evidenced by the discrepancy between the initial and final fidelities in Fig.~\ref{Fide2}{\bf (b)}. However this tendency has an upper limit, after which the intermediate regime begins. This intermediate regime is characterized by a monotonic decrease of the difference $F(0)-F(\tau)$ as the annealing velocity increases. As can be seen in the inset of Fig.~\ref{Fide2}{\bf (b)}, in this intermediate regime the system loses any ability to follow the instantaneous ground state during the superradiant phase, but somehow manages to have a finite probability of remaining in the ground state of the normal phase after the cycle is completed. This can be interpreted as the system being significantly quenched only during its passage to the superradiant phase, in a process that cannot longer be approximated as an \gls{LZM}. In previous works, we have shown that this process is fundamentally a squeezing mechanism in both subsystems, followed by a generation of light-matter entanglement~\cite{AcevedoPRA2015,Acevedo2015NJP}. This process becomes increasingly irrelevant in terms of being able to change the initial state as $\nu$ increases in the intermediate regime, since the system has less and less time to undergo any changes. This explains the monotonic tendency toward reduced hysteresis effects as the sudden quench limit is reached.\\
\\
\begin{figure}[t!]
\begin{center}
\includegraphics[scale=1.1]{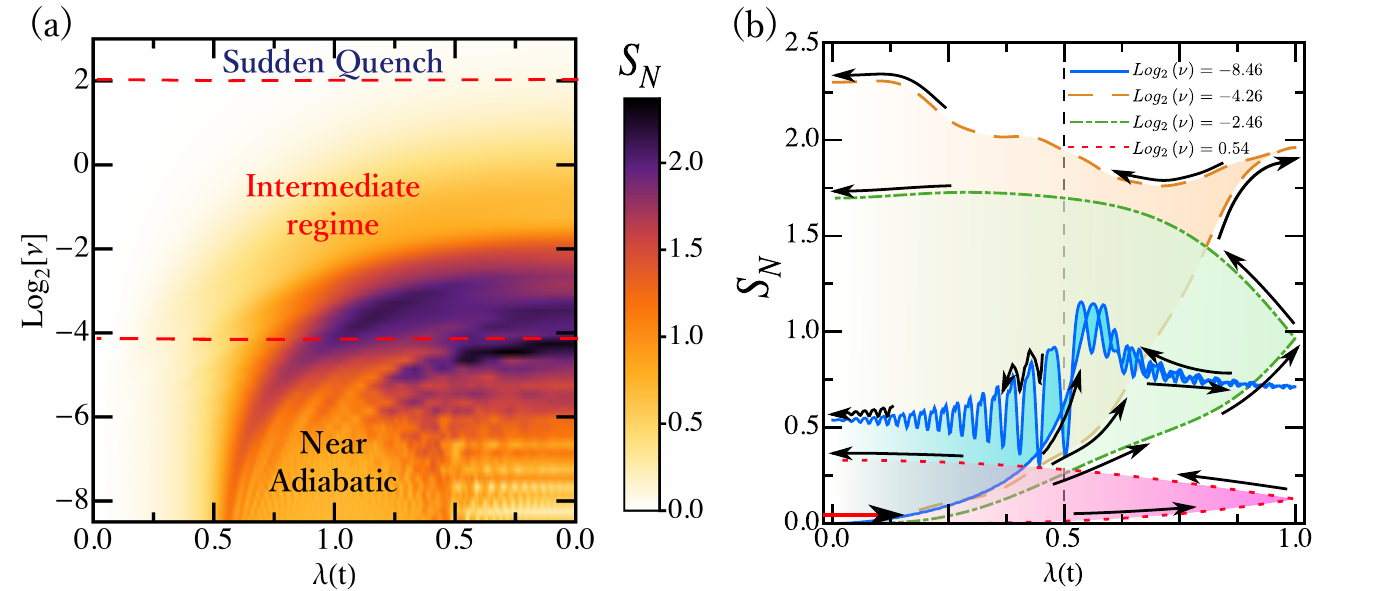}
\caption[ Quantum hysteresis profile analogous to that of Fig.~\ref{Fide2}{\bf (a)} but now depicting the time evolution of the von Neumann entropy $S_N$.] {\label{fentro}  {\bf (a)} Quantum hysteresis profile analogous to that of Fig.~\ref{Fide2}{\bf (a)} but now depicting the time evolution of the von Neumann entropy $S_N$ as defined in Eq.~\eqref{eqentro}. This quantity is an indicator of the entanglement between the light and matter subsystems. The dashed lines are a guide to the eye. {\bf (b)} Quantum hysteresis curves for $S_N$, illustrating the behavior of light-matter entanglement in the different dynamical regimes.}
\end{center}
\end{figure}
In addition to the ground state fidelity, we have analyzed the quantum hysteresis by looking at the light-matter entanglement generated by the cycle in terms of von Neumann entropy. Given a subsystem $A$, the von Neumann entropy is defined as:
\begin{equation} \label{eqentro}
S_N=-\tr{\rhoo_A\log \pap{\rhoo_A }}\:,\:\:\:\rhoo_A=\mathrm{tr}_B\pac{ \left|\psi \right.\rangle \rangle \left.\psi \right|}
\end{equation}
where $B$ is the complementary subsystem and the total system is in a total state $\left|\psi\right.\rangle$ that is pure. When the total system is in such a pure state, the entropy of subsystem $A$ is equal to the entropy of its complementary subsystem $B$, and this quantity $S_N$ is a measure of the entanglement between both subsystems. The natural choice for such a bipartition of the \gls{DM} is where one subsystem is the light (i.e. the radiation mode) and the other subsystem is the matter (i.e. the set of $N$ qubits). Figures~\ref{fentro}{\bf (a-b)} show results for the von Neumann entropy for the cases discussed so far in this chapter for ground state fidelity. Since the \gls{DM} is a closed system (i.e. a pure quantum state with a unitary evolution), the increase of $S_N$ in each subsystem is synonymous with an irreversible interchange of information between the light and matter during the cycle, hence providing a more direct thermodynamical interpretation for the memory effects of the cycle.\\
\\
In Figs.~\ref{fentro}{\bf (a-b)}, the near adiabatic regime shows a new aspect of interest: the von Neumann entropy is not always increasing over time, which means that for slow annealing velocities, information is not always dispersing from light to matter and vice versa. Instead, there is some level of feedback for each subsystem, so that they are still able to retain some of their initial state independence. However, this feedback becomes increasingly imperfect so that at annealing velocities near the boundary with the intermediate regime, the information mixing attains maximal levels. After that, the mixing of information between light and matter is always a monotonic dispersion process, which becomes reduced as the time of interaction is reduced more and more. This establishes a striking difference between the lack of memory effects in the adiabatic and sudden quench regimes: the former's cycle comprises a large but reversible change, while the latter's cycle is akin to a very small but irreversible one. In practice, both mean relatively small changes to the initial condition -- however this is a consequence of two very different properties. This interplay between actual change and its reversibility may explain why the transition between those two regimes is more intricate that might initially have been imagined.\\
\\
We want to emphasize that our numerical results has performed an extensive convergence tests. As a guide to identify the required truncation in the Fock space we show in Fig.~\ref{Conver} the dynamical profile of the radiation reduced density matrix elements $\langle n \lvert \rho_{b}\rvert n \rangle$ for several values of the quenching velocity $\nu$.\\
\\
\begin{figure}[h]
\begin{center}
\includegraphics[scale=0.95]{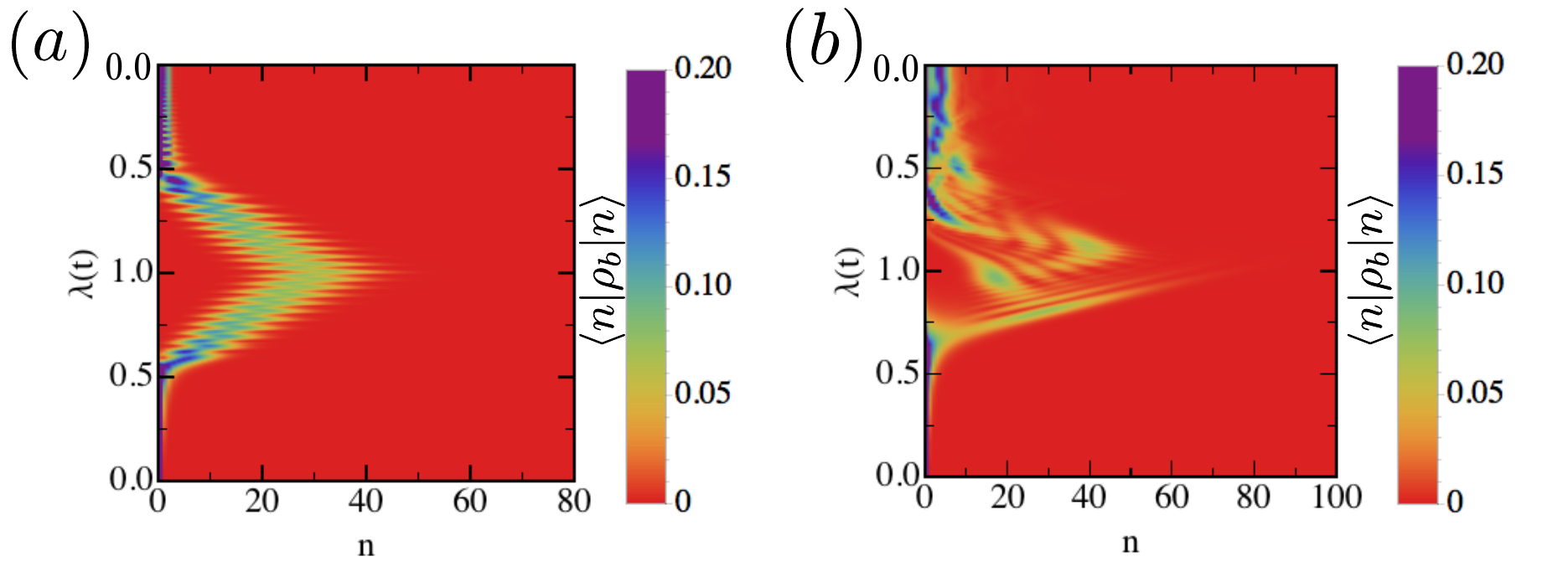}
\caption[Dynamic profile of values the matrix element $\langle n \lvert \rho_{b}\rvert n \rangle$.]{\label{Conver} Dynamic profile of values the matrix element $\langle n \lvert \rho_{b}\rvert n \rangle$ as a function of time dependent radiation-matter parameter $\lambda(t)$, where $\rho_b$ is a density matrix for the bosonic Fock state $\lvert n \rangle$. Left panel $(a)$: $\log_2 \pap{\nu}=-8.46$. Right panel $(b)$: $\log_2 \pap{\nu}=-4.46$.  }
\end{center}
\end{figure}
\\
According to these figures it is evident that the optimum truncation of the Fock space depends on the quenching velocity.

\section{Conclusions}

We have presented a quantum hysteresis analysis of the finite-size \gls{DM} in cycles that cross the \gls{QPT} from the non-interacting to the strong coupling regime and back again. In order to explore and quantify the resulting collective memory effects, we have employed the ground state fidelity and the light-matter entanglement measured through the light and matter von Neumann entropy. The former measure is more oriented to adiabatic quantum control, while the latter is more related to quantum information and quantum thermodynamics. We have identified the entire range of regimes of the cyclic dynamical process: from the adiabatic limit at small annealing velocities, to the sudden quench regime. This revealed that the transition between these two regimes is by no means a trivial one, due to an interplay between the amount of change undergone by the system as compared to the reversible character of that change. Towards the near adiabatic regime, some degree of reversibility is possible despite the system being forced to undergo large changes, which means that information can still go back and forward between the light and matter subsystems. This generates an oscillatory behavior comparable to \gls{LZS} processes. By contrast towards the sudden quench regime, the information exchange between light and matter is always dispersive but gets smaller and smaller as the interaction times are reduced. These two regimes could have their own interesting applications which we leave for future exploration. In particular, the interference occurring in the near adiabatic regime could be important for spectroscopy applications, since it reveals details of the interaction during the hysteresis process. By contrast, characterization of the intermediate regime is important for quantum control, since it improves understanding of the squeezing process that precedes the dynamical generation of light-matter entanglement.
\begin{savequote}[45mm]
``In thinking out the applications of mathematic and physics, it is perfectly natural that mathematics will be useful when large numbers are involved in complex situations"
\qauthor{Richard P. Feynman}
\end{savequote}
\chapter[Entanglement and Schmidt Gap in a Driven \gls{DM}]{Dynamics of Entanglement and the Schmidt Gap in a Driven Dicke Model}
\begin{center}
\begin{tabular}{p{15cm}}
\vspace{0.1cm}
\quad \lettrine{\color{red1}{\GoudyInfamily{I}}}{n} this chapter, we continue to explore the non-equilibrium dynamics in driving light-matter systems. We study the impact of such a pulsed coupling on the light-matter entanglement in the Dicke model as well as the respective subsystem quantum dynamics. Our dynamical many-body analysis exploits the natural partition between the radiation and matter degrees of freedom, allowing us to explore time-dependent intra-subsystem quantum correlations by means of squeezing parameters, and the inter-subsystem Schmidt gap for different pulse duration (i.e. ramping velocity) regimes -- from the near adiabatic to the sudden quench limits. In this way, we show that a pulsed stimulus can be used to generate many-body quantum coherences in light-matter systems of general size. We identify a novel form of dynamically-driven quantum coherence emerging 
 for general $N$ and without having to access the empirically challenging strong-coupling regime. Its properties depend on the {\em speed} of the changes in the stimulus. Non-classicalities arise within each subsystem that have eluded previous analyses. Our findings show robustness to losses and noise, and have potential functional implications at the systems level for a variety of nanosystems, including collections of $N$ atoms, molecules, spins, or superconducting qubits in cavities -- and possibly even vibration-enhanced light-harvesting processes in macromolecules.
\end{tabular}
\end{center}
\begin{center}
\begin{tabular}{p{15cm}}
\vspace{0.1cm}
This chapter is published in references~\cite{Gomez_JPB2017,Gomez_Frontiers2018}:  
\begin{itemize}
\item {\bf F. J. G\'omez-Ruiz},  J. J. Mendoza-Arenas, O. L. Acevedo, F. J. Rodr\'iguez, L. Quiroga, and N. F. Johnson. {\it Dynamics of Entanglement and the Schmidt Gap in a Driven Light-Matter System}. J. Phys. B: At. Mol. Opt. Phys. {\bf 51}, 024001 (2018).
\item  {\bf F. J. G\'omez-Ruiz}, O. L. Acevedo, F. J. Rodr\'iguez, L. Quiroga, and N. F. Johnson. {\it Pulsed generation of quantum coherences and non-classicality in light-matter systems}. Frontiers in Physics, {\bf 6}, 92 (2018).
\end{itemize}
\end{tabular}
\end{center}
\section{Introduction}
The understanding, characterization and manipulation of non-equilibrium correlated many-body systems has benefitted from several remarkable experimental and theoretical advances in recent years~\cite{georgescu2014rmp, Zoller_NJP2011}. Although, by definition, any laboratory sample will necessarily interact with its laboratory environment~\cite{breuer}, modern technologies have succeeded in isolating quantum systems to a significant degree within a large variety of experimental settings~\cite{Lloyd_SC96, Schneider_RP2012, Houck_Nat2012}. Many of these realizations can be regarded as particular cases of an interaction between matter and radiation, or some other form of bosonic excitation field. 
From a theoretical point of view, many of these systems can be modeled to a reasonable approximation by considering the matter subsystem as two-level systems (qubits) and the radiation subsystem as a set of independent harmonic oscillators. Examples of such modeling include cavity \gls{QED}~\cite{nori2011nature,xiang2013rmp} and circuit \gls{QED}~\cite{niemczyk2010nature,peterson2012nature}, impurities immersed in \gls{BEC}s~\cite{ng2008pra,haikka2011pra,sabin2014scirep}, and artificial atoms of semiconductor heterostructures interacting with light~\cite{Sarah_IOP2012} or with plasmonic excitations~\cite{Dzsotjan2010prb}. Since these systems contain various degrees of freedom, their theoretical study has been traditionally approached using approximate perturbative methods~\cite{breuer}. \\
\\
Most of the theoretical treatments to date rely on the assumption that the matter-radiation interaction is static, and either very weak or very strong. However from an empirical perspective, these regimes do not represent any technological boundary -- indeed, the coupling strength in real systems is quite likely to be in between these limits. The potential richness of effects in this intermediate case {\em and} in the regime of non-static coupling, is therefore of significant interest for temporal quantum control in practical quantum information processing and quantum computation. On a more fundamental level, an open-dynamics quantum simulator would be invaluable for shedding new light on core issues at the foundations of physics, ranging from the quantum-to-classical transition and quantum measurement theory~\cite{Zurek1} to the characterization of Markovian and non-Markovian systems~\cite{Bruer_PRL, PRBluis,Cosco2017arxiv}.\\
\\
In fact, the interactions between electronic excitations in matter and quantized collective excitations, lie at the heart of conventional condensed matter physics - in which the focus is on periodic systems - as well as nanostructures which are increasingly being fabricated from materials of common interest to chemists, physicists and biologists. Characterizing how collective quantum behavior can be generated in such systems, and what its properties are, represents a challenging research area -- and also an important technological one, e.g. for quantum information processing -- since each system is ultimately a many-body quantum system embedded in an environment. Of particular interest is the issue of correlations or `coherence' in such systems, which in its purest quantum mechanical form manifests itself as many-body quantum entanglement. Recently, new experimental setups have shown a high degree of control of coherence in scenarios involving elementary boson excitations or confined photons interacting with atoms, molecules or artificial nanostructures in cavities~\cite{Yuan2018,Yuan2017,castroPRL}. Interest in the resulting collective coherences now extends beyond the realm of inorganic materials, to organic and biomolecular systems for which there is an ongoing debate concerning the origin and robustness of such quantum coherences in warm environments ~\cite{1,Yuan2017,Flick}. For example, the recent {\em Nature} review of Scholes et al.~\cite{1} tentatively points toward a surprising ubiquity of coherence phenomena across chemical and biophysical systems that are driven by some external stimulus -- typically a high-power light source which provides time-dependent perturbations that generate vibrational responses on the ultrafast scale~\cite{Tsakmakidiseaaq,1,4,5,Jha1,8,23,33,41}.  It is suspected that many of these coherence phenomena involve some generic form of  quantum mechanical interference between the many-body wave function amplitudes of the system's electronic and vibrational (i.e. boson field) components~\cite{1,5}. Indeed there is a body of evidence~\cite{1,4,5,8,23,33,41,Stockman} suggesting that coherence  phenomena in chemical and biophysical systems of general size can show a surprising level of robustness and extended survival time in the presence of noise. Reference~\cite{1} suggests that these observations are so ubiquitous that focus should be turned toward exploring the connection between coherence and possible biological function. \\
\\
Unfortunately, it is impossible to evaluate the exact quantum evolution of a driven mixed exciton-carrier-vibrational system of arbitrary size. Any theoretical analysis will therefore, by necessity, make approximations in terms of the choice of specific simplifying geometries, the specific number of system components included in the calculation (e.g. $N=1$ dimer as in Ref.~\cite{5}), choices about the coupling between the various excitations of the system, and the manner in which memory effects are averaged over or truncated. While convenient computationally, such approximations have left open the question of the fundamental nature of such coherence phenomena, and how they might possibly be generated as the number $N$ of system components increases towards the tens, hundreds or thousands as in real experimental samples. This highlights the need for calculations that purposely avoid these conventional approximations, albeit while making others, in an effort to better understand the general many-body problem for arbitrary $N$ and arbitrary matter-boson coupling strength.\\
\\
In this chapter, we explore this dynamical regime which is opened up by manipulating the strength of the light-matter coupling in time -- for example using external pulses that generate a coupling that cycles from weak to strong and back again. Specifically, we use a general, time-dependent many-body Hamiltonian, namely the Dicke model, to study the impact of a single pulse in the light-matter coupling, on the quantum correlations at the collective and subsystem levels. Exploiting the natural partition between the radiation and matter degrees of freedom, we explore the time-dependent squeezing parameters of each subsystem, and the entanglement spectrum through the Schmidt gap, for different pulse duration (i.e. ramping velocity) regimes, ranging from the near-adiabatic to the sudden quench limits. The results show that both the inter-subsystem {\em and} and intra-subsystem quantum correlations signal the emergence of the superradiant phase. In addition, in the intermediate ramping regimen, both subsystems remain entangled at the end of the applied pulse, which should be of interest for quantum control schemes.\\
\\
Additionally, we study how many-body quantum coherences (specifically quantum entanglement) can be generated, and perhaps ultimately understood, for a rather generic nanostructure system coupled to a bosonic filed and subject to an external stimulus. Our approach to capturing the effects of a time-dependent field-matter interaction is via the modulation of the matter polarization generated by a time-dependent, externally applied pulse stimulus. While our calculations are not specifically designed to mimic any particular physical nor biochemical nanostructure system, we illustrate our results by referring to a hybrid qubit matter system coupled to a single-mode boson field. While we freely admit that our calculations lack the fine details of other works targeted at specific experimental systems, the generic nature of our calculations allows an examination of what might currently be missed from other calculations that adopt the traditional approximations.\\
\\

\section{Pulsed Generation of Quantum Coherences and Non-classicality in Driven \gls{DM}}\label{sectPG}
We now proceed to characterize the complete dynamical \gls{QPT} profile by focusing on properties of each subsystem, namely the matter subsystem composed by the all-to-all (qubit) spin network, and the radiation mode subsystem. We analyze a wide range of annealing velocities $\upsilon$, and use a logarithmic scale for showing these values of the velocity, defined by $\Gamma=\log_2(\upsilon)$. This range varies from the slow near adiabatic regime,  through the intermediate regime, to the fast sudden-quench regime. 

\begin{figure}[t]
\begin{center}
\includegraphics[scale=0.9]{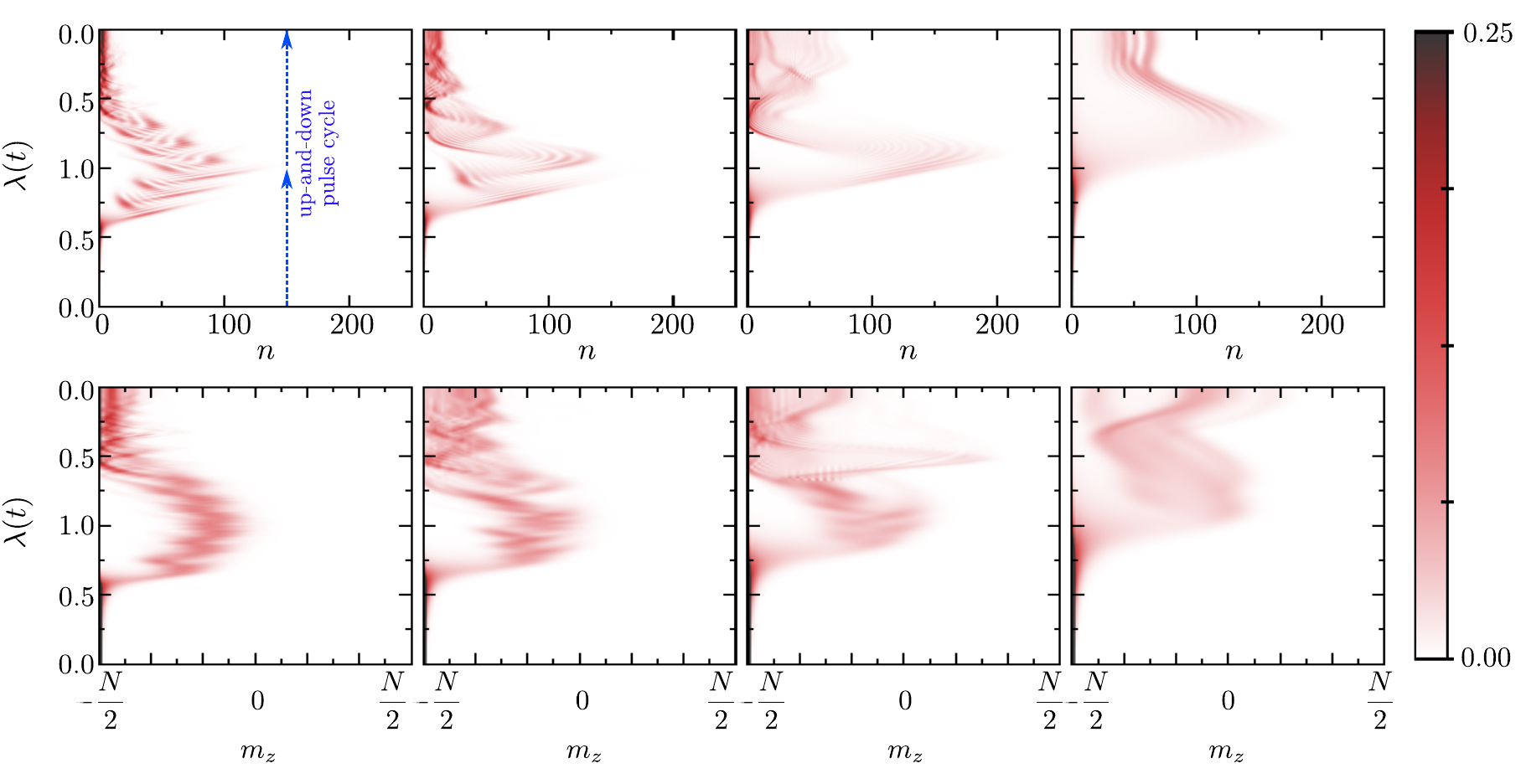}
\caption[Projection of reduced density matrix for each subsystem in driven \gls{DM}.]{Projection of reduced density matrix for each subsystem. Upper panels: $\langle n | \rho_{B}| n\rangle$. Lower panels: $\langle m_z | \rho_{Q}| m_z\rangle$. The values of velocities are, from left to right: $\Gamma=-7.76,\, -5.76,\, -4.76,\, -2.76$. The color scale was adjusted to improve visualization of the results. System size $N=81$}\label{fig_2_6}
\end{center}
\end{figure}

\begin{figure}[t]
\begin{center}
\includegraphics[scale=0.9]{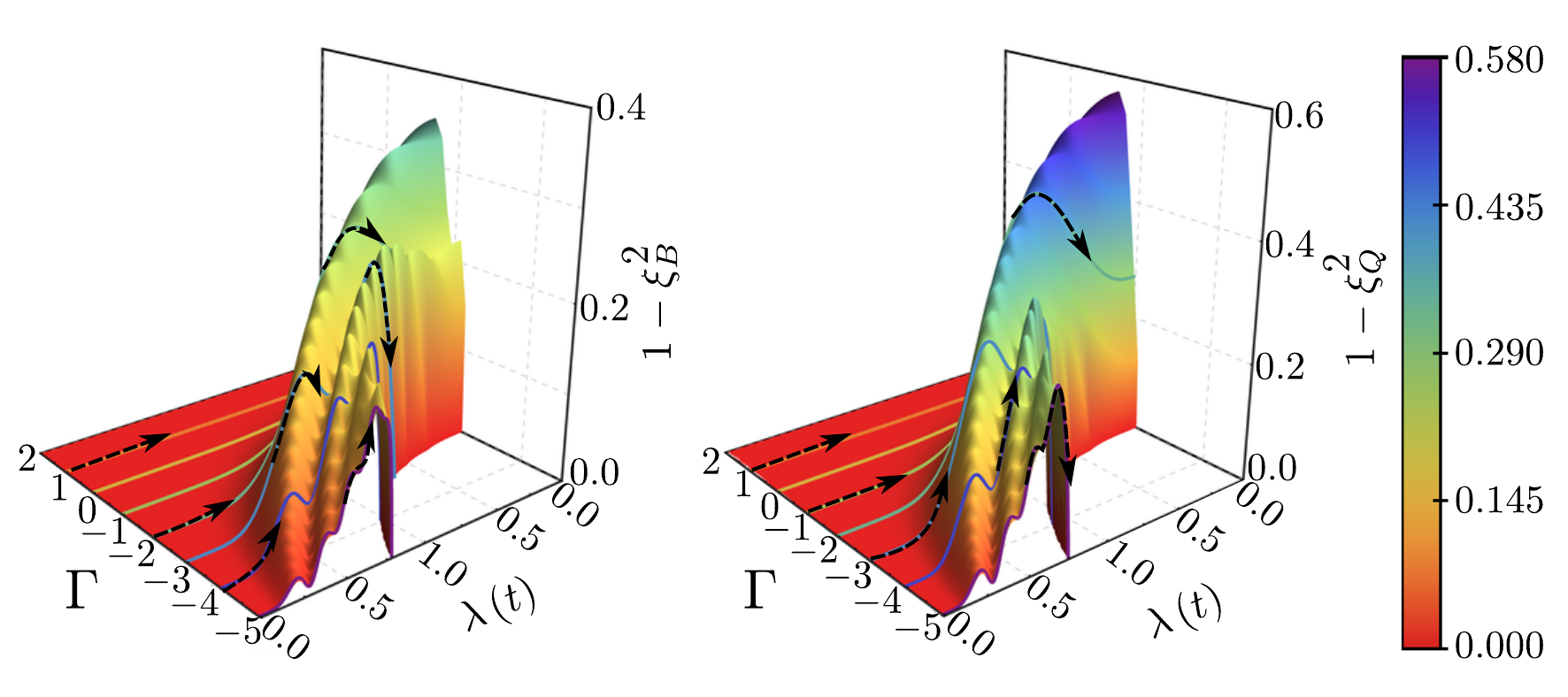}
\caption[Dynamic profiles of squeezing for both subsystem matter and light.]{Dynamic profiles of the matter subsystem in which, for fixed annealing velocities, time varies according to the direction of arrows. Left panel: Evolution $1-\xi_{B}^{2}$, as defined in Eq.~\eqref{Sbos}, whenever it is greater that zero (squeezed radiation). Right: Two qubit concurrence $c_{w}\pap{N-1}=1-\xi_{Q}^{2}$. System size $N=81$}\label{fig_3_6}
\end{center}
\end{figure}

We begin our discussion of the driven dynamics of the Dicke model by considering the diagonal elements of the reduced density matrices of the bipartition. Figure~\ref{fig_2_6} shows the instantaneous projection of the reduced density matrix for the matter and radiation subsystems over the $J_z$ and Fock basis respectively, for several values of the annealing velocity. For the slowest ramping ($\Gamma=-7.76$), both the radiation mode and qubits remain entirely unexcited before crossing the quantum critical point $\lambda_c$ when increasing $\lambda$, with only the respective states of $n=0$ and $m_z=-N/2$ being populated. After crossing $\lambda_c$ the population is transferred to states of larger $n$ and $m_z$, a process that continues up to the time where $\lambda$ starts decreasing. With the reversal of $\lambda$ the population of large values of $m_z$ and $n$ is transferred back to lower values, in a highly-symmetrical form with respect to the turning point. When $\lambda_c$ is crossed again, both the radiation field and the set of qubits become almost completely unexcited, with only the lowest values of $m_z$ and $n$ being populated. Since the reversal of the matter and light dynamics is not completely achieved, this corresponds to a near-adiabatic regime instead of a true adiabatic one for lower velocities). \\
\\
For larger annealing velocities, the dynamic population of states with $m_z>-N/2$ and $n>0$ remains qualitatively similar to that of the near-adiabatic limit during the linear increase of $\lambda$. However two main qualitative differences are observed. First, this population transfer occurs further and further away from $\lambda_c$ as $v$ increases, indicating that the ground-state \gls{QPT} is not being immediately captured. Second, larger values of $n$ and $m_z$ are reached, since a faster ramping velocity provides a stronger excitation to the system. On the other hand, the population dynamics of the $\lambda$ reversal regime is very different to that close to adiabaticity. Even though the population is also transferred back to states of lower quantum numbers, the symmetry with respect to the turning point is lost, and at the end of the dynamics, when the matter and radiation become uncoupled, they are still highly excited. This already indicates that for large annealing velocities, the system gets so excited that it does not simply follow the decrease of $\lambda$, which is of course only expected in the adiabatic limit. Similar asymmetric results are found in the squeezing and entanglement spectrum results shown below.\\   
\\
Now we describe the squeezing parameter for both subsystems, starting with the light degrees of freedom. The squeezing of light states has widely been studied in the literature. A squeezed state of light arises in a simple quantum model comprising non-linear optical processes such as optical parametric oscillation and four-wave mixing. The fundamental importance of the squeezed state is characterized by the property that the variance of the quadrature operator $\hat{x}$ is less than the value $1/2$ associated with the vacuum and coherent state. The squeezing parameter in the field mode $\xi_{B}^{2}$ is expressed in terms of the variance (Var) and covariance (Cov) of the field quadratures as~\cite{Walls}              
 \begin{equation}\label{Sbos}
  \xi_{B}^{2} = {\rm Var}\pap{\hat{x}} + {\rm Var}\pap{\hat{p}} - \sqrt{\pap{{\rm Var}\pap{\hat{x}}-{\rm Var}\pap{\hat{p}}}^2 + 4 {\rm Cov}\pap{\hat{x},\hat{p}}^2}.
 \end{equation} 
In the left panel of Figure~\ref{fig_3_6}, we present a novel way to generate a photon squeezed state. At $t=0$ the quantum cavity starts in the vacuum state $|n=0\rangle$. As before, the radiation-matter parameter varies as a simple linear up-down pulse, forming a triangular ramping. Our results show the existence of a specific regime of annealing velocities such that while the pulse is applied, the photon squeezing tends to increase (besides small oscillations) even after the reversal ramping of $\lambda(t)$ has started. Furthermore, we note that for this velocity regime, the final state of light has high squeezing when the final radiation-matter parameter is zero. \\
   
\begin{figure}[h!]
\begin{center}
\includegraphics[scale=0.75]{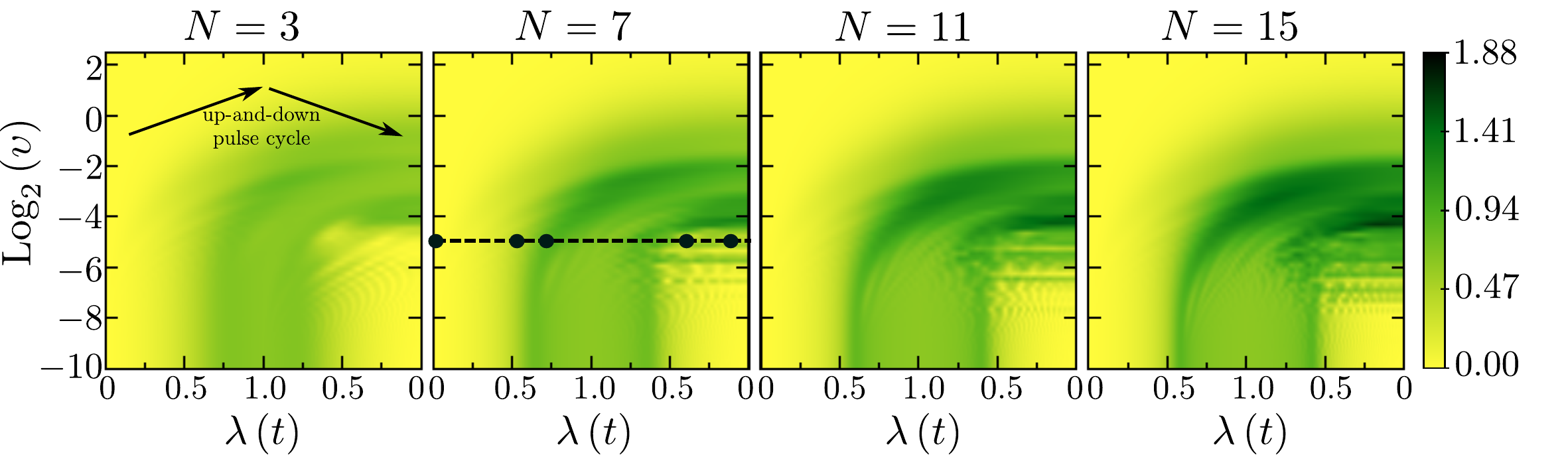}
\caption[Collective quantum coherence generated by a simple up-and-down pulse.]{Collective quantum coherence generated by a simple up-and-down pulse (i.e. triangular $\lambda(t)$ indicated in first panel), as measured by the von Neumann entropy which quantifies the quantum entanglement between the electronic and bosonic (e.g. vibrational) subsystems. By the end of just one up-and-down cycle for a broad range of intermediate return trip times, a substantial amount of quantum coherence is generated in the $N$-component system for general $N$. If the external perturbation is then turned off, for example because the pulse has ended, the generated coherence will survive as long as the built-in decoherence/dephasing mechanisms in the sample allow it to last. The darker the color, the larger the quantum coherence (see color bar).  The larger the $\upsilon$, the less negative the logarithm (i.e. higher on the vertical scale), and the shorter the return trip time. Since these results look qualitatively similar for any $N\geq 3$, they offer insight into the ubiquity of coherences observed empirically in chemical and biophysical systems \cite{1}. Increasing $N$ simply increases the numerical value of the peak value, while choosing a smaller $\lambda(t)$ maximum just reduces the magnitude of the effect. The five points indicated along the horizontal dashed line for $N=7$, correspond to the five specific values of time at which the sub-system Wigner functions are evaluated in the next section. }\label{fig3_6}
\end{center}
\end{figure}

Now we discuss the dynamics of spin squeezing, which has also been the object of intense research in the past few decades. For example, the natural idea of transferring squeezing from light to atoms has been attracting attention both theoretically and experimentally. The notion of spin squeezing has arisen mainly from two considerations: the study of particle correlations and entanglement~\cite{Wang_PRA2003, Zoller, Nick}, as well as the improvement of measurement precision in experiments~\cite{Wineland}. The experimental proposals for transferring squeezing from light to atoms include placing the latter in a high-Q cavity so they interact repeatedly with a single-field (not squeezed) mode~\cite{Ueda}, and illuminating bichromatic light on atoms in a bad cavity~\cite{Klaus}. The intrinsic spin squeezing in a large atomic radiating system was studied in Ref.~\cite{yukalov}, where spin-squeezed states were generated by means of strong interatomic correlations induced by photon-exchange. Spin squeezing can also be produced via a squeezing exchange between motional and internal degrees of freedom of atoms~\cite{Saito}.  For a detailed review, we refer to Ref.~\cite{Jian}.\\  
\\
The definition of spin-squeezing is not unique~\cite{Jian}. For our propose we use the definition given in Ref.~\cite{Wang_PRA2003}, in which a relation between entanglement for a two-qubit subsystem as measured by the Wootters concurrence $c_{w}$~\cite{wootters} and the spin squeezing parameter $\xi_{Q}$ was established, namely
\begin{equation} \label{squee_concu}
\xi_{Q}^{2}=1-\pap{N-1}c_{w}.
\end{equation}
Since each qubit is equally entangled with each other, the monogamic character of entanglement is manifested in Eq.~\eqref{squee_concu} by the $N-1$ factor.\\
\\
In the right panel of Fig.~\ref{fig_3_6} we show the spin squeezing for a wide range of velocities. We find a regime of intermediate annealing velocities for which the squeezing is large at the end of the pulse, which coincides with the velocity regime for which the photonic squeezing is magnified. In previous works by some of us, we showed that the intermediate velocity regime allows for the generation of entanglement~\cite{Gomez_Entropy2016,Acevedo2015NJP}; this is manifested in the generation of squeezing in both light and matter. A fundamental and novel feature of our results is that there is no need of ultra-strong coupling to have squeezing in both light and matter. In addition, we note that the squeezing after the pulse widely exceeds the values that would be achieved through a near-adiabatic evolution.\\
\\
In this way, we are interested in the system's quantum coherences and non-classicality following pulsed perturbations. For general ramping velocity $\upsilon$, the amplitude of being either in the ground or the collected excited states, accumulates a dynamical phase with these  channels interfering with each other and hence forming the oscillatory patterns.  At low ramping velocities, the near-adiabatic regime has a general tendency to show an increase in memory effects as the cycles get faster. However for a broad range of intermediate ramping velocities (Fig.~\ref{fig3_6}) a new regime emerges which is characterized by large quantum coherence between the bosonic (e.g. vibrational) and electronic subsystems. This process would represent a squeezing mechanism in both the electronic and vibrational subsystems, followed by the generation of electronic-vibrational coherence in the form of genuine quantum mechanical  entanglement~\cite{AcevedoPRA2015, Acevedo2015NJP, Gomez_Entropy2016}. As the annealing velocity is further increased, the system has less and less time to undergo any changes.\\
    
The collective coherence in Fig.~\ref{fig3_6} is purely quantum in nature  (i.e. entanglement); it involves an arbitrary number $N$ of components ($N\geq 3$); and it is achieved using any up-and-down $\lambda(t)$ and {\em without} the need to access the strong matter-bosonic field (e.g. electron-vibrational) coupling limit. This is important in practical terms since strong coupling can be hard to generate and control in a reliable way experimentally. Instead, as illustrated in Fig.~\ref{fig3_6} for each value of $N$, we find that the same macroscopic coherence is generated by choosing intermediate ramping velocities and undergoing a return trip, as shown. Moreover the same qualitative result as Fig.~\ref{fig3_6} holds for any $N\geq 3$ and becomes stronger with $N$. Hence we have shown that  by the end of just one up-and-down cycle for a broad range of intermediate return trip times, a substantial amount of quantum coherence will have been generated in the $N$-component system for general $N$. This enhanced entanglement region can be seen as bounded by a maximum ramping velocity $\ups_{\mathrm{max}}$ above which the sudden quench approximation is valid, and a minimum ramping velocity $\ups_{\mathrm{min}}$ below which the adiabatic condition is fulfilled. $\ups_{\mathrm{min}}$ does not depend on the maximum value of $\lambda(t)$ reached, which is to be expected since the ground state in the ordered phase has an asymptotic of $S_N \rightarrow \log 2$ and the adiabatic condition should only depend on the system size $N$. The scaling $\ups_{\mathrm{min}} \propto N^{-1}$ that emerges, comes from a relation for the minimal energy gap at the critical threshold~\cite{AcevedoPRA2015}. The upper bound $\ups_{\mathrm{max}}$ does not depend on system size. In the near adiabatic regime, the von Neumann entropy is not always increasing with time, which means that for slow annealing velocities, information is not always dispersing from the vibrational subsystem to the molecular subsystem and vice versa. Instead, there is some level of feedback for each subsystem, so that they are still able to retain some of their initial state independence. However, this feedback becomes increasingly imperfect so that at annealing velocities near the boundary with the intermediate regime, the information mixing attains maximal levels. After that, the mixing of information between vibrational and electronic subsystems is always a monotonic dispersion process, which becomes reduced as the time of interaction is reduced more and more. This establishes a striking difference between the lack of memory effects in the adiabatic and sudden quench regimes: the former's cycle comprises a large but reversible change, while the latter's cycle is akin to a very small but irreversible one. In practice, both mean relatively small changes to the initial condition -- however this is a consequence of two very different properties. This interplay between actual change and its reversibility may explain why the transition between those two regimes is more intricate that might have otherwise been imagined.
\begin{figure}[t!]
\begin{center}
\includegraphics[scale=0.8]{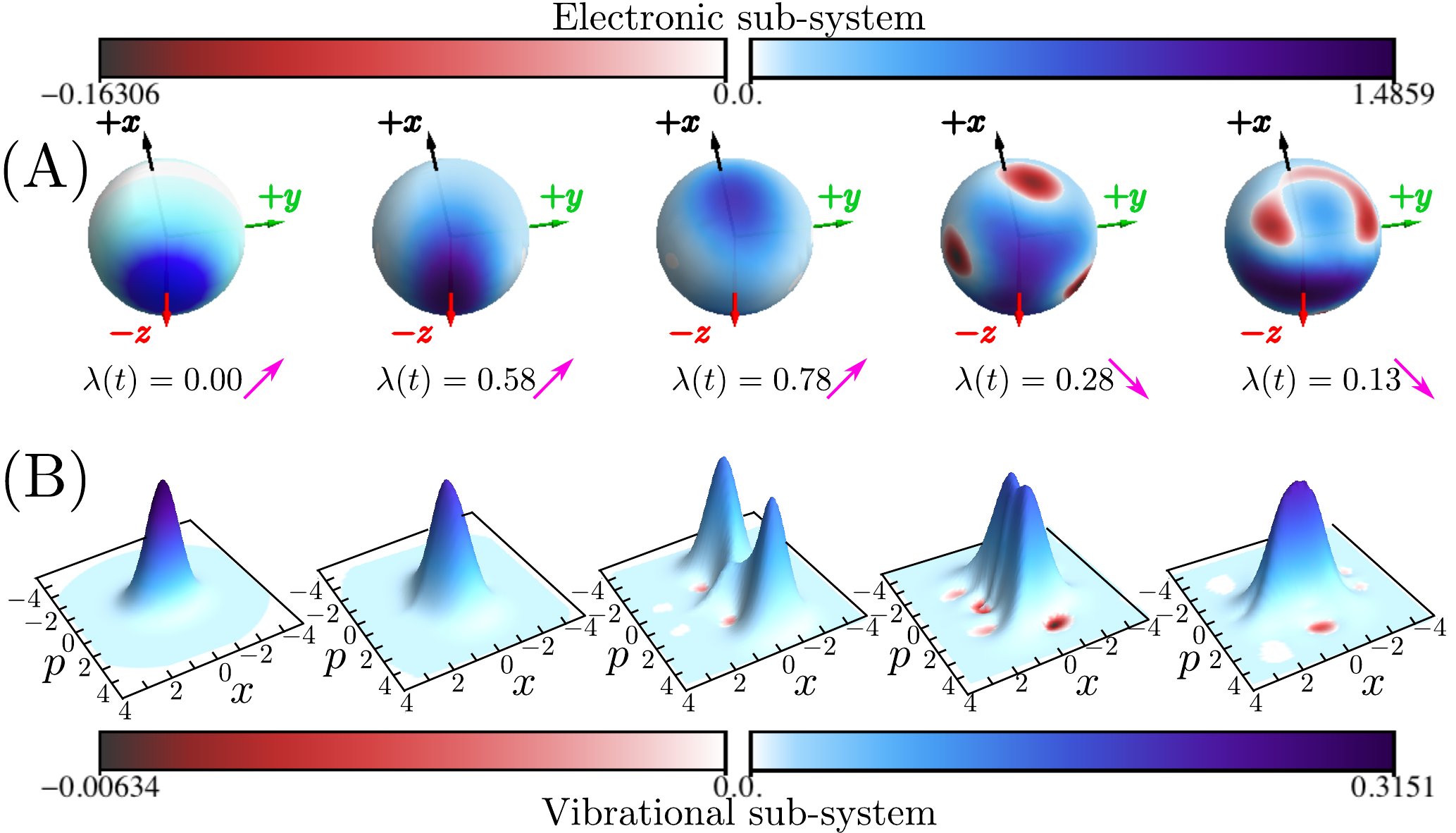}
\caption[Wigner Functions for both light-matter subsystems.]{(A) Electronic sub-system Agarwal-Wigner Functions $W_q$ and (B) boson field/vibrational sub-system Wigner functions $W_b$, shown at two values of $\lambda(t)$ in each portion of an up-and-down pulse cycle. The pulse cycle is depicted as the horizontal dashed line in Fig.~\ref{fig3_6} for $N=7$. The $\upsilon$ value is purposely chosen not to be the optimal one producing the strongest coherence $\pap{\log_{2}\pap{\upsilon}=-5}$, because we want to illustrate the type of non-classicality that can be achieved for broader values of $\upsilon$. Most importantly, by the end of just one up-and-down cycle, both $W_q$ and $W_b$ develop complex non-classical patterns for a broad range of intermediate return trip times and general $N\geq 3$ (see Fig.~\ref{fig3_6}). $W_q$ and $W_b$ are phase space representations. Though positive portions may be quantum mechanical or classical, the negative portions (red and black) that appear demonstrate unambiguous non-classicality. In (A), opposite Bloch hemispheres are not shown because of symmetry: $W_q(\theta,\phi+\pi)=W_q(\theta,\phi)$. In (B), $W_b$ is represented in the $x-p$ plane of position (vertical) and momentum (horizontal) quadrature.}\label{fig4_6}
 \end{center}
 \end{figure}

Our system shows the novel feature of demonstrating non-classicality in both the vibrational {\em and} the electronic subsystems for arbitrary $N$.
Specifically, Fig.~\ref{fig4_6} shows  this non-classicality generated separately within each subsystem during the up-and-down $\lambda(t)$ cycle, and is represented by the Agarwal-Wigner-Function and Wigner quasi-distributions for the electronic and vibrational subsystems respectively. As $\lambda(t)$ increases from zero, the Wigner Function exhibits squeezing, with the Wigner function then splitting along the $x$ and $-x$ directions and no longer concentrated around the initial state. Increasing $\lambda(t)$ further leads to appearance of negative scars (see red portions) which are uniquely non-classical phenomena -- though we stress that even  positive portions of $W_q$ and $W_b$ can exhibit quantum mechanical character. Both $W_q$ and $W_b$ not only develop multiple negative regions which are a marker of non-classical behavior, but they also contain so-called sub-Planckian structures which have been related to quantum chaos. Most importantly, by the end of just one up-and-down cycle, both $W_q$ and $W_b$ have developed complex non-classical patterns, with a blend of regular and chaotic character.

 \subsection{Schmidt gap in light-matter systems} \label{gap_section}
The observation several years ago of the fundamental role of entanglement on quantum criticality led to intense research on characterizing \gls{QPT}s by means of different measures such as entanglement entropy and concurrence~\cite{amico2002nature,gu2003pra,latorre2003prl,wu2004prl,laflorencie2006prl,zanardi2006njp,buonsante2007prl,amico2008rmp,jjma2010pra,pino2012pra,hofmann2014prb}. Shortly after, it was shown that the entanglement spectrum, i.e. the set of eigenvalues of the reduced density matrix of one subsystem resulting from a bipartition, provides valuable information on the properties of topological phases~\cite{haldane2008prl}, and remarkably even more than the entanglement entropy. Since then, several works have analyzed the behavior of the entanglement spectrum, and in particular of the Schmidt gap (the difference between the two largest eigenvalues) close to criticality for different scenarios. These include zero-temperature \gls{QPT}s~\cite{chiara2012prl,lepori2013prb,bayat2013nat}, where the Schmidt gap has been suggested as an order parameter, many-body localization~\cite{Gray2017arxiv}, and dynamical crossings of \gls{QPT}s at different speeds~\cite{canovi2014prb,torlai2014jstat,Qijun2015prb,francuz2016prb,coulamy2017pre}. The latter situation, corresponding to our point of interest in the present work, has been mostly studied for quantum spin chains. Now we discuss the dynamics of the Schmidt gap of the non-equilibrium Dicke model.\\   
\\
In contrast to several condensed-matter systems, the Dicke model immediately suggests a bipartition which allows for a direct study of the physical properties of subsystems of different nature, i.e.  the set of qubits and the radiation field. Thus we calculate the entanglement spectrum for this bipartition. In general, the dynamical state of the total system $\left|\psi\pap{t}\right.\rangle$ is represented by the bipartite form in Eq.~\ref{state}. A standard singular-value-decomposition therefore allows us to rewrite this state as 
\begin{equation} \label{schmidt}
\left|\psi\pap{t}\right.\rangle=\sum_{\alpha=1}^{\Xi}S_{\alpha}\pap{t}  \left|\Phi_{\alpha}^{\pas{m_{z}}}\pap{t}\right.\rangle\otimes \left|\Phi_{\alpha}^{\pas{n}}\pap{t}\right.\rangle,
\end{equation}
with
\begin{eqnarray*}
\left|\Phi_{\alpha}^{\pas{m_{z}}}\pap{t}\right.\rangle&=\sum_{m=-N/2}^{N/2} U_{m,\alpha}\pap{t} \left|m_{z}\right.\rangle,\qquad\left|\right.\Phi_{\alpha}^{\pas{n}}\pap{t}\rangle&=\sum_{n=0}^{\chi} V_{\alpha,n}\pap{t} \left|n\right.\rangle,
\end{eqnarray*}
and where the unitary matrices ${\bf U}$ and ${\bf V}$ are defined on the corresponding subspaces $\mathcal{H}_{m_z}$ and $\mathcal{H}_{n}$ of the set of qubits and radiation field respectively. The new orthonormal states $\pac{\left| \right.\Phi_{\alpha}^{\pas{m_{z}}}\pap{t}}$ and $\pac{\left|\right. \Phi_{\alpha}^{\pas{n}}\pap{t}\rangle}$ are known as Schmidt states. The diagonal elements $S_{\alpha}\geq 0$ in the expansion of Eq.~\eqref{schmidt} are the Schmidt coefficients, which satisfy $\sum_{\alpha}S_{\alpha}^{2}=1$ due to the normalization of the state and are assumed to be arranged in descending order with $\alpha$. Finally $\Xi=\min(N+1,\chi+1)$ is the Schmidt rank, which corresponds to the total number of coefficients in the decomposition.\\

The reduced density matrices for the two subsystems, $\rho_{m_z}\pap{t}={\rm tr}_{n}\pap{\left|\psi\pap{t} \right.\rangle \langle\left. \psi\pap{t}\right|}$ and $\rho_{n}\pap{t}={\rm tr}_{m_z}\pap{ \left| \psi\pap{t}\right.\rangle \langle\left.\psi\pap{t}}\right|$, follow directly from the Schmidt decomposition of Eq.~\eqref{schmidt} and are given by
\begin{eqnarray*}
 \rho_{m_z}\pap{t}&=\sum_{\alpha=1}^{\Xi} S_{\alpha}^{2}\pap{t}\left|\right.\Phi_{\alpha}^{\pas{m_{z}}}\pap{t}\rangle \langle \Phi_{\alpha}^{\pas{m_{z}}}\pap{t}\left.\right|\\
\rho_{n}\pap{t}&=\sum_{\alpha=1}^{\Xi} S_{\alpha}^{2}\pap{t} \left|\right.\Phi_{\alpha}^{\pas{n}}\pap{t}\rangle \langle\Phi_{\alpha}^{\pas{n}}\pap{t}\left.\right|,
 \end{eqnarray*}
 which immediately shows that both $\rho_{m_z}\pap{t}$ and $\rho_{n}\pap{t}$ are diagonal in their respective Schmidt basis and have identical spectra. As a result, the Schmidt gap $\Delta_S$ is defined as
 \begin{equation}
 \Delta_S \equiv \left| S^2_{2}-S^2_{1}\right|,
 \end{equation}
corresponding to the difference between the two largest eigenvalues of the reduced density matrix of any of the two subsystems, and is thus a property shared by both. In the following we describe the behavior of the Schmidt gap as the \gls{QPT} of the Dicke model is crossed with the triangular ramping at different annealing velocities $\upsilon$.\\

\begin{figure}[h!]
\begin{center}
\includegraphics[scale=0.9]{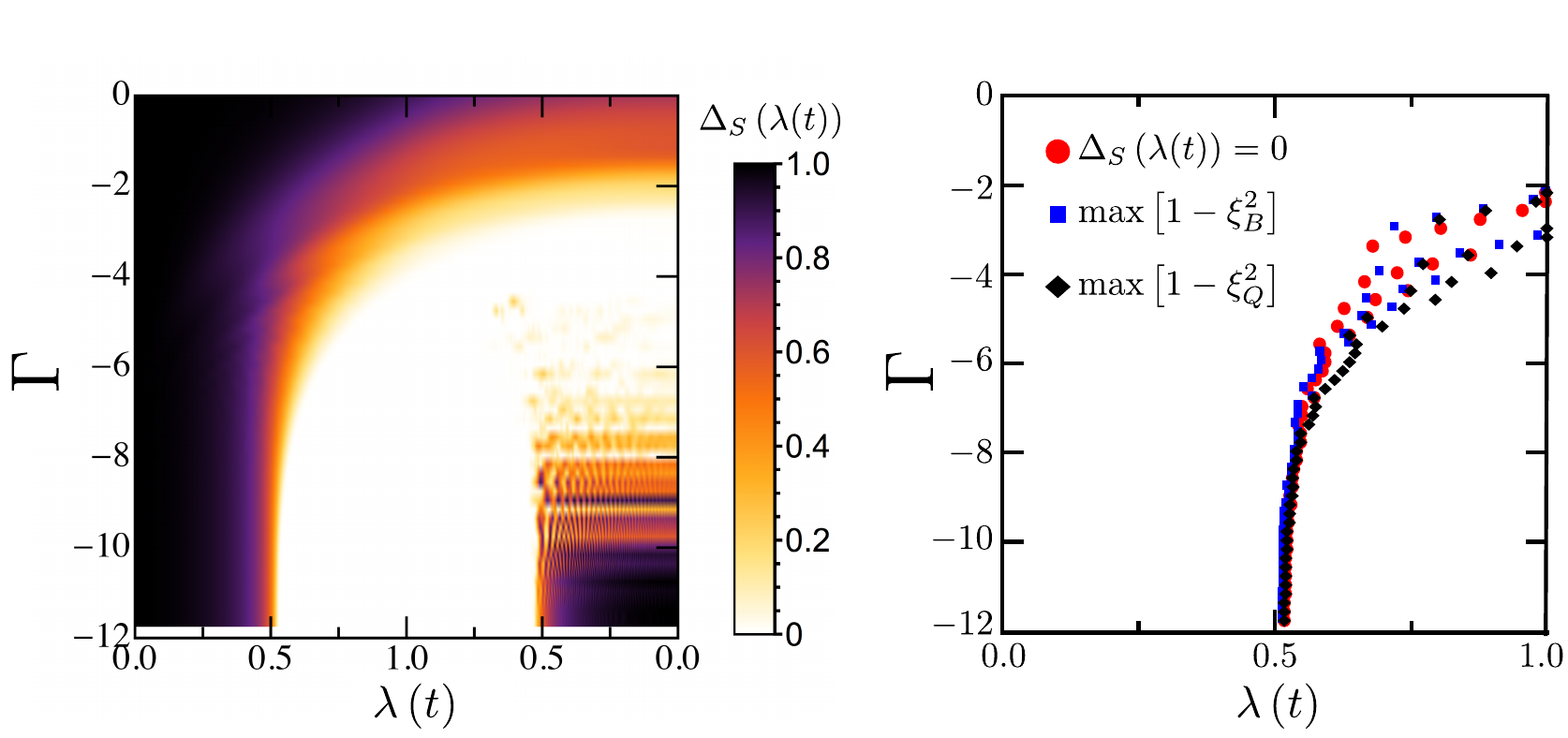}
\caption[Schmidt gap as a function of the annealing velocity.]{Left panel. Schmidt gap $\Delta_S$ as a function of the annealing velocity $\Gamma=\log_2(\upsilon)$ and the time-dependent light-matter interaction $\lambda(t)$. Right panel. Comparison of the $\lambda$ values where the Schmidt gap vanishes, and where the maximal squeezing of both qubits and photons takes place, for the same annealing velocities of the left panel and the ramping of increase of $\lambda$. System size $N=81$}\label{fig_6_6}
\end{center}
\end{figure}

We first consider the crossing of the quantum critical point $\lambda_c=1/2$ during the linear increase of $\lambda(t)$, corresponding to the $0\rightarrow1$ regime in the left panel of Fig~\ref{fig_6_6}. Since the initial state $\left|\psi(t=0)\right.\rangle$ is simply a product, only $S_1(t=0)=1$ is finite, while the other Schmidt coefficients are zero; thus $\Delta_S(t=0)=1$. During the subsequent dynamics $\Delta_S$ monotonically decreases, at a $\Gamma-$dependent rate. In the near adiabatic regime ($\Gamma\lesssim-10$) $S_1$ and $S_2$ cross and the Schmidt gap closes slightly above $\lambda_c$, which suggests that it actually captures the \gls{QPT} between normal and superradiant states. This is similar to previous results of adiabatic dynamical crossings of \gls{QPT}s in spin chains~\cite{canovi2014prb,torlai2014jstat,Qijun2015prb}, where the gap closes near to the corresponding transition. However, in contrast to these cases where the Schmidt coefficients separate and continue crossing during the subsequent dynamics, here $\Delta_S$ remains being zero. As the annealing velocity increases up to the intermediate regime, the Schmidt gap maintains the same qualitative decay with $\lambda$, closes further away from $\lambda_c$ similarly to dynamical crossings on spin chains, and remains zero afterwards. However for even faster ramping processes, in the sudden quench regime ($-3\lesssim\Gamma$), the decay of the gap is so slow that it remains finite when the reversal of $\lambda$ begins.\\
\\
Now we discuss the dynamical crossing when $\lambda$ is reversed, depicted in the $1\rightarrow0$ regime of the left panel of Fig~\ref{fig_6_6}. The main feature of the near adiabatic ramping is that slightly above $\lambda_c$ the Schmidt gap becomes finite again, signaling the return of the system to the normal phase. Moreover, the system actually goes back to the initial product state $\left|\psi(0)\right.\rangle$, since the Schmidt gap reaches the value $\Delta_S=1$ when $\lambda=0$. For higher annealing velocities ($-10\lesssim\Gamma\lesssim-7$) the gap shows an initial fast non-monotonic growth, after which it tends to saturate to a finite value following an oscillatory dynamics. This indicates that even though the qubit and radiation subsystems become disconnected at the end of the pulse, the total state is not just a simple product but an entangled configuration. Thus this intermediate far-from-adiabatic triangular ramping could be exploited as a protocol for preparing entangled states of non-interacting subsystems. For larger annealing velocities but before the sudden quench regime, where the Schmidt gap became zero before starting the light-matter coupling reversal ($-7\lesssim\Gamma\lesssim-5$), it emerges again before crossing $\lambda_c$ but exhibits complex  dynamics including more points of closure. For somewhat higher velocities we observe a scenario where the gap remains finite during the first stage of the driving, but since the dynamics is not so slow it still becomes zero shortly after the start of the reversal stage, before crossing $\lambda_c$ for the second time ($-5\lesssim\Gamma\lesssim-3$). This no longer occurs in the sudden quench regime, where due to the very slow dynamics the Schmidt gap never closes.

 \section{Discussion about quantum coherences and non-classicality in Driven \gls{DM} } \label{sect_discussion}
The results presented in Section~\ref{sectPG}, in particular the similar qualitative profiles of the squeezing parameters and the Schmidt gap as a function of $\Gamma$ and $\lambda$, suggest that both might serve as indicators of the same non-equilibrium phenomenon. Now we briefly discuss this connection, along with a simple approach to the problem, and a possible future application. 

 \subsection{Squeezing functions and Schmidt gap}
In the right panel of Fig.~\ref{fig_6_6} we show, for each annealing velocity considered, the value of $\lambda$ at which the Schmidt gap becomes zero during its increase ramping. As previously discussed, the gap vanishes at higher values of $\lambda>\lambda_c$ as the velocity increases, moving away from the near adiabatic limit. Close to the sudden quench regime ($-5\lesssim\Gamma\lesssim-2$) this general trend continuous, even though the increase is non-monotonic as the dynamics (and thus determining the exact closing point) becomes more involved. In spite of this behavior we find that remarkably, the closure of the gap coincides (quite well for low velocities, approximately for high velocities) with the points in which the maximal squeezing parameters of both qubits and photons take place. This is also shown in the right panel of Fig.~\ref{fig_6_6}, where the different scenarios are plotted simultaneously.\\
\\
Furthermore this also agrees with the values of $\lambda$ in which the qubit and radiation order parameters become finite (see Ref.~\cite{Acevedo2015NJP}). Thus the Schmidt gap can be considered as a complementary quantity to the order parameters of the Dicke model~\cite{Qijun2015prb}, as the former is finite when the latter are zero and vice versa~\cite{Acevedo2015NJP}. These results suggest that both the Schmidt gap and the squeezing parameters are indicators of the emergence of the superradiant state when dynamically crossing the \gls{QPT}, even at high velocity.\\ 
\\
The behavior of these quantities is far more complex during the reversal stage. Due to the strongly-oscillating behavior at low velocities and the more erratic dynamics at high velocities, determining correctly the vanishing point of the Schmidt gap is much more complicated. However the qualitative form of the squeezing parameters depicted in Fig.~\ref{fig_3_6} suggests that the connection between both types of quantities remains valid.\\

\subsection{Landau-Zener-Stuckelberg approach}
A common theme running through our results is the appearance of large quantum correlations in the regime of intermediate pulse duration in the variation in $\lambda(t)$, or equivalently intermediate ramping velocity. A full many-body theory of this dynamical generation of quantum correlations is not possible at the present, and would likely require a novel theoretical technique for treating Eq.~\ref{Hdicke} in a non-perturbative way. However as a first step towards understanding the complex dynamics discussed here, we consider the simplest version of what happens to a quantum system when it crosses a quantum critical point driven by a time-dependent Hamiltonian. Specifically we provide a heuristic treatment by appealing to the phenomenon of Landau-Zener-Stuckelberg interferometry, by means of which possible trajectories of a quantum system interfere with each other when a transition between energy levels at an avoided crossing (a Landau-Zener transition) is crossed. As discussed in detail in Ref.~\cite{georgescu2014rmp}, when a two-level system is subject to periodic driving with sufficiently large amplitude, a sequence of transitions occurs. The phase accumulated between transitions (commonly known as the Stuckelberg phase) may result in constructive or destructive interference.

Following this heuristic approach, we imagine that we can approximate the complex energy-level diagram of this many-body light-matter system as simply a ground state and a excited-state manifold, separated by some minimum energy gap $\Delta$ during the driving process. During the up-sweep alone, there is a single pass through the avoided crossing (i.e. remnant of the critical point) and so the probability that the system then ends up in this excited state manifold is given by $P_+=P_{\rm LZ}={\rm exp}(-2\pi \Delta^2/4v)$~\cite{georgescu2014rmp}. A similar result follows for the down-sweep alone. However since a pulse involves the double-passage through the avoided crossing region, the resulting probability is given by $P_+=4P_{\rm LZ}(1-P_{\rm LZ}) {\rm sin}^2 \Phi$, where $\Phi$ is the sum of two separate phase contributions: one through the quasi-adiabatic portion and one through the non-adiabatic portion.  Averaging over these phases, and hence averaging over the fine-scale oscillations seen in our results, the probability that the system ends up in the excited state manifold following the pulse is given by ${\overline P_+}=2P_{\rm LZ}(1-P_{\rm LZ})$. As a crude approximate energy scale we set $\Delta=0.5$, which is the value of $\lambda$ at which a purely static \gls{QPT} occurs in Eq.~\ref{Hdicke}. As $v$ increases, ${\overline P_+}$ rises from zero to a maximum and then decays back to zero. Its maximum value is $0.5$ which corresponds to the maximum entropy scenario in a simple two-level system. We then obtain numerically that ${\overline P_+}$ starts decaying from its maximum when $\upsilon\approx 1$. This suggests that the correlation features that we observe should also fall off for $\upsilon\rightarrow 1$ ($\Gamma\rightarrow0$), as observed.

\subsection{Future application: system-environment entanglement}
Our findings are also relevant in an entirely different way: if we consider the matter subsystem as the system of interest, and the radiation subsystem as the environment, then our results provide new insight into how a system and its environment become entangled over time, as the system-environment interaction varies. To explore this in the future, instead of considering a single pulse as we do here, the system-environment interaction could be chosen to be a sequence of such pulses which may arrive randomly (e.g. following a Poisson distribution) or become correlated in terms of their arrival times. As such, our model and analysis can provide a first step toward a better understanding of environmental decoherence -- and its flip side, quantum control -- over time. This is important since a primary goal of quantum control is to reliably manipulate quantum systems while preserving advantageous properties such as coherence, entanglement, and purity. Instead of the complex interaction between the system (e.g. matter) and its surroundings (e.g. radiation) being assumed to hamper the system's evolution, it is possible that a suitable sequence of corrective pulses might be used to provide positive feedback to the system and hence maintain its quantum coherences. We leave this for future work.\\

 \begin{figure}[h!]
\begin{center}
\includegraphics[scale=0.8]{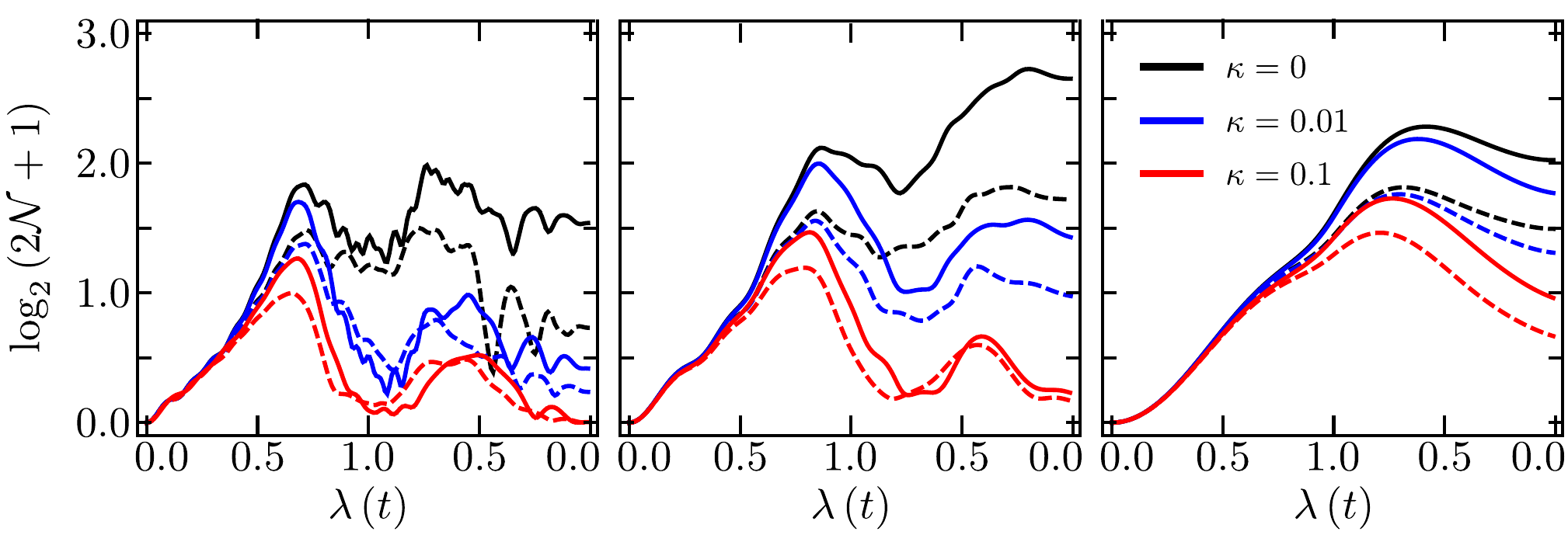}
\caption[Quantum logarithmic negativity]{We show evidence of the robustness of the many-body electronic-vibrational entanglement, as witnessed by quantum logarithmic negativity, to decoherence/losses. Results are shown for three representative, intermediate duration up-and-down pulses (i.e. the annealing velocities $\upsilon$ for left, middle, right panels are $\log_{2}\pap{\upsilon}=-4.64,\, -3.32,\, -1.32$ respectively).  Results are shown for $N=5$ (dashed lines) and $N=11$ (solid lines) and for several values of decoherence $\kappa$.}\label{fig5_6}
 \end{center}
 \end{figure}

\section{Impact of losses and noise in driven \gls{DM}}
Following the density matrix approach of Ref.~\cite{AcevedoPRA2015}, we have investigated numerically how the presence of decoherence/losses to the environment in the chemical or biophysical system will affect the dynamics discussed above, as illustrated in Fig.~\ref{fig5_6}. The widely-accepted best entanglement measurement in an open quantum system is the quantum negativity $\mathcal{N}\left(\rho\right)=\frac{1}{2}\left(\left| \rho^{\Gamma}_{q}\right|_{1} - 1 \right)$ where $ \rho^{\Gamma}_{q}$ is the partial transpose of $\rho$ with respect to the electronic subsystem, and $\left|\hat{\mathcal{O}}\right|_{1}\equiv \mathrm{tr}\pac{\sqrt{\hat{\mathcal{O}}^{\dagger}\hat{\mathcal{O}} }} $  is the trace norm. The electronic-vibrational density matrix $\rho\pap{t}$ evolves as~\cite{breuer}:
\begin{equation}
\frac{d}{d t}\hat{\rho} = -\ii \pas{\Ho,\hat{\rho}} +2\kappa \pap{\bar{n}+1} \mathcal{L}\pap{\hat{\rho};\hat{a}}+2\kappa \bar{n} \mathcal{L}\pap{\hat{\rho};\hat{a}^{\dagger}}
\label{eqME}
\end{equation}
where the Lindblad superoperator $\mathcal{L}\pap{\rho;\hat{\mathcal{O}}}$ for the arbitrary operator $\hat{\mathcal{O}}$ is defined as $\hat{\mathcal{O}}\rho\,\hat{\mathcal{O}}^{\dagger}-\frac{1}{2}\pac{\hat{\mathcal{O}}^{\dagger}\hat{\mathcal{O}},\rho}$ and $\pac{\bullet,\bullet}$ is the traditional anti-commutator. Moreover, $\kappa$ is the damping rate and $\bar{n}$ is the thermal mean photon number.
All our main results survive well if the decoherence term through interaction with the environment, is anywhere up two orders of magnitudes lower than the main energy scale. Furthermore, even if dissipation is at values of just an order of magnitude below, spin squeezing effects remain highly robust, with increasing noise resistance with system size. Vibrational field squeezing surprisingly survives to dissipation regimes comparable to the Hamiltonian dynamics itself. On the other hand, detailed features of the chaotic stage (such as order parameter oscillations, negative regions, and sub-Planck structures) are far more sensitive to decoherence. These very sensitive features could be used as tools for measuring very weak forces. In our analysis, we have found that introducing small but finite values of the average number of phonons $\bar{n}$ (such as those typical at the ultra-low temperatures in most experimental realizations) does not change qualitatively the conclusions; it just slightly intensifies the process of decoherence.
 
\section{Driven system of arbitrary size}
\begin{figure}[h!]
\begin{center}
\includegraphics[scale=0.7]{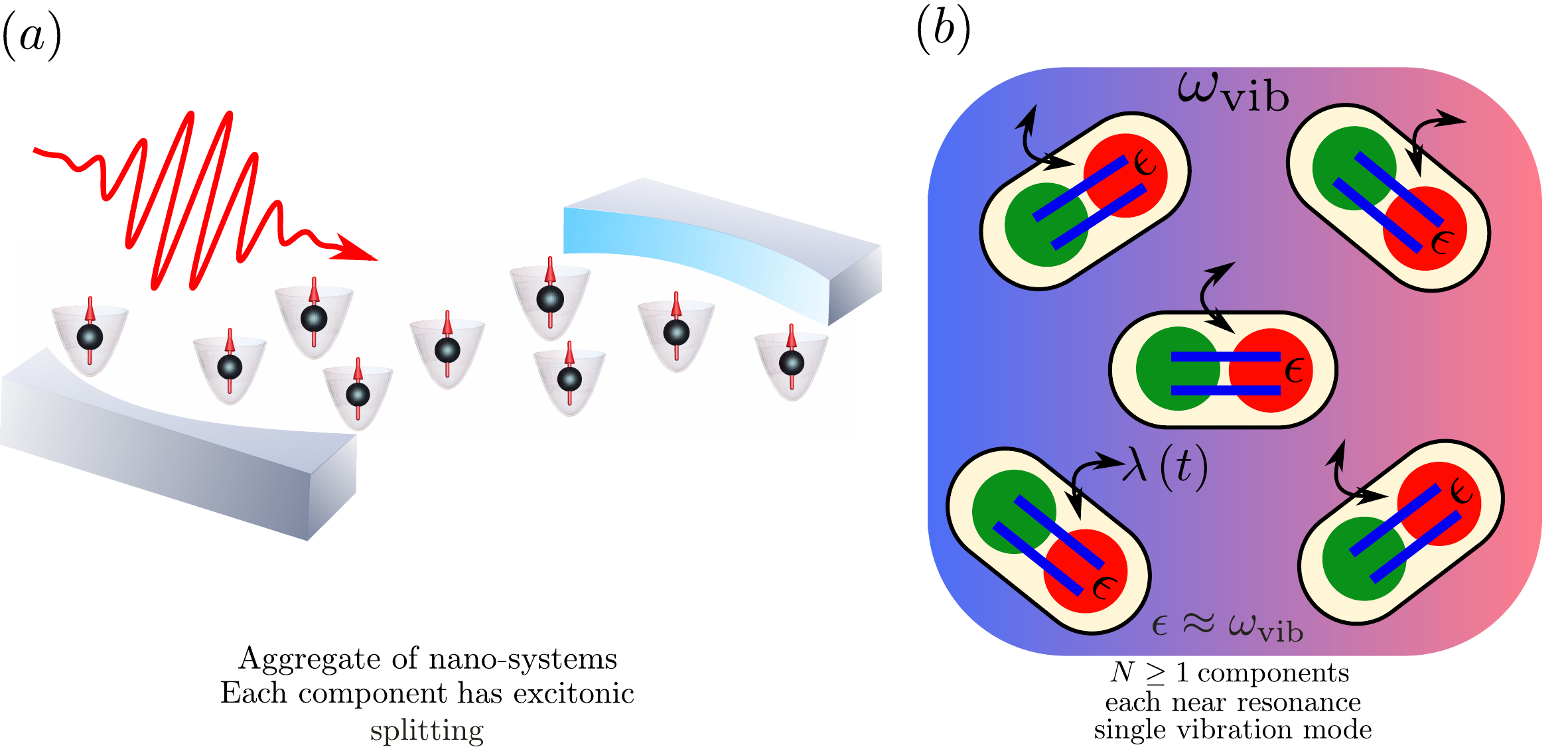}
\caption[Prototype system of arbitrary size.]{(A) {\bf Prototype system.} As an illustration of how our general model and results might be applied in the future, this schematic diagram indicates the type of system that could mimic the dynamics that we analyze for $N$ qubits immersed in a single-mode bosonic environment. We stress that similar implementations have already been built experimentally within the atomic physics community. (B) Schematic representation of the single mode, resonant version of our model (Eq.~\eqref{hdic}). Though it is not our intention to accurately describe any one implementation, we note that in the possible setting of light harvesting/processing  in biochemical systems, each qubit or dimer in Fig.~\ref{fig_6_7}(A) comprises two split excitonic energy levels with separation ${\epsilon}$ which can be regarded as the basic two-level component in our $N$-component system. The coupling ($\lambda(t)$) is time-dependent in order to capture the complex swathe of additional non-equilibrium, anharmonic interactions that can be generated in the system by an external pulsed stimulus.}\label{fig_6_7}
\end{center}
\end{figure}

A general nanosystem that contains an arbitrary number of components and is driven by some external perturbation beyond linear response, will have a time-dependent Hamiltonian that resembles the following schematic form:
\begin{equation}\label{E1}
\begin{split}
{H_{\rm gen}}(t)&=\sum_{\{{\rm vib}\}} {a_{\rm vib}}^{\dagger}{a_{\rm vib}} + \sum_{\{{\rm ext-field}\}} {b_{\rm ext-field}}^{\dagger}{b_{\rm ext-field}} \\
&+ \sum_{\{{\rm elect}\}} {c_{\rm elect}}^{\dagger}{c_{\rm elect}}+ \Lambda(\{\{{a_{\rm vib}}\},\{{b_{\rm ext-field}}\},\{{c_{\rm elect}}\}\}; t)
\end{split}
\end{equation}
where $\{{a_{\rm vib}}\}$, $\{{b_{\rm ext-field}}\}$ and $\{{c_{\rm elect}}\}$ represent the set of all vibrational, externally-generated and electronic modes respectively, i.e. $\{{a_{\rm vib}}\}$ includes all delocalized (phonons) and localized (vibrational) modes, $\{{b_{\rm ext-field}}\}$ includes the quantization of the external field which may, for example, be photonic but is not necessarily black-body or weak. The operators $\{{c_{\rm elect}}\}$ account for every electron and hence hole excitation, including those that form excitons, and therefore when the interaction $\Lambda(\dots)$ is included, they can describe any exciton as well as free carriers, and any coupling between them. Our implementation of Eq.~\eqref{E1} accounts for an arbitrary number $N$ of nanosystem components (e.g. $N$ identical dimers from Fig.~\ref{fig_6_7}) whose excitonic levels become coupled to particular vibrational modes of the system, as in  Fig.~\ref{fig_6_7}(b). The coupling between the electronic and vibrational components is enhanced by dynamical fields that can be created inside the system as a result of a strong external driving field (e.g. pulsed light). Since we are interested in nonlinear measurement techniques in which out-of-equilibrium effects arise, this induced dynamical coupling can be regarded as arising from the anharmonic interactions that tend to cycle up and down in time as the molecular system distorts in response to  perturbations (e.g. external light pulse).
Our approach makes the reasonable assumption that the driven system with anharmonic interactions can be described phenomenologically using classical fields instead of full quantum field operators. This has the effect of averaging over some higher-order quantum fluctuations while including others rather precisely. Specifically, we replace quadratic operator terms by a $c$-number with a time-dependent coefficient which acts as an effective pump on the remaining excitations. Reference \cite{PRBluis} further demonstrates the reasonability of this approximation for the explicit case of control of non-Markovian effects in the dynamics of polaritons generated in semiconductor microcavities at high laser-pumping pulse intensities -- however the same approach can be applied to any other Hamiltonian of the general form of Eq.~\eqref{E1}. As a result, an effective classical intensity sets the coupling strength which becomes time-dependent. The resulting Hamiltonian can be written as that describing a time-dependent generalized Dicke-like model for any $N\geq 1$ \cite{Acevedo2014PRL}:
\begin{equation}\label{hdic}
{H_N}(t)=\sum_\beta\omega_\beta {a_\beta}^{\dagger}{a_\beta} + \sum_{i=1}^{N}\sum_{\alpha_i\in i}\frac{\epsilon_{\alpha_i}}{2}{\sigma}_{z,\alpha_i}^{i} +
\sum_\beta  \sum_{i=1}^{N}\sum_{\alpha_i\in i}
\frac{\lambda_{\alpha_i,\beta}^{i}(t)}{\sqrt{N}}\left({a_\beta}^{\dagger}+{a_\beta}\right){\sigma}_{x,\alpha_i}^{i}
\end{equation}
where ${\sigma}_{p,\alpha_i}^{i}$ denotes the Pauli operators for excitation $\alpha_i$ on each component (e.g. dimer, Fig.~\ref{fig_6_7}(b)) $i$ with $p=x,z$. The first term is the set of vibrational modes $\{\beta\}$ which may or may not be localized around certain locations. The second term represents the electronic excitations $\{\alpha_{i}\}$ localized on each of the components  $i=1,\dots,N$ (e.g. dimers, Fig.~\ref{fig_6_7}(b)). The two electronic states on each component may be hybrid excitonic states, e.g. $|X\rangle$ and $|Y\rangle$ in Ref.~\cite{5}. The third term gives the coupling between the electronic and vibrational terms, by means of which energy and quantum coherence can be transferred back and forth between these molecular components $\{\alpha_{i}\}$ and the vibrational modes $\{\beta\}$. We stress that our choice of $N$ components in Eq.~\eqref{hdic} does not mean that this is necessarily the total number of molecular units in the system under study: It may happen that in practice only some portion of the macromolecular system is probed by the experiment, hence $N$ can in principle be tailored to account for this.\\
\\
Our focus here is on near resonant conditions since these are the most favorable for generating large coherences. Hence we assume for the moment that each component $i$ has one multi-electron energy level separation that is approximately the same as one of the possible vibrational energies, and is also approximately the same for all $N$ components. All other electronic excitation and vibrational modes will be off resonance: including them would modify the quantitative values in our results but the main qualitative findings would remain.\\
\\
Figures~\ref{fig_1}(a) and~\ref{fig_1}(b) provide a motivation for the components in our model inspired by Ref.~\cite{5}, and a schematic of the resonant version of our model (Eq.~\eqref{hdic}) comprising $N$ dimer pairs, where each has two hybridized excitonic states which are energy-split by $\epsilon$. Figure~\ref{fig_6_7}(a) shows the example of a possible application of our model, which features a single LHCII complex~\cite{5}  that is ubiquitous in light-harvesting antennae of cyanobacteria, cryptophyte algae and higher plants~\cite{5}. Indeed LHCII is probably the most ubiquitous light-harvesting complex on the planet. There are three candidate components in each LHCII complex as shown in Fig.~\ref{fig_6_7}(a). Each comprises two hybridized excitonic states with energy splitting $\epsilon=\sqrt{\Delta^2+4V^2}$ where $\Delta$ is the energy difference between the two chromophores' individual exciton states and $V$ is the dipole-dipole coupling strength that provides the inter-chromophore coupling and hence hybridization~\cite{5}. As noted in Ref.~\cite{5}, the Chlb$_{b-a}$ pair has a mode around 750cm$^{-1}$ that is coupled to the electronic dynamics, and this energy is also close to the frequency of the pyrrole in-plane deformations -- meaning that if driven anharmonically, it could in principle generate time-dependent couplings $\lambda(t)$ as in Eq.~\eqref{hdic}. We also note that even this single-mode resonance assumption can be generalized by matching up different  excitation energies $\epsilon'$, $\epsilon''$, etc. to the nearest vibrational energies $\omega'$, $\omega''$ etc. and then solving Eq.~\eqref{hdic} in the same way for each subset $(\epsilon',\omega')$ etc. For example, if the $N$ components are partitioned into $n$ subpopulations, where each subpopulation has the same resonant energy and vibrational mode but where these values differ between subpopulations, the total Hamiltonian will approximately  decouple into $H^{(1)}\oplus H^{(2)}\oplus H^{(3)}\dots \oplus H^{(N)}$. Any residual coupling between these subpopulations might then be treated as noise, as discussed later.\\
\\ 

\subsection{Temporal coupling from driving field}
We now justify the claim that memory effects can arise in the exciton-vibration (XV) dynamics due to the interaction with a (controllable) exterior field, and hence justify the use of a time-dependent $\lambda(t)$. For quantum systems embedded in complex environments, where extra degrees of freedom modulate the interaction between the quantum system of interest and a large reservoir, effective non-Markovian behaviors in the quantum system dynamics arise even though the reservoir itself can be described within a Markovian approximation~\cite{budini}. In our model, the memory effects are due to the parametric pulsed coupling between the exciton and the vibration modes which is represented by the time-dependent XV coupling. Although the phase imprinted by the excitation laser is lost during the first steps of electron-exciton relaxation from the high energy sector to the XV region, this is not a sufficient reason to exclude any coherent-like behavior in the relaxing XV dynamics. Indeed it can be shown that for a wide class of phase-mixed states of the pump modes, results for the signal population can be obtained that are identical to those for a coherent population of those modes. In order to clarify this critical point, we now show that our basic premise is justified for a variety of reasons. According to the extensive literature concerning previous versions of a generic Hamiltonian such as that given by Eq.~\eqref{E1}, in the classical limit the system is equivalent to two coupled harmonic oscillators. This information is enough to gain analytical insight into the solution of the resulting quadratic system. The driven system in this limit is described by two coupled harmonic oscillators with a time-dependent coupling frequency. Consequently for this purpose, we will consider a simplified model for the parametric process that contains just $3$ boson modes (for the sake of simplicity we describe now the $N$ dimer subsystem in the low excitation limit as an effective boson $b$ mode), as described by the Hamiltonian:
\begin{eqnarray}\label{E3}
\hat{H}=\omega_a \hat{a}^{\dagger}\hat{a}+\chi \left(\hat{a}^{\dagger}\hat{a}\right)^2+\omega_b \hat{b}^{\dagger}\hat{b}+\omega_c \hat{c}^{\dagger}\hat{c}+g\left(\hat{a}^{\dagger}\hat{b}^{\dagger}\hat{c}^2+\hat{a}\hat{b}\hat{c}^{\dagger 2}\right)
\end{eqnarray}
where the operators $\hat{a}$, $\hat{b}$ and $\hat{c}$ correspond to the vibration, exciton and high energy controllable exciton modes as employed in the general Hamiltonian in Eq.~\eqref{E1}. Note that we allow for anharmonic terms of strength $\chi$ for the vibration mode.
We now consider the effect of the pump state on the dynamics of this simple, but representative, model. In particular, we consider the excitation of high energy electron states, which indirectly feeds (through relaxation process) an effective pump reservoir which follows the applied radiation pulse shape.  We assume in Eq.~\eqref{E3} that $\left(\hat{a}^{\dagger}\hat{b}^{\dagger}\hat{c}^2+\hat{a}\hat{b}\hat{c}^{\dagger 2}\right) = h(t)(\hat{a}^{\dagger}\hat{b}^{\dagger}+\hat{a}\hat{b})$ where $h(t)$ represents the applied pulse shape. It is usually argued that the expectation value $\langle \hat{c}^{\dagger 2}\rangle$ ($\langle\hat{c}^{2}\rangle$) is different from zero only if the high energy reservoir states have coherent populations. Since the laser pulse excites electrons at a higher energy, the excess energy might be expected to relax into the exciton region giving rise to a coherent interaction. However this is not necessarily the case: after non-resonant excitation, the phase imprinted by the excitation laser is generally lost. The appearance of a well-defined phase is often regarded as the true characteristic feature of a coherent state. A careful analysis of unitary dynamics from mixed states, such as those produced by incoherent relaxation processes, shows that coherent-like behaviors can indeed often be obtained. In order to justify this last claim we compute the time evolution of XV observables under two kind of pump initial states: (i) A pure initial state like $\left|\right.\Psi\rangle=\left|\right.0_a\rangle \left|\right.0_b\rangle \left|\right.\alpha_c\rangle$, denoting the vacuum state for both XV modes, and a pump coherent state. (ii) A statistical mixed state with no phase information at all, given by a density matrix $\hat{\rho}_P=\int_0^{2\pi}d\theta P(\theta)\hat{\Pi}_p(\theta)\left|\right.\Psi\rangle\langle\Psi\left|\right.\hat{\Pi}_p^{-1}(\theta)$, where $\hat{\Pi}_p(\theta)=e^{i\hat{N}\theta}$, with $\hat{N}=\hat{a}^{\dagger}\hat{a}+\hat{b}^{\dagger}\hat{b}+\hat{c}^{\dagger}\hat{c}$, denotes a phase smearing operator, given the fact that it takes a pump coherent state $\left|\right.\alpha_c\rangle$ to a different phase coherent state $\left|\right. e^{i\theta}\alpha_c\rangle$, leaving the XV modes in the vacuum state. The function $P(\theta)$ fixes the pump phase smearing effect with $P(\theta)\geq 0$ and $\int_0^{2\pi}d\theta P(\theta)=1$. Since $\left[\hat{H},\hat{N}\right]=0$, it follows that the time-evolution operator $\hat{U}(t)=e^{-i\hat{H} t}$ commutes with the phase smearing operator $\hat{\Pi}_p(\theta)$. It is now an easy task to obtain for any XV observable like $\hat{a}^{\dagger k}\hat{a}^{l}$, the time-evolution as
\begin{equation}
\begin{split}\label{Eq4}
\langle\hat{a}^{\dagger k}\hat{a}^{l}\rangle_P=& Tr\{\hat{a}^{\dagger k}\hat{a}^{l} \hat{\rho}_P(t) \}\\
=&\int_0^{2\pi}d\theta P(\theta)Tr\{\hat{U}^{-1}(t)\hat{\Pi}_p^{-1}(\theta)\hat{a}^{\dagger
k}\hat{a}^{l} \hat{\Pi}_p(\theta)\hat{U}(t)|\Psi\rangle\langle\Psi|\}\ .
\end{split}
\end{equation}
Since $ \hat{\Pi}_p^{-1}(\theta)\hat{a}^{\dagger k}\hat{\Pi}_p(\theta)=e^{-ik\theta}\hat{a}^{\dagger k}$ and $ \hat{\Pi}_p^{-1}(\theta)\hat{a}^{l}\hat{\Pi}_p(\theta)=e^{i l\theta}\hat{a}^{l}$ it follows that
\begin{eqnarray}\label{E5}
\langle\hat{a}^{\dagger k}\hat{a}^{l}\rangle_P=\langle\hat{a}^{\dagger k}\hat{a}^{l}\rangle_0\int_0^{2\pi}d\theta
P(\theta)e^{-i(k-l)\theta}
\end{eqnarray}
where $\langle\hat{a}^{\dagger k}\hat{a}^{l}\rangle_0$ corresponds to the initial state with the pump in a coherent state. From Eq.~\eqref{E5} it is evident that the population dynamics of the vibrational subsystem ($k=l=1$) is fully insensitive to this class of phase smearing in the pump state, $\langle\hat{a}^{\dagger}\hat{a}\rangle_P=\langle\hat{a}^{\dagger}\hat{a}\rangle_0$, as well as other vibrational correlations as long as the pump phase smearing probability $P(\theta)$ remains practically constant.\\
\\
The main physical ingredients of the general, complex XV system as given by Eq.~\eqref{E1} are captured by this simple $3$-mode Hamiltonian. Therefore we can conclude that for a wide class of coherent pump-plus-relaxation process conditions, our main results on the non-Markovian evolution of the vibrational population are indeed meaningful. Hence the replacement of $\hat{c}$-pump operators by complex numbers -- which consequently yields a time-dependent XV coupling strength $\lambda(t)$ -- is justified. Also, the range of validity of our assumption is the same as the usual one for the parametric approximation which requires a highly populated coherent state, $|\alpha_c|\gg 1$, and short times, $gt\ll 1$. These conditions are precisely identical to those under which we show our model fits with previous studies of XV coherence generation: high excitation and a rapid relaxation dynamics. Therefore, there is indeed a formal justification for reducing the last terms in Eq.~\eqref{E3} to $g h(t)(\hat{a}^{\dagger}\hat{b}^{\dagger}+\hat{a}\hat{b})$ where $h(t)$ represents the applied pulse shape -- hence justifying the time-dependent interaction $\lambda(t)\sim g h(t)$ in Eq.~\eqref{hdic}.\\
\\
We have so far assumed a single $\epsilon,\omega$ pair are close to each other in energy. In the limit that other pairs are also near resonance but these resonances have very different energies from $\epsilon,\omega$, a similar dynamical coherence can develop within each of these subspaces of the full Hamiltonian (Eq.~\eqref{hdic}). Each pair will have its own up-and-down return trip time (and hence ramping velocity $\upsilon$) for which the coherence is maximal. Since the full Hamiltonian can then be written approximately as a sum of these separate subspaces, the full many-body wave function will include a product of the coherent wavefunctions $\Psi_{\epsilon',\omega'}(t)$ for these separate $\{\epsilon',\omega'\}$ subspaces. In the more complex case where several pairs are close together in energy, they will each tend to act as noise for each other. Suppose that the coherence for pair $\epsilon,\omega$ is described by $\Psi_{\epsilon,\omega}(t)$ and it is perturbed by noise from two pairs $\{\epsilon'',\omega''\}$ and $\{\epsilon''',\omega'''\}$ which happen to be nearby in energy. The fact that they are dynamically generated in the same overall system due to the same incident pulses, means that they will likely represent correlated noise. Such correlated noise from various sources can actually help maintain the coherence of $\Psi_{\epsilon,\omega}(t)$ over time. To show this, consider the following simple example (though we stress that there are an infinite number of other possibilities using other numbers and setups, see Ref.~\cite{Lee}) in which we treat $\Psi_{\epsilon,\omega}(t)$ for the pair $\epsilon,\omega$ as a two-level system. The two subspaces $\{\epsilon'',\omega''\}$ and $\{\epsilon''',\omega'''\}$ each generate decoherence of $\Psi_{\epsilon,\omega}(t)$ in the form of discrete stochastic phase-damping kicks. Such phase kicks are a purely quantum mechanical mechanism for losing coherence, as opposed to dissipation.
The probability distributions of the kicks from these two subspaces are $P_A, P_B$. In addition, the kicks are such that the kick of $\Psi_{\epsilon,\omega}(t)$, described by the rotation angle $\theta_2$ is correlated to the previous rotation angle ($\theta_1$):
\begin{equation}
\begin{split}
P_A(\theta_2|\theta_1)&=\begin{cases}
\frac{1}{3}[ \delta(\theta_2) + \delta(\theta_2+\frac{\pi}{2}) + \delta(\theta_2-\frac{\pi}{2})], &\qquad\text{if $\theta_1 \in \{-\frac{\pi}{2}, 0, \frac{\pi}{2} \}$,}\\
\delta(\theta_2), & \qquad\text{otherwise}
\end{cases}\\
 P_B(\theta_2|\theta_1) &=\begin{cases}
 \frac{1}{3}[ \delta(\theta_2-\epsilon)  +\delta(\theta_2+\frac{3 \pi}{4}) + \delta(\theta_2-\frac{\pi}{4} ) ],&\text{if $\theta_1 \in \{-\frac{3 \pi}{4}, \epsilon, \frac{\pi}{4} \}$}\\
 \delta(\theta_2-\epsilon),&\text{otherwise}
 \end{cases}
\end{split}
\end{equation}
with similar conditions holding for all subsequent pairs $\theta_i$ and $\theta_{i-1}$ (see Ref. \cite{Lee} for general discussion). The specific choice of angles may be generalized. The parameter $\epsilon$ is small, and its presence just acts as a memory of which probability distribution was selected in the previous step. If $P_A$ represents the only noise-source applied, and assuming the initial angle of rotation is $0$ (i.e. $\theta_1=0$) then $P_A(\theta_n, \ldots, \theta_1)= \prod_{i=2}^{n} P_A(\theta_{i}| \theta_{i-1})= (\frac{1}{3})^{n-1}$. Hence if under the influence of subspace $\{\epsilon'',\omega''\}$ (and hence $P_A$), the density matrix for $\Psi_{\epsilon,\omega}(t)$ will have off-diagonal elements (which correspond to the decoherence) that decrease by a factor $\frac{1}{3}$ after each phase-kick.
Similar arguments hold if $P_B$ is the only noise-source applied to the system and if we assume $\theta_1 =\epsilon$. Combining the two noise-sources (i.e. probability distributions) at random means that the angles of rotation can take on seven values, $\{ -\pi/3 , -\pi/2, 0,  \epsilon, \pi/3, \pi/2, \pi \}$. The decay factor now becomes exactly $2/3$ in the  limit of $\epsilon\rightarrow 0$. This means that the {\em combination} of the noise sources causes a slower decoherence of $\Psi_{\epsilon,\omega}(t)$ than each on their own. Hence it is  possible that the quantum coherence of $\Psi_{\epsilon,\omega}(t)$ due to a near resonance of $\epsilon,\omega$ as studied in detail in this section is actually favored by having competing coherence processes in the same system.\\
\\
For completeness, we also now clarify how a time-dependent pulse of  incident light can produce a time-dependent pulse of excitonic-vibrational coupling. An incident electromagnetic (light) field ${\vec E}$ with any pulse shape will generate an internal polarization field ${\vec P}$ within the material, given exactly by Maxwell's Equations. The equation describing the time-domain behavior in a general, anisotropic and nonlinear medium subject to a general time and position-dependent light field $\vec E$ is given by:
$$ \nabla \times \nabla \times \vec E + \mu_0\sigma\frac{\partial \vec E}{\partial t} +\mu_0\frac{\partial^2 {\vec{\vec\epsilon}}\cdot\vec E}{\partial^2 t}=-\mu_0 
\frac{\partial^2 {\vec P}}{\partial^2 t}$$
in which the standard symbols have their well-known meaning from electromagnetic theory (e.g. $\vec{\vec\epsilon}$ is a complex second-order tensor). If the medium is lossless then $\sigma=0$ and so this equation can be rewritten as:
$$[ \nabla \times (\nabla \times)  + \frac{1}{\epsilon_0 c^2} 
\frac{\partial^2}{\partial t^2} {\vec{\vec\epsilon}} \cdot ] {\vec E}=-\frac{1}{\epsilon_0 c^2}\frac{\partial^2 {\vec P}}{\partial^2 t} \ .$$
Though nonlinear and anisotropic in general, the presence of $\partial^2/{\partial t^2}$ terms for ${\vec E}$ and ${\vec P}$ in both equations means that a pulse in ${\vec E}$ will generate a similar pulse in ${\vec P}$, and hence a pulse in the internal electric field dynamics coupling the electronic and vibrational systems (i.e. a pulse in $\lambda(t)$ as in our model Hamiltonian).
\section {Conclusions}
We have presented theoretical results for the quantum correlations that develop in a many-body light-matter system, as a result of dynamically manipulating the strength of the light-matter coupling -- specifically, in the form of a single pulse. Our approach was to solve numerically a general, time-dependent many-body Hamiltonian, and exploit the natural partition between radiation and matter degrees of freedom. Specifically, we presented results on intra-subsystem quantum correlations, namely the time-dependent matter and radiation squeezing parameters, and the inter-subsystem Schmidt gap for different pulse duration (i.e. ramping velocity) regimes, from the near adiabatic to the sudden quench limits. The results reveal that both types of quantities signal the emergence of the superradiant state when the quantum critical point is dynamically crossed, by the maximal value of the squeezing parameters and the vanishing of the gap. It is also observed that beyond the near adiabatic limit, the light and matter subsystems remain entangled even when they become uncoupled at the end of the pulse, which could be exploited as a protocol to engineer entangled states of non-interacting systems. Thus our results should also be of interest for temporal control schemes in practical quantum information processing and quantum computation. On a more fundamental level, our results may be helpful for the development of an open-dynamics quantum simulator, for shedding new light on core issues at the foundations of physics, including the quantum-to-classical transition and quantum measurement theory~\cite{Zurek1}, and characterization of Markovianity in quantum systems~\cite{Bruer_PRL, PRBluis,Cosco2017arxiv}. Our findings could also help shed light on system-environment entanglement, if we view the matter subsystem as the system of interest and the radiation subsystem as the environment, and if the system-environment interaction is chosen to be a sequence of pulses with different correlation properties.\\
\\ 
However, our study has of course many limitations: the most obvious perhaps being its lack of specific chemical and biological details and hence the apparent difficulty in saying anything specific about a particular chemical or biophysical system in which coherence has been observed. But just as the physics of critical phenomena has been able to obtain predictions about systems-level behaviors of wide classes of chemically distinct materials near critical points {\em without} including all these details, so too it is possible that the phenomenon of quantum coherence is also, to a certain degree, detail-independent. \\
\\
With this in mind, our findings predict that nanosystems of fairly general size and driven by pulses (e.g. due to a high power external light source or some other applied field) can show surprisingly strong quantum coherence and non-classicality without necessarily passing to the strong coupling regime, but instead through its dynamics -- in particular, the {\em speed} of the dynamical changes that are induced. As we show in Fig.~\ref{fig3_6}, the resulting coherence builds up during the up-and-down ramping associated with an external driving pulse (e.g. light pulse) and is large at the end of it. If this ramping is then turned off, for example because the pulse has ended, the generated coherence will survive as long as the built-in decoherence/dephasing mechanisms allow it to last. Our calculations show that it could remain for a significant time if the noise is not too large Our approach complements existing work in that we avoid the usual type of approximations prevalent in the coherence literature~\cite{1} and instead present results that in principle apply to general $N\geq 3$. The Hamiltonian that we consider is purposely simpler and more generic than many studied to date in order that we can focus attention on understanding the conditions under which optimal coherence can be generated and hence become available for functional use. Though we considered the coupling $\lambda$ to be taken to a relatively modest value ($\sim 1$) and returned, even lower maximum values will give qualitatively similar effects.\\
\\
What about the functional advantage of such coherence? A functional advantage for $N=1$ has already been discussed in Ref.~\cite{5}: in particular, the exciton energy can be transferred coherently from a pure exciton state in a single dimer component, to a mixed excitonic-vibrational state as shown by the individual components in Fig.~\ref{fig_6_7}(a). Given that our results apply in principle to an arbitrary number $N$ of components, and these components may in principle have significant spatial separations, our results  suggest a new {\em systems-level} functional advantage in terms of being able to transfer energy and information coherently throughout the entire $N$-body collective. In particular, since each component (i.e. dimer in Fig.~\ref{fig_6_7}) contributes to an important energy transfer pathway towards exit sites, as discussed in Ref.~\cite{5}, our finding of emergent quantum coherence underpinned by sub-system non-classicalities, implies a systems-level benefit, as opposed to the local advantage for $N=1$~\cite{5}.\\
\\
There should also a wider range of interest in these findings, e.g. aggregates of real or artificial atoms in cavities and superconducting qubits~\cite{ExpDicke1,ExpDicke2}, as well as trapped ultra-cold atomic systems~\cite{bloch2012nat,schneider2012rpp,georgescu2014rmp}, the collective generation and propagation of entanglement~\cite{amico2002nature, wu2004prl, RomeraPLA2013, Reslen2005epl, Acevedo2015NJP, AcevedoPRA2015}, the development of spatial and temporal quantum correlations~\cite{sun2014pra,FernandoPRB2016}, critical universality~\cite{Acevedo2014PRL}, and finite-size scalability~\cite{CastanoPRA2011, CastanoPRA2012}. Hence the effects described in this chapter may be accessible under current experimental realizations in a broad class of systems of interest to physicists. As a result, our findings should be of interest for quantum control protocols which are in turn of interest in quantum metrology, quantum simulations, quantum computation, and quantum information processing~\cite{Gernot2006,Rey2007,Dziarmaga10,Hardal_CRP2015,NiedenzuPRE2015}. 
\begin{savequote}[45mm]
``In its efforts to learn as much as possible about nature, modern physics has found that certain things can never be ``known” with certainty. Much of our knowledge must always remain uncertain. The most we can know is in terms of probabilities."
\qauthor{Richard P. Feynman}
\end{savequote}
\chapter[Inhomogeneous Kibble-Zurek mechanism.]{Inhomogeneous Kibble-Zurek mechanism}
\begin{center}
\begin{tabular}{p{15cm}}
\vspace{0.1cm}
\quad \lettrine{\color{red1}{\GoudyInfamily{T}}}{he}  Kibble-Zurek mechanism is a highly successful paradigm to describe the dynamics of both thermal and quantum phase transitions. It is one of the few theoretical tools that provide an account of non-equilibrium behavior in terms of equilibrium properties. It predicts that in the course of a phase transition topological defects are formed. In this chapter, we elucidate the emergence of adiabatic dynamics in an inhomogeneous quantum phase transition. We show that the dependence of the density of excitations with the quench rate is universal and exhibits a crossover between the standard \gls{KZM} behavior at fast quench rates, and a steeper power-law dependence for slower ramps.  Local driving of quantum critical systems thus leads to a much more pronounced suppression of the density of defects, that constitute a testable prediction amenable to a variety of platforms for quantum simulation including cold atoms in optical gases, trapped ions and superconducting qubits. Our results establish the universal character of the critical dynamics across an inhomogeneous quantum phase transition, that we proposed for favoring adiabatic dynamics.\\
\\
This chapter is published in reference~\cite{Gomez_PRL2019}:  {\bf F. J. G\'omez-Ruiz}, A. del Campo. {\it Universal Dynamics of  Inhomogeneous Quantum Phase Transitions: Suppressing Defect Formation}. Phys. Rev. Lett. {\bf 122}, 080604 (2019). 
\end{tabular}
\end{center}
\newpage
\section{Introduction}
The development of new methods to induce or mimic adiabatic dynamics is essential to the progress of quantum technologies. In many-body systems, the need to develop new methods to approach adiabatic dynamics is underlined for their potential application to quantum simulation and adiabatic quantum computation.\\
\\
The Kibble-Zurek mechanism (\gls{KZM}) is a paradigmatic theory to describe the dynamics across both classical continuous phase transitions and quantum phase transitions~\cite{Kibble76a,Kibble76b, Zurek96a,Zurek96b,Zurek96c, Dziarmaga10,Polkovnikov11}. The system of interest  is assumed to be driven by a quench of an external control parameter $h(t)=h_c(1-t/\tau_Q)$ in a finite-time $\tau_Q$ across the critical value $h_c$. The mechanism exploits the divergence of the relaxation time $\tau(\epsilon)=\tau_0/|\epsilon|^{z\nu}$ (critical slowing down) as a function of the dimensionless distance  to the critical point  $\epsilon=(h_c-h)/h_c=t/\tau_Q$, to estimate the time scale, known as the freeze-out time $\hat{t}$, in which the dynamics ceases to be adiabatic.  The dynamics is therefore controlled by the quench time $\tau_Q$ and by  $z$ and $\nu$,  which are referred to as   the dynamic and correlation-length critical exponent, respectively. The central prediction of the \gls{KZM} is the estimate of the size of the domains in the broken symmetry phase using the equilibrium value of the correlation length $\xi(\epsilon)=\xi_0/|\epsilon|^{\nu}$, at the value $\epsilon(\hat{t})=\hat{\epsilon}$. As a result, the average domain size exhibits a universal power-law scaling dictated by $\xi(\hat{t})=\xi_0(\tau_Q/\tau_0)^{\nu/(1+z\nu)}$.  At the boundary between domains, topological defects form. In turn, the density of defects is set by $d=\xi(\hat{t})^{-1}\sim\tau_Q^{-\beta_{\rm KZM}}$ with $\beta_{\rm KZM}=\nu/(1+z\nu)$. The \gls{KZM} constitutes a negative result for the purpose of suppressing defect formation, given that in an arbitrarily large system, defects will be formed no matter how slowly the phase transition is crossed. This has motivated  a variety of approaches to circumvent the \gls{KZM} scaling law and favor adiabatic dynamics, including  nonlinear protocols \cite{Diptiman08,Barankov08},   optimal control \cite{Doria11,Rahmani11,DeChiara13}, shortcuts to adiabaticity \cite{delcampo12,Saberi14,Campbell15}, and the simultaneous tuning of multiple parameters of the system \cite{SauSengupta14},  to name some relevant examples \cite{DS15}.\\
\\
Test-beds for the experimental demonstration of universal dynamics at criticality are often inhomogeneous, and it is this feature which paves the way to defect suppression. Under a finite-rate quench of an external control parameter, the system does not reach the critical point everywhere at once. Rather,  a choice of the broken symmetry  made locally at the critical front can influence the subsequent symmetry breaking across the system, diminishing the overall number of defects. In this scenario, the paradigmatic \gls{KZM} fails, and should be extended to account for the inhomogeneous character of the system~\cite{ZD08, Zurek09, DM10, DM10b, ions1, DRP11}. An Inhomogeneous Kibble-Zurek mechanism (\gls{IKZM}) has been formulated in classical phase transitions~\cite{Zurek09, ions1, DRP11} following the early insight by Kibble and Volovik~\cite{KV97}. The current understanding is summarized in~\cite{DKZ13,DZ14}. Its key predictions are a suppression of the net number  of excitations with respect to the homogeneous scenario, and an enhanced power-law scaling of the residual density of excitations as a function of the quench rate.\\
\\
In classical systems, numerical evidences in favor of the \gls{IKZM} have been reported ~\cite{ions1}. Three experimental groups have reported an enhanced dependence of the density of kinks with the quench rate  across a structural continuous phase transition in trapped Coulomb crystals ~\cite{EH13,Ulm13,Pyka13}.  However, a related experiment testing soliton formation during Bose-Einstein condensation of a trapped atomic cloud  under forced evaporative cooling was consistent with the standard \gls{KZM} in a homogeneous setting~\cite{Lamporesi13}. In addition,  a verification of the power-law in both numerical studies and experiments has been limited by the range of testable quench rates and defect losses. Defect suppression induced by causality has also been shown to play a role in  inhomogeneous quantum systems, that have so far been explored by numerics and adiabatic perturbation theory~\cite{CK10,DM10,DM10b,Rams16,Nishimori18,Mohseni18}.\\ 
\\
In this chapter, we establish the universal character of the critical dynamics across an inhomogeneous quantum phase transition and the validity of the \gls{IKZM} in the quantum domain. We show that the dependence of the density of excitations with the quench rate  is universal and exhibits a crossover between the standard \gls{KZM} at fast quench rates, and a steeper power-law dependence for slower ramps, that favors defect suppression. 
\section{Dynamics of an inhomogeneous quantum phase transition}
\begin{figure}[t]
\begin{center}
\includegraphics[scale=0.85]{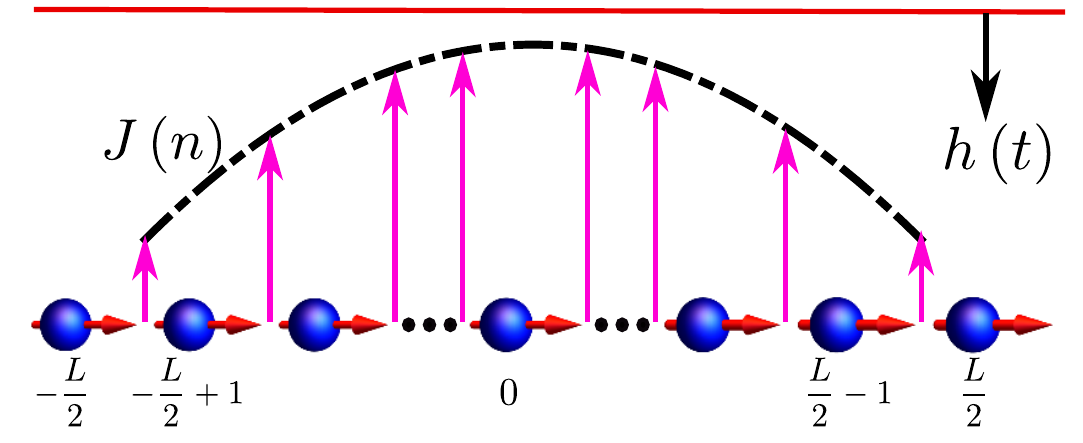}
\includegraphics[scale=0.85]{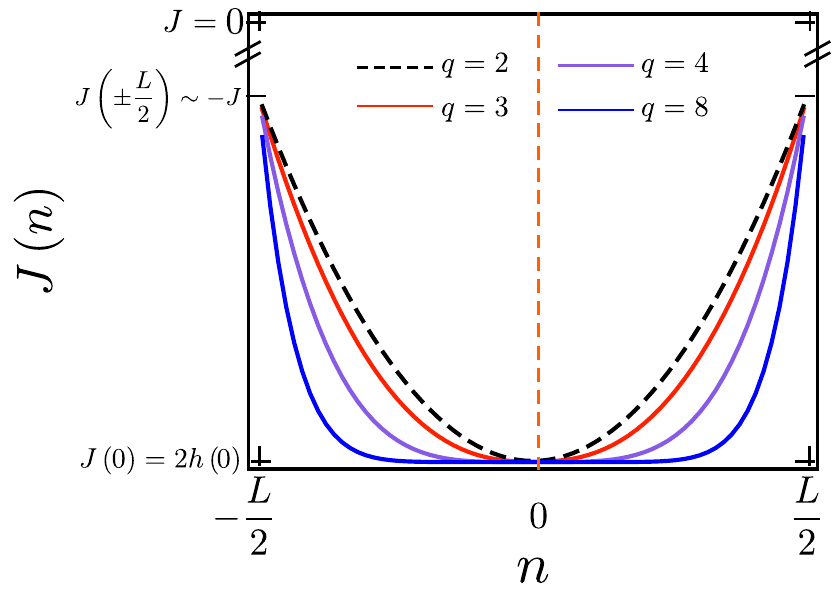}
\caption[Inhomogeneous quantum phase transition.]{{\it Left panel:} Schematic illustration of a one-dimensional transverse-field quantum Ising chain with a symmetric spatial modulation of the tunneling amplitude  $J\pap{n}$ (vertical arrow). As the homogeneous magnetic field $h(t)$ (red line) is decreased, the critical point is first crossed locally at the center of the chain. Subsequently, the critical front spreads sideways at a speed $v_F(n)$ that can be controlled by the quench rate. {\it Right panel:} Schematic illustration of a symmetric spatial modulation of the tunneling amplitude  $J\pap{n}$ given by Eq.~\eqref{Smod_J}  for differences  values of $q$. 
}\label{fig_1_7}
\end{center}
\end{figure}

The one-dimensional inhomogeneous quantum Ising model in a transverse magnetic field $h$ describes a chain of $L$ spins with the Hamiltonian
\beqa
\hat{H}_0=-\sum_{n=1}^{L-1} J(n)\hat{\sigma}_{n}^{z}\hat{\sigma}_{n+1}^{z}-\sum_{n=1}^{L}h\pap{t} \hat{\sigma}_{n}^{x}.
\label{H_Ising}
\eeqa
The setup~\eqref{H_Ising} is schematically represented in Figure~\ref{fig_1_7}. \\
\\
Its homogeneous version ($J(n)=J$) is a paradigmatic model to study quantum phase transitions~\cite{sachdev},  and its quantum simulation in the laboratory is at reach in a variety of quantum platforms including superconducting circuits \cite{Barends2016a}, Rydberg atoms \cite{Labuhn2016} and trapped ions \cite{Monroe17}. 
The homogeneous transverse-field Ising model (H-\gls{TFQIM}) exhibits a quantum phase transition at $h_c = \pm J$ between a paramagnetic phase ($|h|>J$) and ferromagnetic phase ($|h|<J$).  Therefore, it is convenient to introduce the reduced parameter $\varepsilon=(J-h)/J$. The gap between the ground and excite state closes as $\Delta= 2|h-J|$, so the relaxation time $\tau=\hbar/\Delta=\tau_0/|\varepsilon|$ diverges as the system approaches the critical point (critical slowing down). The equilibrium healing length reads $\xi=2J/\Delta=1/|\varepsilon|$ in units of the lattice spacing. 
The critical dynamics of a  H-TFIM is well-described by the standard \gls{KZM}\cite{Dziarmaga10,Polkovnikov11}. The nonadiabatic dynamics results in the creation of topological defects. In the classical case, the latter are formed at the boundary between adjacent domains in the broken symmetry phase and are known as ($\mathbb{Z}_2$) kinks. These excitations involve coherent quantum superpositions and are generally  delocalized \cite{Dziarmaga12}. This is particularly the case in translationally invariant systems \cite{Dziarmaga05}.
The quantum \gls{KZM} sets the average distance between kinks by the equilibrium value of the correlation length at the instant when the dynamics ceases to be adiabatic~\cite{Polkovnikov05,Damski05,Dziarmaga05,ZDZ05}. This time scale known as the freeze-out time can be estimated by equating the relaxation time to the time elapsed after the critical point, $\tau(\hat{t})=|\varepsilon/\dot{\varepsilon}|$, 
whence it follows that $\hat{t}=\sqrt{\tau_0\tau_Q}$. The density of topological defects $d\sim\xi(\hat{t})^{-1}$ scales then as 
\beqa\label{KZM}
d_{\text{KZM}}=\frac{1}{\sqrt{2J\tau_Q/\hbar}} .
\eeqa

An exact calculation shows that $d=\frac{1}{2\pi}d_{\text{KZM}}$~\cite{Dziarmaga05}. We wish to investigate how this paradigmatic scenario is modified  in inhomogeneous quantum phase transitions, extending in doing so the \gls{IKZM} to the quantum domain. We consider a smooth spatial modulation of the  tunneling amplitude $J(n)$ with maximum at $n=0$ and refer to  \eqref{H_Ising} by I-\gls{TFQIM} in this case.
\beqa
\label{Smod_J}
J(n)=J(0)\left(1-\alpha_q |n|^q\right).
\eeqa
To keep the same end values of $J(n)$ at $n=\pm L/2$ for different values of $q$, the constant $\alpha_q$ includes a dependence on $q$. For two different values of   $q=q_1,q_2$, the corresponding constants are related by $\alpha_{q_1}=\alpha_{q_2}|L/2|^{q_2-q_1}$. In right panel of the figure~\ref{fig_1_7}, we show the symmetric spatial modulation considered in Eq.~\eqref{Smod_J}. The choice of $\alpha_q$ is such that the  interaction  coupling at the edges of the chain recovers the value in the homogeneous Ising model $J(n)=J\sim1$, that we use as a reference. Further, we let  the quench of the  magnetic field to be homogeneous and with constant rate $\tau_Q$, 
\beqa\label{rampeq}
h(t)=J(0)\left(1-\frac{t}{\tau_Q}\right),
\eeqa 
during the time interval  $t\in[-\tau_Q,\tau_Q]$. 
Alternatively one could consider the driving of a homogeneous system with a spatially-dependent magnetic field. We introduce the dimensionless control parameter  $\varepsilon(n,t)=\frac{h(t)-J(n)}{J(n)}$ that provides a notion of local  distance to the critical point. It takes values $\varepsilon(x,t)>0$ in the high symmetry (paramagnetic)  phase, reaches $\varepsilon(x,t)=0$ at the critical point, and the broken-symmetry phase for $\varepsilon(x,t)<0$ (ferromagnetic phase). In what follows, we consider the case in which the  system is initially prepared deep in the ground state of the paramagnetic phase such that $\varepsilon(n,t)>0$ everywhere in the chain.\\
\\
 As a result of the spatial modulation of $J(n)$, we introduce an effective quench rate with a spatial dependence,  $\tau_Q(n)=\tau_Q\frac{J(n)}{J(0)}=\tau_Q\pap{1-\alpha_q |n|^q}$. From the condition  $\varepsilon(x_F,t_F)=0$, the time at which criticality is reached at the site $n_F=n$ is  $t_F=\tau_Q\alpha_q |n|^q$ and thus $\varepsilon(n,t)=[t-t_F(n)]/\tau_Q(n)$.  This expression determines the trajectory of the critical front  and yields the following estimate for the local front velocity
\beqa
v_F=\frac{1}{\alpha_q q\tau_Q|n|^{q-1}}.
\eeqa
The divergence of the front velocity in the neighborhood  of $n=0$ is consistent with the fact that the modulation of $J(n)$  for $q>1$ is flattened and the quench is locally homogeneous ($v_F\rightarrow\infty$)  in this region. Topological defects may only form in regions where the front velocity $v_F$ surpasses the speed of sound $s(n)=2J(n)/\hbar$. Assuming that this condition is only satisfied around $n=0$, the speed of sound can be approximated by the constant $s(n)\approx s(0)$ and the  the half-size of such region can be estimated as
\beqa
|\hat{n}|<\left(\frac{\hbar}{2\alpha_q q\tau_QJ(0)}\right)^{\frac{1}{q-1}},
\eeqa 

The effective size of the system for kink formation is simply $2|\hat{n}|$ and one can thus expect a suppression of the total number of defects by a factor $2|\hat{n}|/L$ with respect to the homogeneous scenario ($J(n)=J(0)$).  The net density of defects is estimated to be given by
\beqa
d\sim\frac{2|\hat{n}|}{L\xi(\hat{t})}.
\eeqa

whenever $q>1$. Assuming $\alpha_q |\hat{n}|^q\ll1$, one can use the estimate of the homogeneous \gls{KZM} for the average distance between kinks $\xi(\hat{t})$. As a result, the estimate of the \gls{IKZM} for the net number of defects in the I-\gls{TFQIM} reads

\beqa
\label{qdikzm}
d_{\rm IKZM}&=\frac{2}{L}\left(\frac{1}{\alpha_q q}\right)^{\frac{1}{q-1}}\left(\frac{\hbar}{2J(0)\tau_Q}\right)^{\frac{q+1}{2q-2}}.
\eeqa
The density of defects \eqref{qdikzm} thus scales as a  power-law  $d_{\rm IKZM}\sim\tau_{Q}^{-\beta_{\rm IKZM}(q)}$ in which the power-law exponent  $\beta_{\rm IKZM}(q)=\frac{q+1}{2q-2}$ depends explicitly on $q$. We note that for large values of $q$, the power-law exponent $\beta_{\rm IKZM}(q)$ reduces to that in homogeneous systems. The power law in Eq.~\eqref{qdikzm} is the analogue in spatially inhomogeneous systems of the expression for  the density of defects generated in the passage through a quantum critical point induced by a nonlinear quench in the time domain \cite{Diptiman08,Barankov08}. Specifically, the latter scenario concerns a homogeneous system driven by a homogeneous field such that $\epsilon= |t/\tau_Q|^r$, and leads to density of defects $d\propto\tau_Q^{-\frac{r \nu}{1+r z\nu}}$.\\
\\
We further note that inhomogeneous driving results in a power-law suppression of the density of defects with respect to the homogeneous scenario
\beqa
\frac{d_{\rm IKZM}}{d_{\rm KZM}}=\frac{2}{L}\left(\frac{\hbar}{2\alpha_q q\tau_QJ(0)}\right)^{\frac{1}{q-1}}.
\eeqa
From experimental point view, the parabolic spatial modulation $q=2$ often arises  in trapped systems using the local density approximation \cite{DKZ13} and it is accessible in quantum simulators \cite{Barends2016a,Labuhn2016,Monroe17}. Easily, the experimental limit can be achieved as a smooth spatial modulation. Using a Taylor series expansion, from Eq. \eqref{Smod_J} can be locally approximated by  a quadratic function  of the form $J(n)=J(0)(1-\alpha n^2)+\mathcal{O}(n^3)$. \\
\\
 As a result, the estimate of the IKZM for the net number of defects in the I-\gls{TFQIM} reads
\beqa\label{inhomo2}
d_{\text{IKZM}}=\frac{1}{\alpha L}\bigg[\frac{\hbar}{2J(0)\tau_Q}\bigg]^{\frac{3}{2}}.
\label{dikzm}
\eeqa
We note  the three-fold enhancement of  the power-law exponent, an easily testable prediction that we demonstrate numerically in what follows. The condition $\alpha\hat{n}^2\ll 1$ is not sufficient to test the \gls{IKZM} scaling law  Eq. \eqref{dikzm}. When  $2\hat{n}/\hat{\xi}\sim1$ the applicability of the \gls{KZM} can be called into question. Numerical simulations \cite{ions1} and several experiments \cite{EH13,Ulm13,Pyka13} have reported a steepening of the scaling at the onset of adiabatic dynamics in the course of classical phase transitions. To avoid running into this regime, we  demand $2\hat{n}/\hat{\xi}>1$.\\
\\
Further,  $2\hat{n}$ should be large enough so that the power law scaling can be observed, without saturation at fast quench rates. Hence, we are led to consider slow quenches in large system sizes with small $\alpha$, and $\hbar\alpha/(4J(0))\ll\tau_Q<\hbar/(2J(0)\alpha^{2/3})$.\\
\\
On the other hand, in the limit case  with $q<1$ is not without interest as 
\beqa
v_F=\alpha_q q\tau_Q|n|^{1-q},
\eeqa
this is, it increases with $\tau_Q$ and away from the center of the chain $n=0$.
The condition for topological defect formation is then fulfilled in two disconnected regions, $[-L/2,-\hat{n}]$ and $[\hat{n},L/2]$,
where
\beqa
|\hat{n}|>\left(\frac{2J(n)}{\hbar\alpha_q q\tau_Q}\right)^{\frac{1}{1-q}}.
\eeqa
Thus, the spatial distribution of topological defects is expected to  be concentrated at the edges of the chain as opposed to its center.
In turn, the predicted density of topological defects  no longer follows a simple power-law scaling
\beqa
d_{\rm IKZM}&=&\frac{2(L/2-|\hat{n}|)}{L\hat{\xi}}\\
&=&d_{\rm KZM}-\frac{2}{L}\left(\frac{4J(n)J(0)}{\hbar^2\alpha_q q}\right)^{\frac{1}{1-q}}\left(\frac{\hbar}{2J(0)\tau_Q}\right)^{\frac{3-q}{2-2q}}.\nonumber
\eeqa
being governed by the combination of two different power-laws.\\
\\
The lack of a power-law scaling can also occur for $q>1$ whenever defect formation is not restricted to the neighborhood of $n=0$. To appreciate this, it is required to take into account the spatial modulation of $s(n)$ and note that the condition  $v_F>s(n)$ can then be satisfied both in the proximity of $n=0$ as well as near $n=\pm L/2$.
The condition for defect formation, $v_F>s(n)$, then yields
\beqa
\label{effsize}
|n|^{q-1}(1-\alpha_{q}|n|^q)<\frac{\hbar}{2\alpha q\tau_QJ(0)}
\eeqa
The preceding analysis  follows from disregarding the second term in the left hand side, that leads to deviations from \eqref{qdikzm}.
\section{Numerical Approach}
\begin{figure}[h!]
\begin{center}
\includegraphics[scale=1.2]{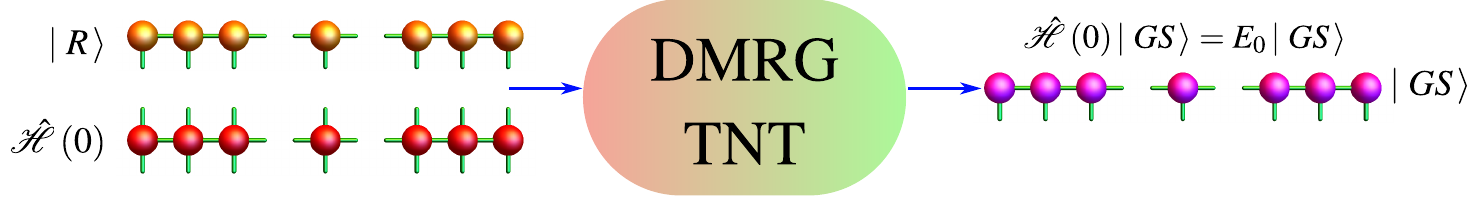}
\caption[Schematic illustration of numerical \gls{DMRG} protocol.]{Schematic illustration of numerical \gls{DMRG} protocol used to calculate the instantaneous ground state energy $\left|GS\right\rangle$ and  the corresponding ground-state energy $E_0$. The numerical routine uses as an input a random matrix-product state $\left| R\right\rangle$ and the matrix-product operator corresponding to Eq.~\eqref{H_Ising} at $t=-\tau_Q$.}
\label{Sfig_2}
\end{center}
\end{figure}

In order to provide a quantitative evidence of the \gls{IKZM}, we perform numerical simulations based on tensor-network algorithms~\cite{tnt}. Specifically at $t=-\tau_Q$,  we first calculate the ground state for~\eqref{H_Ising}  by means of the \gls{DMRG} algorithm~\cite{white1992prl, white1993prb}, using a matrix-product state and matrix-product  operator description of the system with open boundary conditions~\cite{schollwock2011ann}. In  Fig.~\ref{Sfig_2}, we depict a schematic representation of DMRG algorithm. The input for the \gls{DMRG} algorithm is a random MPS represented graphically by $\left|R\right.\rangle$ and the matrix product operator corresponding to Eq.~\eqref{H_Ising} at $t=-\tau_Q$. The algorithm uses a traditional \gls{DMRG} minimization process and yields the matrix product state for the  ground-state  and  the corresponding ground-state energy.

\begin{figure}[h!]
\centering \includegraphics[scale=0.8]{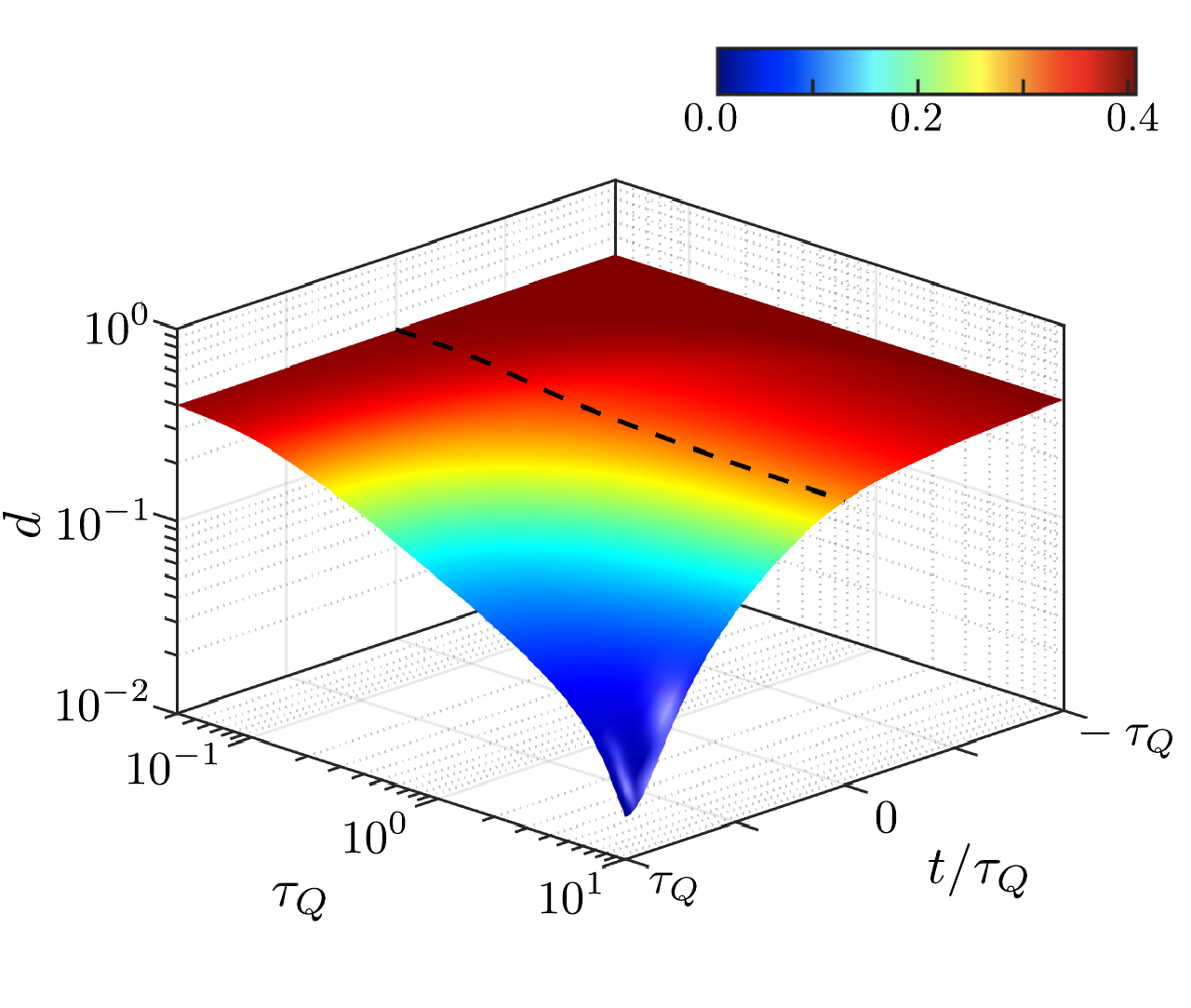}
		\caption[Generation of topological defects under inhomogeneous driving.]{\label{fig_2_7} Density of  kinks $d$ as a function of  the quench rate $\tau_Q$ and the scaled time of evolution $t/\tau_Q$. The dashed black line represent the time $t_c \pap{0}$ at which criticality is reached at the center of the linear  chain ($L=50$). The initial magnetic field is $h(-\tau_Q)=10J$ so that the initial state is deep in the paramagnetic phase, with  $J(0)=5J$ and  $J(\pm L/2)=J$ ($q=2$).}
\end{figure}

The time evolution induced by  the ramp across the critical point \eqref{rampeq} is implemented by the 
 \gls{TEBD} algorithm, suited  for one-dimensional many-body systems. 
\begin{equation}\label{kinks}
d\pap{t}\equiv \frac{1}{2L}\sum_{n=1}^{L-1}\langle \left.\psi(t) \right| \pap{\hat{I}-\hat{\sigma}_{n}^{z}\hat{\sigma}_{n+1}^{z}}\left|\psi(t)\right.\rangle  
\end{equation} 
where $\left|\psi(t)\right.\rangle$ is the instantaneous evolution state. Using the direct evaluation of Eq.~\eqref{kinks}, Figure~\ref{fig_2_7} shows the dynamics of the density of kinks as a function of the time of evolution $t/\tau_Q$ during the crossing of the phase transition for different quench rates $\tau_Q$. In particular, we consider a linear chain of $L=50$ spins with open boundary conditions, described by matrix product states with bond dimension up to $\chi=1500$. The dashed  line signals the time at which the phase transition is reached in the center of  the  chain. 

\begin{figure}[h!]
\begin{center}
\includegraphics[scale=0.7]{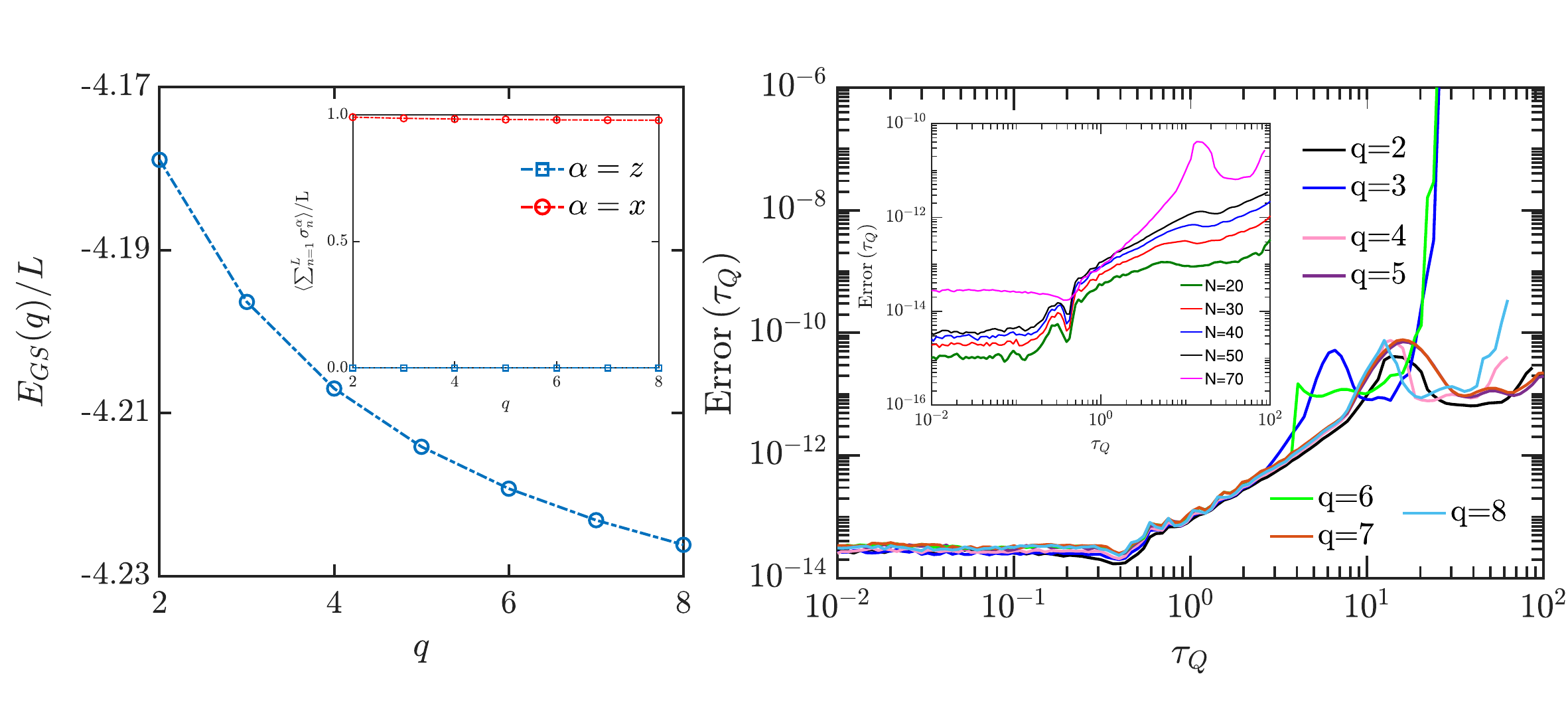}
\caption[Numerical test for Inhomogeneous Kibble-Zurek Mechanism.]{Left: The main panel show the expectation value of ground state energy as a function of $q$. In the inset, the expectation value of magnetization is shown along $x$ and $z$ direction. The system size is $L=70$. Right: In the main panel, the final truncation error  is shown as a function of $\tau_Q$, for severals values of $q$. The inset  shows the final truncation error for different systems sizes for $q=2$.}
\label{Sfig_3}
\end{center}
\end{figure}

Figure~\ref{Sfig_3} shows the ground state energy per particle as a function of $q$, as well as the initial magnetization along $x$ and $z$ direction. For this \gls{DMRG} process, we consider matrix product state  dimension up to $\chi=500$.  During the simulation of non-equilibrium dynamics,  correlations in the system generally increase with the time of evolution and   convergence is checked by increasing the  matrix product state dimension $\chi$. In the main right panel of Fig.~\ref{Sfig_3}, we show the final truncation error,  calculated as the sum of the truncation errors of each singular value decomposition performed. In this way, we calculate the expectation value for the operator of the density of kinks  
  
\newpage
Figure~\ref{Sfig_2} shows the scaling of the density of defects as a function of the quench time $\tau_Q$ for different values of $q$ governing the spatial dependence of the tunneling matrix elements $J(n)$. In all cases, the density of defects saturates to a constant value in the limit of the fast quenches. For slower values, a scaling law is first observed that is well described by the \gls{KZM} for homogeneous systems. In this regime the velocity fo the front $v_F$ exceeds the relevant speed of sound everywhere in the system making the transition effectively homogeneous, in spite of the spatial dependence  of $J(n)$. For even slower quenches, a second scaling regime is observed where the power-law exponent is in agreement with the prediction in Eq. \eqref{qdikzm}. This is the main quantitative prediction of the \gls{KZM} extension to inhomogeneous systems in both classical and quantum domains.  Interestingly, the crossover between the homogeneous and inhomogeneous scenario is clearer for even values of $q=4,6,8$ when the density of defects drops a fine value across the crossover. The onset of adiabatic dynamics is found for even slower values of the quench time. We emphasize that deviations from the scaling regimes illustrated here can be expected whenever defect formation not restricted to the center of the chain, in view of Eq. \eqref{effsize}. The fitted power-law exponents are collected in Table \ref{qtable} for both the homogeneous and inhomogenous power-laws.
While the exponent $\beta_{\rm KZM}$ approaches the constant theoretical value  $\beta_{\rm KZM}=1/2$, the larger exponent $\beta_{\rm IKZM}$ exhibits a dependence on the value of $q$ in agreement with the prediction $\beta_{\rm IKZM}=(q+1)/(2q-2)$.\\

\begin{table}[h!]
\begin{center}
\begin{tabular}{| c | c | c || c | c || c |}\hline\hline
$q$ & $\beta_{\rm KZM}\pm\Delta\beta_{\rm KZM}$& $r^{2}_{\rm KZM}$ &  $\beta_{\rm IKZM}\pm\Delta\beta_{\rm IKZM}$& $r^{2}_{\rm IKZM}$ & $\beta_{\rm IKZM}(q)$\\ \hline\hline
$2$ & $0.52\pm 0.03$ & $0.9991$ & $1.51\pm 0.03$ & $0.9993$ &3/2\\\hline 
$3$ & $0.56\pm 0.03$ & $0.9990$ & $1.08\pm 0.10$ & $0.9993$ &1\\\hline
$4$ & $0.58\pm 0.02$ & $0.9991$ & $0.77\pm 0.07$ & $0.9993$ &5/6\\\hline 
$5$ & $0.57\pm 0.02$ & $0.9990$ & $0.70\pm 0.03$ & $0.9999$&3/4\\\hline
$6$ & $0.55\pm 0.02$ & $0.9990$ & $0.65\pm 0.05$ & $0.9991$&7/10\\\hline
$7$ & $0.62\pm 0.02$ & $0.9996$ & $0.64\pm 0.02$ & $0.9997$&2/3\\\hline
$8$ & $0.57\pm 0.03$ & $0.9992$ & $0.61\pm 0.01$ & $0.9996$&9/14\\\hline\hline
\end{tabular}
\end{center}
\caption{{\bf Numerical power-law exponents.} The density of defects generated across an inhomogeneous phase transition exhibits a crossover from a power-law  $d_{\rm KZM}\sim\tau_Q^{-\beta_{\rm KZM}}$ characteristic of truly homogeneous systems to a second power-law of the form  $d_{\rm IKZM}\sim\tau_Q^{-\beta_{\rm IKZM}}$, describing slower quenches with larger values of the quench time $\tau_Q$. We show the numerical results for the fitted corresponding power-law exponents $\beta_{\rm KZM}$ and $\beta_{\rm IKZM}$ for severals values of $q$ and system size $N=70$. The power-law exponent $\beta_{\rm KZM}$ is approximately independent of the value of $q$ and takes slightly higher values than the theoretical prediction $\beta_{\rm KZM}=1/2$ for the Ising model, in agreement with previous theoretical and numerical studies. The second power law is characterized by an exponent in good agreement with the theoretical prediction $\beta_{\rm IKZM}=(q+1)/(2q-2)$.}\label{qtable}
\end{table}
\begin{figure}[t]
\begin{center}
\includegraphics[scale=0.87]{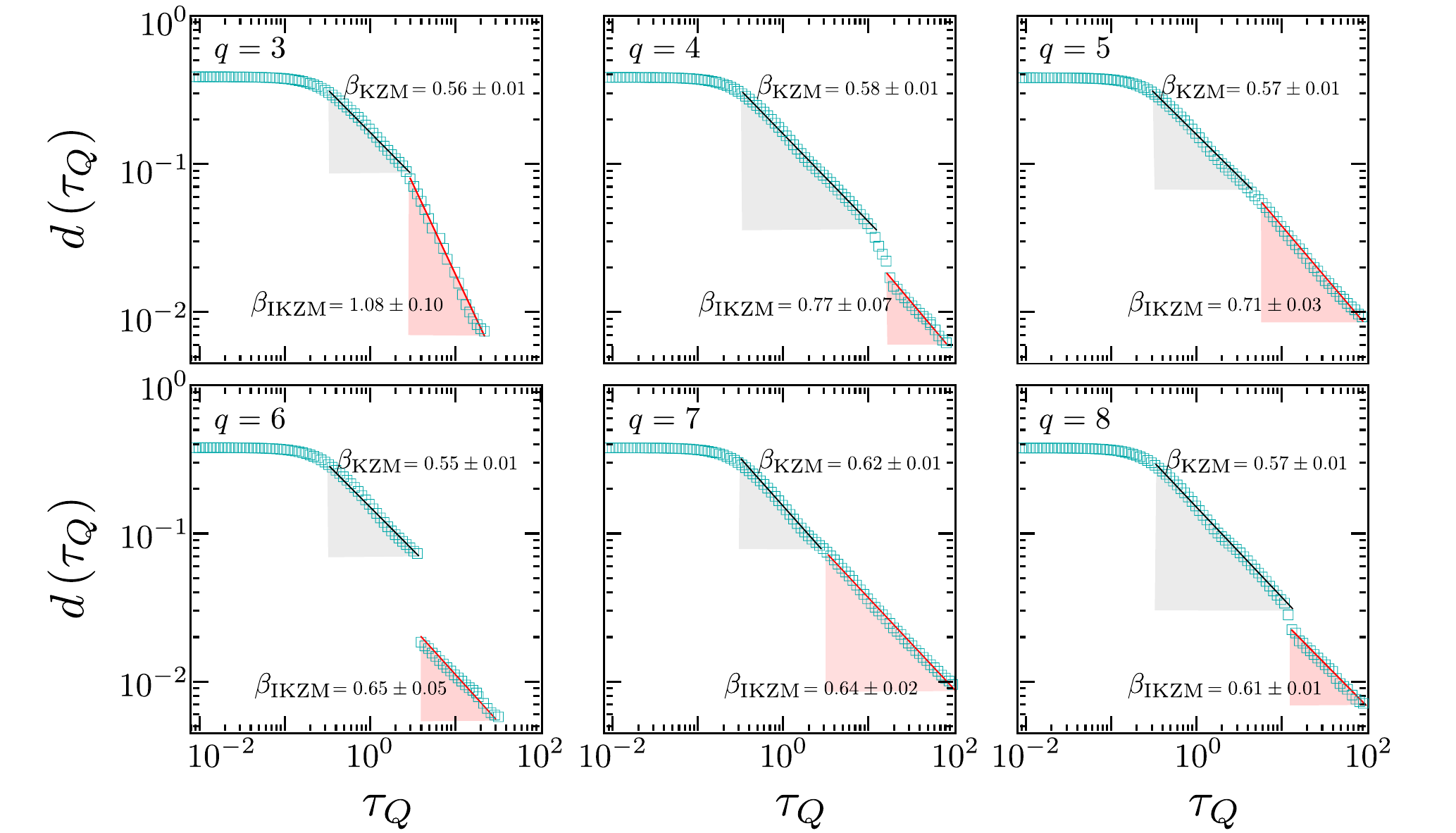}
\caption[Density of topological defect with  a non-quadratic tunneling matrix element $J(n)$.]{The density of  defects is plotted as a function of the quench time in a doubly logarithmic scale for different values of $q$ and system size $N=70$. In all cases a transition is observed from a regime governed by homogeneous \gls{KZM} at fast quenches and an enhanced suppression of defects for slow quenches. The latter arises from the interplay of the velocity of the critical front $v_F$ and the speed of sound $s$ and is the key prediction of the \gls{IKZM}. }\label{Sfig_4}
\end{center}
\end{figure}
\begin{figure}[t!]
\begin{center}
\includegraphics[scale=0.7]{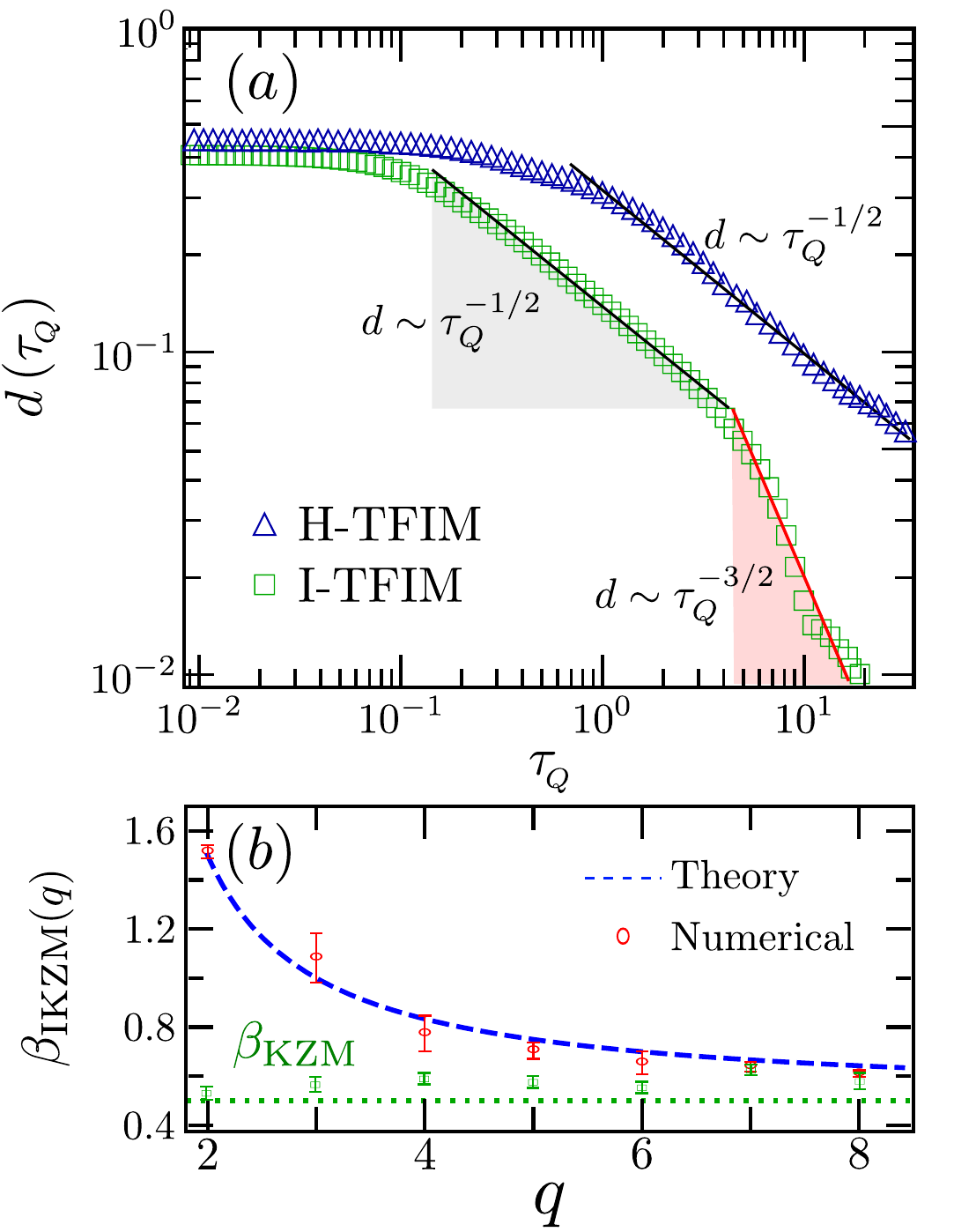}
\caption[Density of topological defects $d$ as a function of the quench time $\tau_Q$.]{(a) In a homogeneous  transition (H-\gls{TFQIM}, denoted by $\triangle$) the dependence of $d$ on $\tau_{Q}$ is described by the original \gls{KZM}, up to a saturation of $d$ at fast quench rates and the onset of adiabatic dynamics for slow quenches. When the critical point is crossed locally (I-\gls{TFQIM}, denoted by  $\Box$), the dependence is no longer described by a single-power law and exhibits a crossover between two universal regimes, described by Eqs.~\eqref{KZM} and  Eq.~\eqref{dikzm}. The symbols correspond to the numerical results for $L=50$, and the solid lines are the linear fits ($q=2$). (b) Comparison between the theoretical and numerical power-law exponents for different values of $q$ in both homogenous and inhomogeneous  scaling regimes. Results corresponding to the homogenous scaling ($q\to\infty$) are shown in green.} \label{fig_3_7}
\end{center}
\end{figure}

In Figure~\ref{fig_3_7}(a), we depict the power-law dependence for the density of excitations upon completion of the phase transition ($t=\tau_Q$) as a function $\tau_Q$ for both H-\gls{TFQIM} and I-\gls{TFQIM} cases. The homogeneous case \cite{Dziarmaga05} is governed by the universal  \gls{KZM} scaling $d\sim \tau_{Q}^{-0.51\pm0.03}$, with regression coefficient $0.9994$, and is plotted as a reference. The numerical data is well-described by the  \gls{KZM} prediction $d\sim 1/\sqrt{\tau_Q}$. The density of defects   also exhibits  a saturation at fast quenches  as well as deviations at the onset of adiabatic dynamics in the limit of slow driving. These features  in the extreme of very fast and slow quenches are  also shared by the inhomogeneous crossing of the phase transition.  In addition, numerical simulations in Fig.~\ref{fig_3_7}(a) establish that for intermediate quench rates the density of defects exhibits a crossover between two different universal regimes, dictated by the \gls{KZM} (fast rate quench) and \gls{IKZM} (slow rate quench) scalings derived in Eqs.~\eqref{KZM} and~\eqref{inhomo2} respectively. We report in the I-\gls{TFQIM} a critical exponent for \gls{KZM} of $d\sim \tau_{Q}^{-0.52\pm0.03}$ with regression coefficient $0.9991$.  Therefore, fast quench rates the ingomogeneous critical dynamics is  described by the \gls{KZM} prediction $d\sim 1/\sqrt{\tau_Q}$. Further, the scaling for slower quench rates is described by $d\sim \tau_{Q}^{-1.51\pm0.03}$ with regression coefficient $0.9992$, in  agreement with the theoretical prediction $d\sim \tau_Q^{-3/2}$. This \gls{IKZM} scaling holds for quench rates between the onset fo adiabatic dynamics and he characteristic quench rate $\tau_Q^*$ at which the crossover occurs. The latter  can be estimated by setting the effective system size  equal to the physical system size, i.e., $2\hat{n}=L$. This yields $\tau_Q^*=\hbar/(2\alpha J(0)L)$ and we have verified numerically its inverse linear dependence of the  value of $\tau_Q^*$  observed numerically with the system size  $L$ and  and  $\alpha$. Figure~\ref{fig_3_7} and~\ref{Sfig_4} shows the agreement between the theoretical an analytical power-law exponents in the inhomogeneous scaling regime when different values of  $q$  are considered (other than $q=2$), governing the modulation of $J(n)=J(0)(1-\alpha_q |n|^q)$. \\
\\
We want to emphasize that, the thermodynamics of inhomogeneous systems is subtle. In this limit, the density fo topological defects is described by the \gls{IKZM} (instead of the conventional \gls{KZM}) for slow quenches (satisfying $\tau_Q>\tau_Q^*$). Moreover, the dependence $\alpha_q(L)$  was not part of our focus as we are mainly concerned with signatures of universality of the power-law scaling of the density of defects on the quench time. Inhomogeneous systems in the laboratory are currently of very small size though (tens of particles in ion chains, for instance), so we prefer not to emphasize  the thermodynamic limit. Regarding the homogeneous character of the dynamics in the center of the systems this is {\it always } a feature of the \gls{IKZM} as discussed in last discussion  (whenever $q>1$). Further, the size of the domains in the  central region of the spin system can be estimated with the homogeneous \gls{KZM} as we also explicitly discussed in this section. However, the density of defects  is not computed locally, but globally with respect to the total system size $L$. Within the \gls{IKZM} treatment, defects form only in the central part of the system of size 
$L_{\rm eff}=2|\hat{n}|$  where $L_{\rm eff}=L_{\rm eff}(\tau_Q)$ is itself a function of the quench time (being determined by the interplay between the front velocity $v_F$ and the sound velocity ${s}$, i.e., by the condition $v_F>s$). Additionally,  in our discussions, we are interested in the density of defects in the {\it total} system. For a homogenous system this is
\begin{equation}
d=\frac{1}{L}\langle n\rangle=\frac{1}{L}\frac{L}{\hat{\xi}}=\frac{1}{\hat{\xi}},
\end{equation}
where the estimate of the mean number of topological defects $\langle n\rangle =L/\hat{\xi}$ is computed as the ratio of the total system size $L$ and the domain size $\hat{\xi}$.
For an inhomogeneous system this is modified to
\begin{equation}
d=\frac{1}{L}\langle n\rangle=\frac{1}{L}\frac{L_{\rm eff}(\tau_Q)}{\hat{\xi}}=\frac{2|\hat{n}|}{L}\frac{1}{\hat{\xi}},
\end{equation}
where $\hat{\xi}$ is the size of the domains in the central region of the chain and the effective system size is generally smaller than $L$. Whenever $L_{\rm eff}/L\ll 1$, the size of the domains can be estimated by the homogeneous KZM as we discuss in the text, i.e., $\hat{\xi}=\xi_0(\tau_Q/\tau_0)^{\frac{\nu}{1+z\nu}}\sim\sqrt{2J(0)\tau_Q/\hbar}$.\\
\\
In particular, the physics behind the case $q \to \infty$ is discussed by considering  the finite but increasing values $q=2,4,8$. We note that for $q=8$ the modulation of $J(n)$ is already very steep. But defect formations occurs only in the effective system size $L_{\rm eff}$. This should be taking into account to correct the estimate by the referee. The density of defects is not simply given by $d \sim 1/(J(0) \tau_Q)^{1/2}$ but by 
$d \sim (L_{\rm eff}/L)1/(J(0) \tau_Q)^{1/2}~\tau_Q^{-3/2}$.\\
\\
Further, the case  $q \to 1$ is singular in that the front velocity no longer diverges. Depending on its finite value it can be larger or small that the sound velocity. If larger, defect formation is governed by the conventional KZM. If $v_F<s$ then the dynamics is essentially adiabatic. The case $q=1$ has been addressed in the literature using adiabatic perturbation theory.\\
\\
 \section[Universal scaling of the density of kinks]{Density of defects with arbitrary spatial dependence of the critical front and critical exponents $\nu$ and $z$}
The preceding section can be generalized for an arbitrary critical system characterized by a correlation length critical exponent $\nu$ and a dynamic critical exponent $z$. As discussed in the last section, the expression for the relevant sound velocity, is then
\beqa 
\label{hats}
\hat s =  \frac {\hat \xi} {\hat \tau} = \frac {\xi_0} {\tau_0} \bigg[\frac {\tau_0} {\tau_Q(n)} \bigg]^{\frac {\nu(z-1)} {1+\nu z}}.
\eeqa  
Assuming defect formation to be restricted to the neighborhood of $n=0$, one can set  $\tau_{Q}(n)\approx\tau_Q(0)=\tau_Q$ in this expression.
The condition $v_F>\hat s $ leads to the estimate of (half) the effective size for defect formation
\beqa
|n|<|\hat{n}|=\left(\frac{1}{\alpha_q q \xi_0}\right)^{\frac{1}{q-1}}\bigg(\frac {\tau_0} {\tau_Q} \bigg)^{\frac {\nu(z-1)} {(1+\nu z)(q-1)}}.
\eeqa
Using the KZM estimate for the size of the domains, $\hat{\xi}=\xi_0(\tau_Q/\tau_0)^{\nu/(1+\nu z)}$, the density of defects becomes
\beqa
d_{\rm IKZM}=\frac{2|\hat{n}|}{L\hat{\xi}}=\frac{2}{L\xi_0}\left(\frac{1}{\alpha_q q \xi_0}\right)^{\frac{1}{q-1}}\bigg(\frac {\tau_0} {\tau_Q} \bigg)^{\frac {1+q\nu} {(1+\nu z)(q-1)}}.
\eeqa
Further, one can generally write 
\beqa
d_{\rm IKZM}\sim \hat{X}d_{\rm KZM}, 
\eeqa
this is, the IKZM scaling follows from the paradigmatic KZM result for homogeneous systems, taking into account the effective fraction of the system $\hat{X}=2\hat{n}/L$ that depends on the quench rate as $\hat{X}\sim 1/\tau_Q$.
\section{Conclusions} 
In summary, we have explored the effect of local driving in the universal dynamics across a quantum phase transition using the paradigmatic quantum Ising chain as a testbed. A local crossing of the critical point can result from inhomogeneities  in  the system or the spatial modulation in the external fields that drive the transition.  As the critical point is reached locally, there is an interplay between the speed of sound and the velocity of propagation of the critical front.  The effective system size in which topological defects can form acquires then a dependence on the quench rate.  For fast quenches, the residual density of defects is well described by a power law in agreement with the original Kibble-Zurek mechanism. As the quench rate decreases there exists a crossover  to a novel power-law scaling behavior of the density of defects, that is characterized by a larger exponent, higher than that predicted by the Kibble-Zurek mechanism.  Local driving  thus leads to a much more pronounced  suppression of the density of defects, that constitute a testable prediction amenable to a variety of platforms for quantum simulation including cold atoms in optical gases, trapped ions and superconducting qubits. Our results should prove useful in a variety of contexts including the preparation of  phases of matter in quantum simulators and the engineering of inhomogeneous schedules in quantum annealing.

\begin{savequote}[45mm]
``It doesn't matter how beautiful your theory is, it doesn't matter how smart you are. If it doesn't agree with experiment, it's wrong."
\qauthor{Richard P. Feynman}
\end{savequote}
\chapter[\gls{STD} from critical dynamics in a trapped-ion quantum simulator]{Statistics of topological defects from critical dynamics in a trapped-ion quantum simulator}
\begin{center}
\begin{tabular}{p{15cm}}
\vspace{0.1cm}
\quad \lettrine{\color{red1}{\GoudyInfamily{T}}}{he} crossing of a quantum phase transition leads to the formation of topological defects whose full counting statistics has been predicted to be universal. 
Using a trapped-ion quantum simulator, we experimentally probe the the kink distribution resulting from driving  a one-dimensional quantum Ising chain through the paramagnet-ferromagnet quantum phase transition. Quasiparticles are shown to obey a Poisson binomial distribution. All cumulants of the kink number distribution are nonzero and scale with a universal power-law as a function of the quench time in which the transition is crossed. We experimentally verified this scaling for the first cumulants and report deviations due to the dephasing-induced anti-Kibble-Zurek mechanism. Our results establish that  the universal character of  the critical dynamics can be extended beyond the paradigmatic Kibble-Zurek mechanism, that accounts for  the mean kink number, to characterize the full  probability distribution of topological defects.\\
\\
This chapter is submitted in reference~\cite{Cui_Arxiv2019}: J-M. Cui, {\bf F. J. G\'omez-Ruiz}, Y-F. Huang, C-F. Li, G-C. Guo, A. del Campo. {\it Testing quantum critical dynamics beyond the Kibble-Zurek mechanism with a trapped-ion simulator}. arXiv:1903.02145 (2019). 
\end{tabular}
\end{center}
\newpage
\section{Introduction}
The Kibble-Zurek mechanism (\gls{KZM}) is a highly successful paradigm to describe the dynamics of both thermal and quantum phase transitions~\cite{Kibble76a,Kibble76b,Zurek96a,Zurek96c}. It is one of the few theoretical tools that provide an account of nonequilibrium behavior in terms of equilibrium properties~\cite{Polkovnikov05,Damski05,Dziarmaga05,ZDZ05}. It predicts that in the course of a phase transition topological defects are formed. Since its conception, the \gls{KZM} has had immense impact on the scientific community. This influence spans over three decades and has spurred uninterrupted research since the mid 70s. It has been broadly used in a cosmological setting to discuss structure formation and the quest for cosmic strings. In condensed matter it has spurred research on liquid crystals, colloidal monolayers, superfluid helium, superconducting rings, ion chains, among a long list of examples~\cite{Ulm13,Pyka13,EH13,Keim15,Lamporesi13,Navon15}. AMOP experiments have tested the theory by studying the formation of soliton, vortex and spin textures in Bose-Einstein condensates. This activity has advanced our understanding of critical dynamics, e.g. by extending the \gls{KZM} to inhomogeneous systems~\cite{DKZ13}. In the quantum domain, experimental progress has been more limited and led by the use of quantum simulators in a variety of platforms~\cite{Xu2014,Cui16,Wu16}.\\
\\
The central prediction of \gls{KZM} is that the mean number of topological defects $\la n\ra$, formed when a system is driven through a critical point in a time scale $\tau_Q$, is given by a universal power-law $\la n\ra\sim \tau_Q^{-\beta}$. The power-law exponent $\beta=D\nu/(1+z\nu)$ is fixed by the dimensionality of the system $D$ and a combination of the equilibrium correlation-length and dynamic critical exponents, denoted by $\nu$ and $z$, respectively. Essentially, the \gls{KZM} is a statement about the breakdown of the adiabatic dynamics across a critical point. As such, it provides useful heuristics for the preparation of ground-state phases of matter in quantum simulation as well as for adiabatic quantum computation \cite{Suzuki09b}.\\
\\
In this context, the work present in this chapter provides a paradigm shift. The efforts to study the formation of topological defects have been biased by the available theoretical framework, focusing exclusively on the mean number of defects. Our work provides the first experimental evidence of the universal character of the distribution of the number of these defects. By shifting the emphasis to the probability distribution for defect formation, our work opens new directions of inquiry and paves the ways to solve open question in physics, such as the apparent absence of cosmic strings in the observable universe. \\
\\
\section{Full counting statistics of topological defects}\label{Sfull}

Beyond the focus of the \gls{KZM}, the full counting statistics encoded in the probability distribution $P(n)$ can be expected to  shed further light. The number distribution of topological defects $P(n)$  has become available in recent experiments~\cite{Lukin17}.  In addition, the distribution of kinks has recently been explored theoretically in the one-dimensional transverse-field quantum Ising model (\gls{TFQIM})~\cite{delcampo18}, a paradigmatic testbed of quantum phase transitions~\cite{sachdev}.  The distribution $P(n)$ has been predicted to be universal and fully determined by the scaling theory of phase transitions~\cite{delcampo18}. \\
\\
We shall be interested in experimentally probe the the kink distribution resulting from driving  a one-dimensional quantum Ising chain through the paramagnet-ferromagnet quantum phase transition. In this way, we consider a one-dimensional Ising model (see Eq.~\eqref{hami_xy} in the limit case $\gamma=1$). The \gls{TFQIM} is described by the Hamiltonian
\beqa
\hat{H}\pap{t}=-J\sum_{n=1}^L(\hat{\sigma}_n^z\hat{\sigma}_{n+1}^z+g\pap{t} \hat{\sigma}_n^x),
\label{H_Ising}
\eeqa
We show that in the static case ($g\pap{t}=g$), the hamiltonian~\eqref{H_Ising} can be written as a free fermion model, making use of the Jordan-Wigner, Fourier and Bogoliubov-de Gennes transformations. As a result, a new set of quasiparticle operators $\gamma_k$ that diagonalize the Hamiltonian as:
 \begin{eqnarray}
\hat{\mathcal{H}}=\sum_{k>0}\hat{H}_{k}=\sum_{k>0}\epsilon_{k}\pap{g}\pap{\hat{\gamma}_{k}^{\dagger}\hat{\gamma}_{k}+\hat{\gamma}_{-k}^{\dagger}\hat{\gamma}_{-k}-1},
\end{eqnarray}
where $\hat{\gamma}_{k}$ are quasiparticle operators, with $k$ labeling each mode and taking values $k=\frac{\pi}{N}\pap{2m-1}$ with $m=-\frac{N}{2}+1,\ldots,\frac{N}{2}$. The energy $\epsilon_k$ of the $k$-th mode is $\epsilon_{k}\pap{g}=2J\sqrt{\pap{g-\cos k}^2 +\sin^2 k}$. Other  physical observables can as well be expressed in both the spin and momentum representations. We are interested in characterizing the number of kinks. As conservation of momentum  dictates that kinks are formed in pairs,  we focus on the kink-pair number operator $\hat{\mathcal{N}}~\equiv~\sum_{m=1}^{N}\left(\hat{1}-\hat{\sigma}_{m}^{z}\hat{\sigma}_{m+1}^{z}\right)/4$. The later  can be equivalently written as $\hat{\mathcal{N}}=\sum_{k>0}\hat{\gamma}_{k}^{\dagger}\hat{\gamma}_{k}$, where $\hat{\gamma}_{k}^{\dagger}\hat{\gamma}_{k}$ is a Fermion number operator, with eigenvalues $\{0,1\}$. As different $k$-modes are decoupled,  this representation paves the way to simulate the dynamics of the phase transition in the TFQIM in ``momentum space'': the dynamics of each mode can be simulated with an ion-trap qubit, in which the  expectation value of $\hat{\gamma}_{k}^{\dagger}\hat{\gamma}_{k}$ can be measured. To this end, we consider the quantum critical dynamics induced by a ramp of the magnetic field
\begin{eqnarray}
g(t)=\frac{t}{\tau_{\rm Q}}+g(0),\label{gt}
\end{eqnarray}
in a time scale $\tau_{\rm Q}$ that we shall refer to as the quench time. We further choose $g(0)<-1$ in the paramagnetic phase. In momentum space, driving the phase transition is equivalently described by an ensemble of Landau-Zener crossings. This observation proved key in establishing the validity of the \gls{KZM} in the quantum domain~\cite{Dziarmaga05,Xu2014,Cui16}: the average  number of kink pairs $\langle\hat{\mathcal{N}}\rangle=\langle n\rangle$ after the quench scales as 
\begin{eqnarray}
\langle n\rangle_{{\rm KZM}}=\frac{N}{4\pi}\sqrt{\frac{\hbar}{2J\tau_{\rm Q}}},\label{navkzm}
\end{eqnarray}
in agreement with the universal power law $\langle n\rangle_{{\rm KZM}}\propto\tau_{\rm Q}^{-\frac{\nu}{1+z\nu}}$ with critical exponents $\nu=z=1$ and one spatial dimension ($D=1$).\\
\\
The full counting statistics of topological defects is encoded in the probability $P(n)$ that a given number of kink pairs $n$ is obtained as a measurement outcome of the observable $\hat{\mathcal{N}}$. To explore the implications of the scaling theory of phase transitions, we focus on the characterization of the probability distribution $P(n)$ of the  kink-pair number in the final nonequilibrium state upon completion of the crossing of the critical point induced by~\eqref{gt}. Exploiting the equivalence
between the spin and momentum representation, the dynamics in each mode leads to two possible outcomes, corresponding to the mode being found in the excited state (e) or the ground state (g), with probabilities $p_{\rm e}=p_{k}$ and $p_{\rm g}=1-p_{k}$, respectively. Thus, one can associate with each mode $k>0$ a discrete random variable, with excitation probability $p_{k}=\langle\hat{\gamma}_{k}^{\dagger}\hat{\gamma}_{k}\rangle$. The excitation probability of each mode is that of Bernoulli type.\\ 
\\
The kink-pair number distribution $P(n)$ describes the probability of observing $n$ pairs of kinks. The number of kink pairs is measured
by the observable 
\begin{eqnarray}
\hat{\mathcal{N}}=\frac{1}{4}\sum_{m=1}^{N}(1-\hat{\sigma}_{m}^{z}\hat{\sigma}_{m+1}^{z})=\sum_{k\geq0}\hat{\gamma}_{k}^{\dagger}\hat{\gamma}_{k},
\end{eqnarray}
with eigenvalues $n=0,1,2,\dots$ The kink-pair number distribution is thus defined as the quantum expectation value of the projector
operator $\delta[\hat{\mathcal{N}}-n]$ on the subspace with $n$ pairs of kinks 
\begin{eqnarray}
P(n)={\rm tr}\pas{\hat{\rho}\,\delta[\hat{\mathcal{N}}-n]},
\end{eqnarray}
where the expectation value is taken over the final state $\hat{\rho}$ upon completion of the quench. The Fourier transform of $P(n)$ is
the characteristic function 
\begin{eqnarray}
\widetilde{P}(\theta) & = & \int_{-\pi}^{\pi}d\theta P(n)e^{in\theta}=\mathbb{E}\left[e^{in\theta}\right].\label{charfunc}
\end{eqnarray}
The cumulants $\kappa_{q}$ of the distribution $P(n)$ are defined
via the expansion of the cumulant-generating function, which is the logarithm
of the characteristic function, 
\begin{eqnarray}
\log\widetilde{P}(\theta)=\sum_{q=1}^{\infty}\frac{(i\theta)^{q}}{q!}\kappa_{q}.
\end{eqnarray}

We next focus on the characterization of $P(n)$ for the state that
results from the nonadiabatic crossing of the phase transition in
the TFQIM and derive the exact expression for
the first few cumulants $\{\kappa_{q}\}$. To this end, we note that
the nonadiabatic crossing of the phase transition results in the production
of quasi-particles (and kinks) in pairs. Specifically, due to the
conservation of momentum, quasiparticles with wavectors $+k$ and
$-k$ are excited jointly. The distribution of the number of pairs
of kinks is a Poisson binomial distribution associated with $N/2$
Bernoulli trials each with probability $p_{k}$ for $k>0$. Its characteristic
function reads 
\begin{eqnarray}
\widetilde{P}(\theta)=\prod_{k>0}\left[1+\pap{e^{i\theta}-1}p_{k}\right],
\end{eqnarray}
with $k\in\{\pi/N,3\pi/N,\ldots,(N-1)\pi/N\}$ for an Ising chain with
period boundary conditions and $p_{k}$ denoting the excitation probability
in the $k$-mode. The Poisson binomial distribution is well studied and
its cumulants are known to be of the form 
\begin{eqnarray}
\kappa_{1} & = & \langle n\rangle=\sum_{k>0}p_{k},\\
\kappa_{2} & = & {\rm Var}(n)=\sum_{k>0}p_{k}(1-p_{k}),\\
\kappa_{3} & = & \langle(n-\langle n\rangle)^{3}\rangle=\sum_{k>0}p_{k}(1-p_{k})(1-2p_{k}).
\end{eqnarray}

Their explicit evaluation can be performed using the Landau-Zener
formula for $p_{k}$. In the continuum limit, this yields 
\begin{align*}
\kappa_{1}\equiv\langle n\rangle & =\sum_{k>0}p_{k}\\
& =\frac{N}{2\pi}\int_{0}^{\pi}dk\exp\left(-\frac{1}{\hbar}2\pi J\tau_{\rm Q}k^{2}\right).
\end{align*}
In terms of the error function, it is found that 
\begin{eqnarray}
\kappa_{1}=\frac{N}{4\pi}\sqrt{\frac{\hbar}{2J\tau_{\rm Q}}}\times{\rm erf}\left(\pi\sqrt{\frac{2\pi J\tau_{\rm Q}}{\hbar}}\right),\label{k1c}
\end{eqnarray}
an expression that is valid even for moderately fast quenches, as
long as described by the Landau-Zener formula but possibly away from
the scaling limit. However, for large $\tau_{\rm Q}>\hbar/(2\pi^{3}J)$,
${\rm erf}\left(\frac{\sqrt{\pi}}{2d}\right)$ approaches unity and
the average number of kink pair simply reads 
\begin{eqnarray}
\langle n\rangle_{{\rm KZM}}=\frac{N}{4\pi}\sqrt{\frac{\hbar}{2J\tau_{\rm Q}}},\label{k1a}
\end{eqnarray}
in agreement with the Kibble-Zurek Mechanism~\cite{Dziarmaga05}.

A similar derivation shows that the variance is given by 
\begin{align*}
\kappa_{2} & \equiv{\rm Var}(n)=\sum_{k>0}p_{k}(1-p_{k})\\
& =\frac{N}{2\pi}\int_{0}^{\pi}dke^{-\frac{2\pi J\tau_{\rm Q}}{\hbar}k^{2}}\left(1-e^{-\frac{2\pi J\tau_{\rm Q}}{\hbar}k^{2}}\right).
\end{align*}
This results in the exact expression 
\begin{eqnarray}
{\rm Var}(n)=\left[{\rm erf}\left(\sqrt{\frac{2\pi^{3}J\tau_{\rm Q}}{\hbar}}\right)-\frac{1}{\sqrt{2}}{\rm erf}\left(\sqrt{\frac{4\pi^{3}J\tau_{\rm Q}}{\hbar}}\right)\right]\langle n\rangle_{{\rm KZM}},\label{k2c}
\end{eqnarray}
that in the scaling limit reduces to 
\begin{eqnarray}
{\rm Var}(n)=\left(1-\frac{1}{\sqrt{2}}\right)\langle n\rangle_{{\rm KZM}}.\label{k2a}
\end{eqnarray}
\begin{figure}[h!]
\begin{center}
\includegraphics[width=0.6\textwidth]{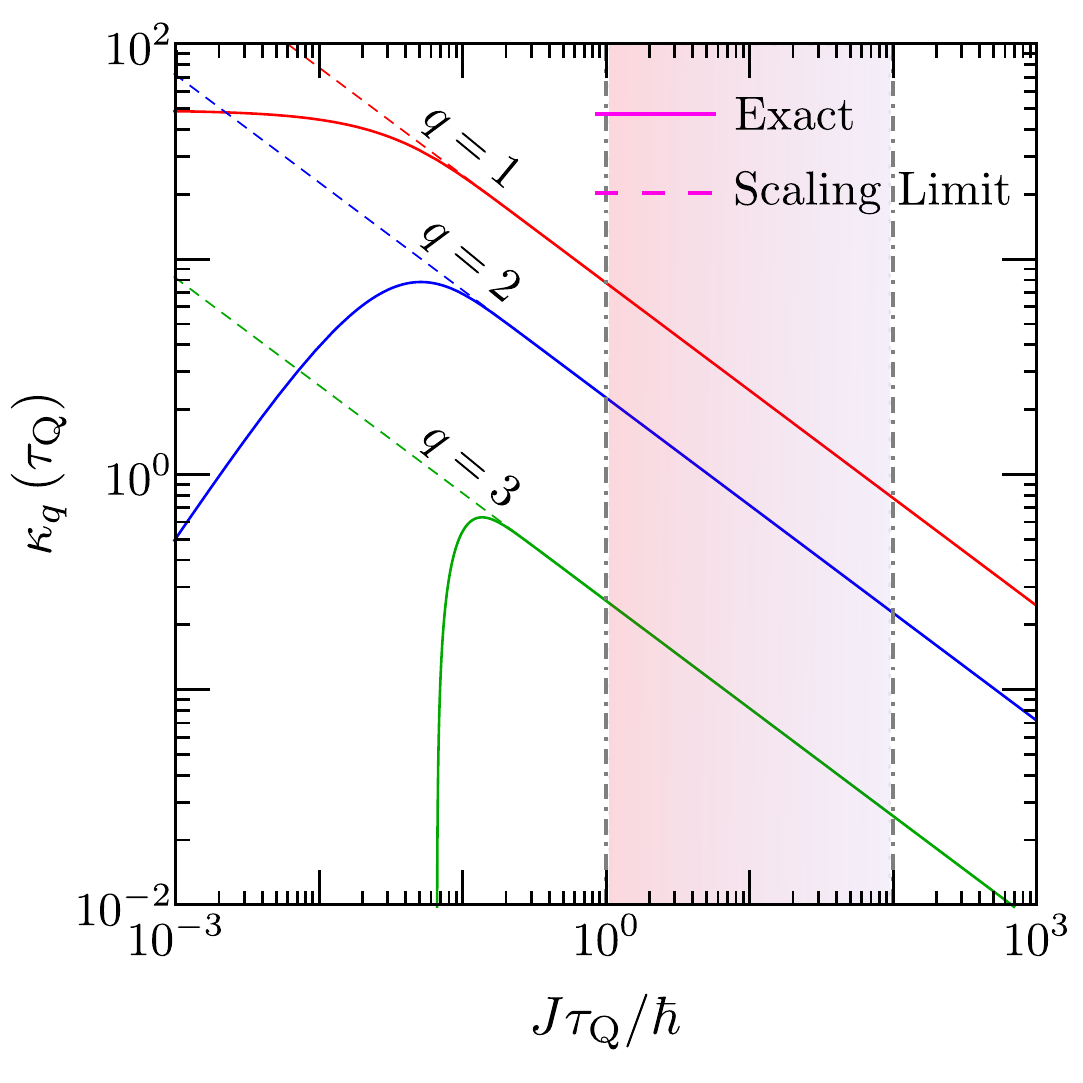} 
\end{center}
\caption[Universal scaling of the cumulants $\kappa_{q}(\tau_{\rm Q})$  of the kink-pair number distribution $P(n)$ as a function of the quench time.]{\textbf{Universal scaling of the cumulants $\kappa_{q}(\tau_{\rm Q})$ 
of the kink-pair number distribution $P(n)$ as a function of the quench time.} 
The solid lines show the exact result for the first three cumulant $\kappa_{q}(\tau_{\rm Q})$ with $q\in{1,2,3}$, 
as predicted by equations~\eqref{k1c},~\eqref{k2c}, and~\eqref{k3c}. The dashed lines show the universal scaling of the first cumulant
according to Eqs.~\eqref{k1a},~\eqref{k2a}, and~\eqref{k3a}. In the experimental regime, marked by the shaded area, the scaling
behavior holds.} 
\end{figure}

To characterize the non-normal features of the distribution of kinks,
we also provide the exact expression for the third cumulant $\kappa_{3}$,
that equals the third centered moment $\mu_{3}=\langle(n-\langle n\rangle)^{3}\rangle$.
In particular, we note that $\kappa_{3}$ is related to the skewness
$\gamma_{1}$ of the distribution, as $\gamma_{1}=\kappa_{3}/\kappa_{2}^{3/2}$.
In the continuum, it reads 
\begin{eqnarray}
\kappa_{3} & \equiv & \langle(n-\langle n\rangle)^{3}\rangle=\sum_{k>0}p_{k}(1-p_{k})(1-2p_{k})\nonumber \\
& = & \frac{N}{2\pi}\int_{0}^{\pi}dke^{-\frac{2\pi J\tau_{\rm Q}}{\hbar}k^{2}}\!\!\left(1-e^{-\frac{2\pi J\tau_{\rm Q}}{\hbar}k^{2}}\right)\!\!\left(1-2e^{-\frac{2\pi J\tau_{\rm Q}}{\hbar}k^{2}}\right)\nonumber \\
& = & \left[{\rm erf}\left(\sqrt{\frac{2\pi^{3}J\tau_{\rm Q}}{\hbar}}\right)-\frac{3}{\sqrt{2}}{\rm erf}\left(\sqrt{\frac{4\pi^{3}J\tau_{\rm Q}}{\hbar}}\right)+\frac{2}{\sqrt{3}}{\rm erf}\left(\sqrt{\frac{6\pi^{3}J\tau_{\rm Q}}{\hbar}}\right)\right]\langle n\rangle_{{\rm KZM}}.\label{k3c}
\end{eqnarray}
This expression is simplified in the scaling limit, performing a power
series expansion with $\tau_{\rm Q}\rightarrow\infty$. To leading order,
we obtain the expression quoted  in the main text 
\begin{eqnarray}
\kappa_{3}=\left(1-\frac{3}{\sqrt{2}}+\frac{2}{\sqrt{3}}\right)\langle n\rangle_{{\rm KZM}}.\label{k3a}
\end{eqnarray}

Note that the scaling expressions Eqs.~\eqref{k1a},~\eqref{k2a} and~\eqref{k3a} are obtained from the more general ones, Eqs.~\eqref{k1c},~\eqref{k2c} and~\eqref{k3c}, whenever the error functions can be replaced by unity. The error function increases monotonically and
swiftly from zero value to unity as the argument is increased from zero value. Indeed, it saturates at unity for fairly moderate values
of the argument, e.g., ${\rm erf}(2)=0.995$. By imposing $\sqrt{\frac{2q\pi^{3}J\tau_{\rm Q}}{\hbar}}>2$,
we predict the onset of the scaling limit for $\kappa_{q}$ at 
\begin{eqnarray}
\tau_{\rm Q}>\frac{2\hbar}{q\pi^{3}J}.
\end{eqnarray}
This condition is satisfied in all the experimental realizations we have performed. As a result, it is justified to consider the scaling
limit, in which it is actually possible to derived a closed form expression
of the characteristic function~\cite{delcampo18} 
\begin{eqnarray}
\widetilde{P}(\theta)=\exp\left[-\langle n\rangle_{{\rm KZM}}\,{\rm Li}_{3/2}(1-e^{i\theta})\right],\label{CGFa}
\end{eqnarray}
in terms of the polylogarithmic function ${\rm Li}_{3/2}(x)=\sum_{p=1}^{\infty}x^{p}/p^{3/2}$.

From this expression, it follows that all cumulants are actually nonzero and proportional to $\langle n\rangle_{{\rm KZM}}$. The kink-pair
number distribution is manifestly non-normal. The Gaussian (normal) approximation discussed in the main text follows from neglecting all
$\kappa_{q}$ with $q>2$. One can also approximate $1-1/\sqrt{2}\approx3/\pi^{2}$,
so that 
\begin{eqnarray}
\widetilde{P}(\theta)\simeq\exp\left[-\langle n\rangle_{{\rm KZM}}\left(i\theta-\frac{3}{2\pi^{2}}\theta^{2}\right)\right],
\end{eqnarray}
and thus 
\begin{eqnarray}\label{Gausspn}
P(n)\simeq\frac{1}{\sqrt{6\langle n\rangle_{{\rm KZM}}/\pi}}\exp\left[-\frac{\pi^{2}(n-\langle n\rangle_{{\rm KZM}})^{2}}{6\langle n\rangle_{{\rm KZM}}}\right].
\end{eqnarray}

In the next section, we will probe the universal distribution of topological defects formed across a quantum phase transition with a trapped-ion quantum simulator.

\section{Experimental test of the universal full counting \gls{STD} created across a quantum phase transition with a trapped-ion quantum simulator}
\subsection{Experimental design}
\begin{figure}[h!]
\centering \includegraphics[width=15cm]{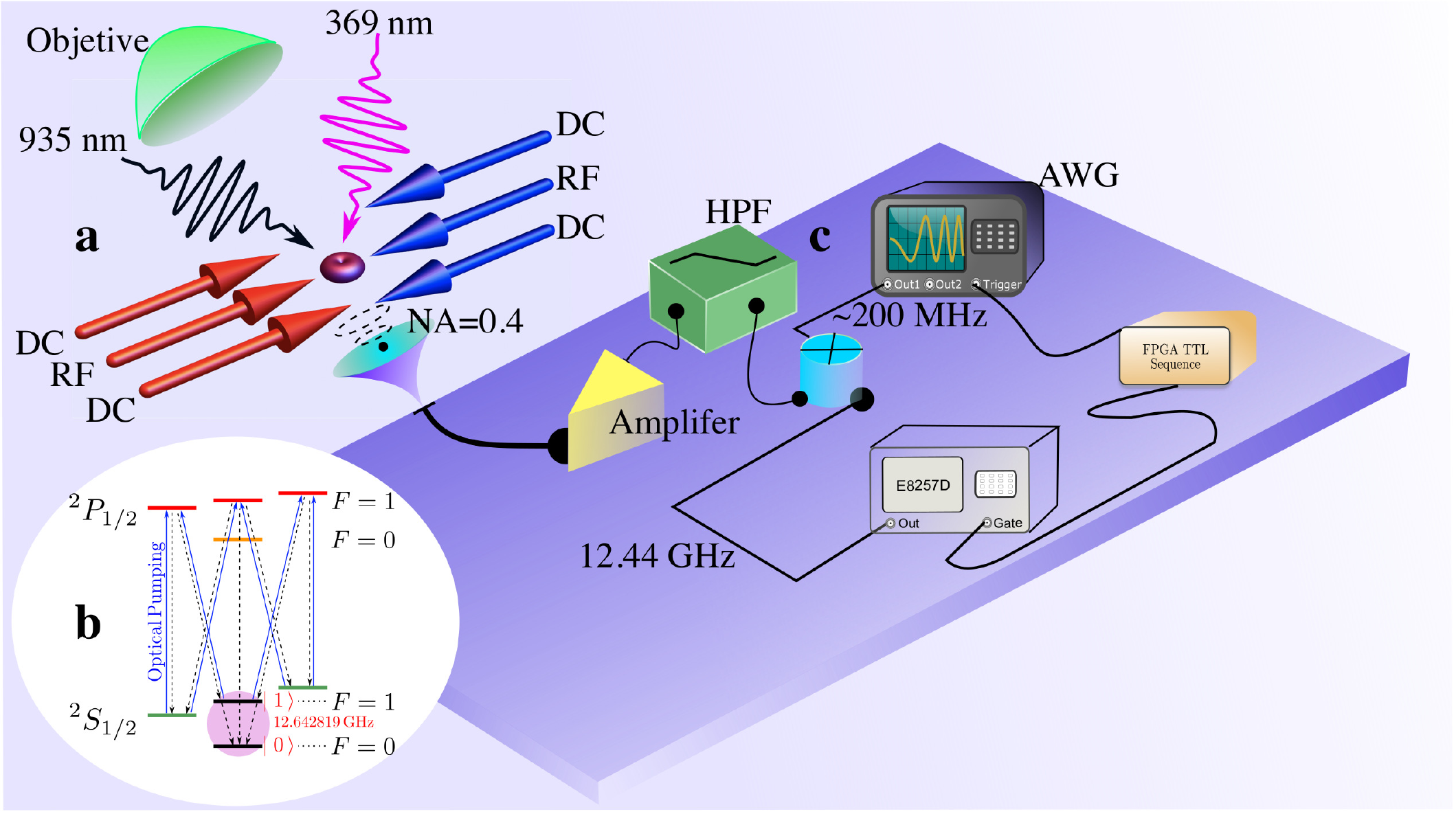} 
\caption[Experiment setup of the universal full counting \gls{STD} created across a quantum phase transition with a trapped-ion quantum simulator]{\label{fig_C8_1}{\bf Experiment setup.} \textbf{a} A single $^{171}{\mathrm{Yb}}^{+}$ is trapped in a needle trap, which consists of six needles on two perpendicular planes. \textbf{b}  The qubit energy levels are denoted by $\left|0\right\rangle$ and $\left|1\right\rangle $, which is the hyperfine clock transition of the trapped ion. \textbf{c} The qubit is driven by a microwave field, generated by a mixing wave scheme in the high pass filter (HPF).  Operations on the qubit are implemented by programming the arbitrary wave generator (AWG). The quantum critical dynamics of the one-dimensional transverse-field quantum Ising model is detected by measuring corresponding Landau-Zener crossings governing the dynamics in each mode.}
\end{figure}
Experimentally, the excitation probabilities $\{p_{k}\}$ can be measured by probing the dynamics in each mode, that is simulated with the ion-trap qubit. The dynamics of a single mode $k$ is described by a Landau-Zener crossing, that  is implemented with a $^{171}\mathrm{Yb^{+}}$ ion confined in a Paul trap consisting of six needles placed on two perpendicular planes, as shown in Fig.~\ref{fig_C8_1}{\bf a}. The hyperfine clock transition in the ground state $S_{1/2}$ manifold is chosen to realize the qubit, with energy levels denoted by $\left|0\right\rangle \equiv\left|F=0,\,m_{F}=0\right\rangle $ and $\left|1\right\rangle \equiv\left|F=1,\,m_{F}=0\right\rangle $, as shown in Fig.~\ref{fig_C8_1}{\bf b}.

\subsection{State preparation and manipulation} 
At zero static magnetic field, the splitting between $\left|0\right\rangle $ and $\left|1\right\rangle $ is 12.642812 GHz. We applied a static magnetic field of 4.66 G to define the quantization axis, which changes the $\left|0\right\rangle $ to $\left|1\right\rangle $ resonance frequency to 12.642819 GHz, and creates a 6.5 MHz Zeeman splittings for $\mathrm{^{2}S_{1/2},\,F=1}$. In order to manipulate the hyperfine qubit with high control, coherent driving is implemented by a wave mixing method, see the scheme in Fig.~\ref{fig_C8_1}{\bf c}. First, an arbitrary wave generator (AWG) is programmed to generate signals around 200 MHz. Then, the waveform is mixed with a 12.442819 GHz microwave (generated by Agilent, E8257D) by a frequency mixer. After the mixing process, there will be two waves around 12.242 GHz and 12.642 GHz, so a high pass filter is used to remove the 12.224 GHz wave. Finally, the wave around 12.642 GHz is amplified to 2\,W  and used to irradiate the trapped ion with a horn antenna.

\subsection{Measurement} 
For a typical experimental measurement in a single mode, Doppler cooling is first applied to cool down the ion~\cite{Olmschenk2007}. The ion qubit is then initialized in the $\left|0\right\rangle $ state, by applying a resonant light at 369nm to excite ${S_{1/2}},\,F=1$ to ${P_{1/2}},\,F=1$. Subsequently, the programmed microwave is started to drive the ion qubit. Finally, the population of the bright state $\left|1\right\rangle $ is detected by fluorescence detection with another resonant light at 369nm, exciting ${S_{1/2}},\,F=1$ to ${P_{1/2}},\,F=0$. Fluorescence of the ion is collected by an objective with 0.4 numerical aperture (NA). A 935nm laser is used to prevent the state of the ion to jump to metastable states \cite{Olmschenk2007}.  The initialization process can prepare the $\left|0\right\rangle $ state with fidelity $>99.9\%$. The total error associated with the state preparation and measurement is measured as 0.5\% ~\cite{Maunz2017NC}.

We can rewrite the explicit Hamiltonian of the $k$-th mode as a linear combination of a single two-levels system~\cite{Dziarmaga05}. In this way, experimentally, the Hamiltonian of a single $k$-mode can be explicitly mapped to a qubit Hamiltonian describing a two-level  system driven by a chirped  microwave pulse, 
\begin{equation}\label{TLS_H}
\hat{H}_{k}^{\rm TLS} =\frac{1}{2}\hbar\pap{\Delta_k(t)\hat{\sigma}_{z}+\Omega_{\rm R}\hat{\sigma}_{x}}, 
\end{equation}
where $\Omega_{\rm R}=4J/\hbar$ and  $\Delta_k(t)=4J[g(t)+\cos{k}]/(\hbar\sin{k})$ are the Rabi frequency and the detuning of the chirped pulse, respectively. In the experiment, the Rabi frequency of the qubit simulator is set around 20 kHz, which depends on the driven power of our microwave amplifier. Higher microwave power can  shorten the Rabi time $T_{R}=1/\Omega_{R}$, and reduce the total  operation time. To simulate the quench process, one would like to vary  $g$ from $-\infty$ to 0, an idealized evolution that needs infinite time, which can not be realized in experiment. However, to explore the universality associated with the crossing of the critical point  one can initialize the system sufficiently deep in the paramagnetic phase. According to the \gls{KZM},  it is sufficient to  choose an initial value of the magnetic field such that the corresponding equilibrium relaxation time is much smaller than the time left until crossing the critical point. The system is then prepared out of the ``frozen region'' in the language of the adiabatic-impulse approximation~\cite{Dziarmaga05}.  For the quench time $\tau_{\rm Q}\geq1$, the initial value of $g=-5$ is far out of the frozen time. We can simulate the TFQIM with an initial $g=-5$ and no  excitations, by preparing the  initial state of the qubit  in the ground state of $\hat{H}_{k,i}^{s}=\hat{H}_{k}^{s}(g=-5)$
before the quench process. The initial state can be derived by solving the eigenvectors of Eq.~\eqref{TLS_H}, which is
\begin{equation}
\left|\psi_{-}\right\rangle _{k,i}=\cos\frac{\theta_{k,i}}{2}\left|0\right\rangle -\sin\frac{\theta_{k,i}}{2}\left|1\right\rangle ,\label{eq:phi_i}
\end{equation}
where $\theta_{k,i}=-\arctan\frac{\sin\left(ka\right)}{5+\cos\left(ka\right)}$.
\begin{figure}[h!]
	\centering \includegraphics[width=0.8\textwidth]{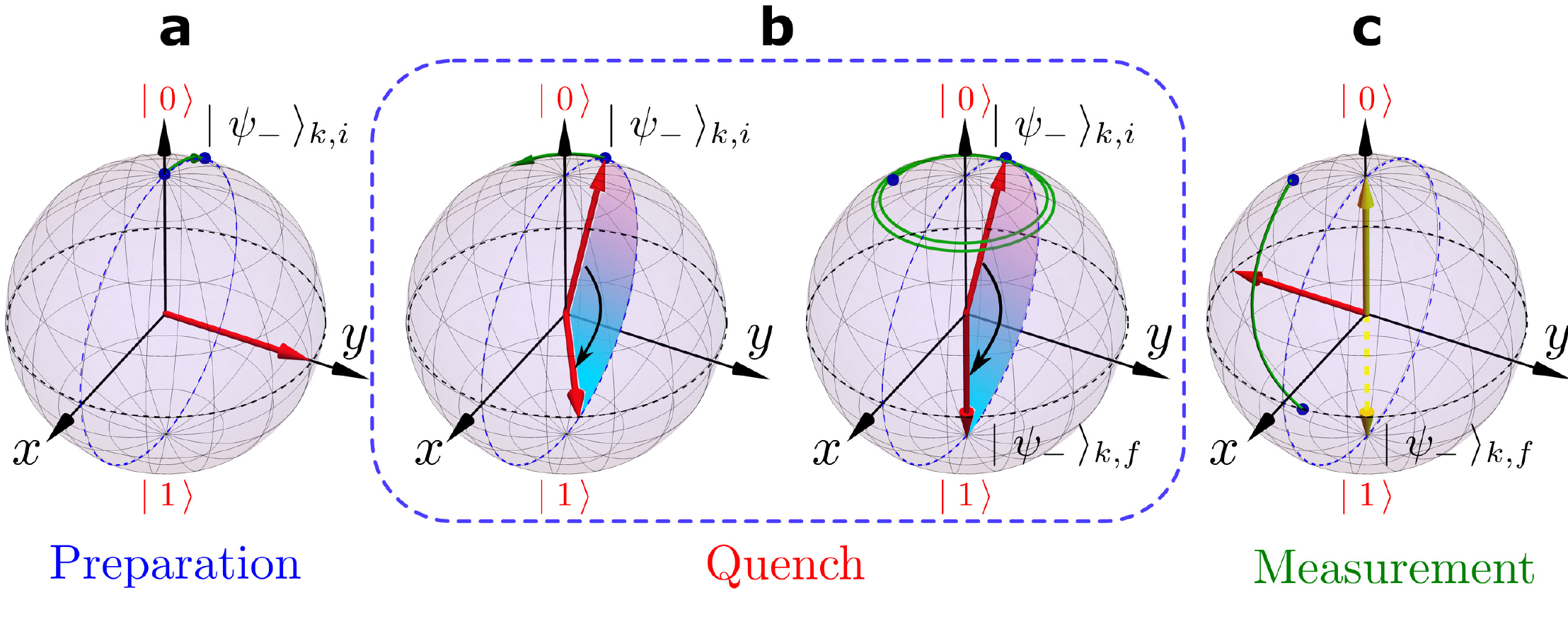} 
	\caption[Scheme to measure the excitation probability with a trapped-ion quantum simulator]{\textbf{Scheme to measure the excitation probability.} The quantum critical dynamics of the one-dimensional transverse-field quantum Ising model is detected
			by measuring corresponding Landau-Zener crossings governing the dynamics
			in each mode. For each mode, a typical process to measure the excitation
			probability in three stages is shown in \textbf{a}, \textbf{b} and
			\textbf{c}.}
	\label{fig_block}
\end{figure}
The scheme to detected the quantum critical dynamics of the one-dimensional TFQIM in each mode by using a single qubit is shown in Fig.~\ref{fig_block}.  The whole process can be divided to three steps. Before the process, the qubit has been pumped to $\left|0\right\rangle $ state by using a 369nm laser to excite transition ${S_{1/2}},\,F=1\rightarrow\ensuremath{{P_{1/2}},\,F=1}$. In the first stage, the ion-trap qubit  is prepared into the state $\left|\psi_{-}\right\rangle _{k,i}$ by a resonant microwave pulse. The second stage is the quench process; the Hamiltonian is time dependent and varies from $\hat{H}_{k,i}^{s}\equiv \hat{H}_{k}^{s}(g=-5)$ to $\hat{H}_{k,f}^{s}\equiv \hat{H}_{k}^{s}(g=0)$.  This is implemented   by driving a chirped pulse into the qubit. The chirped pulse is in the form of  $\Delta_{k}(t)=4J\frac{g\pap{t}-\cos\left(ka\right)}{\hbar\sin\left(ka\right)}$ with $g(t)=5(t/T_{p}-1)$, where $T_{p}=5J\tau_{\rm Q}\sin(ka)/\left(\hbar\Omega_{R}\right)$ is the pulse length. The third stage is to measure the excitation probability after
the quench, which is to measure the occupation probability on the excited eigenstate of $\hat{H}_{k,f}^{s}$. The excited eigenstate is $\left|\psi_{-}\right\rangle _{k,f}=\sin\theta_{k,f}\left|0\right\rangle +\cos\theta_{k,f}\left|1\right\rangle $, where $\theta_{k,f}=-ka$. As the fluorescence detection on $^{171}\mathrm{Yb}^{+}$ can only discriminate $\left|0\right\rangle $ and $\left|1\right\rangle $ sates, we need to rotate the state $\left|\psi_{-}\right\rangle _{k,f}$ to $\left|1\right\rangle $ and then detect the bright state probability.  We thus use a qubit rotation and a fluorescence detection  in this
stage. The calculated waveform to set the AWG to perform the pre-rotation, quench and post-rotation in the three stages will be discussed in
the next subsection. 
\subsection{Driving Waveform with an Arbitrary Wave Generator}
We consider the driving microwave from AWG is $A\cos(\omega_{c}t+\varPhi(t))$,
where $A$ is the amplitude, $\omega_{c}$ is the carrier frequency
and $\varPhi(t)$ is an arbitrary phase function. The carrier frequency
can be mixed up with some frequency $\omega_{0}$ from a local oscillator
(LO) by using a frequency mixer. The waveform at the output of the
mixer is
\[
A\cos(\omega_{c}t+\varPhi(t))\cos(\omega_{o}t)=A/2\left(\cos((\omega_{o}+\omega_{c})t+\varPhi(t))+\cos((\omega_{o}-\omega_{c})t+\varPhi(t))\right).
\]
When the wave passes through the high pass filter, the wave form is filtered as $A/2\cos((\omega_{o}+\omega_{c})t+\varPhi(t))$. In the
experiment, we choose the frequency $\omega_{o}+\omega_{c}$ resonant
with the qubit transition $\omega_{Q}=\omega_{o}+\omega_{c}$, so the
magnetic field of the final driving waveform is $\mathbf{B}(t)=\mathbf{B_{0}}\cos(\omega_{Q}t+\varPhi(t))$.
The interaction between the magnetic field and spin is $H_{I}=-\mathbf{\mu}\cdot\mathbf{B}$,
where $\mu$ is the magnetic dipole of the spin qubit. The Hamiltonian
can be further simplified after the rotating wave approximation in the
interaction frame
\begin{equation}
\hat{H}_{I}(t)=\frac{\hbar}{2}\left[\Omega_{R}\left|1\right>\left\langle 0\right|e^{i\varPhi(t)}+\Omega_{R}^{*}\left|0\right>\left\langle 1\right|e^{-i\varPhi(t)})\right],\label{H4E}
\end{equation}
where the Rabi frequency $\Omega_{R}=-\left\langle 0\right|\mathbf{\mu}\cdot\mathbf{B}\left|1\right>/\hbar$.
The phase function $\varPhi(t)$ corresponds to the azimuthal angle
in the Bloch sphere. Equation~\eqref{TLS_H} can be also expressed
in interaction frame 
\[
H_{I,k}^{s}=\frac{\hbar}{2}\left(\Omega_{R}\left|1\right>\left\langle 0\right|\exp\left(i\int\Delta(t)dt\right)+\Omega_{R}^{*}\left|0\right>\left\langle 1\right|\exp\left(-i\int\Delta(t)dt\right)\right),
\]
so in the quench stage, the phase function is 
\[
\varPhi(t)=\intop_{0}^{T_{P}}\Delta(t)dt=\frac{\Omega_{R}}{\sin ka}\left(\frac{\Omega_{R}}{\tau_{\rm Q}\sin ka}t^{2}-\left(5+\cos ka\right)t\right).
\]
 \begin{figure}[t!]
\centering \includegraphics[width=9cm]{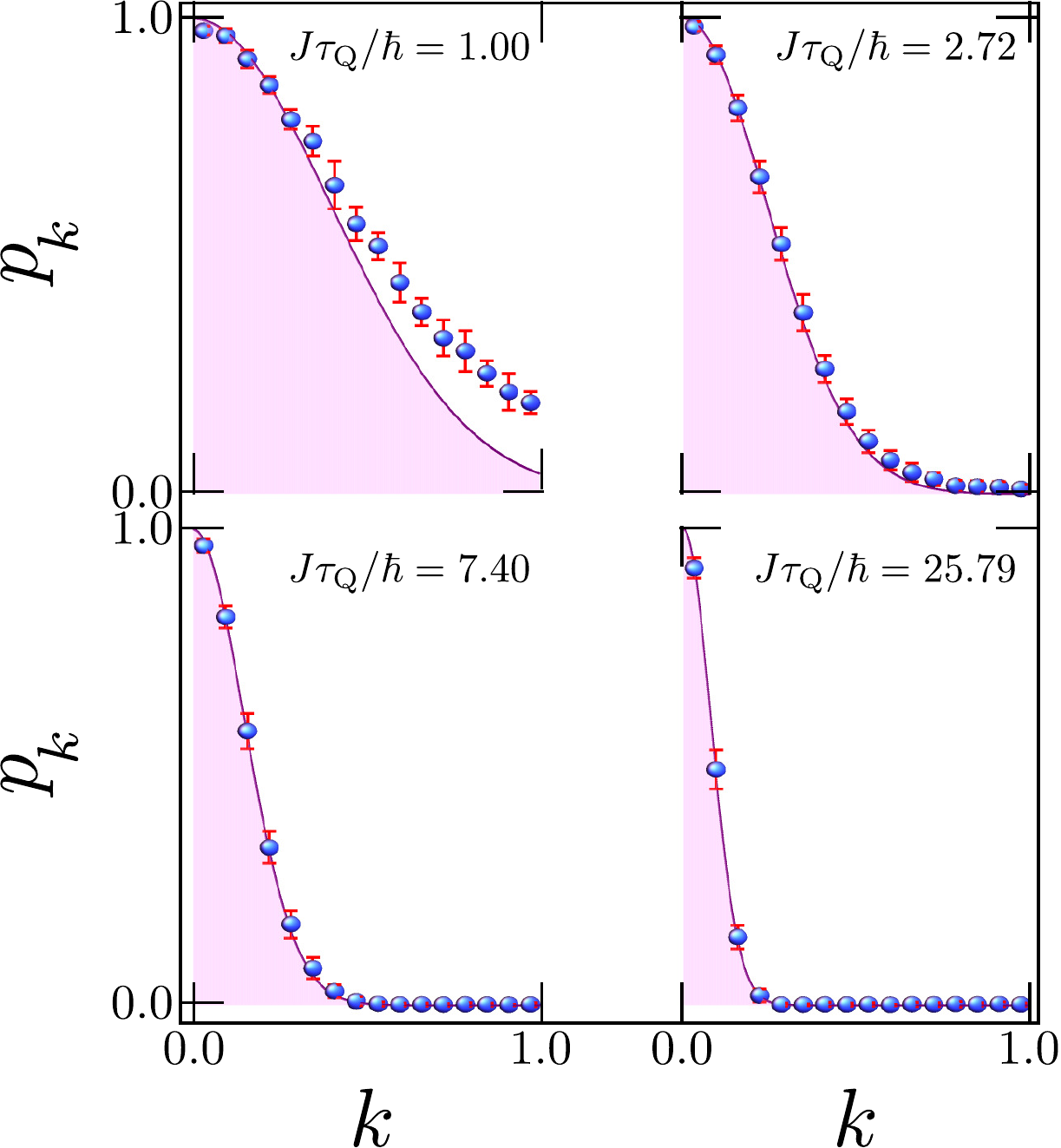} 
\caption[Excitation probability in the ensemble of Landau-Zener crossings.]{\label{C8_fig_2}\textbf{Excitation probability in the ensemble of Landau-Zener crossings.} The quantum critical dynamics of the one-dimensional transverse-field quantum Ising model is accessible in an ion-trap quantum simulator, which implements the corresponding Landau-Zener crossings governing the dynamics in each mode, labelled by wavector $k$. The Rabi frequency $\Omega_{\rm R}=4J/\hbar$ was set to $2\pi\times 20$ kHz in experiment. For each value of the wavector $k$, the excitation probability is estimated from $10000$ measurements. The shaded region describes the excitation probability predicted by  the Landau-Zener formula. Deviations from the later become apparent for fast quench times, especially for  large values of $k$. Error bars indicate the standard-deviation over $10000$ measurements.}
\end{figure}

We rotate the state along a vector in the equatorial plane to prepare
a state from $\left|0\right>$, or measure state to $\left|1\right>$,
which means $\varPhi(t)$ is constant in these two stages. The pulse
length for the preparation and measurement stages is determined by
the polar angle of the ground state at the beginning and end of the quench
process, $t=\theta/\Omega_{R}$, respectively. The whole expression
of $\varPhi(t)$ in the three stages can be derived as
\[
\varPhi(t)=\begin{cases}
\frac{\pi}{2}, & (0,\,t_{1})\\
\frac{\Omega_{R}}{\sin ka}\left(\frac{\Omega_{R}}{\tau_{\rm Q}\sin ka}(t-t_{1})^{2}-\left(5+\cos ka\right)(t-t_{1})\right), & (t_{1},\,t_{2})\\
\phi_{f}-\frac{\pi}{2}, & (t_{2},t_{3})
\end{cases}
\]
where $t_{1}=(2\pi-\theta_{k,i})/\Omega_{R}$, $t_{2}=t_{1}+T_{p}$, $t_{3}=t_{2}+(2\pi+\theta_{k,f})/\Omega_{R}$ and $\phi_{f}=-5\tau_{\rm Q}(2.5+\cos ka)$.\\
\\
\paragraph*{Qubit preparation and detection error}

There are some limitations in the preparation and measurement of the
qubit. We measured the error through a preparation and detection experiment.
First, we prepare the qubit in the $\left|0\right>$ state by optical
pumping method, and detect the ion fluorescence. Ideally, no photon
can be detected as the ion is in the the dark state. However, there
are dark counts of the photon detector  as well as photons scattered
from the environment. We repeat the process $10^{5}$ times, and estimate
the detected photon numbers. We also repeat the process 10000 times
by preparing the qubit in the bright state. Histograms for the photon
number in the dark and bright states are shown in the Supplementary
Figure.~\ref{fig_SPM_error}. In the experiment, the threshold is
selected as 2: the state of the qubit is identified as the bright
state when the detected photon number is greater or equal to 2. Obviously,
there is a bright error $\epsilon_{B}$ for the dark state above the threshold,
and a dark error $\epsilon_{D}$ for the bright state below the threshold.
The total error can be take as $\epsilon=(\epsilon_{B}+\epsilon_{D})/2$.
\begin{figure}[h!]
	\centering \includegraphics[width=0.8\textwidth]{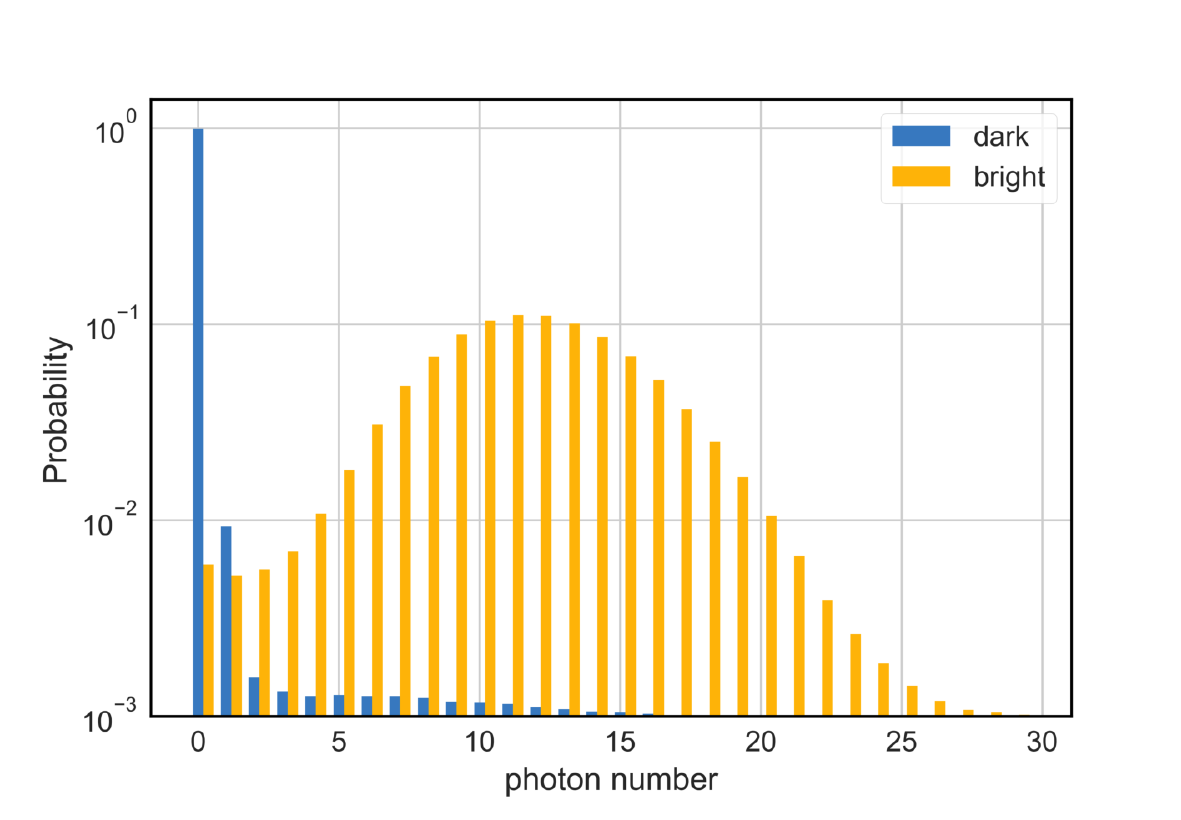} 
	\caption[Histogram for photon counts in state preparation and detection
			experiments.]{\textbf{Histogram for photon counts in state preparation and detection
			experiments.} The distribution of photon counts is shown when the
		qubit state is prepared in $\left|0\right>$ (dark state) and $\left|1\right>$
		(bright state). }
	\label{fig_SPM_error} 
\end{figure}

Following the steps in last subsections, we can measure the excitations probability after a finite ramp with normalized quench parameter $A=J\tau_{\rm Q}/\hbar$, we can vary $g(t)$ from -5 to 0. For this purpose, a chirped pulse with length of  $T_p=5A T_{\rm R}\sin{k}$ and $g(t)=5t/T_p-5$ is used in the experiment, with $T_{\rm R}=1/\Omega_{\rm R}$ denoting the Rabi time. A typical operation process driven by the programmed microwave is illustrated on the Bloch sphere, see Methods for details. First, the qubit is prepared in the ground state of $\hat{H}_{k}^{\rm TLS}$ at $g(0)=-5$. Then, $g(t)$ is ramped from $-5$ to 0 with quenching rate $1/\tau_{\rm Q}$. Finally, the ground state of the final $\hat{H}_{k}^{\rm TLS}$ is rotated to qubit $\left|0\right>$, so that the excited state is mapped to $\left|1\right>$,  which can be detected as a bright state. The measured excitation probability for different quench times is shown in Fig.~\ref{C8_fig_2}.
\subsection{Measurement of full probability distribution of topological defects} 
\begin{figure}[t!]
\centering \includegraphics[width=14cm]{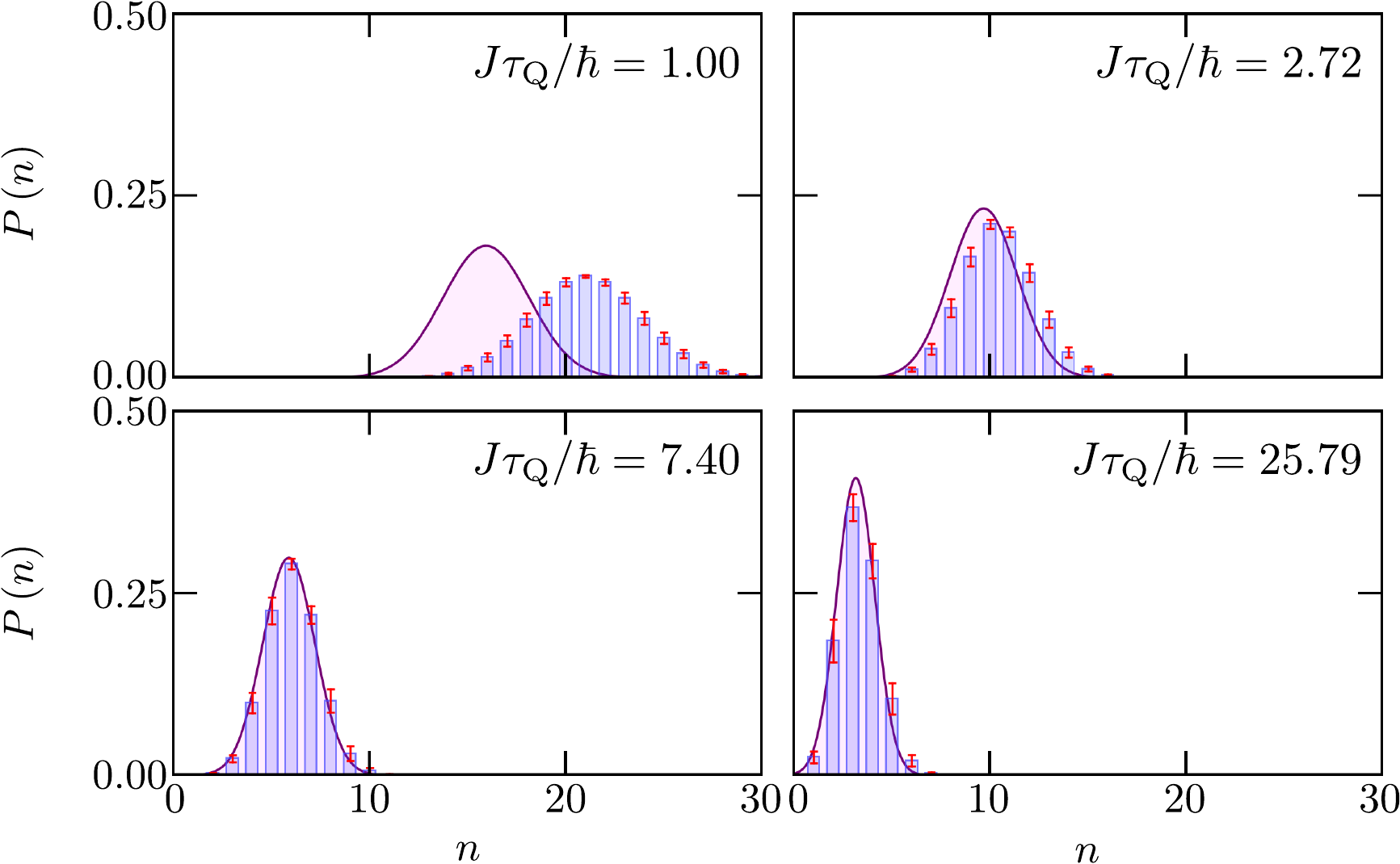} 
\caption[Probability distribution of the number of kink pairs $P(n)$ generated as a function of the quench time.]{\label{C8_fig_3}\textbf{Probability distribution of the number of kink pairs $P(n)$ generated as a function of the quench time.} The experimental kink-pair number distribution for a transverse-field quantum Ising model with $N=100$ spins is compared with the Gaussian approximation derived in the scaling limit, Eq.~\eqref{Gausspn}, ignoring high-order cumulants. The mean and width of the distribution are reduced as the quench time is increased. The experimental $P(n)$ is always broader and shifted to higher kink-pair numbers than the theoretical prediction. Non-normal features of $P(n)$ are enhanced near the sudden-quench limit and at the onset of adiabatic dynamics. Error bars indicate the standard-deviation over $10000$ measurements.}
\end{figure}

From the experimental data, the distribution $P(n)$ of the number of kink pairs is obtained using the characteristic function~\eqref{charfunc} of a Poisson binomial distribution in which the probability $p_{k}$ of each Bernoulli trial is set by the experimental value of the excitation probability upon completion of the corresponding Landau-Zener sweep. This result is compared with the theoretical prediction valid in the scaling limit --the regime of validity of the \gls{KZM}-- in which $P(n)$ approaches the normal distribution, away from the adiabatic limit~\cite{delcampo18}.  A comparison between $P(n)$ and Equation~\eqref{Gausspn} is shown in Fig.~\ref{C8_fig_3}. The matching between theory and experiment is optimal in the scaling limit far away from the onset of the adiabatic dynamics or fast quenches, for which non-universal corrections are expected. We further note that the onset of adiabatic dynamics enhances non-normal features of the experimental $P(n)$ that cannot be simply accounted for by a truncated Gaussian distribution, that takes into account the fact that the number of kinks $n=0,1,2,3,\dots$
 \begin{figure}[t!]
\centering \includegraphics[width=8.5cm]{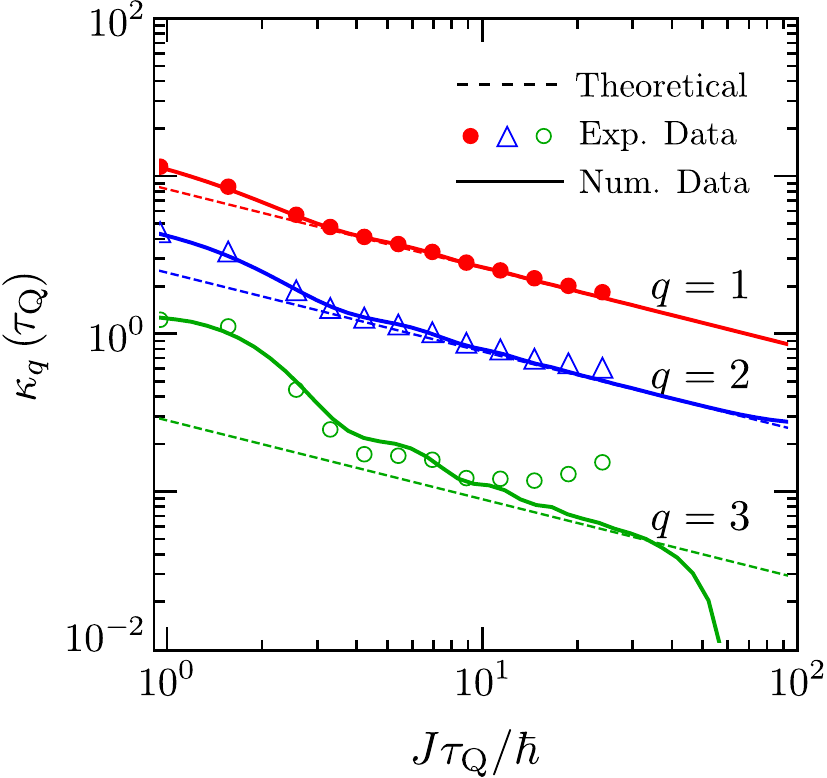}
 \caption[Universal scaling of the cumulants $\kappa_{q}(\tau_{\rm Q})$ of the kink-pair number distribution $P(n)$ as a function of the quench time.]{\label{C8_fig_4}\textbf{Universal scaling of the cumulants $\kappa_{q}(\tau_{\rm Q})$ of the kink-pair number distribution $P(n)$ as a function of the quench time.} The experimental data (symbols) is compared with the scaling prediction (dashed lines) and the numerical data (solid lines) for the first three cumulants with $q=1,2,3$. The universal scaling of the first cumulant $\kappa_{1}(\tau_{\rm Q})=\langle n\rangle$ is predicted
by the \gls{KZM} according to Eq.\eqref{navkzm}. Higher-order cumulants are also predicted to exhibit a universal scaling with the quench time. All cumulants of the experimental $P(n)$ exhibit deviations from the universal scaling at long quench times consistent with a dephasing-induced anti-\gls{KZM} behavior. Further deviations from the scaling behavior are observed at fast quench times. The range of quench times characterized by universal behavior is reduced for high-order cumulants as $q$ increases. The error bars are much smaller than the size of the symbols used to depict the measured points.}
\end{figure}

 To explore universal features in $P(n)$, we shall be concerned with the scaling of its cumulants $\kappa_{q}$ ($q=1,2,3\dots$) as a function of the quench time. We focus on the first three. The first one is set by the \gls{KZM} estimate for the mean  number of kink pairs $\kappa_{1}(\tau_{\rm Q})=\langle n\rangle_{{\rm KZM}}$. The second one equals the variance and as shown in the Section~\ref{Sfull} is set by $\kappa_{2}(\tau_{\rm Q})={\rm Var}(n)=(1-1/\sqrt{2})\langle n\rangle_{{\rm KZM}}$. Finally, the third cumulant is given by the third centered moment $\kappa_{3}(\tau_{\rm Q})=\langle(n-\langle n\rangle)^{3}\rangle=(1-3/\sqrt{2}+2/\sqrt{3})\langle n\rangle_{{\rm KZM}}$. Indeed, all cumulants are predicted to be nonzero and proportional to $\langle n\rangle_{{\rm KZM}}$~\cite{delcampo18}, with  the scaling of the first cumulant $\langle n\rangle_{{\rm KZM}}\propto\tau_{\rm Q}^{-\frac{1}{2}}$ being dictated by the \gls{KZM}.  We compared this theoretical prediction with the cumulants of the experimental $P(n)$ in Fig.~\ref{C8_fig_4}, represented by dashed lines and symbols, respectively.  The deviations from the power law occur at very fast quench times, satisfying $\tau_{\rm Q}<\hbar/(\pi^{3}J)$. However,  this regime  is not probed in the experiment, as all data points are  taken for larger values of $\tau_{\rm Q}$. Within the parameters explored,  deviations at fast quenches with $J\tau_{\rm Q}/\hbar\sim1$ are due to the fact that the excitation probability $p_{k}$ in each mode is not accurately described by the Landau-Zener formula, as shown in Fig.~\ref{C8_fig_2}.\\
 \\ 
For moderate ramps, the power-law scaling of the cumulants is verified for $q=1,2$.  The power-law scaling for $\kappa_{q}(\tau_{\rm Q})$ with $q>1$ establishes the universal character of critical dynamics beyond the \gls{KZM}. The later is explicitly verified for $\kappa_{2}(\tau_{\rm Q})$, as shown in Fig.~\ref{C8_fig_4}. Beyond the fast-quench deviations shared by the first cumulant, the power law scaling of the variance of  the number of kink pairs  $\kappa_{2}(\tau_{\rm Q})$ extends to all the larger values of the quench time explored, with barely noticeable deviations.
By contrast, for  the third cumulant $\kappa_{3}(\tau_{\rm Q})$ the experimental data is already dominated by non-universal contributions both at short quench times, away from the scaling limit. For slow ramps, the third cumulant exhibits an onset of adiabatic dynamics due to finite size effects around $J\tau_{\rm Q}/\hbar=10^2$. Finite-size effects are predicted to lead to a sharp suppression of the third cumulant. However, the experimental data shows that $\kappa_3$ starts to increase with   $\tau_{\rm Q}$. This is reminiscent of the anti-Kibble-Zurek behavior that has been reported in the literature for the first cumulant in the presence of heating sources~\cite{Griffin12,Dutta16} and we attribute it to noise-induced dephasing of the trapped-ion qubit. The nature of these deviations is significantly more pronounced in high-order cumulants,  reducing the regime of applicability of the universal scaling.

\section{Conclusions}
In summary, using a trapped-ion quantum simulator we have probed the full counting statistics of topological defects in the quantum Ising chain, the paradigmatic model of quantum phase transitions. 
The statistics of kinks has been shown to be described by the Poisson binomial distribution, with cumulants obeying a universal power-law with the quench time in which the phase transition is crossed. Our findings demonstrated that the scaling theory associated with a  critical point rules the formation of topological defects beyond the scope of the Kibble-Zurek mechanism, that is restricted to the average number. Our work could be extended to probe systems with topological order in which defect formation has been predicted to be anomalous~\cite{Bermudez09}. We anticipate that the universal features of the full counting statistics of topological defects may prove useful in the error analysis of adiabatic quantum annealers, where the Kibble-Zurek mechanism already provided a useful heuristics~\cite{Suzuki09b}. \\
\\
In addition, our work raises new questions as to whether a similar scaling theory can be derived to characterize the probability distribution of other observables in statistical mechanics and condensed matter physics.  The equilibrium distribution of the magnetization has broad application from hydrodynamics to memory devices. Is there a universal nonequilibrium distribution for it as well? Is such distribution universal in complex systems such as spin glasses or chaotic systems?

\begin{savequote}[45mm]
``Work as hard and as much as you want to on the things you like to do the best. Don't think about what you want to be, but what you want to do."
\qauthor{Richard P. Feynman}
\end{savequote}
\chapter[General Conclusions]{General Conclusions}
\quad \lettrine{\color{red1}{\GoudyInfamily{U}}}{nderstanding} the far-from equilibrium dynamics of many-body systems, the breakdown of adiabaticity and its control, is key to the development of quantum technologies. At the same time, progress on the experimental realization of quantum simulation in a variety of platforms continues to sharpen the formulation of fundamental principles in non-equilibrium physics. A clear example of this symbiosis is provided by the study of the dynamics in a spontaneous symmetry-breaking scenario. In this thesis, we study the non-equilibrium properties in spin-photon systems and we obtained relevant results about their. In the next items, we summarized the main finds in this thesis.\\
\\
We show how to detect the presence of quantum phase transitions by means of local different-time correlations and violations of Leggett-Garg inequalities. We show, by looking at a quite general spin$-1/2$ chain, that there is a relationship between discontinuities, at fixed time, of the derivatives of the correlations and of the ground state energy as functions of a parameter in the system, and, focusing on two paradigmatic examples, that the methods numerically work.\\
\\
In the same line of temporal correlations, we extend our finds to Majorana Fermion context. In this way, a fundamental problem in many-body quantum physics is the assessing of quantum phase transitions, in general, and in particular topological phase transitions are becoming increasingly studied from both the theoretical and experimental points of view. A large number of physical properties acting as indicators of such transitions have been proposed to this end, and a particularly important subset are those based on surface or edge physical quantities. Especially in the last few years there has been a surge of experimental interest in highly efficient ways of distinguishing between possible alternative explanations of what has been believed to be a cornerstone of the detection of Majorana fermions in condensed matter setups such as a zero bias peak. A fundamental open problem was to find an optimal smoking gun signature of zero mode Majorana modes in semiconductor-superconductor wires or equivalently edge-chain localized fermionic elementary excitations. Our work provides a novel candidate protocol based on temporal rather than spatial two-point correlations, namely local and non-local two-time correlations, and does so by exploring clear distinguishing features in each phase, trivial and topological phases, for the three relevant temporal domains of short-time, intermediate and long time behaviors of the temporal correlations, a distinctive feature of our work with respect to any other previously known protocol.\\
\\
On the other hand, we deals with the time-dependent driving of the celebrated Dicke-model across the superradiant transition and back to the normal phase. We are interested in the finite-time effects of driving across a quantum phase transition twice. To this end they consider the unitary/closed Dicke model with no dissipation. As the second order phase transition is approached, the system becomes effectively a two-level system. For this reason, we analyze the transition in terms of a modification of the Landau-Zener model in which one drives the system forward and backward through the avoided crossing. One observes oscillations in the transition probabilities referred to as St\"uckelberg oscillations. We analyze the ground state fidelity and the light-matter entanglement measured using the von Neumann entropy. We find that for slow driving, in the so-called \textquotedblleft near-adiabatic\textquotedblright regime, both quantities show a strong hysteresis: the system does not trace back its evolution.\\
\\
To expand further, we also want to emphasize that a major contribution of our results is that it purposely avoids many of the approximations typically employed in the analysis of coherent effects in many-body quantum systems. In this way, we are able to show for first time that vibronic coherence can be generated in exciton-vibration coupled systems without accessing the strong coupling regime; The same theoretical framework can be used to describe electron-phonon coupling in molecular/nanoparticle aggregates as well as to address the impact of an applied time-varying (pulsed) radiation field; We are able to uncover relationships between the pulsed fields and the generation of possible non-classical matter states.\\
\\ 
In the third part, we discuss the onset of the Kibble-Zurek Mechanism in the context of inhomogeneous systems, when a quantum phase transition is crossed as a consequence of a slow variation of one Hamiltonian parameter. We show that inhomogeneities may generally reduce the impact of defects formation, since a non-homogeneous system can locally exhibit a critical behavior, which is absent in other regions. More specifically, it is possible to find out a crossover time which separates a region behaving as in the original \gls{KZM}, from another region where the power-law exponent for the scaling of defects is larger. This scenario is explicitly tested in a quantum Ising ring in a transverse field, where the couplings are parabolically modulated in space. The study of adiabatic dynamics and \gls{KZM} in a many-body context is certainly receiving a great deal of interest in the last years, due to the advances in quantum simulations. This is also witnessed by the increasing number of papers that are appearing, which are focused on the dynamics across phase transitions.\\
\\
Finally, we describes the defect formation of a transverse quantum Ising chain when driven through a paramagnetic-ferromagnetic phase transition with a finite rate (here driven by the external magnetic field). Using the Jordan-Wigner transformation, not an Ising-chain of spin-$1/2$ fermions is investigated but an ensemble of fermionic creation and annihilation operations is investigated in experiment. Ramping the
magnetic field is equivalent to an ensemble of Landau-Zener crossings of a two-state quantum systems. This is realized by Yb-ions in a Paul trap where a static magnetic field creates a Zeemann splitting between $F=0$ and $F=1$. Fluorescence microscopy gives the population of the bright state. We argue that our experimental results, about full counting statistic of kinks distribution, have a broad impact as our experiment has manifold applications. Knowledge of the distribution of defects proves useful in the characterization in adiabatic quantum computers.  Assessing the statistics of topological defects and its universal character is bound to motivate new experiments across the plethora of experimental platforms where topological defect formation has been explored, e.g., in the light of the Kibble-Zurek mechanism. These include liquid crystals, superconducting qubits, optical lattices, trapped BECs, colloidal monolayers, trapped ion chains and Coulomb crystals, etc. For each platform, new experiments can be envisioned to collect the full counting statistics by improving and developing novel measurement techniques.  Quantum simulators aiming at the preparation of novel phases of matter can be guided by the possibility of tailoring the distribution of topological defects. New protocols in optimal control can be conceived that aim not only at reducing the average number of excitations but as well at engineering the fluctuations.

\bibliography{Mybib_Thesis}{}
\bibliographystyle{apsrev4-1}

\appendix
\begin{savequote}[45mm]
\end{savequote}
\chapter[Exact diagonalization \gls{AXY-TFIM}]{Exact diagonalization the  anisotropic XY Ising model in a transversal magnetic field}\label{Apend1}
\lettrine{\color{red1}{\GoudyInfamily{W}}}{e} consider a general model: the anisotropic XY Ising model in  a transverse field (AXY). In one dimension the AXY has been introduced by Leib et al [1]. $(h=0)$ y por Katsura [2]. They considered a chain of $N$ spins governed by the Hamiltonian
\begin{equation}
\hat{H}=-J\sum_{i=1}^{N}\left[\left(\frac{1+\gamma}{2}\right)\sx_{i}\sx_{i+1}+\left(\frac{1-\gamma}{2}\right)\sy_{i}\sy_{i+1}\right]-h\sum_{i=1}^{N}\sz_{i}
\end{equation} 
where the operators $\sx_{i}$, $\sy_{i}$ and $\sz_{i}$ are spins $1/2$ operators represented by Pauili matrices and $\gamma$ is a parameter characterizing the degree of anisotropy of the interactions in the $XY$ plane. The $\gamma=1$ case corresponds to the Ising model in a transverse field (ITF), while the $\gamma=0$ case give the $XY$ model in a transversal field (XYTF).\\
\\
\noindent The Hamiltonian can be mapped exactly on a free fermion model, consisting of an assembly of non interacting Fermi-Dirac oscillators. then the analytical process of diagonalization of the Hamiltonian is shown. If you have a two level system (TLS), is to identify with the state $\ketC$ the ground state and the excited state $\ket1$. 
\begin{equation*}
\ketC =\begin{pmatrix}
1\\
0
\end{pmatrix}\qquad\ket1 =\begin{pmatrix}
0\\
1
\end{pmatrix}
\end{equation*}
\noindent as the Pauli matrices are defined in the form   
\begin{align*}
\sx&=\begin{pmatrix}
0 & 1\\
1 & 0
\end{pmatrix} & \sy &=\begin{pmatrix}
0 & -i\\
i & 0
\end{pmatrix} & \sz &=\begin{pmatrix}
1 & 0\\
0 & -1
\end{pmatrix} 
\end{align*}
\noindent It is noteworthy that
\begin{align*}
\sz\ket1&=-\ket1\\
\sz\ketC&=\ketC\\
\end{align*}
\noindent $b_{i}$ and $b_{i}^{\dagger}$ operators are fermionic creation and annihilation operators. For TLS is defined from
\begin{align*}
b^{\dagger}\ketC&= \ket1 & b^{\dagger}\ket1&=0\\
b\ket1&=\ketC & b\ketC&=0
\end{align*}
\noindent then 
\begin{align*}
\left(2b^{\dagger}b-I\right)\ketC&=-\ketC\\
\left(2b^{\dagger}b-I\right)\ket1&=\ket1
\end{align*}
therefore
\begin{align*}
\left(2b^{\dagger}_{i}b_{i}-I\right)&=-\sz_{i}
\end{align*}
\noindent Now the form of the operator $b_{i}$ are found in the representation of eigenvectors of $\sz_{i}$
\begin{align*}
\begin{pmatrix}
b_{11} & b_{12}\\
b_{21} & b_{22}
\end{pmatrix}\begin{pmatrix}
1\\
0
\end{pmatrix}=\begin{pmatrix}
0\\
0
\end{pmatrix}
\end{align*}
\noindent from which it can be concluded that $b_{11}=b_{21}=0$, now to the $b$ operator acting on the excited state must  
\begin{align*}
\begin{pmatrix}
0 & b_{12}\\
0 & b_{22}
\end{pmatrix}\begin{pmatrix}
0\\
1
\end{pmatrix}=\begin{pmatrix}
1\\
0
\end{pmatrix}
\end{align*}
then 
\begin{align*}
b=\begin{pmatrix}
0 & 1\\
0 & 0
\end{pmatrix}
\end{align*}
therefore
\begin{align*}
b^{\dagger}=\begin{pmatrix}
0 & 0\\
1 & 0
\end{pmatrix}
\end{align*}
\noindent according to the form of $b$ and $b^{\dagger}$ fermionic operators can be identified with the ladder operators of the TLS $\hat{\sigma}^{\dagger}$ and $\hat{\sigma}^{-}$ respectively. However, this analogy is valid for only one site of the chain of spins; this can not be extended to multiple sites for the fermion operators anticommute in different locations, while the ladder operators for TLS commute. therefore to satisfy the relation of fermionic anticomutacion $\left\{b_{i},b_{j}^{\dagger}\right\}=\delta_{ij}$ the operators $b_{i}$ and $b_{i}^{\dagger}$ are defined as
\begin{align*}
b_{i}&=\prod_{m<i}\left(\sz_{m}\right)\sigma_{i}^{\dagger}\\
b_{i}^{\dagger}&=\prod_{m<i}\left(\sz_{m}\right)\sigma_{i}^{-}
\end{align*}
\noindent inverse transformation is given by
\begin{align*}
\sigma_{i}^{\dagger} &= \prod_{m<i}\left(1-2b_{m}^{\dagger}b_{m}\right)b_{i},\\
\sigma_{i}^{-}&= \prod_{m<i}\left(1-2b_{m}^{\dagger}b_{m}\right)b_{i}^{\dagger}. 
\end{align*}
this new redefinition of the fermionic operators, allows anticommutation and commutation relations for fermionic operators are met. This transformation is known as the Jordan-Wigner transformation. The following commutation and anticommutation rules were quickly verified:
\begin{align*}
\left\{b_{i},b_{j}^{\dagger}\right\}&=\delta_{ij}, & \left\{b_{i},b_{j}\right\}=\left\{b_{i}^{\dagger},b_{j}^{\dagger}\right\}&=0\\
\left[\sigma_{i}^{\dagger},\sigma_{j}^{-}\right]&=\delta_{ij}\sz_{i} & \left[\sz_{i},\sigma_{j}^{\pm}\right]&=\pm2\delta_{ij}\sigma_{i}^{\pm}  
\end{align*}
where the curly brackets represent anticommutation and square brackets are commutators. Now, as we know that the ladder operators for a TLS satisfy the following relations with the Pauli matrix
\begin{align*}
\sigma_{i}^{x} &=\sigma_{i}^{\dagger}+\sigma_{i}^{-}\\
\sigma_{i}^{y} &=i\left(\sigma_{i}^{-}-\sigma_{i}^{\dagger}\right)
\end{align*}
replacing the Jordan-Wigner transformations, we obtain
\begin{align*}
\sigma_{i}^{x} &=\prod_{m<i}\left(1-2b_{m}^{\dagger}b_{m}\right)b_{i}+\prod_{m<i}\left(1-2b_{m}^{\dagger}b_{m}\right)b_{i}^{\dagger}\\
\sigma_{i}^{x} &=\prod_{m<i}\left(1-2b_{m}^{\dagger}b_{m}\right)\left(b_{i}+b_{i}^{\dagger}\right)\\
\sigma_{i}^{y} &=i\left[\prod_{m<i}\left(1-2b_{m}^{\dagger}b_{m}\right)b_{i}^{\dagger}-\prod_{m<i}\left(1-2b_{m}^{\dagger}b_{m}\right)b_{i}\right]\\
\sigma_{i}^{y} &=i\left[\prod_{m<i}\left(1-2b_{m}^{\dagger}b_{m}\right)\left(b_{i}^{\dagger}-b_{i}\right)\right]
\end{align*}
Now, is explicitly calculated in terms of the Hamiltonian
\begin{align*}
\sigma_{i}^{x}\sigma_{i+1}^{x}&=\left[\prod_{m<i}\left(1-2b_{m}^{\dagger}b_{m}\right)b_{i}+\prod_{m<i}\left(1-2b_{m}^{\dagger}b_{m}\right)b_{i}^{\dagger}\right]\times\\
&\quad \left[\prod_{n<i+1}\left(1-2b_{n}^{\dagger}b_{n}\right)b_{i+1}+\prod_{n<i+1}\left(1-2b_{n}^{\dagger}b_{n}\right)b_{i+1}^{\dagger}\right]
\end{align*}
\begin{align*}
\sigma_{i}^{x}\sigma_{i+1}^{x}&= \left[ \prod_{m<i}\left(1-2b_{m}^{\dagger}b_{m}\right)b_{i}\right] \left[\prod_{n<i}\left(1-2b_{n}^{\dagger}b_{n}\right)\left(1-2b_{i}^{\dagger}b_{i}\right)b_{i+1}\right]+\\
&\quad\left[\prod_{m<i}\left(1-2b_{m}^{\dagger}b_{m}\right)b_{i}\right]\left[\prod_{n<i}\left(1-2b_{n}^{\dagger}b_{n}\right)\left(1-2b_{i}^{\dagger}b_{i}\right)b_{i+1}^{\dagger}\right]+\\
&\quad\left[\prod_{m<i}\left(1-2b_{m}^{\dagger}b_{m}\right)b_{i}^{\dagger}\right]\left[\prod_{n<i}\left(1-2b_{n}^{\dagger}b_{n}\right)\left(1-2b_{i}^{\dagger}b_{i}\right)b_{i+1}\right]+\\
&\quad\left[\prod_{m<i}\left(1-2b_{m}^{\dagger}b_{m}\right)b_{i}^{\dagger}\right]\left[\prod_{n<i}\left(1-2b_{n}^{\dagger}b_{n}\right)\left(1-2b_{i}^{\dagger}b_{i}\right)b_{i+1}^{\dagger}\right]
\end{align*}
\begin{align*}
\sigma_{i}^{x}\sigma_{i+1}^{x}&= \left[ \prod_{m<i}\left(1-2b_{m}^{\dagger}b_{m}\right)\right] \left[\prod_{n<i}\left(1-2b_{n}^{\dagger}b_{n}\right)\right]\left[b_{i}\left(1-2b_{i}^{\dagger}b_{i}\right)b_{i+1}\right]+\\
&\quad\left[\prod_{m<i}\left(1-2b_{m}^{\dagger}b_{m}\right)\right]\left[\prod_{n<i}\left(1-2b_{n}^{\dagger}b_{n}\right)\right]\left[b_{i}\left(1-2b_{i}^{\dagger}b_{i}\right)b_{i+1}^{\dagger}\right]+\\
&\quad\left[\prod_{m<i}\left(1-2b_{m}^{\dagger}b_{m}\right)\right]\left[\prod_{n<i}\left(1-2b_{n}^{\dagger}b_{n}\right)\right]\left[b_{i}^{\dagger}\left(1-2b_{i}^{\dagger}b_{i}\right)b_{i+1}\right]+\\
&\quad\left[\prod_{m<i}\left(1-2b_{m}^{\dagger}b_{m}\right)\right]\left[\prod_{n<i}\left(1-2b_{n}^{\dagger}b_{n}\right)\right]\left[b_{i}^{\dagger}\left(1-2b_{i}^{\dagger}b_{i}\right)b_{i+1}^{\dagger}\right]
\end{align*}
the indices n and m are as dumb, the products
\begin{align*}
\left[\prod_{m<i}\left(1-2b_{m}^{\dagger}b_{m}\right)\right]\left[\prod_{n<i}\left(1-2b_{n}^{\dagger}b_{n}\right)\right]&=1
\end{align*}
then 
\begin{align*}
\sigma_{i}^{x}\sigma_{i+1}^{x}&= b_{i}\left(1-2b_{i}^{\dagger}b_{i}\right)b_{i+1}+b_{i}\left(1-2b_{i}^{\dagger}b_{i}\right)b_{i+1}^{\dagger}+ b_{i}^{\dagger}\left(1-2b_{i}^{\dagger}b_{i}\right)b_{i+1}+b_{i}^{\dagger}\left(1-2b_{i}^{\dagger}b_{i}\right)b_{i+1}^{\dagger}\\
&=b_{i}b_{i+1}-2b_{i}b_{i}^{\dagger}b_{i}b_{i+1}+b_{i}b_{i+1}^{\dagger}-2b_{i}b_{i}^{\dagger}b_{i}b_{i+1}^{\dagger}+b_{i}^{\dagger}b_{i+1}-2b_{i}^{\dagger}b_{i}^{\dagger}b_{i}b_{i+1}+b_{i}^{\dagger}b_{i+1}^{\dagger}-2b_{i}^{\dagger}b_{i}^{\dagger}b_{i}b_{i+1}^{\dagger}
\end{align*}
Now as $b_{i}b_{i}=0$ and $b_{i}^{\dagger}b_{i}^{\dagger}=0$, then the nonzero terms are
\begin{equation}
\boxed{\sigma_{i}^{x}\sigma_{i+1}^{x}=b_{i}b_{i+1}+b_{i}b_{i+1}^{\dagger}+b_{i}^{\dagger}b_{i+1}+b_{i}^{\dagger}b_{i+1}^{\dagger}}
\end{equation}
Now also is necessary to know the form of $\sigma_{i}^{y}\sigma_{i+1}^{y}$, thus calculating explicitly
\begin{align*}
\sigma_{i}^{y}\sigma_{i+1}^{y}&=\left[\prod_{m<i}\left(1-2b_{m}^{\dagger}b_{m}\right)b_{i}-\prod_{m<i}\left(1-2b_{m}^{\dagger}b_{m}\right)b_{i}^{\dagger}\right]\times\\
&\quad\left[\prod_{n<i+1}\left(1-2b_{n}^{\dagger}b_{n}\right)b_{i+1}^{\dagger}-\prod_{n<i+1}\left(1-2b_{n}^{\dagger}b_{n}\right)b_{i+1}\right]\\
&= \left[\prod_{m<i}\left(1-2b_{m}^{\dagger}b_{m}\right)b_{i}\right]\left[\prod_{n<i+1}\left(1-2b_{n}^{\dagger}b_{n}\right)b_{i+1}^{\dagger}\right]\\
&\quad -\left[\prod_{m<i}\left(1-2b_{m}^{\dagger}b_{m}\right)b_{i}\right]\left[\prod_{n<i+1}\left(1-2b_{n}^{\dagger}b_{n}\right)b_{i+1}\right]\\
&\quad -\left[\prod_{m<i}\left(1-2b_{m}^{\dagger}b_{m}\right)b_{i}^{\dagger}\right]\left[\prod_{n<i+1}\left(1-2b_{n}^{\dagger}b_{n}\right)b_{i+1}^{\dagger}\right]\\
&\quad + \left[\prod_{m<i}\left(1-2b_{m}^{\dagger}b_{m}\right)b_{i}^{\dagger}\right]\left[\prod_{n<i+1}\left(1-2b_{n}^{\dagger}b_{n}\right)b_{i+1}\right]\\
\sigma_{i}^{y}\sigma_{i+1}^{y}&=\left[\prod_{m<i}\left(1-2b_{m}^{\dagger}b_{m}\right)b_{i}\right]\left[\prod_{n<i}\left(1-2b_{n}^{\dagger}b_{n}\right)\left(1-2b_{i}^{\dagger}b_{i}\right)b_{i+1}^{\dagger}\right]\\
&\quad -\left[\prod_{m<i}\left(1-2b_{m}^{\dagger}b_{m}\right)b_{i}\right]\left[\prod_{n<i}\left(1-2b_{n}^{\dagger}b_{n}\right)\left(1-2b_{i}^{\dagger}b_{i}\right)b_{i+1}\right]\\
&\quad -\left[\prod_{m<i}\left(1-2b_{m}^{\dagger}b_{m}\right)b_{i}^{\dagger}\right]\left[\prod_{n<i}\left(1-2b_{n}^{\dagger}b_{n}\right)\left(1-2b_{i}^{\dagger}b_{i}\right)b_{i+1}^{\dagger}\right]\\
&\quad + \left[\prod_{m<i}\left(1-2b_{m}^{\dagger}b_{m}\right)b_{i}^{\dagger}\right]\left[\prod_{n<i}\left(1-2b_{n}^{\dagger}b_{n}\right)\left(1-2b_{i}^{\dagger}b_{i}\right)b_{i+1}\right]\\
&=\left[\prod_{m<i}\left(1-2b_{m}^{\dagger}b_{m}\right)\right]\left[\prod_{n<i}\left(1-2b_{n}^{\dagger}b_{n}\right)\right]\left[b_{i}\left(1-2b_{i}^{\dagger}b_{i}\right)b_{i+1}^{\dagger}\right]\\
&\quad -\left[\prod_{m<i}\left(1-2b_{m}^{\dagger}b_{m}\right)\right]\left[\prod_{n<i}\left(1-2b_{n}^{\dagger}b_{n}\right)\right]\left[b_{i}\left(1-2b_{i}^{\dagger}b_{i}\right)b_{i+1}\right]\\
&\quad -\left[\prod_{m<i}\left(1-2b_{m}^{\dagger}b_{m}\right)\right]\left[\prod_{n<i}\left(1-2b_{n}^{\dagger}b_{n}\right)\right]\left[b_{i}^{\dagger}\left(1-2b_{i}^{\dagger}b_{i}\right)b_{i+1}^{\dagger}\right]\\
&\quad + \left[\prod_{m<i}\left(1-2b_{m}^{\dagger}b_{m}\right)\right]\left[\prod_{n<i}\left(1-2b_{n}^{\dagger}b_{n}\right)\right]\left[b_{i}^{\dagger}\left(1-2b_{i}^{\dagger}b_{i}\right)b_{i+1}\right]
\end{align*}
the indices n and m are as dumb, the products
\begin{align*}
\left[\prod_{m<i}\left(1-2b_{m}^{\dagger}b_{m}\right)\right]\left[\prod_{n<i}\left(1-2b_{n}^{\dagger}b_{n}\right)\right]&=1
\end{align*}
then 
\begin{align*}
\sigma_{i}^{y}\sigma_{i+1}^{y}&=b_{i}\left(1-2b_{i}^{\dagger}b_{i}\right)b_{i+1}^{\dagger}-b_{i}\left(1-2b_{i}^{\dagger}b_{i}\right)b_{i+1}-b_{i}^{\dagger}\left(1-2b_{i}^{\dagger}b_{i}\right)b_{i+1}^{\dagger}+b_{i}^{\dagger}\left(1-2b_{i}^{\dagger}b_{i}\right)b_{i+1}\\
&=b_{i}b_{i+1}^{\dagger}-2b_{i}b_{i}^{\dagger}b_{i}b_{i+1}^{\dagger}-b_{i}b_{i+1}+2b_{i}b_{i}^{\dagger}b_{i}b_{i+1}-b_{i}^{\dagger}b_{i+1}^{\dagger}+2b_{i}^{\dagger}b_{i}^{\dagger}b_{i}b_{i+1}^{\dagger}+b_{i}^{\dagger}b_{i+1}-2b_{i}^{\dagger}b_{i}^{\dagger}b_{i}b_{i+1}
\end{align*}
Now as $b_{i}b_{i}=0$ and $b_{i}^{\dagger}b_{i}^{\dagger}=0$, then the nonzero terms are
\begin{equation}
\boxed{\sigma_{i}^{y}\sigma_{i+1}^{y}=b_{i}b_{i+1}^{\dagger}-b_{i}b_{i+1}-b_{i}^{\dagger}b_{i+1}^{\dagger}+b_{i}^{\dagger}b_{i+1}}
\end{equation}
replacing each of the terms in the previously calculated one has Hamiltonian AXY
\begin{align*}
\hat{H}&=-\frac{J}{2}\sum_{i=1}^{N}\left[\left(1+\gamma\right)\sx_{i}\sx_{i+1}+\left(1-\gamma\right)\sy_{i}\sy_{i+1}\right]-h\sum_{i=1}^{N}\sz_{i}\\
&=-\frac{J}{2}\left\{\left[\sum_{i=1}^{N}\left(1+\gamma\right)\sx_{i}\sx_{i+1}\right]+\left[\sum_{i=1}^{N}\left(1-\gamma\right)\sy_{i}\sy_{i+1}\right]\right\}-h\sum_{i=1}^{N}\sz_{i}\\
&=-\frac{J}{2}\left\{\left[\sum_{i=1}^{N}\left(1+\gamma\right)b_{i}b_{i+1}+b_{i}b_{i+1}^{\dagger}+b_{i}^{\dagger}b_{i+1}+b_{i}^{\dagger}b_{i+1}^{\dagger}\right]\right.\\
&\left.+\left[\sum_{i=1}^{N}\left(1-\gamma\right)b_{i}b_{i+1}^{\dagger}-b_{i}b_{i+1}-b_{i}^{\dagger}b_{i+1}^{\dagger}+b_{i}^{\dagger}b_{i+1}\right]\right\}-h\sum_{i=1}^{N}\left(1-2b_{i}^{\dagger}b_{i}\right)\\
\end{align*} 
have explicitly calculating
\begin{eqnarray}\label{HXY}	
\boxed{H=-Nh-J\sum_{i=1}^{N}\left[b_{i}^{\dagger}b_{i+1}+b_{i}b_{i+1}^{\dagger}+\gamma b_{i}^{\dagger}b_{i+1}^{\dagger}+\gamma b_{i}b_{i+1}\right]+2h\sum_{i=1}^{N}b_{i}^{\dagger}b_{i}}
\end{eqnarray}

The diagonalization of (\ref{HXY}) for $h$ independent of $t$ is completed by using two more transformations: \emph{(i)} Fourier transformation, and \emph{(ii)} Bogoliubov transformation. We can still carry out the Fourier transform. Define
\begin{align*}
b_{j}^{\dagger}&=\frac{1}{\sqrt{N}}\sum_{p}e^{-i j\phi_{p}}a^{\dagger}_{p},\\
b_{j}&=\frac{1}{\sqrt{N}}\sum_{p}e^{i j\phi_{p}}a_{p},\\
\end{align*} 
The above transformed Hamiltonian is already quadratic in the operator and it is diagonalisable. Let us consider fermion in momentum space. where $j\phi_{p}= j R_{j}$, where the complete set of wavevector is $j=2\pi m /N$,
\begin{align*}
	m&=-\left(N-1\right)/2,\ldots, -1/2, 1/2,\ldots,\left(N-1\right)/2\qquad\text{(for $N$ even)}\\
	m&=N/2,\ldots,0,\ldots,N/2\qquad\text{(for $N$ odd)}
\end{align*}
 from the above transformation are calculated explicitly each of the terms of the Hamiltonian
 
\begin{eqnarray}
\boxed{\sum_{j=1}^{N}b_{j}^{\dagger}b_{j}=\frac{1}{N}\sum_{p}a_{p}^{\dagger}a_{p}}
\end{eqnarray}
 
\begin{align*}
 \sum_{j=1}^{N}b_{j}^{\dagger}b_{j+1}&=\frac{1}{N}\sum_{j=1}^{N}\sum_{p}e^{-i j \phi_{p}}a_{p}^{\dagger}\sum_{m}e^{ i (j+1)\phi_{m}}a_{m}\\
 &=\frac{1}{N}\sum_{m,p} a_{p}^{\dagger}a_{m}e^{ i\phi_{m}}  \sum_{j=1}^{N} e^{-i j \phi_{p}} e^{ i j \phi_{m}}\\
 \sum_{j=1}^{N}b_{j}^{\dagger}b_{j+1}&=\frac{1}{N}\sum_{m,p}a_{p}^{\dagger}a_{m}e^{ i\phi_{m}}\delta_{p,m}
 \end{align*}
\begin{eqnarray}
\boxed{\sum_{j=1}^{N}b_{j}^{\dagger}b_{j+1} =\frac{1}{N}\sum_{p}a_{p}^{\dagger}a_{p}e^{ i\phi_{p}}}
\end{eqnarray}
\begin{eqnarray}
\boxed{\sum_{j=1}^{N}b_{j}b_{j+1}^{\dagger}=\frac{1}{N}\sum_{p}a_{p}^{\dagger}a_{p}e^{-i\phi_{p}}}
\end{eqnarray}
\begin{align*}
\sum_{j=1}^{N}b_{j}^{\dagger}b_{j+1}^{\dagger}&=\frac{1}{N}\sum_{m,p}a_{p}^{\dagger}a_{m}^{\dagger}e^{-i\phi_{m}}\sum_{j=1}^{N}e^{-i j \phi_{p}}e^{-i j \phi_{m}}\\
&=\frac{1}{N}\sum_{m,p}a_{p}^{\dagger}a_{m}^{\dagger}e^{-i\phi_{m}}\delta_{p,-m}\\
&=\frac{1}{N}\sum_{p}a_{p}^{\dagger}a_{-p}^{\dagger}e^{i\phi_{p}}=\frac{1}{N}\sum_{p}a_{-p}^{\dagger}a_{p}^{\dagger}e^{-i\phi_{p}}\\
&=\frac{1}{2N}\sum_{p}\left(a_{p}^{\dagger}a_{-p}^{\dagger}e^{i\phi_{p}}+a_{-p}^{\dagger}a_{p}^{\dagger}e^{-i\phi_{p}}\right)\\
&=\frac{1}{2N}\sum_{p}\left(a_{p}^{\dagger}a_{-p}^{\dagger}e^{i\phi_{p}}-a_{p}^{\dagger}a_{-p}^{\dagger}e^{-i\phi_{p}}\right)
\end{align*}
\begin{eqnarray}
	\boxed{\sum_{j=1}^{N}b_{j}^{\dagger}b_{j+1}^{\dagger}=-\frac{i}{N}\sum_{p}\sin\left(\phi_{p}\right)a_{p}^{\dagger}a_{-p}^{\dagger}}
\end{eqnarray}
\begin{eqnarray}
	\boxed{\sum_{j=1}^{N}b_{j}b_{j+1}=-\frac{i}{N}\sum_{p}\sin\left(\phi_{p}\right)a_{p}a_{-p}}
\end{eqnarray}
replacing the Hamiltoniano (\ref{HXY})
\begin{align*}
H&=-\frac{J}{N}\sum_{p}\left[a_{p}^{\dagger}a_{p}e^{ i\phi_{p}}+a_{p}^{\dagger}a_{p}e^{-i\phi_{p}}-i\gamma \sin\left(\phi_{p}\right)a_{p}^{\dagger}a_{-p}^{\dagger} - i\gamma \sin\left(\phi_{p}\right)a_{p}a_{-p} \right]\\
&\qquad+\sum_{p}\left(\frac{2h}{N}a_{p}^{\dagger}a_{p}-\frac{h}{N}\right)\\
&=-\frac{J}{N}\sum_{p}\left[a_{p}^{\dagger}a_{p}\left[\left(e^{i\phi_{p}}+e^{-i\phi_{p}}\right)-\frac{2h}{J}\right]-i\gamma\sin\left(\phi_{p}\right)\left(a_{p}^{\dagger}a_{-p}^{\dagger}+a_{p}a_{-p}\right)+\frac{h}{J}\right]\\
H&=-\frac{J}{N}\sum_{p}\left[a_{p}^{\dagger}a_{p}\left[2\cos\left(ap\right)-\frac{2h}{J}\right]-i\gamma\sin\left(ap\right)\left(a_{p}^{\dagger}a_{-p}^{\dagger}+a_{p}a_{-p}\right)+\frac{h}{J}\right]
\end{align*}
and $N$ is the total number of sites.
\begin{align*}
H&=-\frac{J}{N}\sum_{p}\left[2a_{p}^{\dagger}a_{p}\left[\cos\left(ap\right)-\frac{h}{J}\right]-i\gamma\sin\left(ap\right)\left(a_{p}^{\dagger}a_{-p}^{\dagger}+a_{p}a_{-p}\right)+\frac{h}{J}\right]\\
\end{align*}
\begin{eqnarray*}	
\boxed{H=-\frac{J}{N}\sum_{k}\left[\left(\cos\left(a k\right)-\frac{h}{J}\right)\left(a_{k}^{\dagger}a_{k}+a_{-k}^{\dagger}a_{-k}\right)-i\gamma\sin\left(ak\right)\left(a_{k}^{\dagger}a_{-k}^{\dagger}+a_{k}a_{-k}\right)+\frac{h}{J}\right]}
\end{eqnarray*}
Next, we use the Bogoliubov transformation to map into a new set of fermionic operators $(\gamma_{k})$ whose number is conserved. These new operators are defined by a unitary transformation on the pair $a_{k}$, $a_{-k}^{\dagger}$:
\begin{align*}
	\gamma_{k}= u_{k}a_{k}-i v_{k}a_{-k}^{\dagger}
\end{align*}
where $u_{p}$, $v_{p}$ are real numbers satisfying $u_{k}^{2}+v_{k}^{2}=1$, $u_{-k}=u_{k}$, and $v_{-k}=-v_{k}$. It can be checked that canonical fermion anticomutation relation for the $a_{k}$ imply that the same relations are also satisfied by the $\gamma_{k}$, that is
	\[\left\{\gamma_{k},\gamma_{k'}^{\dagger}\right\}=\delta_{k,k'},\qquad\left\{\gamma_{k}^{\dagger},\gamma_{k'}^{\dagger}\right\} =\left\{\gamma_{k},\gamma_{k'}\right\}=0
\]
we also note the inverse transformation
\begin{eqnarray*}
	a_{k}=u_{k}\gamma_{k} + i v_{k}\gamma_{-k}^{\dagger}.
\end{eqnarray*}
therefore
\begin{align*}
	a_{-k}&=u_{k}\gamma_{-k} - i v_{k}\gamma_{k}^{\dagger}\\
	a_{k}^{\dagger}&=u_{k}\gamma_{k}^{\dagger} - i v_{k}\gamma_{-k}\\
	a_{-k}^{\dagger}&=u_{k}\gamma_{-k}^{\dagger} + i v_{k}\gamma_{k}
	\end{align*}
using the relationships found above are calculated explicitly each of the terms of the Hamiltonian
\begin{align*}
a_{k}^{\dagger}a_{k}&=u_{k}^{2} \gamma_k^{\dagger}\gamma_k + i u_k v_k \gamma_k{}^{\dagger }\gamma_{-k}^{\dagger}-i u_k v_k \gamma_{-k}\gamma_k + v_k^2\gamma _{-k}\gamma_{-k}{}^{\dagger } 
\\
a_{-k}^{\dagger}a_{-k}&= u_k^2 \gamma_{-k}^{\dagger}\gamma_{-k} - i u_k v_k\gamma_{-k}^{\dagger}\gamma_k^{\dagger}+i u_k v_k \gamma_k\gamma_{-k} + v_k^2 \gamma_k\gamma_k^{\dagger}\\
a_{k}^{\dagger}a_{-k}^{\dagger}&=u_{k}^{2} \gamma _{k}^{\dagger}\gamma_{-k}^{\dagger}+i u_k v_k \gamma _{k}^{\dagger}\gamma_k - i u_k v_k \gamma_{-k}\gamma_{-k}^{\dagger}+v_k^2\gamma_{-k}\gamma_k\\
a_{k}a_{-k}&= u_k^2 \gamma_k \gamma_{-k}+i u_k v_k \gamma_{-k}^{\dagger}\gamma_{-k} - i u_k v_k \gamma_k\gamma _k^{\dagger}+v_k^2 \gamma_{-k}^{\dagger}\gamma_k^{\dagger}
\end{align*}
using the relationships found above are calculated explicitly each of the terms of the Hamiltonian
\begin{align*}
H&=-\frac{J}{N}\sum_{k}\left[\left(\cos\left(a k\right)-\frac{h}{J}\right) \left( u_{k}^{2} \gamma_k^{\dagger}\gamma_k + i u_k v_k \gamma_k{}^{\dagger }\gamma_{-k}^{\dagger}-i u_k v_k \gamma_{-k}\gamma_k + v_k^2\gamma _{-k}\gamma_{-k}{}^{\dagger}\right.\right.\\
&\qquad \left.+ u_k^2 \gamma_{-k}^{\dagger}\gamma_{-k} - i u_k v_k\gamma_{-k}^{\dagger}\gamma_k^{\dagger}+i u_k v_k \gamma_k\gamma_{-k} + v_k^2 \gamma_k\gamma_k^{\dagger} \right)\\
&\qquad -i\gamma\sin\left(ak\right)\left(u_{k}^{2} \gamma _{k}^{\dagger}\gamma_{-k}^{\dagger}+i u_k v_k \gamma _{k}^{\dagger}\gamma_k - i u_k v_k \gamma_{-k}\gamma_{-k}^{\dagger}+v_k^2\gamma_{-k}\gamma_k\right.\\
&\qquad \left.\left.+u_k^2 \gamma_k \gamma_{-k}+i u_k v_k \gamma_{-k}^{\dagger}\gamma_{-k} - i u_k v_k \gamma_k\gamma _k^{\dagger}+v_k^2 \gamma_{-k}^{\dagger}\gamma_k^{\dagger}\right)+\frac{h}{J}\right]
\end{align*}
using anticommutation relations and factoring like terms we obtain
\begin{align*}
H&=-\frac{J}{N}\sum_{k}\left[2\gamma_{k}^{\dagger}\gamma_{k}\left(u_{k}^{2}\left(\cos\left(a k\right)-\frac{h}{J}\right)+\gamma \sin\left(ak\right)u_{k}v_{k}\right)\right.\\
&\qquad +2\gamma_{k}\gamma_{k}^{\dagger}\left(\left(\cos\left(a k\right)-\frac{h}{J}\right)v_{k}^{2}-\gamma\sin\left(ak\right)u_{k}v_{k}\right)\\
&\qquad -i\gamma_{k}\gamma_{-k}\left(\gamma\sin\left(ak\right)\left\{u_{k}^{2}-v_{k}^{2}\right\}+2\left(\cos\left(a k\right)-\frac{h}{J}\right)u_{k}v_{k}\right)\\
&\qquad-i\gamma_{k}^{\dagger}\gamma_{-k}^{\dagger}\left(\gamma\sin\left(ak\right)\left\{u_{k}^{2}-v_{k}^{2}\right\}+2\left(\cos\left(a k\right)-\frac{h}{J}\right)u_{k}v_{k}\right)+\left.\frac{h}{J}\right]
\end{align*}
using the relationships found above are calculated explicitly each of the terms of the Hamiltonian and using the terms $\gamma^{\dagger}_{k}\gamma^{\dagger}_{-k}$ and $\gamma_{k}\gamma_{-k}$ are zero for the Hamiltonian diagonalizes and implementing the relationship $u_{k}=\cos\left(\theta_{k}/2\right)$ and $v_{k}=\sin\left(\theta_{k}/2\right)$, relationship is then to be met angles, so
\begin{align}\label{angul}
\gamma\sin\left(ak\right)\left\{u_{k}^{2}-v_{k}^{2}\right\}+2\left(\cos\left(a k\right)-\frac{h}{J}\right)u_{k}v_{k}&=0\notag\\
\gamma\sin\left(ak\right)\left\{u_{k}^{2}-v_{k}^{2}\right\}&=-2\left(\cos\left(a k\right)-\frac{h}{J}\right)u_{k}v_{k}\notag\\
\gamma\sin\left(ak\right)\left\{\cos^{2}\left(\theta_{k}/2\right)-\sin^{2}\left(\theta_{k}/2\right)\right\}&=-2\left(\cos\left(a k\right)-\frac{h}{J}\right)\cos\left(\theta_{k}/2\right)\sin\left(\theta_{k}/2\right)\notag\\
\gamma\sin\left(ak\right)\cos\theta_{k}&=\left(\frac{h}{J}-\cos\left(a k\right)\right)\sin\theta_{k}\notag\\
\tan\theta_{k}&=\frac{\gamma\sin\left(ak\right)}{\left(\frac{h}{J}-\cos\left(a k\right)\right)}
\end{align}
and $N$ is the number of spins in the chain, you can define the interaction Hamiltonian by place of occupation of each spin. such that
\begin{align*}
H_{I}&=-J\sum_{k}\left[2\gamma_{k}^{\dagger}\gamma_{k}\left(u_{k}^{2}\left(\cos\left(a k\right)-\frac{h}{J}\right)+\gamma \sin\left(ak\right)u_{k}v_{k}\right)\right.\\
&\qquad +2\gamma_{k}\gamma_{k}^{\dagger}\left(\left(\cos\left(a k\right)-\frac{h}{J}\right)v_{k}^{2}-\gamma\sin\left(ak\right)u_{k}v_{k}\right)+\left.\frac{h}{J}\right]
\end{align*}
Now using the anticommutation relation, the Hamiltonian can be written
 \begin{align*}
H_{I}&=-J\sum_{k}\left[2\gamma_{k}^{\dagger}\gamma_{k}\left(u_{k}^{2}\left(\cos\left(a k\right)-\frac{h}{J}\right)+\gamma \sin\left(ak\right)u_{k}v_{k}\right)\right.\\
&\qquad +2\left(1-\gamma_{k}^{\dagger}\gamma_{k}\right)\left(\left(\cos\left(a k\right)-\frac{h}{J}\right)v_{k}^{2}-\gamma\sin\left(ak\right)u_{k}v_{k}\right)+\left.\frac{h}{J}\right]
\end{align*}  
It expanded term by term, we obtain
\begin{align*}
H_{I}&=\sum_{k}\left[-2 J u_{k}^2 \gamma _{k}^{\dagger}\gamma _k \left(\cos(ak)-\frac{h}{J}\right)+2 J v_{k}^2 \gamma _{k}^{\dagger}\gamma_{k} \left(\cos(a k)-\frac{h}{J}\right)\right.\\
&\qquad \left.-4\gamma J u_{k} v_{k} \sin(a k)\gamma _{k}^{\dagger}\gamma_{k} -2 J v_{k}^2 \left(\cos(a k)-\frac{h}{J}\right)+2
\gamma J u_{k} v_{k} \sin(a k)-h\right]
\end{align*}
arranging terms it must be
\begin{align*}
H_{I}&=\sum_{k}\left[\left\{2 J \left(\frac{h}{J}-\cos(ak)\right)\left(u_{k}^2 -v_{k}^2\right)+4\gamma J u_{k} v_{k} \sin(a k)\right\}\gamma _{k}^{\dagger}\gamma _k\right.\\
&\qquad \left. +2 J v_{k}^2 \left(\cos(a k)-\frac{h}{J}\right)-2
\gamma J u_{k} v_{k} \sin(a k)-h\right]
\end{align*}
In order to simplify, we use the definitions of amplitud $u_{k}=\cos\pap{\theta_{k}/2}$ and $v_{k}=\sin\pap{\theta_{k}/2}$; therefore, using the trigonometric relations $u_{k}^{2}-v_{k}^{2}=\cos\pap{\theta_k}$, $2u_{k}v_{k}=\sin\pap{\theta_k}$, and $v_{k}^{2}=\frac{1-\cos\pap{\theta_k}}{2}$. In this way, we can rewritten the hamiltonian  
\begin{align*}
H_{I}&=\sum_{k}\left[\left\{2 J \left(\frac{h}{J}-\cos(ak)\right)\cos\pap{\theta_{k}}+2\gamma J \sin\pap{\theta_k} \sin(a k)\right\}\gamma _{k}^{\dagger}\gamma _k\right.\\
&\qquad \left. +2 J \pap{\frac{1-\cos\pap{\theta_k}}{2}}\left(\frac{h}{J}-\cos(a k)\right)-
\gamma J \sin\pap{\theta_k} \sin(a k)-h\right]
\end{align*}
We can simplify a little more
\begin{align*}
H_{I}&=\sum_{k}\left[\left\{2 J \left(\frac{h}{J}-\cos(ak)\right)\cos\pap{\theta_{k}}+2\gamma J \sin\pap{\theta_k} \sin(a k)\right\}\gamma _{k}^{\dagger}\gamma _k\right.\\
&\qquad \left.  -J \left(\frac{h}{J}-\cos(ak)\right)\cos\pap{\theta_{k}}-\gamma J \sin\pap{\theta_k} \sin(a k)\right]+\sum_{k}\cos\pap{ak}
\end{align*}
given that $\sum_{k}\cos\pap{ak}=0$ therefore:
\begin{align*}
H_{I}&=\sum_{k}\left[\left\{2 J \left(\frac{h}{J}-\cos(ak)\right)\cos\pap{\theta_{k}}+2\gamma J \sin\pap{\theta_k} \sin(a k)\right\}\gamma _{k}^{\dagger}\gamma _k\right.\\
&\qquad \left.  -J \left(\frac{h}{J}-\cos(ak)\right)\cos\pap{\theta_{k}}-\gamma J \sin\pap{\theta_k} \sin(a k)\right]
\end{align*}
Now, 
\begin{align*}
H_{I}&= \sum_{k}2J \pap{\left(\frac{h}{J}-\cos(ak)\right)\cos\pap{\theta_{k}}+\gamma  \sin\pap{\theta_k} \sin(a k)} \pac{\gamma _{k}^{\dagger}\gamma _k-\frac{1}{2}}
\end{align*}

On the other hand, given the relation~\eqref{angul}, we can evaluate each trigonometric function as:
\begin{align}
\sin\pap{\theta_{k}}&=\frac{\gamma\sin\pap{ak}}{\sqrt{\pap{\frac{h}{J} -\cos\pap{ak}}^2+\gamma^2\sin^2\pap{ak}}} &&\cos\pap{\theta_{k}}=\frac{\pap{\frac{h}{J}-\cos{ak}}}{\sqrt{\pap{\frac{h}{J} -\cos\pap{ak}}^2+\gamma^2\sin^2\pap{ak}}}
\end{align}
   can be demonstrated that the system is mapped to a system of non-interacting fermions with Hamiltonian given by
\begin{eqnarray*}
	\boxed{H=\sum_{k}\lambda\left(\theta_{k}\right)\left\{\gamma_{k}^{\dagger}\gamma_{k}-\frac{1}{2}\right\}}
\end{eqnarray*}
where $\lambda\left(\theta_{k}\right)=2J\sqrt{\left(\frac{h}{J}-\cos \left(a k\right)\right)^{2} +\gamma^{2}\sin^{2}\left(ak\right)}$, with the ground state defined $\gamma_{k}\left|\textbf{0}\right\rangle=0$. The groundstate free energy has the form
\begin{eqnarray*}
	\boxed{\epsilon_{0}\equiv\frac{E_{0}}{NJ}=\frac{1}{2\pi}\int_{-\pi}^{\pi}dk \lambda\left(\theta_{k}\right)}
\end{eqnarray*}
\begin{savequote}[45mm]
\end{savequote}
\chapter{Published Papers}
\lettrine{\color{red1}{\GoudyInfamily{I}}}{n} the following pages we have decided to add the first page of every papers that have been published chronological during the author's doctoral research. We stress that all these papers are fundamentally published forms of the doctoral research this thesis is about, and are included with the consent of the other authors.
\begin{enumerate}
\item[6.] {\bf F. J. G\'omez-Ruiz}, A. del Campo
    \textquotedblleft \href{https://link.aps.org/doi/10.1103/PhysRevLett.122.080604}{Universal dynamics of inhomogeneous quantum phase transitions: suppressing defect formation.}\textquotedblright  \emph{Phys. Rev. Lett.}  {\textbf 112}, 080604 (2019).
\item[5.]  \textbf{F. J. G\'omez-Ruiz}, J. J. Mendoza-Arenas, F. J. Rodríguez, C. Tejedor, and L. Quiroga, \textquotedblleft \href{https://link.aps.org/doi/10.1103/PhysRevB.97.235134}{Universal two-time correlations, out-of-time-ordered correlators and Leggett-Garg inequality violation by edge Majorana fermion qubits.}\textquotedblright \emph{Phys. Rev. B}, {\textbf 97}, 235134 (2018). 
 \item[4.]  \textbf{ F. J. G\'omez-Ruiz},  L. Acevedo,  F. J. Rodríguez, L. Quiroga and N. F. Johnson, \textquotedblleft \href{https://doi.org/10.3389/fphy.2018.00092}{Pulsed Generation of Quantum Coherences and Non-classicality in Light-Matter Systems.}\textquotedblright {\em Front. Phys.}, \textbf{ 6}, 92 (2018).  
  \item[3.]   \textbf{ F. J. G\'omez-Ruiz}, J. J. Mendoza-Arenas, O. L. Acevedo,  F. J. Rodríguez, L. Quiroga and N. F. Johnson,  \textquotedblleft \href{https://doi.org/10.1088/1361-6455/aa9a92}{Dynamics of Entanglement and the Schmidt Gap in a Driven Light-Matter System.}\textquotedblright {\em J. Phys. B: At. Mol. Opt. Phys.}, \textbf{ 51}, 024001 (2018).
   \item[2.]  \textbf{ F. J. G\'omez-Ruiz}, O. L. Acevedo, L. Quiroga, F. J. Rodr\'iguez and N. F. Johnson, \textquotedblleft \href{http://www.mdpi.com/1099-4300/18/9/319}{Quantum Hysteresis in Coupled Light-Matter Systems.}\textquotedblright \emph{Entropy}, \textbf{18}, 319 (2016).	
    \item[1.] \textbf{ F. J. G\'omez-Ruiz}, J. J. Mendoza-Arenas, F. J. Rodríguez, C. Tejedor, and L. Quiroga, \textquotedblleft \href{https://journals.aps.org/prb/abstract/10.1103/PhysRevB.93.035441}{Quantum phase transitions detected by a local probe using time correlations and violations of Leggett-Garg inequalities.}\textquotedblright \emph{Phys. Rev. B}, \textbf{ 93}, 035441 (2016). 
\end{enumerate}
\newpage
\thispagestyle{empty}
\begin{figure}[h!]
 \centering 
 \includegraphics[scale=0.85]{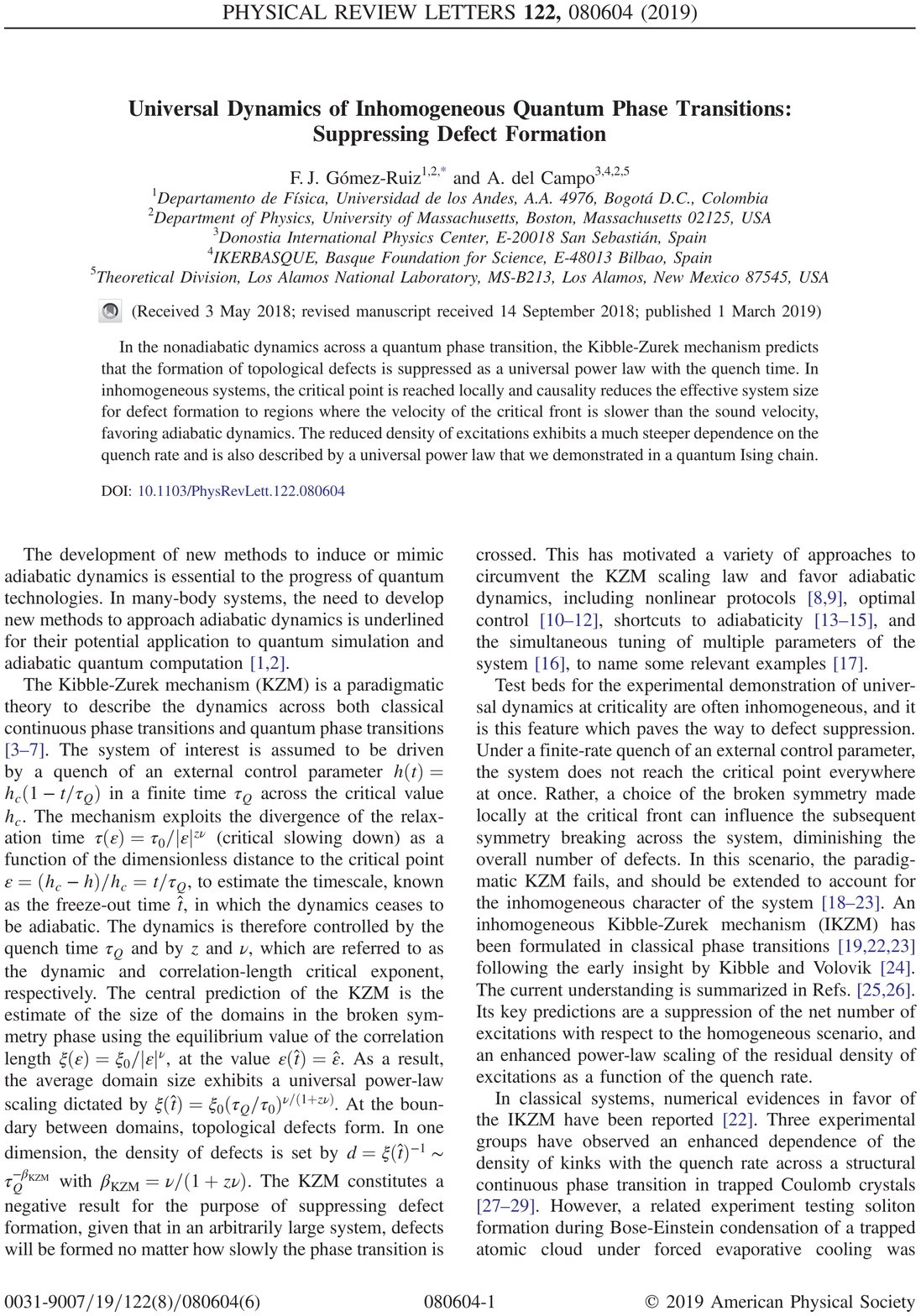}
\end{figure}
\newpage
\thispagestyle{empty}
\begin{figure}[h!]
 \centering 
 \includegraphics[scale=0.85]{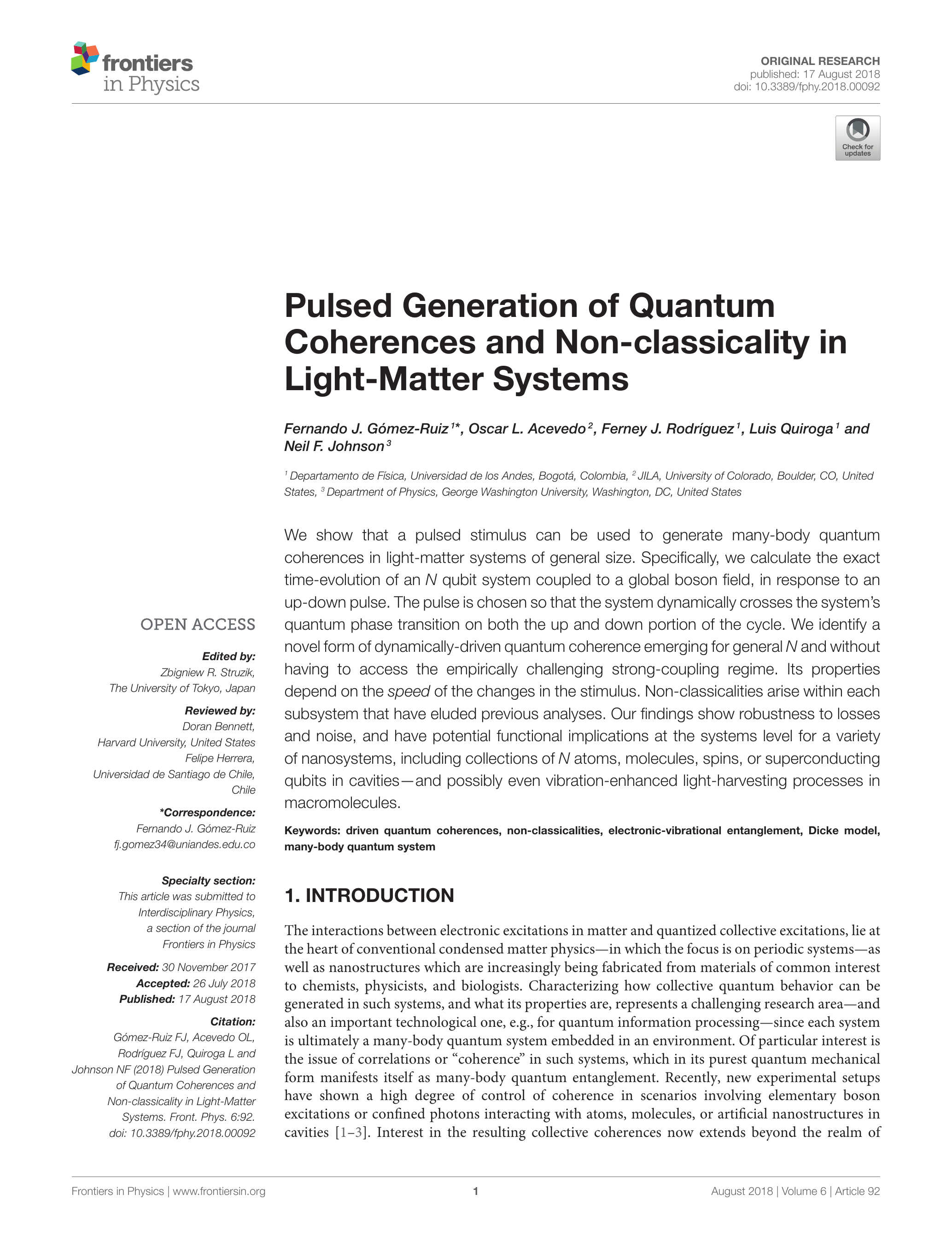}
\end{figure}
\newpage
\thispagestyle{empty}
\begin{figure}[h!]
 \centering 
 \includegraphics[scale=0.85]{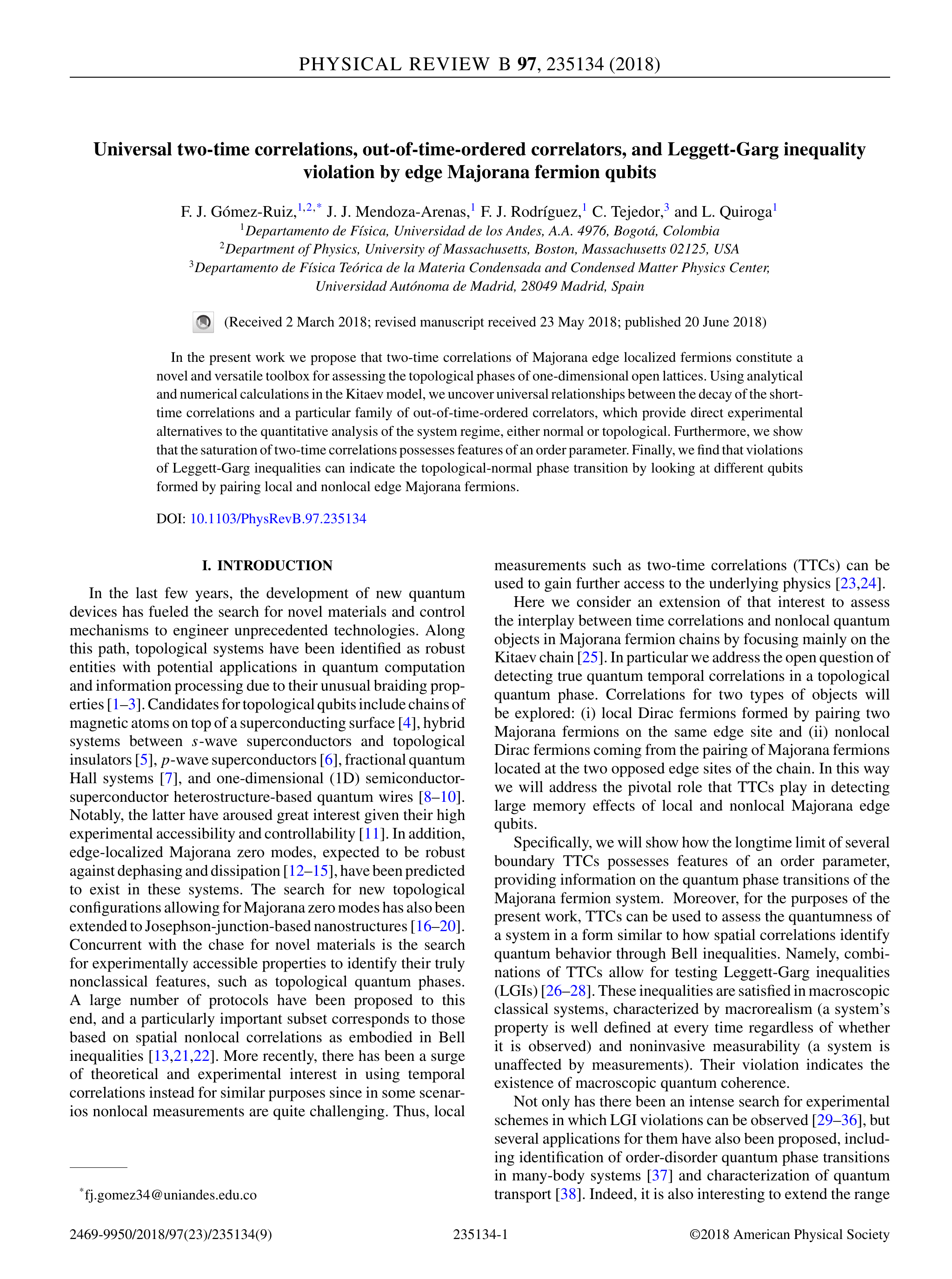}
\end{figure}
\newpage
\thispagestyle{empty}
\begin{figure}[h!]
 \centering 
 \includegraphics[scale=0.85]{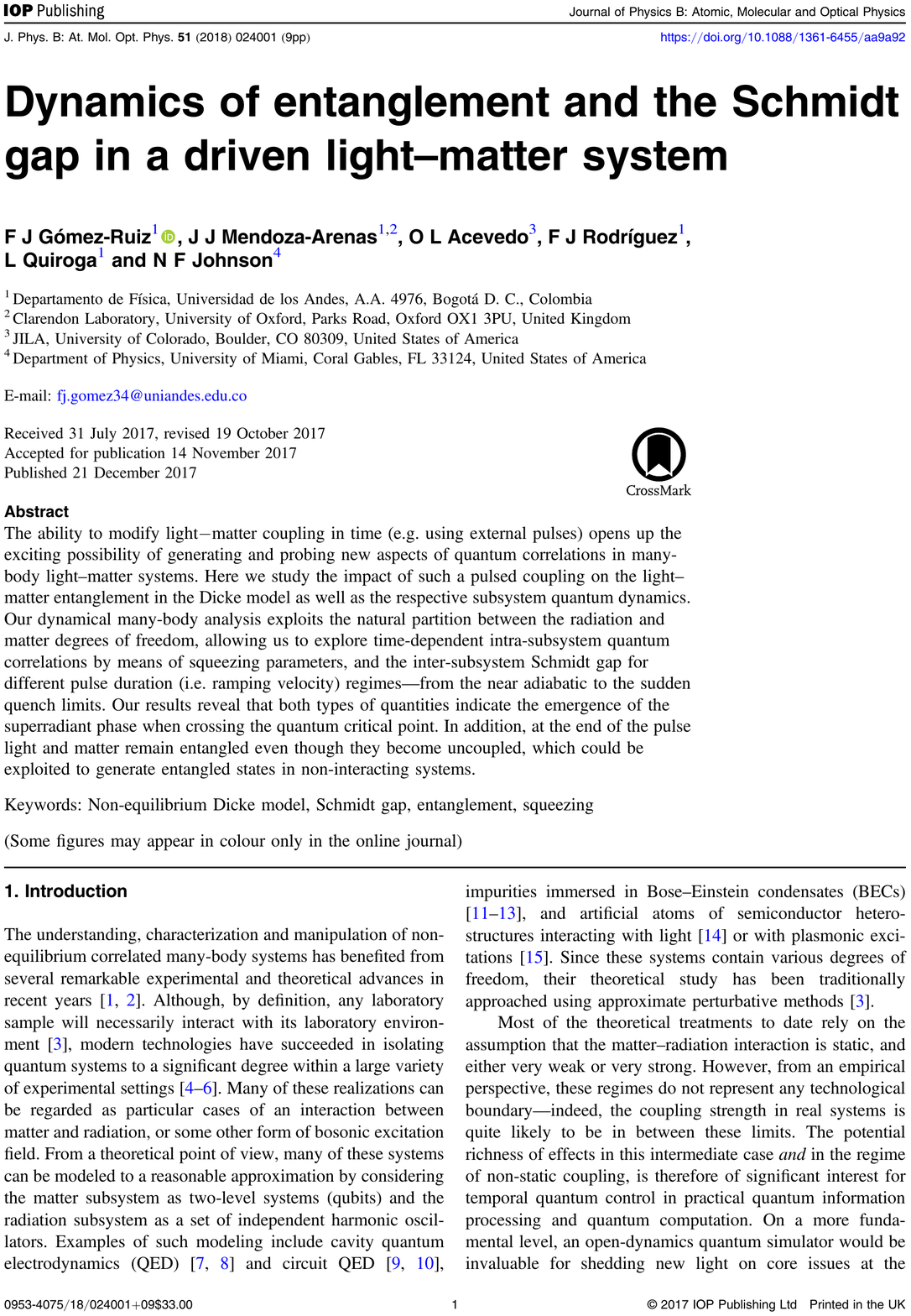}
\end{figure}
\newpage
\thispagestyle{empty}
\begin{figure}[h!]
 \centering 
 \includegraphics[scale=0.85]{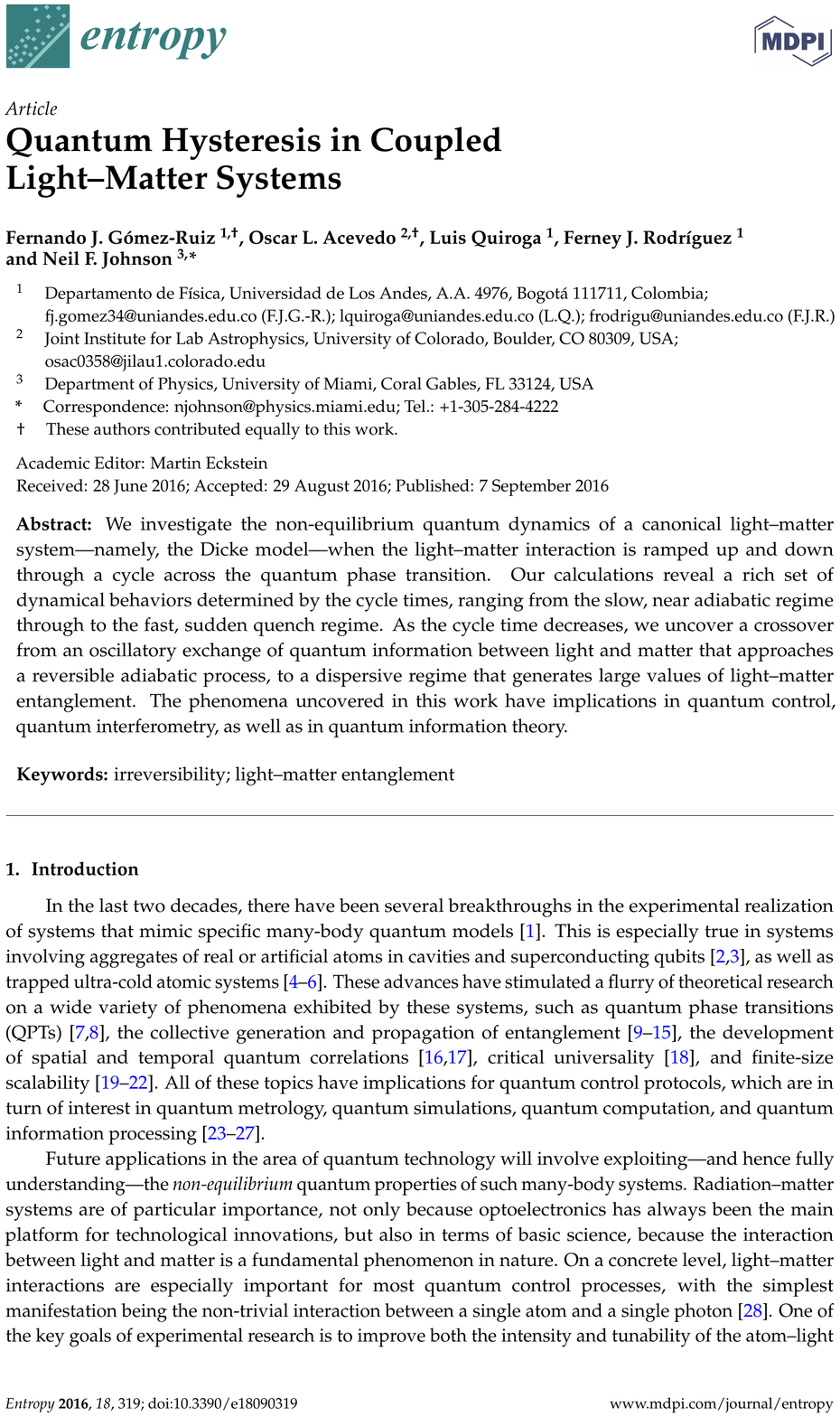}
\end{figure}
\newpage
\begin{figure}[h!]
 \centering 
 \includegraphics[scale=0.85]{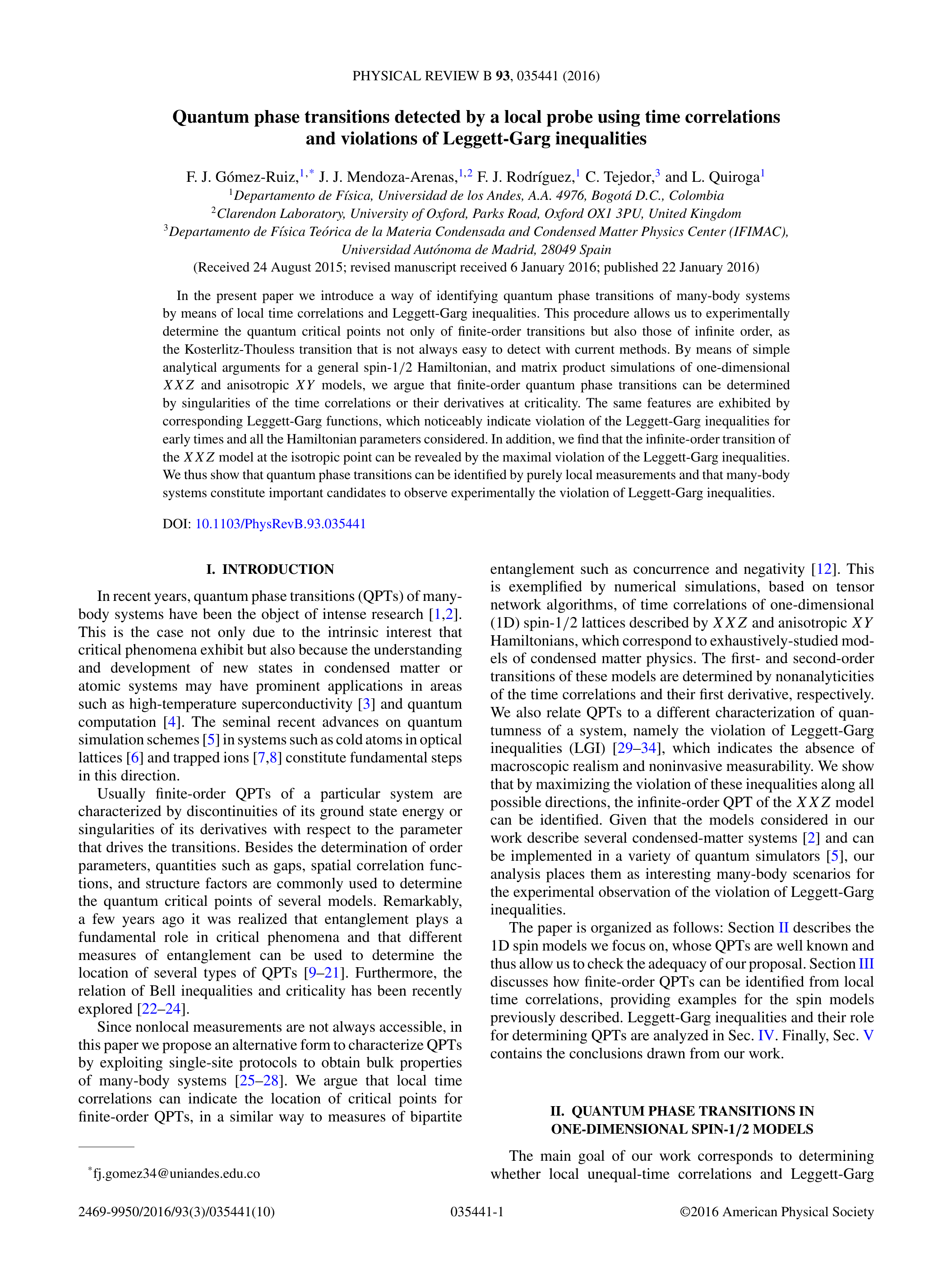}
\end{figure}

\begin{savequote}[45mm]
\end{savequote}
\chapter{Submitted Papers}
\lettrine{\color{red1}{\GoudyInfamily{I}}}{n} the following pages we have decided to add the first page of every preprint that have been submitted chronological during the author's doctoral research. We stress that all these papers are fundamentally published forms of the doctoral research this thesis is about, and are included with the consent of the other authors.
\begin{enumerate}
\item[1.] J-M. Cui, \textbf{F. J. G\'omez-Ruiz}, Y-F. Huang, C-F. Li, G-C. Guo, \& A. del Campo.  \textquotedblleft \href{https://arxiv.org/abs/1903.02145}{Testing quantum critical dynamics beyond the Kibble-Zurek mechanism with a trapped-ion simulator}\textquotedblright \emph{ArXiv:quant-ph} 1903.02145 (2019).
\end{enumerate}
\newpage
\thispagestyle{empty}
\begin{figure}[h!]
 \centering 
 \includegraphics[scale=0.85]{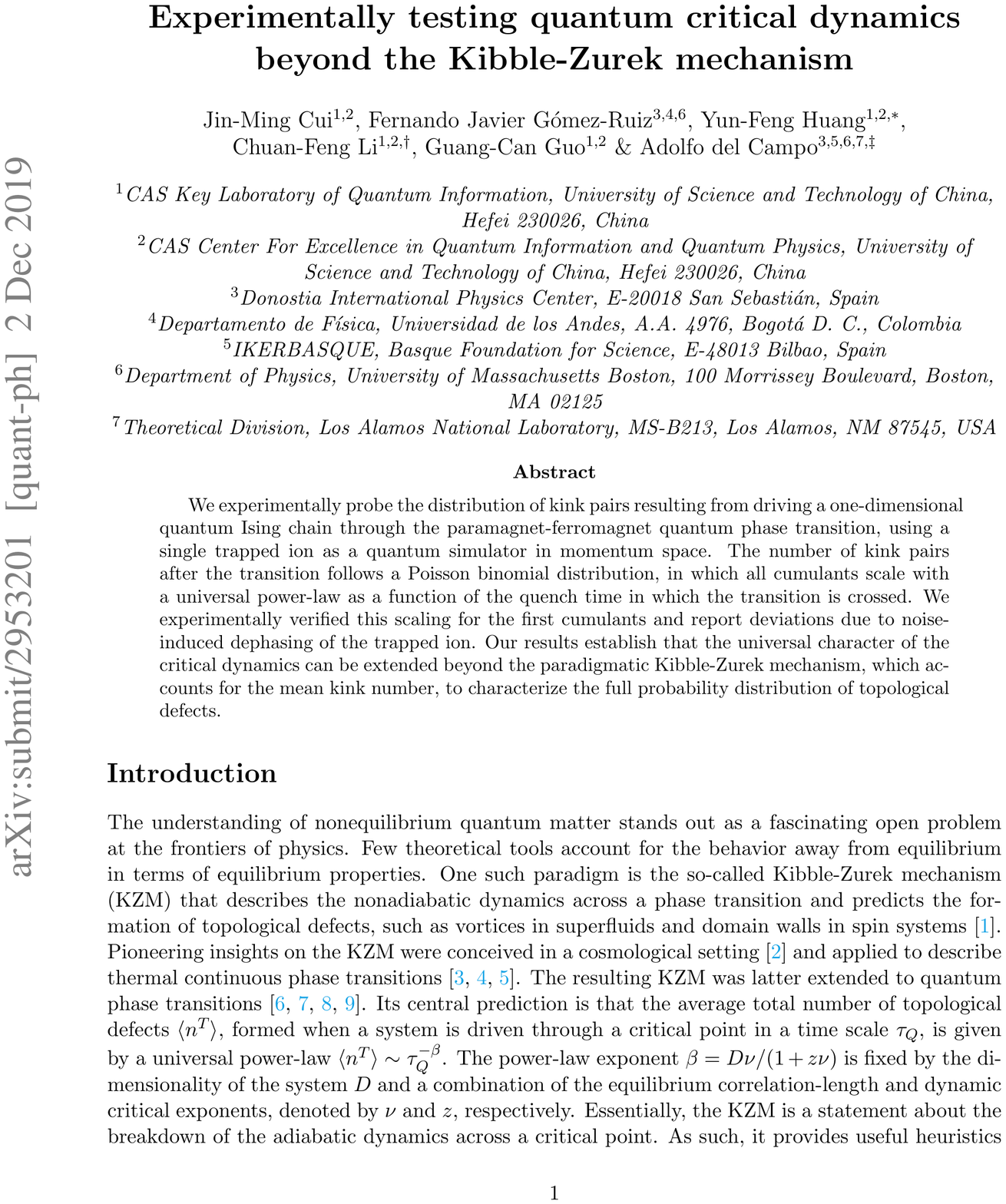}
\end{figure}
\newpage
\begin{savequote}[45mm]
\end{savequote}
\chapter{Additional published and submitted papers}
\lettrine{\color{red1}{\GoudyInfamily{I}}}{n} the following pages we have decided to add first  page of every papers that have been published and/or submitted during the author's doctoral research period but not discuses in these thesis.
\begin{enumerate}
\item[3.] J. J. Mendoza-Arenas, \textbf{F. J. G\'omez-Ruiz}, F. J. Rodr\'iguez, \& L. Quiroga.\textquotedblleft\href{https://doi.org/10.1038/s41598-019-54121-1}{Enhancing violations of Leggett-Garg inequalities in nonequilibrium correlated many-body systems by interactions and decoherence}\textquotedblright \emph{Sci. Rep.} {\bf 9}, 17772 (2019).
\item[2.] N.F. Johnson, \textbf{F.J. G\'omez-Ruiz}, F.J. Rodr\'iguez, \& L. Quiroga.  \textquotedblleft \href{https://arxiv.org/abs/1901.08873}{Quantum Terrorism: Collective Vulnerability of Global Quantum Systems}\textquotedblright \emph{ArXiv:quant-ph} 1901.08873 (2019).  
\item[1.] J. J. Mendoza-Arenas, \textbf{F. J. G\'omez-Ruiz}, M. Eckstein, D. Jaksch,
\& S. R. Clark \textquotedblleft \href{https://doi.org/10.1002/andp.201700024}{Ultra-Fast Control of Magnetic Relaxation in a Periodically Driven Hubbard Model}\textquotedblright Annalen Der Physik, \textbf{529}, 1700024 (2017).
\end{enumerate}
\newpage
\thispagestyle{empty}
\begin{figure}[h!]
 \centering 
 \includegraphics[scale=0.85]{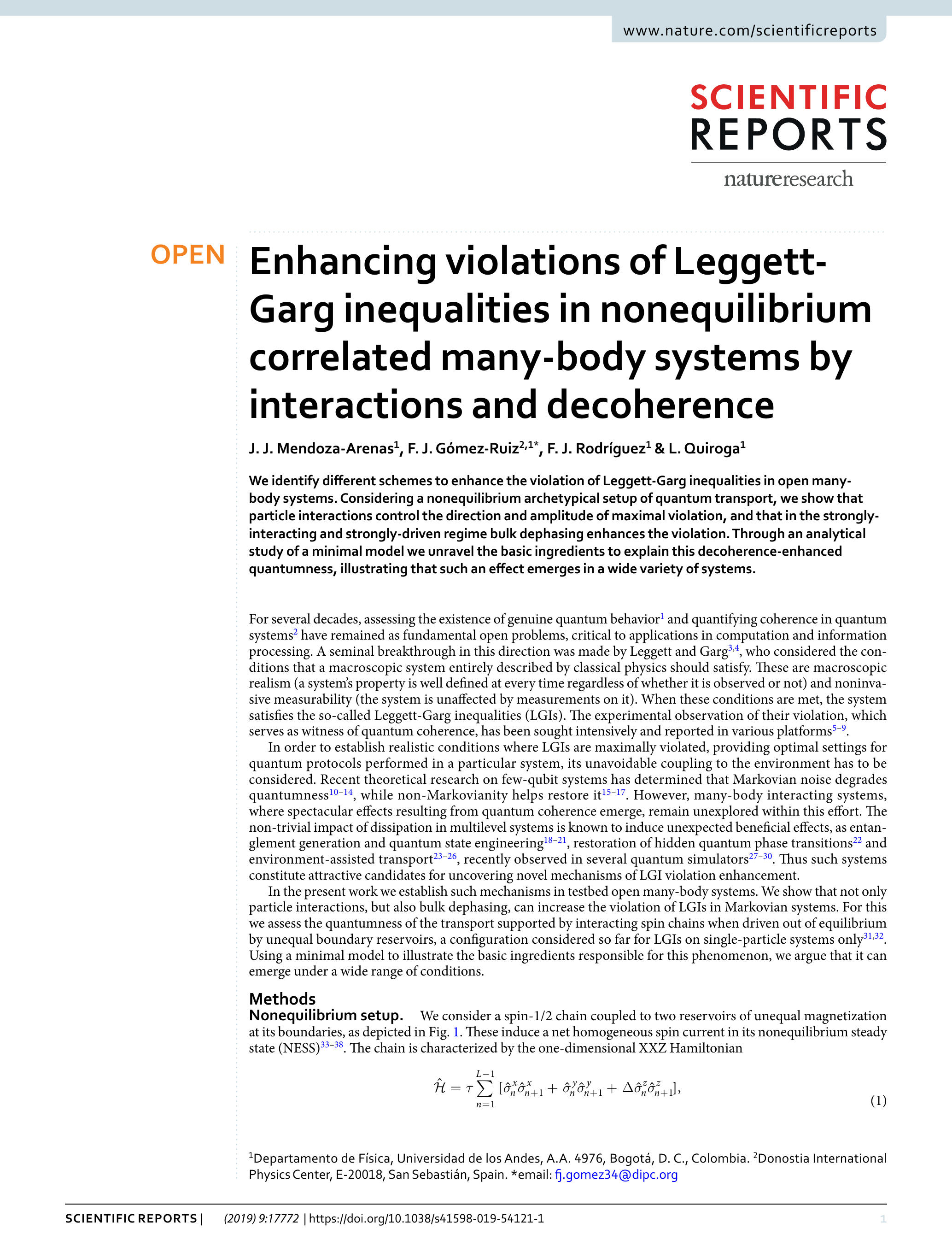}
\end{figure}
\thispagestyle{empty}
\begin{figure}[h!]
 \centering 
 \includegraphics[scale=0.85]{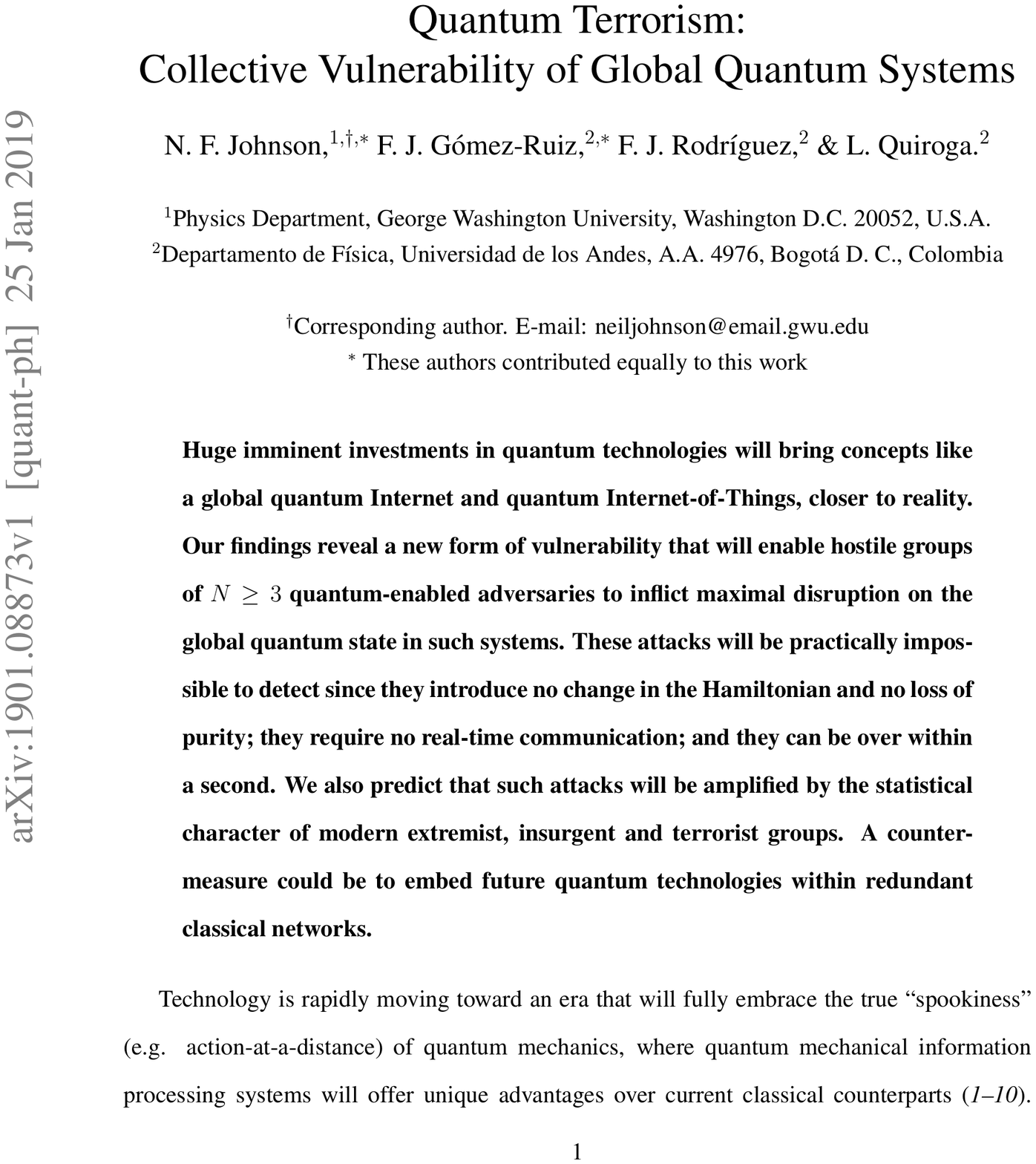}
\end{figure}
\newpage
\thispagestyle{empty}
\begin{figure}[h!]
 \centering 
 \includegraphics[scale=0.85]{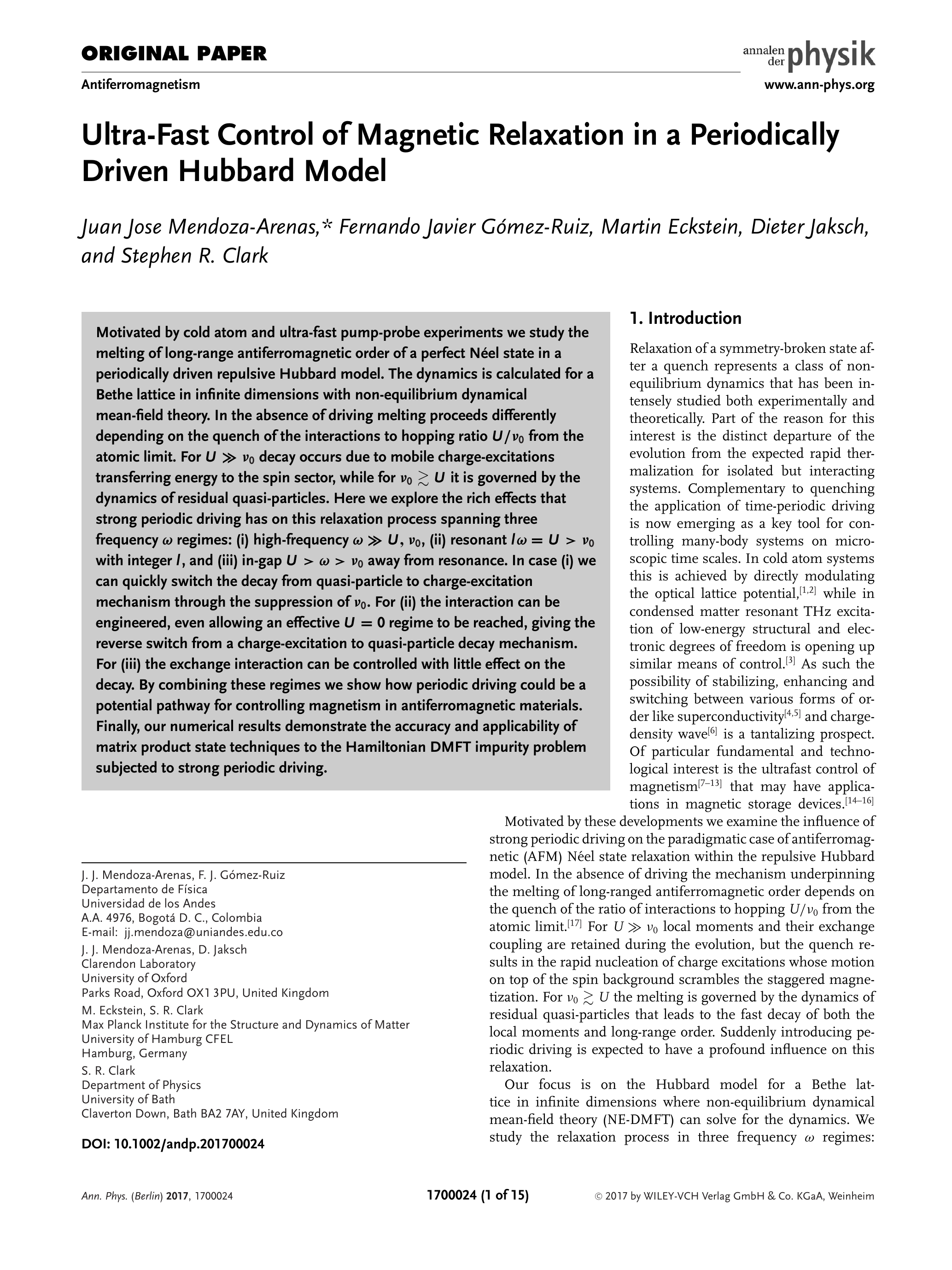}
\end{figure}

\begin{savequote}[45mm]
\end{savequote}
\chapter[Ethic considerations]{Ethic considerations}\label{Apend2}
\lettrine{\color{red1}{\GoudyInfamily{I}}}{n} this case, due to the characteristics of this  thesis (no living beings will be used in any type of experiments), there is no need to make an evaluation with the department ethics committee. Instead, the protocol for the data management will be described: As mentioned in the introduction and chapter 2, once the simulation ends running the data will be downloaded, and only that data will be used to expose the results thesis. Furthermore, if in any case the results of another person or research group were used, they will be duly referenced following the procedure stipulated by the university.

\end{document}